\documentclass[twoside,b5paper,10pt,english]{book}

\usepackage{graphicx}
\graphicspath{ {figures/} }
\usepackage{array}
\usepackage{watermark}
\usepackage{fancybox}
\usepackage[dvipsnames]{xcolor}
\usepackage{babel}[english]
\usepackage[Lenny]{fncychap}
\usepackage[pagestyles]{titlesec}
\usepackage{bookmark}
\usepackage{afterpage}
\usepackage{xstring}
\usepackage{yhmath}
\usepackage{enumerate}  
\usepackage{scrextend}
\addtokomafont{labelinglabel}{\sffamily\bfseries}
\usepackage[b5paper,left=2.5cm,right=1.5cm,top=2.5cm]{geometry}
\makeatletter
\def\@footnotecolor{black} 
\define@key{Hyp}{footnotecolor}{%
 \HyColor@HyperrefColor{#1}\@footnotecolor%
}
\def\@footnotemark{%
    \leavevmode
    \ifhmode\edef\@x@sf{\the\spacefactor}\nobreak\fi
    \stepcounter{Hfootnote}%
    \global\let\Hy@saved@currentHref\@currentHref
    \hyper@makecurrent{Hfootnote}%
    \global\let\Hy@footnote@currentHref\@currentHref
    \global\let\@currentHref\Hy@saved@currentHref
    \hyper@linkstart{footnote}{\Hy@footnote@currentHref}%
    \@makefnmark
    \hyper@linkend
    \ifhmode\spacefactor\@x@sf\fi
    \relax
  }%
\makeatother
\usepackage[a4,center,pdflatex]{crop} 
\usepackage{hyperref}
\hypersetup{
    colorlinks = true,
    linkcolor = blue,
    citecolor = teal, 
    filecolor = magenta,
    footnotecolor = black
}

\makeatletter
\def\p@part{\itshape\color{violet}}
\def\p@section{\itshape\color{violet}}
\def\p@subsection{\itshape\color{violet}}
\def\p@chapter{\itshape\color{violet}}
\makeatother

\usepackage{amsmath,amssymb,amsfonts,bm,cancel,physics,nicefrac}
\usepackage{csquotes}
\usepackage[backend=biber,style=phys,biblabel=brackets]{biblatex}
\usepackage{pdfpages}
\usepackage[font=footnotesize,labelfont=scriptsize]{caption}
\usepackage{subcaption}
\usepackage{tocbibind}

\newpagestyle{mystyle_empt}{%
\sethead[][][]{}{}{}}

\newpagestyle{mystyle}{%
\sethead[\thepage][][Chapter\ \thechapter. \chaptertitle]
{Section \thesection. \sectiontitle}{}{\thepage}
\setheadrule{1pt}
}
\newpagestyle{mystyle2}{%
\sethead[\thepage][][ \chaptertitle]
{\chaptertitle}{}{\thepage}
\setheadrule{1pt}
}
\newpagestyle{mystyle3}{%
\sethead[\thepage][][ Appendix\ \thechapter.  \chaptertitle]
{Appendix\ \thechapter.  \chaptertitle}{}{\thepage }
\setheadrule{1pt}
}
\newpagestyle{mystyle4}{%
\sethead[\thepage][][ Chapter\ \thechapter.  \chaptertitle]
{Chapter\ \thechapter.  \chaptertitle}{}{\thepage }
\setheadrule{1pt}
}

\addto{\captionsenglish}{}

\newcommand{\gs}{\text{gs}}
\newcommand{\eq}{\text{eq}}
\newcommand{\zc}{\text{ZC}}
\newcommand{\Buckled}{\text{B}}
\newcommand{\BuckledStable}{\text{B+}}
\newcommand{\BuckledUnstable}{\text{B}-}
\newcommand{\confSpinElastic}{\mathbf{u},\mathbf{p},\bm{\sigma}}
\newcommand{\hamSpinElastic}{\mathcal{H}(\confSpinElastic)}
\newcommand{\etal}{\emph{et al.}\ }
\newcommand{\ie}{\emph{i.e.}\ }
\newcommand{\eg}{\emph{e.g.}\ }
\newcommand{\const}{\text{const}}
\newcommand{\PDF}{p}
\newcommand{\prob}{P}
\newcommand{\targetx}{x_T}
\newcommand{\survival}{Q}
\newcommand{\FPTpdf}{\phi} 
\newcommand{\FPTpdfLaplace}{\Phi}
\newcommand{\FPT}[1]{%
    \IfInteger{#1} 
    {\ifnum #1=1%
            \mathcal{T}%
        \else%
            \mathcal{T}^{\, #1}%
        \fi
    }
    {\mathcal{T}^{\, #1}}
}
\newcommand{\brackets}[1]{\left[ #1 \right]}
\newcommand{\parenthesis}[1]{\left(#1\right)}
\renewcommand{\braket}[1]{\left\langle #1 \right\rangle}
\renewcommand{\braces}[1]{\left\{#1\right\}}
\renewcommand{\d}[1]{\text{d}#1 \,}
\newcommand{\laplTransformOper}[2]{\mathcal{L}\braces{#1(#2)}}
\newcommand{\laplTransformTilde}[2]{\widetilde{#1}(#2)}
\newcommand{\cdev}[3]{\dfrac{\text{d}^{#1}#2}{\text{d}#3^{#1}}}
\newcommand{\pdev}[3]{\dfrac{\partial^{#1}#2}{\partial#3^{#1}}}

\newcommand{\optMFPT}[1]{\widetilde{#1}}
\newcommand{\optVar}[1]{\widehat{#1}}

\newcommand{\free}{(e)}
\newcommand{\refr}{(r)}
\newcommand{\erfc}{\text{erfc}}
\newcommand{\deltaPhase}[1]{\delta^{#1}}
\newcommand{\sgn}{\text{sgn}}
\newcommand{\disorder}{\eta}
\newcommand{\positiveSol}[1]{#1_+}
\newcommand{\negativeSol}[1]{#1_-}

\DeclareFieldFormat{doilink}{%
  \iffieldundef{doi}{%
    \iffieldundef{url}{%
      \iffieldundef{annotation}{%
          #1%
            }{%
          \href{\thefield{annotation}}{#1}%
            }
        }
        {%
        \href{\thefield{url}}{#1}%
        }%
    }{%
    \href{http://dx.doi.org/\thefield{doi}}{#1}%
    }%
}

\newbibmacro{string+doiurlisbn}[1]{%
  \iffieldundef{doi}{%
    \iffieldundef{url}{%
      \iffieldundef{isbn}{%
        \iffieldundef{issn}{%
          #1%
        }{%
          \href{http://books.google.com/books?vid=ISSN\thefield{issn}}{#1}%
        }%
      }{%
        \href{http://books.google.com/books?vid=ISBN\thefield{isbn}}{#1}%
      }%
    }{%
      \href{\thefield{url}}{#1}%
    }%
  }{%
    \href{http://dx.doi.org/\thefield{doi}}{#1}%
  }%
}

\DeclareFieldFormat{title}{%
  \usebibmacro{string+doiurlisbn}{\mkbibemph{#1}}}
\DeclareFieldFormat
  [thesis]
  {title}{\usebibmacro{string+doiurlisbn}{\mkbibquote{#1\isdot}}}

\DeclareBibliographyDriver{article}{%
\usebibmacro{bibindex}%
\usebibmacro{begentry}%
\usebibmacro{author/translator+others}%
\setunit{\labelnamepunct}\newblock
\usebibmacro{title}%
\newunit\newblock
\printfield{version}%
\newunit\newblock
\printtext[doilink]{%
\usebibmacro{journal+issuetitle}%
\newunit
\usebibmacro{byeditor+others}%
\newunit
\usebibmacro{note+pages}%
\newunit
\printfield[parens]{year}%
}%
\newunit\newblock
\usebibmacro{doi+eprint+url}%
\setunit{\bibpagerefpunct}\newblock
\usebibmacro{pageref}%
\usebibmacro{finentry}}

\begin{document}
\frontmatter
\begin{titlepage}
\begin{center}

\thiswatermark{\put(5,-650){\includegraphics[width=1.1\textwidth]{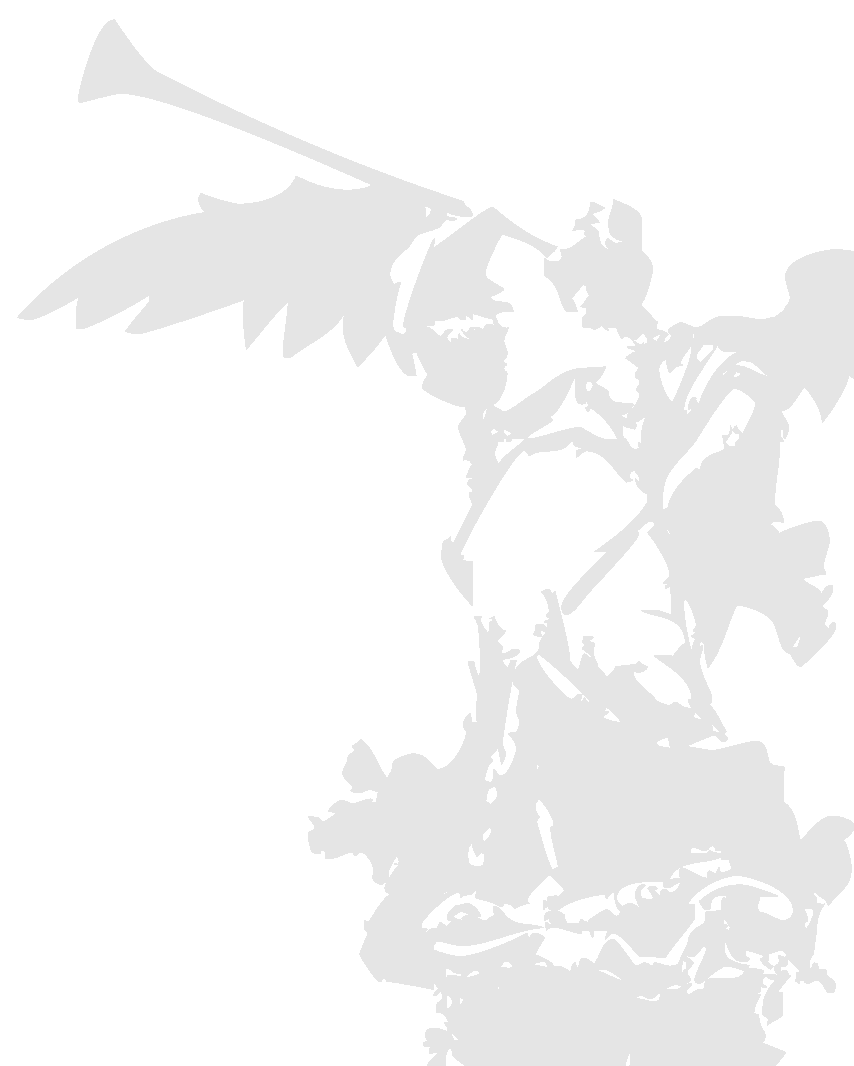}}}

\includegraphics[width=0.45\textwidth]{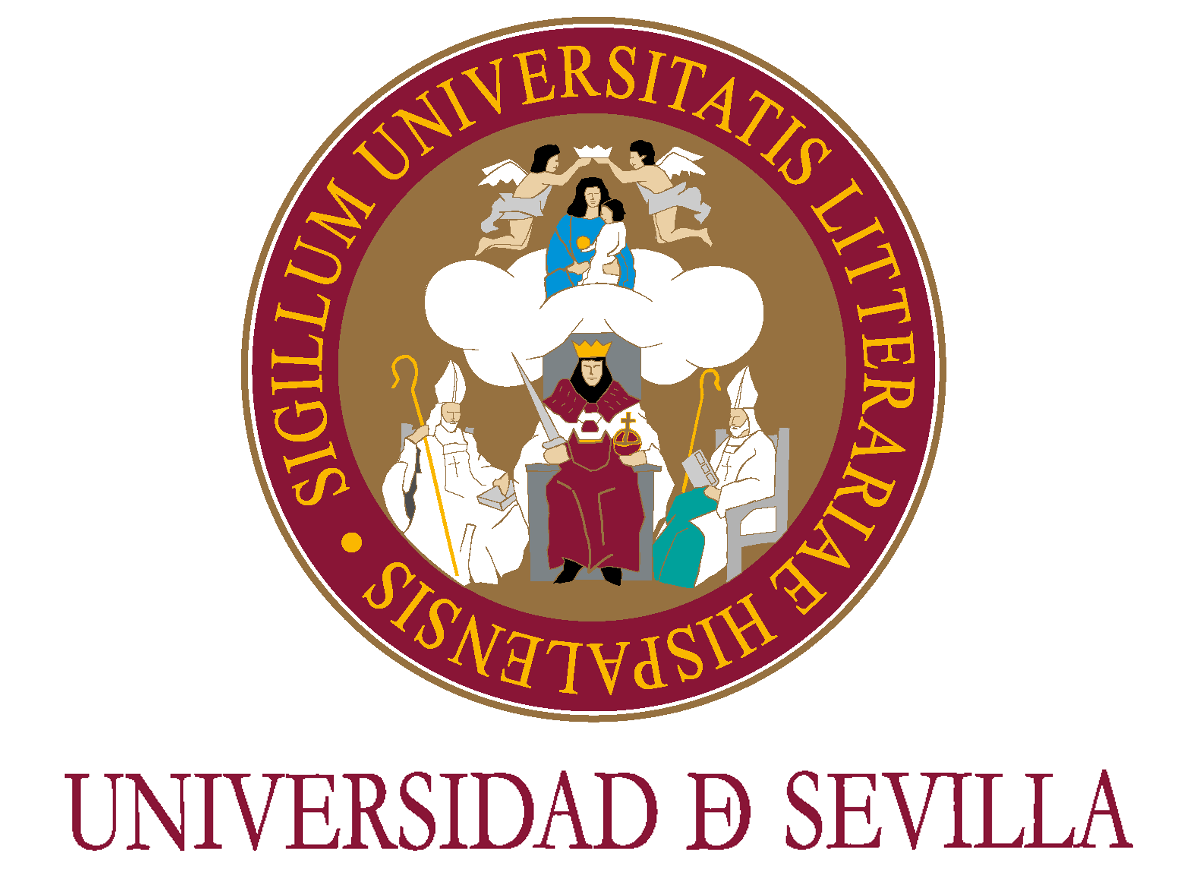}\\[1cm]

 \textsc{\Large Departamento de F\'{\i}sica At\'{o}mica \\Molecular y Nuclear}\\[1.3cm]

 \textsc{\Large Tesis Doctoral}\\[1.3cm]

 \hrulefill \\[0.6cm]
{ \huge \bfseries 
    Simple models for mesoscopic systems: from slender structures to stochastic resetting
 }\\[0.6cm]

 \hrulefill \\[2.0cm]

\begin{minipage}{0.4\textwidth}
\begin{flushleft} \large
\emph{Doctorando:}\\
Gregorio Garc\'{\i}a Valladares \\
$\quad$
\\ $\quad$
\end{flushleft}
\end{minipage}
\begin{minipage}{0.4\textwidth}
\begin{flushright} \large

\emph{Directores:} \\
Carlos Alberto Plata Ramos \\ Antonio Prados Monta\~{n}o
\emph{Tutor:} \\
Antonio Prados Monta\~{n}o
\end{flushright}

\end{minipage}
 
\vfill
 
{\large October, 2025}

\end{center}

\end{titlepage}

\thispagestyle{empty}
\pagestyle{mystyle_empt} 

\cleardoublepage


\chapter*{$\quad$}
\begin{quotation}\begin{flushright}\begin{em}
``The most important step a person can take. It's not the first one, is it?\\
It's the next one. Always the next step, Dalinar.''\\[0.2cm]
\par\end{em}
Dalinar Kholin, from \emph{Oathbringer}, by Brandon Sanderson.
\end{flushright}\end{quotation}

\thispagestyle{empty}


 



{\clearpage \thispagestyle{empty}}
\chapter*{List of publications}

This thesis includes the research contained 
in the following works:
\begin{itemize}
    \item \textbf{Gregorio Garc\'{\i}a Valladares},  Antonio Prados, Carlos A.\ Plata and A.\ Manacorda, ``To reset or not to reset in a finite domain: that is the question'', \href{http://dx.doi.org/10.48550/arXiv.2505.16626}{arXiv:2505.\ 16626} (Phys.\ Rev.\ E in Press, December 2025).  
    \item \textbf{Gregorio Garc\'{\i}a Valladares}, Deepak Gupta,  Antonio Prados and Carlos A.\ Plata, ``Stochastic resetting with refractory periods: pathway formulation and exact results'', \href{http://dx.doi.org/10.1088/1402-4896/ad317b}{Phys.\ Scr.\ \textbf{99}, 045234 (2024)}.  
    \item \textbf{Gregorio Garc\'{\i}a Valladares}, Carlos A.\ Plata, Antonio Prados and A.\ Manacorda, ``Optimal resetting strategies for search processes in heterogeneous environments'', \href{http://dx.doi.org/10.1088/1367-2630/ad06da}{New J.\ Phys.\ \textbf{25}, 113031 (2023)}.  
    \item \textbf{Gregorio Garc\'{\i}a Valladares}, Carlos A.\ Plata and Antonio Prados,``Buckling in a rotationally invariant spin-string model'', \href{http://dx.doi.org/10.1103/PhysRevE.107.014120}{Phys.\ Rev. E \textbf{107}, 014120 (2023)}. 
\end{itemize}


{\clearpage \thispagestyle{empty}}
\addtocontents{toc}{\protect\setcounter{tocdepth}{-1}}
{
  \hypersetup{linkcolor=black}
  \tableofcontents
}
\addtocontents{toc}{\protect\setcounter{tocdepth}{3}}

{\clearpage \thispagestyle{empty}}
\mainmatter



\chapter{Introduction}	
\label{ch:intro}
\pagestyle{mystyle}  

The search for answers lies at the heart of science. While the approach varies across fields and disciplines, every researcher strives to understand the world around them from their own perspective, using the tools at their disposal. Physicists are no exception. It is impossible for any single person to fully grasp the vast scope of physics---I certainly do not. The range of topics is enormous: motion, forces, energy, matter, heat, light\dots\ are but a few of the many topics that physics covers nowadays. Its boundaries are expanding even further with the rise of multidisciplinary sciences.

When thinking about physics, there is a certain beauty in drawing an analogy to grimdark stories. In this analogy, the universe and its phenomena play the role of the villain. Physicists are those courageous heroes who, in the struggle to understand it, take on the universe, wielding their most powerful weapons: rigorous observation, experimentation, and mathematics as the language of nature. Individually, they are destined to fail, but together, they can achieve great things and discover how the universe works. Among these brave warriors, we can find those well-versed in statistical physics.

Statistical physics is a branch of physics that employs statistical methods to explain and predict the behaviour of systems with a large number of constituents, such as atoms or molecules. We are all familiar with concepts like temperature, pressure, or volume in the everyday life, \ie at the macroscopic level. However, these properties emerge from the collective behaviour of a vast number of particles, each of which following the laws of physics at the microscopic level. Statistical physics was developed to provide the tools to connect these two levels of description. Nevertheless, its applications go beyond: statistical physics has been successfully applied to fields as diverse as biology, economics, computer science, and social sciences, where systems with many interacting components are common. One of the most remarkable aspects of statistical physics is its ability to find the key ingredients that govern the behaviour of complex systems---a special arcane magic to extract simplicity from complexity, by modelling only the essential features of the system under scrutiny.

This thesis attempts to contribute to the grand adventure of understanding the universe by applying statistical physics. Specifically, we fundamentally address two questions: Why does a thin plate for a shield bend when heated? How can we design a search strategy to find as fast as possible a hidden treasure? These quite general questions are the motivation behind the two parts of this work. Herein, we will consider mesoscopic descriptions---nor fully microscopic nor fully macroscopic---to capture the essential interactions and behaviours that qualitatively drive the phenomena we are interested in. To better understand the context and the relevance of the problems under study, in the following we start by introducing the necessary background concepts, \ie the mathematical and physical framework alongside a brief review of the state-of-the-art in the relevant fields. On the one hand, part~\ref{part:part-materials} focuses on studying a novel spin-elastic model to describe the mechanical response of low-dimensional materials. On the other hand, part~\ref{part:resetting} is devoted to the study of stochastic processes under resetting, as feasible strategies to optimise search processes.

\section{Stochastic processes}
\label{sec:stochastic_processes}
This section attempts to provide a brief introduction to the theory of stochastic processes, which is the fundamental mathematical framework used throughout this thesis. We are working in the context of non-equilibrium statistical mechanics, \ie systems far from equilibrium, where the dynamics of the system is crucial. Since we are dealing with statistical systems, stochasticity is inherent to the description, so we will first give a brief introduction to probability theory, applied to stochastic processes.

Since probability theory is extensively studied in different textbooks \cite{book:Feller_IntroductionProbabilityTheory_71,book:Feller_IntroductionProbabilityTheory_91,book:VanKampen_StochasticProcessesPhysics_92,book:Gardiner_HandbookStochasticMethods_83}, we just provide here a brief review of its basics. Let us consider a random variable $X$ taking values $x$
---following the usual notation, we use capital letters to denote random variables, and small letters for their possible values.
A stochastic process is a function $Y_X(t)$ that depends on both a random variable $X$ and the time variable $t$, \ie 
\begin{equation}
    \label{eq:stochastic_process_def}
    Y_X(t) \equiv f(X,t). 
\end{equation}
Each possible value $x$ of the random variable defines a realisation---or trajectory---of the process $Y_x(t)=f(x,t)$. The collection of all the possible realisations gives rise to the ensemble. 

Let us introduce the concept of probability density function (PDF) of a random variable $X$ as $\PDF_X(x)$, \ie the probability that the random variable $X$ takes a value in the interval $(x_1,x_2)$ is
\begin{equation}
    \label{eq:Prob_vs_PDF}
    \prob(x_1 \leq X \leq x_2 ) = \int_{x_1}^{x_2} \d{x} \PDF_X(x), 
\end{equation}
or, in other words, the probability that $X$ belongs in the infinitesimal interval $\brackets{x,x+\d{x}}$ is $\PDF_X(x)\d{x}$.\footnote{Expression~\eqref{eq:Prob_vs_PDF} is also valid for discrete random variables since the PDF may include Dirac-delta contributions.} 
The average value of the stochastic process is obtained by
\begin{equation}
    \label{eq:average_stochastic_process}
    \langle Y(t)\rangle = \int \d{x} Y_x(t) \PDF_X(x) = \int \d{x} f(x,t) \PDF_X(x).
\end{equation}
Here, we consider that whenever one has an integral without limits, we are integrating over the whole space, \ie $(-\infty,\infty)$ in the one-dimensional case. By extension, the general way to compute higher-order moments is 
\begin{equation}
    \langle Y(t_1)Y(t_2)\ldots Y(t_n)\rangle = \int \d{x} Y_x(t_1)Y_x(t_2)\ldots Y_x(t_n) \PDF_X(x).
\end{equation}
Lastly, a stochastic process is said to be stationary if all its moments are invariant under time translation, \ie
\begin{equation}
    \langle Y(t_1)Y(t_2)\ldots Y(t_n)\rangle = \langle Y(t_1+\Delta t)Y(t_2+\Delta t)\ldots Y(t_n+\Delta t)\rangle, \, \forall \Delta t\in\mathbb{R},\, \forall \braces{t_1,t_2,\ldots,t_n}.
\end{equation}

\subsection{Markov processes. Chapman-Kolmogorov equation}
\label{subsec:introduction_Markov_processes}

The one-time PDF of the stochastic process $Y_X(t)$ is computed as 
\begin{equation}
    \label{eq:PDF_stochastic_process}
    \PDF_1(y,t) = \int \d{x} \PDF_X(x) \delta(y - Y_x(t)),
\end{equation}
where the subscript indicates the number of times we are considering and $\delta(\cdot)$ is the Dirac-delta distribution. Let us choose $n$ different finite times $t_1,t_2,\ldots,t_n$. For a given realisation, the stochastic process takes values $y_i=Y_x(t_i)$ for a set of given time instants $\{t_1,t_2,\ldots,t_n\}$. The joint PDF of $\{y_1,y_2,\ldots,y_n\}$ is given by
\begin{equation}
    \label{eq:joint_PDF_stochastic_process}
    \PDF_n(y_1,t_1;y_2,t_2;\ldots;y_n,t_n) = \int \d{x}\; \PDF_X(x) \prod_{i=1}^n \delta(y_i - Y_x(t_i)).
\end{equation}
If two times are equal, \eg $t_1=t_2$, then 
\begin{equation}
    \PDF_n(y_1,t_1;y_2,t_2=t_1;\ldots;y_n,t_n) = \delta(y_2 - y_1) \PDF_{n-1}(y_2,t_2;\ldots;y_n,t_n),
\end{equation}
which expresses the fact that $y_2=y_1$ for any realisation of the stochastic process, \ie for any value $x$ of the random variable $X$. Furthermore, a general $n$-th order moment is given by
\begin{equation}
    \label{eq:general_moment_stochastic_process}
    \langle Y(t_1)\ldots Y(t_n)\rangle = \int \d{y_1}\ldots \d{y_n} y_1\ldots y_n\,\PDF_n(y_1,t_1;y_2,t_2;\ldots;y_n,t_n).
\end{equation}
This procedure defines an infinite hierarchy of PDFs $\PDF_n$ that verify Kolmogorov axioms \cite{book:VanKampen_StochasticProcessesPhysics_92}:
\begin{itemize}
    \item $\PDF_n\geq0$, \ie they are always positive.
    \item They are symmetric under the exchange of any two times $(y_i,t_i)$ and $(y_j,t_j)$. 
    \item The marginal PDF is obtained by integrating over any time and letting the rest of them fixed, \ie $\int \d{y_n} \PDF_n(y_1,t_1;\ldots,y_n,t_n) = \PDF_{n-1}(y_1,t_1;\ldots,y_{n-1},t_{n-1})$.
    \item The normalisation condition holds, $\int \d{y_1} \PDF_1(y_1,t_1) = 1$, $\forall t$. 
\end{itemize}

Once we have the hierarchy of PDFs for different times, we may define the conditional PDF using Bayes's theorem, 
\begin{equation}
    \label{eq:conditional_probability_general}
    \PDF_{l|k}(y_{k+1},t_{k+1};\ldots;y_{k+l},t_{k+l}|y_1,t_1;\ldots;y_k,t_k) = \frac{\PDF_{k+l}(y_1,t_1;\ldots;y_{k+l},t_{k+l})}{\PDF_k(y_1,t_1;\ldots;y_k,t_k)}.
\end{equation}
The former expression stands for the PDF for the subensemble of trajectories that have passed through the fixed values $(y_1,t_1)$ $,\ldots,(y_k,t_k)$. Thus, the subscript $l|k$ indicates that we are considering the PDF of $l$ variables for $k$ given values. The normalisation for this conditional PDF imposes that 
\begin{equation}
    \label{eq:conditional_probability_normalisation}
    \int \d{y_{k+1}}\ldots\int \d{y_{k+l}} \PDF_{l|k}(y_{k+1},t_{k+1};\ldots;y_{k+l},t_{k+l}|y_1,t_1;\ldots;y_k,t_k) = 1.
\end{equation}

So far, we have not considered the time evolution of the stochastic process. Let us consider $n$ ordered times $t_1 < t_2 < \ldots < t_n$, then a stochastic process is considered \emph{Markovian} if
\begin{equation}
    \PDF_{1|n-1}(y_n,t_n|y_1,t_1;\ldots;y_{n-1},t_{n-1}) = \PDF_{1|1}(y_n,t_n|y_{n-1},t_{n-1}), 
\end{equation}
which holds $\forall n$, $\forall\braces{t_1,t_2,\ldots,t_n}$, $\forall\braces{y_1,y_2,\ldots,y_n}$. This property is known as the Markov property and states that the conditional probability of the stochastic process at any time $t_n$ only depends on its previous time $t_{n-1}$. For this reason, a Markov process is also called a \emph{memoryless} process, since the future evolution is determined only by the current state of the system, and not on its full previous history. Thus, a Markovian stochastic process is completely characterised by its conditional PDF, or \emph{transition probability}, $\PDF_{1|1}$ and the one-time PDF $\PDF_1$. The general hierarchy of PDFs is then simplified to
\begin{align}
    \label{eq:Markov_PDFs}
    \PDF_n(y_1,t_1;\ldots;y_n,t_n) &= \PDF_1(y_1,t_1) \prod_{i=1}^{n-1} \PDF_{1|1}(y_{i+1},t_{i+1}|y_{i},t_{i}).
\end{align}

When a stochastic process is Markovian, it satisfies two important consistency equations. The first one is the \emph{Chapman-Kolmogorov equation}, which states that the transition probability fulfils\footnote{The proof involves integrating the three-time equation~\eqref{eq:Markov_PDFs} over the intermediate time $t_2$ and applying Bayes's theorem~\eqref{eq:conditional_probability_general}.}
\begin{equation}
    \label{eq:Chapman_Kolmogorov_equation}
    \PDF_{1|1}(y_3,t_3|y_1,t_1) = \int \d{y_2} \PDF_{1|1}(y_3,t_3|y_2,t_2)\PDF_{1|1}(y_2,t_2|y_1,t_1), \quad t_1 \leq t_2 \leq t_3.
\end{equation} 
The second consistency equation is
\begin{equation}
    \label{eq:Markovian_consistency_property}
    \PDF_1(y_2,t_2) = \int \d{y_1} \PDF_{1|1}(y_2,t_2|y_1,t_1)\PDF_1(y_1,t_1).
\end{equation}
Any two positive and normalised PDFs $\PDF_1$ and $\PDF_{1|1}$ that verify the Chapman-Kolmogorov equation~\eqref{eq:Chapman_Kolmogorov_equation} and the consistency property~\eqref{eq:Markovian_consistency_property} may be used to define a Markovian stochastic process, which is unique for those two PDFs. 

We have already introduced that a stochastic process is stationary if all its moments (or equivalently the hierarchy of PDFs) are invariant under global time translation. In the case of a Markov process, the stationary condition is guaranteed if (i) $\PDF_{1}(y,t) = \PDF_{1}(y)$, and (ii) $\PDF_{1|1}(y_2,t_2|y_1,t_1)$ only depends on the time difference $\Delta t = t_2 - t_1$. If only the second condition is fulfilled, the process is called \emph{homogeneous in time}. In this case, the transition probability can be conveniently written as 
\begin{equation}
    \label{eq:transition_probabilities_homogeneous}
    T_{\Delta t} (y_2|y_1) \equiv \PDF_{1|1}(y_2,t_1+\Delta t|y_1,t_1), \quad
    T_{\Delta t=0} (y_2|y_1) = \delta(y_2-y_1), 
\end{equation}
so the Chapman-Kolmogorov equation~\eqref{eq:Chapman_Kolmogorov_equation} reads
\begin{align}
    \label{eq:Chapman_Kolmogorov_homogeneous}
    T_{\Delta t+\Delta t'} (y_3|y_1) = \int \d{y_2} T_{\Delta t}(y_3|y_2) T_{\Delta t'}(y_2|y_1), \quad \forall \Delta t, \Delta t'\geq0, 
\end{align}    
while the consistency property~\eqref{eq:Markovian_consistency_property} becomes
\begin{equation}
    p_1(y_2,t_1+\Delta t) = \int \d{y_1} T_{\Delta t}(y_2|y_1) p_1(y_1,t_1).
    \label{eq:Markovian_consistency_property_homogeneous}
\end{equation}

\subsection{Master equation}
\label{subsec:introduction_Master_equation}
The Chapman-Kolmogorov equation~\eqref{eq:Chapman_Kolmogorov_homogeneous} and the second consistency condition~\eqref{eq:Markovian_consistency_property_homogeneous} are integral equations that are very difficult to solve in general. From the practical point of view, they are hard to manipulate although they are still useful to check the consistency of the solutions. However, it is possible to derive a differential equation, called the \emph{master equation}, that describes the time evolution of the PDFs. 

Let us suppose a homogeneous in time Markov process, where the transition probability is $T_{\Delta t}(y_2|y_1)$, as given by~\eqref{eq:transition_probabilities_homogeneous}. If the time increment $\Delta t$ is small enough, $\Delta t\to 0^+$, then we can expand the transition probability as 
\begin{equation}
    T_{\Delta t}(y_2|y_1) = A(y_1,\Delta t)\delta(y_2-y_1) + \Delta t\, W(y_2|y_1) + \mathcal{O}(\Delta t^2), \quad \Delta t \to 0^+.
    \label{eq:transition_probabilities_expansion}
\end{equation}
The new quantity $W(y_2|y_1)$ represents a \emph{transition rate}, \ie a transition probability (density function) per unit time. In~\eqref{eq:transition_probabilities_expansion}, $A(y_1,\Delta t)$ is a function that is computed by imposing the normalisation of $T_{\Delta t}(y_2|y_1)$. Taking into account such normalisation,
\begin{equation}
    T_{\Delta t}(y_2|y_1) = \brackets{1-\Delta t\int \d{y'} W(y'|y_1)}\delta(y_2-y_1) + \Delta t W(y_2|y_1) + \mathcal{O}(\Delta t^2). 
\end{equation}
The factor in front of the delta has a clear interpretation, it is the probability of remaining in the same state $y_1$ after a short time, which equals one minus the total probability of leaving $y_1$ for any other state $y'$. If we substitute the former expression for $T_{\Delta t}(y_3|y_1)$, considering $\Delta t\to0^+$, into the Chapman-Kolmogorov equation~\eqref{eq:Chapman_Kolmogorov_homogeneous} and change $\Delta t$ to $t$, we obtain the master equation
\begin{equation}
    \label{eq:Master_equation_Transitions}
    \partial_{t} T_{t}(y_3|y_1) = \int \d{y_2} \brackets{W(y_3|y_2) T_{t}(y_2|y_1)-W(y_2|y_3) T_{t}(y_3|y_1)},
\end{equation}
with the initial condition $T_{t=0}(y_3|y_1) = \delta(y_3-y_1)$. Equivalently, the second consistency equation~\eqref{eq:Markovian_consistency_property_homogeneous} gives the master equation for the PDFs $\PDF(y,t)=\PDF_1(y,t)$,
\begin{equation}
    \label{eq:Master_equation_PDFs}
    \partial_t \PDF(y,t) = \int \d{y'} \brackets{W(y|y')\PDF(y',t)-W(y'|y)\PDF(y,t)}.
\end{equation}
This equation describes the time evolution of a continuous Markovian stochastic process, if it were a discrete process, the master equation would take the form 
\begin{equation}
    \partial_t\PDF_n(t) =\sum_{n'} \brackets{W_{nn'}\PDF_{n'}(t)-W_{n'n}\PDF_n(t)}.
\end{equation}
The physical interpretation of the master equation is its providing a gain-loss balance: the first terms $W(y|y')\PDF(y',t)$ ($W_{nn'}\PDF_{n'}(t)$) represent the incoming flux from any state $y'$ ($n'$) to $y$ ($n$), whereas the second terms $W(y'|y)\PDF(y,t)$ ($W_{n'n}\PDF_n(t)$) are the outgoing flux from the state $y$ ($n$) to any other. 
Since the transition probabilities and the one-time PDFs verify the master equation, \ie~\eqref{eq:Master_equation_Transitions} and~\eqref{eq:Master_equation_PDFs}, we will only write hereinafter the one-time PDF $p(y,t)$---they are related by~\eqref{eq:Markovian_consistency_property_homogeneous}. However, we have to take into account they have different initial conditions: $T_{t=0}(y|y') = \delta(y-y')$ for the transition probability, and a given $\PDF(y,t=0)=\PDF_0(y)$ for the one-time PDF.

The modelling of the process is encoded in the transition rate $W_{nn'}$ (or the continuous version $W(y|y')$). The difference between one stochastic process and another stochastic process lies on the form of the $W$'s, which must be calculated from the microscopic dynamics with a short-time analysis (\eg using the Fermi's golden rule \cite{book:Fermi_NuclearPhysicsCourse_50}) or proposed by plausibility arguments (as done in this thesis for our mesoscopic models). 

\subsection{Ensemble description. Fokker-Planck equation}
\label{subsec:introduction_FokkerPlanck}
In the continuous case~\eqref{eq:Master_equation_PDFs}, the master equation is an integro-differential equation for the ensemble of the stochastic process. When possible, it is handy to approximate the master equation by a partial differential equation (PDE) called the \emph{Fokker-Planck equation} (FPE). 

In many physical contexts, one expects that the possible jumps are small enough, in the sense that the PDF changes very little over the interval in which the transition rate is different from zero. Then, we can assume that the transition rates $W(y|y')$ are sharp functions of the difference $\Delta y= y-y'$, so they vanish for large enough $|\Delta y|$.
Consequently, we can transform the master equation~\eqref{eq:Master_equation_PDFs} into the \emph{Kramers-Moyal expansion} \cite{book:VanKampen_StochasticProcessesPhysics_92,book:Gardiner_HandbookStochasticMethods_83}
\begin{equation}
    \label{eq:Kramers_Moyal_expansion_PDE}
    \pdev{}{\PDF(y,t)}{t} = \sum_{\nu = 1}^\infty \frac{(-1)^{\nu}}{\nu!} \pdev{\nu}{}{y}\brackets{a_\nu(y) \PDF(y,t)},
\end{equation}
where its coefficients (jump moments) are defined as
\begin{equation}
    a_\nu(y)= \int \d{y'} (y'-y)^\nu W(y'|y).
    \label{eq:jump_moments_definition}
\end{equation}
The Kramers-Moyal expansion (KME) is an infinite order PDE that has different meanings depending on the truncation. The lowest order indicates the deterministic motion, which leads to the \emph{Liouville equation} in Hamiltonian mechanics. If we truncate the KME at second order, we introduce the fluctuations, so we have a stochastic contribution that leads to the \emph{forward Fokker-Planck equation},
\begin{equation}
        \label{eq:Fokker_Planck_equation}
        \partial_t \PDF(y,t) = -\partial_y \brackets{a_1(y) \PDF(y,t)} + \frac{1}{2}\partial_y^2 \brackets{a_2(y) \PDF(y,t)}.
\end{equation}
Higher contributions stand for corrections to the FPE that we will not consider in this thesis.

Let us suppose that the initial value of the stochastic process at time $t_0$ is $y_0$. The Fokker-Planck equation that describes the time evolution to reach $y$ at time $t$ reads
 \begin{equation}
    \partial_t \PDF_{1|1}(y,t|y_0,t_0) = -\partial_y \brackets{a_1(y) \PDF_{1|1}(y,t|y_0,t_0)} + \frac{1}{2}\partial_y^2 \brackets{a_2(y) \PDF_{1|1}(y,t|y_0,t_0)}.
    \label{eq:Fokker_Planck_equation_conditional}
\end{equation}
Herein, we have used the notation of the conditional probability density $\PDF_{1|1}(y,t|y_0,t_0)$ to emphasise that the initial condition is $p_{1|1}(y,t_0|y_0,t_0)=\delta(y-y_0)$. The solution of this FPE is also known as the \emph{propagator} of the process,\footnote{The propagator terminology stems from Green's function theory. In that context, our Green's function would be $T_{t-t_0}(y|y_0)$, which is related to the solution of the FPE by~\eqref{eq:Markovian_consistency_property_homogeneous}.} since it describes how the system evolves from the initial state $(y_0,t_0)$ to the final state $(y,t)$. Recall that we are working with homogeneous in time processes, so propagators only depend on the time difference $t-t_0$, \ie $\PDF_{1|1}(y,t|y_0,t_0) = \PDF_{1|1}(y,t-t_0|y_0,0)$. Henceforward we will employ the notation $\PDF_{1|1}(y,t|y_0) \equiv \PDF_{1|1}(y,t|y_0,0)$ for the sake of simplicity.

Another interesting approach to study the time evolution of the ensemble is through the \emph{backward Fokker-Planck equation}. Our aim is to know how the initial condition $(y_0,t_0)$ influences the final state $(y,t)$, \ie the time evolution is analysed backwards, from $t$ to $t_0$. To obtain the backward Fokker-Planck equation, we use the Chapman-Kolmogorov equation~\eqref{eq:Chapman_Kolmogorov_equation} for the three times $t_0 \leq t_1 \leq t$, then we differentiate with respect to the intermediate time $t_1$,
\begin{align}
    0 = \partial_{t_1} \PDF_{1|1}(y,t|y_0,t_0) = \int \d{y_1} \bigg[ \partial_{t_1} \PDF_{1|1}(y,t|y_1,t_1) \PDF_{1|1}(y_1,t_1|y_0,t_0) 
    \nonumber\\
    + \PDF_{1|1}(y,t|y_1,t_1) \partial_{t_1} \PDF_{1|1}(y_1,t_1|y_0,t_0)\bigg]. 
\end{align}
If we apply the forward Fokker-Planck equation~\eqref{eq:Fokker_Planck_equation} for $\PDF_{1|1}(y_1,t_1|y_0,t_0)$, after performing integration by parts and evaluate at the end for $t_1=t_0$,\footnote{This procedure is true for any $y_1$ provided that the boundary terms from integration by parts vanish at the contours---either because normalisation if there are no boundaries or due to the boundary conditions associated to each particular situation.} we obtain the backward Fokker-Planck equation
\begin{equation}
    -\partial_{t_0} \PDF_{1|1}(y,t|y_0,t_0) = \brackets{a_1(y_0)\partial_{y_0} + \frac{1}{2} a_2(y_0)\partial_{y_0}^2}\PDF_{1|1}(y,t|y_0,t_0).
    \label{eq:backward_Fokker_Planck_equation}
\end{equation}
And, because we are working with homogeneous in time processes, 
we are able to rewrite the previous PDE making the change of variable $t'=t-t_0$, so 
\begin{equation}
    \partial_{t'} \PDF_{1|1}(y,t'|y_0) = \brackets{a_1(y_0)\partial_{y_0} + \frac{1}{2} a_2(y_0)\partial_{y_0}^2}\PDF_{1|1}(y,t'|y_0),
    \label{eq:backward_Fokker_Planck_equation_time_difference}
\end{equation}
with the initial condition $\PDF_{1|1}(y,t'=0|y_0) = \delta(y-y_0)$. For the sake of simplicity, we omit the subscript along this thesis to denote the propagators, so propagators will be referred as $\PDF(y,t|y_0,t_0)=\PDF(y,t-t_0|y_0)$ from here on.

\subsection{Trajectory description. Langevin equation}

Another possibility to analyse a stochastic process consists of describing the time evolution of individual realisations of the stochastic process. Just as the ensemble description is studied by the master equation or the Fokker-Planck equation for the PDF, it is possible to describe the stochastic process at the trajectory level by using a stochastic differential equation (SDE) called the \emph{Langevin equation}.

Let us consider the non-linear Langevin equation with additive noise \cite{book:VanKampen_StochasticProcessesPhysics_92,book:Gardiner_HandbookStochasticMethods_83}
\begin{equation}
    \label{eq:Langevin_equation_general}
    \dot{y}(t) = A(y) + C \xi(t),
\end{equation}
where $A(y)$ is a deterministic force and $C \xi(t)$ is a stochastic force. Herein, $\xi(t)$ is the unit Gaussian white noise defined by
\begin{equation}
    \langle \xi(t) \rangle = 0, \quad \langle \xi(t)\xi(t') \rangle = \delta(t-t'),
\end{equation}
and $C$ is a constant that modulates the intensity of the stochastic force. Here, the average $\langle \cdot \rangle$ is taken over the different realisations of the noise.

Let us consider a small time interval $\Delta t$, then the time evolution is obtained by integrating the Langevin equation from $t$ to $t+\Delta t$,
\begin{equation}
    \Delta y = y(t+\Delta t) - y(t) = A(t)\Delta t + C \int_t^{t+\Delta t} \d{t'} \xi(t').
\end{equation}
The second term corresponds to a stochastic integral over the white noise $\xi(t)$. Since $\xi(t)$ is a Gaussian process, the integral 
\begin{equation}
    W(\Delta t,t) = \int_t^{t+\Delta t} \d{t'} \xi(t')
\end{equation}
is a Gaussian random variable fully described by its two first moments. They are
\begin{subequations}
    \label{eq:jump_moments_Gaussian_white_noise}
    \begin{align}
        \langle W(\Delta t,t) \rangle &= \int_{t}^{t+\Delta t} \d{t'} \langle \xi(t') \rangle = 0, \\
        \langle W^2(\Delta t,t) \rangle &= \int_{t}^{t+\Delta t} \d{t'} \int_{t}^{t+\Delta t} \d{t''} \langle \xi(t')\xi(t'') \rangle = \int_{t}^{t+\Delta t} \d{t'} \int_{t}^{t+\Delta t} \d{t''} \delta(t'-t'') = \Delta t.
    \end{align}
\end{subequations}
Therefore, we conclude that $W(\Delta t,t) = \mathcal{N}(0,\Delta t) = \Delta t \mathcal{N}(0,1)$, where $\mathcal{N}(\mu,\sigma^2)$ is a Gaussian random variable with mean $\mu$ and variance $\sigma^2$. 

To establish the connection between the Langevin equation and the Fokker-Planck equation, we can compute the jump moments of the process described by equation~\eqref{eq:Langevin_equation_general}. The jump moments~\eqref{eq:jump_moments_definition} can also be written as 
\begin{equation}
    a_\nu(y) = \lim_{\Delta t \to 0^+} \frac{1}{\Delta t}\int \d{\Delta y} (\Delta y)^\nu T_{\Delta t}(y+\Delta y|y) = \lim_{\Delta t \to 0^+} \frac{1}{\Delta t} \langle (\Delta y)^\nu \rangle, \quad \nu\geq 1,
\end{equation}
where we have used~\eqref{eq:transition_probabilities_expansion}. Taking the average over the realisations of~\eqref{eq:Langevin_equation_general} using~\eqref{eq:jump_moments_Gaussian_white_noise}, we obtain 
\begin{equation}
    a_1(y) = A(y), \quad a_2(y) = C^2.
    \label{eq:jump_moments_Langevin}
\end{equation}
Hence, for the cases we are interested in this thesis, the Fokker-Planck equation~\eqref{eq:Fokker_Planck_equation} is equivalent to the Langevin equation~\eqref{eq:Langevin_equation_general}, since it describes the same stochastic process.\footnote{Note that the Fokker-Planck equation that we have previously introduced is more general than the Langevin equation; since we assume additive noise for the latter, $C$ does not depend on $y$.} 





\section{Mechanical response of low-dimensional systems}
\label{sec:intro_spin_elastic_models}

Mechanics is the field of physics that studies the relationship between the motion of the bodies and the forces that are being applied on them. Concretely, continuum mechanics is concerned with the displacements suffered by continuous media, such as solids or fluids, that causes deformations in their structure or shape. 

In this context, we are interested in some particular solids that we refer as low-dimensional materials. They are physical systems in which one or more spatial dimensions are negligible compared with the others. Within this category, we can classify them as 
\begin{enumerate}
    \item Plates or two-dimensional (2d) materials, solids whose thickness is much smaller than the other two dimensions, \ie the planar dimensions. 
    \item Thin rods or one-dimensional (1d) materials, solids whose cross-section is much smaller than their length. 
\end{enumerate}
These systems have been proven to exhibit rich and intriguing properties, which have motivated extensive research toward their practical exploitation \cite{journalarticle:Cea.etal_NumericalStudyRippling_Phys.Rev.B20,journalarticle:Amnuanpol_BucklingInstabilityRotating_EPL21,journalarticle:Chen.etal_SpontaneousTiltSingleClamped_Phys.Rev.Lett.22,journalarticle:Hanakata.etal_ThermalBucklingSymmetry_ExtremeMech.Lett.21,journalarticle:Hanakata.etal_AnomalousThermalExpansion_Phys.Rev.Lett.22,journalarticle:Jain.etal_CompressioncontrolledDynamicBuckling_Phys.Rev.E21,journalarticle:Poincloux.etal_BendingResponseBook_Phys.Rev.Lett.21,journalarticle:LeDoussal.Radzihovsky_ThermalBucklingTransition_Phys.Rev.Lett.21,journalarticle:Thibado.etal_FluctuationinducedCurrentFreestanding_Phys.Rev.E20,journalarticle:Novoselov.etal_ElectricFieldEffect_Science04,journalarticle:Neto.etal_ElectronicPropertiesGraphene_Rev.Mod.Phys.09,journalarticle:Samy.etal_ReviewMoS2Properties_Crystals21,journalarticle:Sangchap.etal_ExploringPromiseOnedimensional_IntJHydrog.Energy24,journalarticle:Samykano_ProgressOnedimensionalNanostructures_Mater.Charact.21}. To name a few, low-dimensional materials exhibit flexibility, high thermal and electrical conductivity, transparency, low Joule effect. This diversity of properties make low-dimensional solids ideal candidates for several applications. They range from very small electronic devices, such as integrated circuits \cite{journalarticle:Hua.Shen_LowdimensionalNanostructuresMonolithic_Chem.Soc.Rev.24}, healthcare monitoring systems \cite{journalarticle:Huang.etal_GrapheneBasedSensorsHuman_Front.Chem.19,journalarticle:Zhang.etal_ScalablyNanomanufacturedAtomically_SmallStruct.22,journalarticle:Seo.etal_SinglechiralitySinglewallCarbon_Phys.Chem.Chem.Phys.25}, or energy harvesting \cite{journalarticle:Mangum.etal_MechanismsSpontaneousCurvature_Membranes21,journalarticle:Gikunda.etal_ArrayGrapheneVariable_Membranes22,journalarticle:Mamun.etal_RecentReviewElectrospun_Membranes23}, to large-scale structures, like perovskite solar cells \cite{journalarticle:Mohammed.etal_TwodimensionalPureBromine_Comp.Cond.Mat.23,journalarticle:Qamar.etal_CarbonNanotubesPerovskite_Synth.Met.24}, or catalysts for hydrogen production \cite{journalarticle:Shanmughan.etal_ExploringFuture2D_IntJHydrog.Energy23}. 

In this thesis, we focus on the mechanical response of these materials, specifically on their spatial profiles and their dependence on the applied external conditions. Our main goal is to understand the buckling phenomenon, a sudden change of the mechanical structure from a flat to bent (buckled) state, which emerges in a wide range of low-dimensional systems \cite{journalarticle:Hanakata.etal_AnomalousThermalExpansion_Phys.Rev.Lett.22,journalarticle:Singh.etal_RipplingBucklingMelting_Phys.Rev.B15,journalarticle:Plummer.Nelson_BucklingMetastabilityMembranes_Phys.Rev.E20,journalarticle:Shankar.Nelson_ThermalizedBucklingIsotropically_Phys.Rev.E21}. The name of buckling comes from its resemblance with the Euler buckling phenomenon on structural engineering \cite{book:Landau.etal_TheoryElasticityVolume_86,journalarticle:Golubovic.etal_DynamicsEulerBuckling_Phys.Rev.Lett.98}, where slender columns bend under load.

From a theoretical point of view, this mechanical change of shape has been studied using the theory of elasticity, which provides a continuum description of the deformations undergone by solids. Section~\ref{subsec:intro_elasticity} is thus devoted to introduce the basic concepts of the theory of elasticity for low-dimensional systems, where we derive the equilibrium differential equation that the spatial profiles must fulfil. Then, in section~\ref{subsec:intro_buckling}, we review how buckling phenomena have been previously studied in the literature using spin-elastic models. Therein, we discuss (lack of) its consistency with the theory of elasticity. 

\subsection{Theory of elasticity}
\label{subsec:intro_elasticity}

Here, we present a brief introduction to the theory of elasticity following Landau's approach \cite{book:Landau.etal_TheoryElasticityVolume_86}. Our aim is to compute equilibrium profiles by applying calculus of variations. The central idea is to minimise the free energy functional, which may be complemented with external stresses and/or boundary conditions.

Let us consider a solid described as a continuous medium. Under the effect of external forces, the solid undergoes structural changes that alter its shape and volume. Let $\mathbf{x}$ and $\mathbf{x}'$ be the spatial vector before and after the deformation, respectively. Then, the displacement field $\mathbf{u}$ is defined as 
\begin{equation}
    \mathbf{u} = \mathbf{x}' - \mathbf{x},
\end{equation}
or $u_i = x_i'-x_i$ for the three spatial components $i=1,2,3$, where we have a certain some three-dimensional orthonormal basis. The distance between two close points before the deformation is given by
\begin{align}
    \text{d}s = \sqrt{\text{d}x_1^2+\text{d}x_2^2+\text{d}x_3^2}, \quad \text{d}s^2 = \sum_{i=1}^3 \text{d}x_i^2 
\end{align}
For the sake of compactness, we will use the Einstein summation convention to simplify the notation to $\text{d}s^2 = \text{d}x_i^2$, \ie repeated indices indicate summation over them. After the deformation, that distance changes to $\text{d}s'^2 = \text{d}x_i'^2 = (\text{d}x_i+\text{d}u_i)^2$. Conveniently, applying the chain rule, it reads
\begin{equation}
    \text{d}s'^2 = \text{d}s^2 + 2 u_{ik} \text{d}x_i\text{d}x_k, 
    \label{eq:elasticity_deformations_distance}
\end{equation} 
where we have defined the second order strain tensor 
\begin{equation}
    u_{ik} = \frac{1}{2}\parenthesis{
        \pdev{}{u_i}{x_k} + \pdev{}{u_k}{x_i} + \pdev{}{u_l}{x_i}\pdev{}{u_l}{x_k}}. 
    \label{eq:elasticity_strain_tensor}
\end{equation}
Equations~\eqref{eq:elasticity_deformations_distance} and~\eqref{eq:elasticity_strain_tensor} describe the infinitesimal variations of distances in the solid due to the deformation. Obviously, the solid is not deformed for a rigid translation $\mathbf{u} = \mathbf{\const}$.

We are going to focus on the limit of small deformations, where the relative changes of the distances are ``small''. In this regime, although the strain tensor components~\eqref{eq:elasticity_strain_tensor} are also small, the displacement field $\mathbf{u}$ can be large in some situations, \eg in the paradigmatic examples of low-dimensional systems. 

The equilibrium state of a solid is related to the microscopic distribution of the system. In the absence of external agents, the equilibrium corresponds to the free configuration of the system. When the solid is deformed, by the action of external agents, internal forces appear to try to restore the original shape, causing the equilibrium profile to change. Its physical origin stems from the atomistic interactions of the solid, which try to take the system to its minimum energy configuration compatible with the external conditions.

Let us consider a deformed solid that only experiences its own internal stresses. In the continuum limit, the total forces acting on the system can be divided into the sum of the forces acting on each infinitesimal volume element. Let $\phi_i$ be the $i$-th component of the force per unit volume, being $i$ an arbitrary direction. Then, the total force exerted to a section $V$ of the total volume $\Omega$ is $\int_{V} \d{\mathbf{x}} \phi_i$. In the absence of external forces, the action-reaction law states that there exists an equal and opposite force that balances any $\phi_i$, so they cancel in the total resultant. Newton's third law thus ensures the total force acting on a volume $V$ is the sum of forces that the surrounding elements exert on it. Hence, it can be represented as a surface integral. Defining $\sigma_{ij}$ as $\phi_i=\partial_{x_j}\sigma_{ij}$, and applying the divergence theorem, we can write 
\begin{equation}
    \int_{V} \d{\mathbf{x}} \phi_i = \int_{V} \d{\mathbf{x}} \partial_{x_j}\sigma_{ij} = \int_{\partial V} \d{S} \sigma_{ij} n_j,
\end{equation}
where $\mathbf{n}$ is the external normal vector to the surface $\partial V$. The tensor $\sigma_{ij}$ is called the \emph{stress tensor}, which represents the force per unit area in the $i$-th direction that is exerted on a surface whose normal is along the $j$-th direction. 

The free energy density $f$ of the deformed body is obtained by an expansion in powers of the strain tensor components $u_{ij}$. Expanding up to second order, the expression of a deformed isotropic body in equilibrium is 
\begin{equation}
    f = f_0 + \frac{1}{2}u_{ii}^2 + \mu u_{ij}^2,
\end{equation}
being $\lambda$ and $\mu$ the first and second Lamé's coefficient, respectively, and $f_0$ the free energy density of the undeformed solid---let us assume $f_0=0$, since we are not interested in its value. In the small deformation regime, we assume that the stress tensor has a linear dependence on the strain tensor, which is known as Hooke's law. The usual way to write Hooke's law is 
\begin{equation}
    \sigma_{ij} = 2\mu\parenthesis{u_{ij}-\frac{1}{3}u_{kk}\delta_{ij}} + K u_{kk}\delta_{ij},
\end{equation} 
where $\mu$ and $K=\lambda+2\mu/3$ are usually called shear and bulk modulus, respectively. In this approximation, the free energy of the deformed solid is given by 
\begin{equation}
    f = \frac{1}{2}\sigma_{ij}u_{ij}=\mu \parenthesis{u_{ij}-\frac{1}{3}u_{kk}\delta_{ij}}^2 + \frac{K}{2} u_{kk}^2.
    \label{eq:elasticity_free_energy_density}
\end{equation}
The first term on the right-hand size (rhs) stands for \emph{pure shear}, \ie deformations that do not alter the volume of the solid, only the shape. The opposite case, represented by the second term, indicates deformations that only altered the volume, which is called \emph{hydrostatic compression/expansion}.

If at the boundaries we apply a pressure, then the boundary conditions are
\begin{equation}
    \sigma_{ij} n_j = P_i, \quad \forall \mathbf{x}\in \partial\Omega,
    \label{eq:elasticity_boundary_conditions}
\end{equation}
where $P_i$ is the applied pressure on the surface $\partial \Omega$ along the $i$-th direction. 

\subsection{Low-dimensional systems: thin plates}

\begin{figure}
    \centering
    \includegraphics[width=0.6\textwidth]{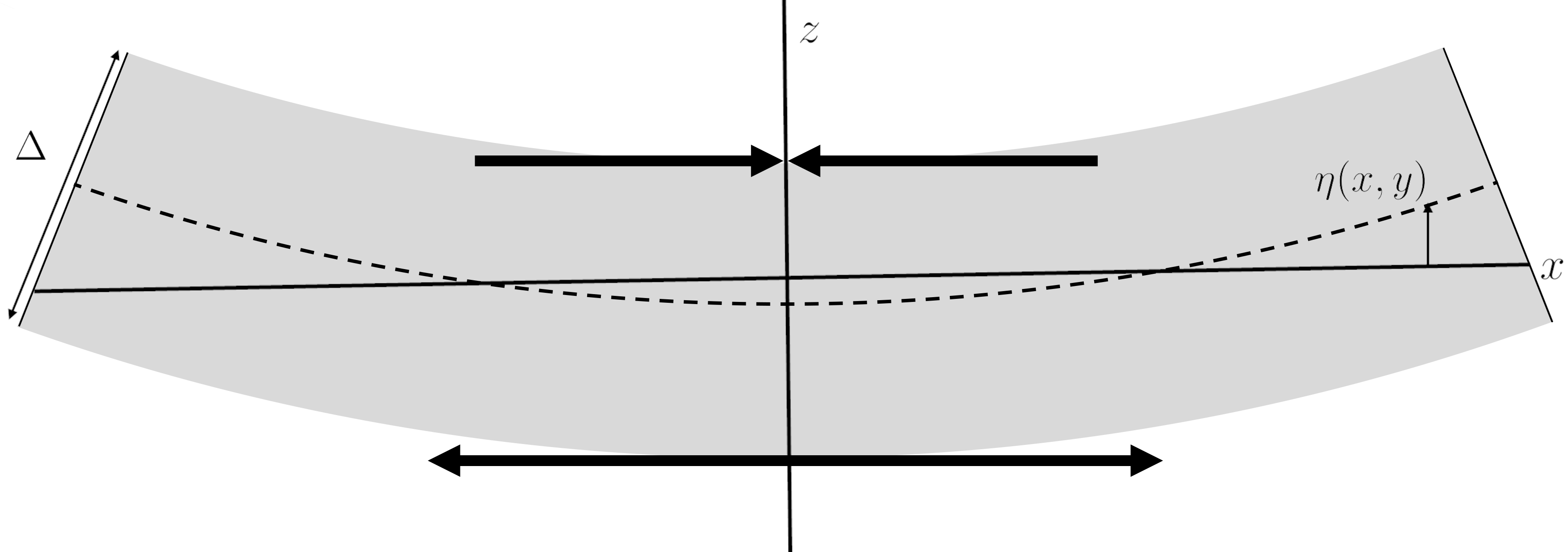}
    \caption{Schematic representation of a thin plate of thickness $\Delta$ that lies on the $xy$-plane. The displacement field $\mathbf{u}=(u_x,u_y,\eta)$ has two in-plane components $u_x$ and $u_y$, and an out-of-plane component $\eta$.}
    \label{fig:intro_elasticity_plate}
\end{figure}

In this section, the previous description is particularised to the case of two-dimensional materials. Let us consider a thin plate of thickness $\Delta \ll d$, where $d$ is a typical spatial length on the plane. As seen in figure~\ref{fig:intro_elasticity_plate}, if the plate is bent, some regions of the plate are compressed (upper surface), whereas others are stretched (lower surface). Inside the plate, we can always find a neutral surface, where the system is neither compressed nor stretched, which is located along the curve defined by the thickness midpoint in figure~\ref{fig:intro_elasticity_plate}. Thus, there is no planar deformations in the neutral surface, so the displacements are given by $\mathbf{u}^{(0)}=(0,0,\eta(x,y))$. Since the plate in thin, we assume it can be bent even if the applied forces are relatively small compared to the internal stresses. Thus, we can neglect the pressures on the boundaries~\eqref{eq:elasticity_boundary_conditions}, leaving $\sigma_{ij}n_j=0$ in $\partial\Omega$. Since we are working with small deformations, the system is slightly bent and we can suppose that normal vector $\mathbf{n}$ is along the $z$-direction. Hence, the stress tensor on both surfaces fulfils $\sigma_{iz}=0$, $i=x,y,z$, $\mathbf{x}$ at $\partial\Omega$. Additionally, since the plate is thin, $\sigma_{iz}$ components must be small compared to the others if they are zero at the surface. Therefore, we may assume that those components are negligible inside the plate, as compared with the remaining ones of the stress tensor, \ie $\sigma_{iz}\simeq 0$, $i=x,y,z$.

Under these assumptions, following Landau's derivations in~\cite{book:Landau.etal_TheoryElasticityVolume_86}, we can write the displacement field components as 
\begin{equation}
    u_x = -z \pdev{}{\eta}{x}, \quad u_y = -z \pdev{}{\eta}{y}, \quad u_z = \eta(x,y).
\end{equation}
The free energy of the system is obtained integrating~\eqref{eq:elasticity_free_energy_density}, which reads 
\begin{subequations}
    \begin{align}
        F &=\iint \d{x}\d{y}\int_{-\Delta/2}^{\Delta/2} \d{z} f = F_1 + F_2,
        \label{eq:elasticity_free_energy_plate_functional}
        \\
        F_1 &=D \iint \d{x}\d{y} \parenthesis{\pdev{2}{\eta}{x}+\pdev{2}{\eta}{y}}^2, 
        \\
        F_2 &= 2D(1-\sigma) \iint \d{x}\d{y} \brackets{
            \parenthesis{\frac{\partial^2\eta}{\partial x\partial y}}^2 - \pdev{2}{\eta}{x}\pdev{2}{\eta}{y}},
    \end{align}
\end{subequations} 
where $D$ and $\sigma$ are constants that depend on the material properties. 

The equilibrium profile is obtained by minimising the free energy functional~\eqref{eq:elasticity_free_energy_plate_functional}. Using calculus of variations \cite{book:Lanczos_VariationalPrinciplesMechanics_70,book:Landau.etal_TheoryElasticityVolume_86}, we derive the Euler-Lagrange equation 
\begin{equation}
    \nabla^4 \eta = 0,
\end{equation}
where $\nabla^4 = \nabla^2\nabla^2$ is the biharmonic operator, and $\nabla^2 = \partial^2/\partial x^2 + \partial^2/\partial y^2$ is the 2d Laplace operator. The previous PDE gives us the equilibrium shape of two-dimensional materials in the absence of external forces. Thus, the term $\nabla^4 \eta$ corresponds to the contribution of the internal forces that the solid exerts to restore its original flat shape after being deformed. 

The calculus of variations also gives us certain boundary terms that must vanish at the contour to ensure the minimisation of the free energy functional, as explained in appendix~\ref{app:variational_principle_higher_order}. The most usual ones correspond to consider either clamped or supported boundary conditions. The supported conditions state that vertical displacement are forbidden at the contour, but the slope is free to vary, so the system can bend at the boundaries. In contrast, clamped conditions impose both the vertical displacements and the slope to be zero at the contour, so the profile is completely fixed at the boundaries.

Let us suppose that the whole plate is subjected to an external force per unit area $P(x,y)$ along the $z$-direction. Then, the extra contribution to the free energy functional~\eqref{eq:elasticity_free_energy_plate_functional} can be proved to be  
\begin{equation}
    F_3= \iint \d{x}\d{y} P(x,y)\, \eta \to \delta F_3 = \iint \d{x}\d{y} P(x,y)\delta \eta.
\end{equation}
Thus, the Euler-Lagrange equation is modified to 
\begin{equation}
    D \nabla^4 \eta = P(x,y).
    \label{eq:elasticity_EL_general_thinplates}
\end{equation}
We must underline that the external stress $P(x,y)$ involves a new source term that breaks the rotational symmetry of the system, \ie any rotation of an equilibrium profile does not have the same energy as the original one. This fact can be easily checked by comparing the contributions from each term in the free-energy functional~\eqref{eq:elasticity_free_energy_plate_functional}: the internal forces $D\nabla^4 \eta$ stem from the term $F_1$, whose free energy density involves the curvature of the profile, $\nabla^2 \eta$, whereas the pressure $P(x,y)$ stems from the term $F_3$, whose free energy density directly involves the profile itself, $\eta$.

\subsection{Graphene and spin-elastic models}
\label{subsec:intro_buckling}

Graphene is one prototypical example of two-dimensional material. In fact, it was one of the precursors of the study of two-dimensional systems and their properties \cite{journalarticle:Novoselov.etal_ElectricFieldEffect_Science04,journalarticle:Neto.etal_ElectronicPropertiesGraphene_Rev.Mod.Phys.09,journalarticle:Amorim.etal_NovelEffectsStrains_Phys.Rep.16}. It is made of a single layer of carbon atoms arranged in a honeycomb lattice. The existence of graphene constituted a major breakthrough in material science, since perfect two-dimensional materials were supposed to be impossible according to the Mermin-Wagner theorem \cite{journalarticle:Mermin.Wagner_AbsenceFerromagnetismAntiferromagnetism_Phys.Rev.Lett.66} and Landau's fluctuation theory, due to the long wavelength fluctuations that destroy the order in two-dimensional systems \cite{book:Landau.Lifshitz_StatisticalPhysicsVolume_13}. Nonetheless, subsequent experiments demonstrated that graphene sheets are not perfectly flat, but they present some out-of-plane displacements due to thermal fluctuations, known as ripples \cite{journalarticle:Meyer.etal_StructureSuspendedGraphene_Nature07,journalarticle:Fasolino.etal_IntrinsicRipplesGraphene_Nat.Mater.07}. 

\begin{figure}
    \centering
    \includegraphics[width=0.8\textwidth]{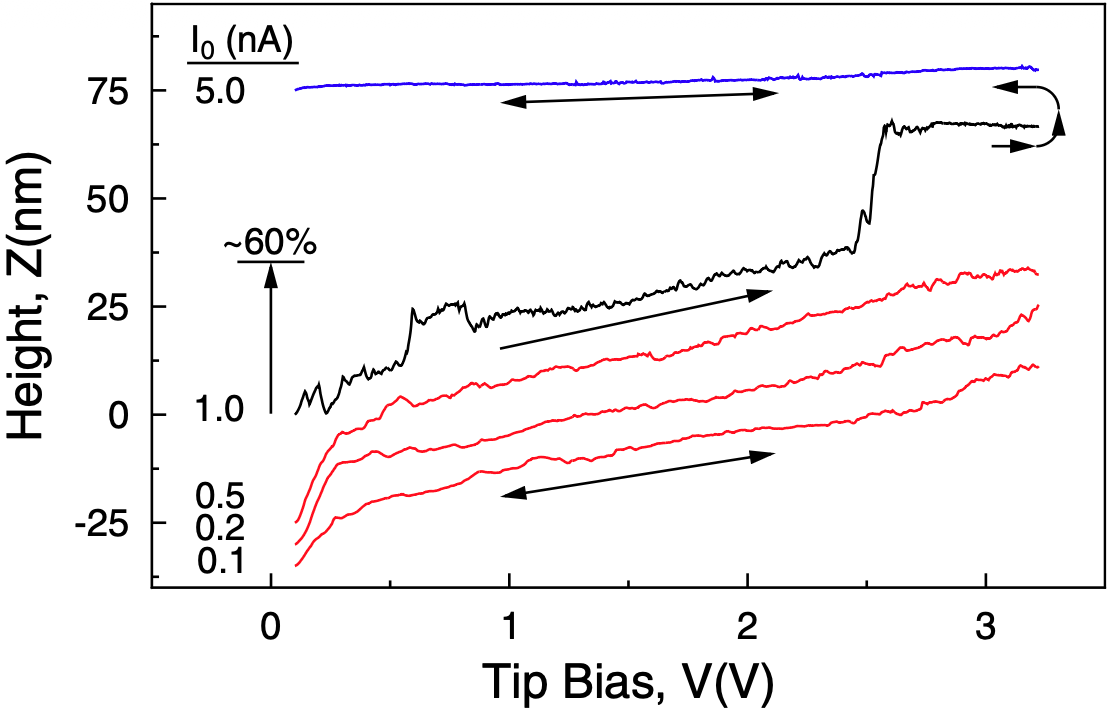}
    \caption{Height of the STM tip as a function of the applied voltage for several constant tunneling currents. Rippled profiles, represented by red curves, are observed for low voltages: the height $Z$ has a smooth and reversible dependence with the voltage. The black curve indicates the critical value, which depends on the local heating---product of the voltage and the current. At this critical voltage, the system undergoes an abrupt transition to a buckled state where the height $Z$ shows an irreversible behaviour: over the blue curve, the system remains buckled even if the voltage is decreased. Curves are slightly offset from each other for the sake of clarity. Image taken from \cite{journalarticle:Schoelz.etal_GrapheneRipplesRealization_Phys.Rev.B15}.}
    \label{fig:introduction_schoelz}
\end{figure}

\begin{figure}
    \centering
    \includegraphics[height=7cm]{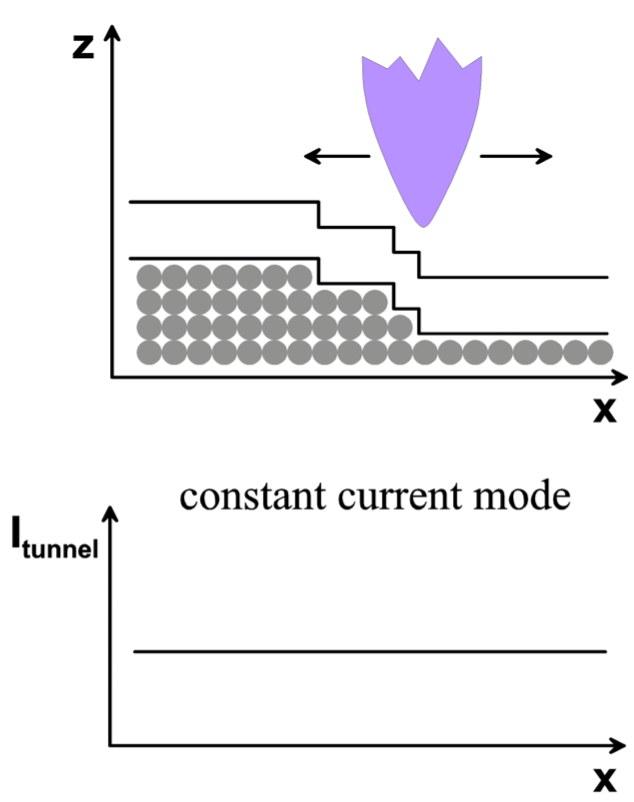}
    \caption{Schematic representation of the current-constant operating mode in scanning tunneling microscopy. The height of the tip $z$ varies in order to keep $I_{\text{tunnel}}$ constant while the tip maps the surface. Image taken from \cite{thesis:Pandelov_InvestigationStructureReactivity_07}.}
    \label{fig:STM}
\end{figure}

The rippling phenomenon has been widely studied using lattice models where the transversal displacements are coupled to electronic degrees of freedom, \ie electron-phonon coupling \cite{journalarticle:Cea.etal_NumericalStudyRippling_Phys.Rev.B20,journalarticle:Fasolino.etal_IntrinsicRipplesGraphene_Nat.Mater.07,journalarticle:San-Jose.etal_ElectroninducedRipplingGraphene_Phys.Rev.Lett.11,journalarticle:Bonilla.Carpio_ModelRipplesGraphene_Phys.Rev.B12,journalarticle:Bonilla.etal_RipplesStringCoupled_Phys.Rev.E12,journalarticle:Ruiz-Garcia.etal_STMdrivenTransitionRippled_Phys.Rev.B16,journalarticle:Ruiz-Garcia.etal_BifurcationAnalysisPhase_Phys.Rev.E17}. Graphene has been observed to experiment a mechanical phase transition that is related to buckling phenomena---the system goes from a flat profile to a bent state. Scanning tunneling microscopy (STM) experiments have shown that, even in the absence of external load, local heating induces a phase transition from a rippled sheet to a buckled membrane \cite{journalarticle:Schoelz.etal_GrapheneRipplesRealization_Phys.Rev.B15,journalarticle:Neek-Amal.etal_ThermalMirrorBuckling_Nat.Commun.14,journalarticle:Lindahl.etal_DeterminationBendingRigidity_NanoLett.12,journalarticle:Eder.etal_ProbingBothSides_NanoLett.13}. STM works by applying a voltage between a sharp tip and a surface. If the tip is close enough to the sample, electrons can tunnel through the vacuum gap between them, producing a tunneling current whose intensity depends on that gap distance and the voltage. Figure~\ref{fig:introduction_schoelz} reports the experiment made in~\cite{journalarticle:Schoelz.etal_GrapheneRipplesRealization_Phys.Rev.B15} using this setup. Therein, an Omicron ultrahigh-vacuum low-temperature STM that operates in \emph{current-constant} mode measures the height $Z$ of a freestanding graphene sample. Fixed the applied voltage $V$, digital images are obtained keeping the tunneling current constant $I_0$ using a feedback loop: electronics adjust the tip height while it scans the surface topography---figure~\ref{fig:STM} illustrates a schematic representation of this operating mode in STM. The parameters of the experiment, the voltage and tunneling current, are directly related to the local heating induced on the graphene sheet via Joule effect. We observe how the height of the tip smoothly increases with the voltage, following a reversible process for low currents. A critical change is observed when considering higher currents. After that critical value, the system undergoes an abrupt transition, turning to an irreversible buckled state.

Previous works have proposed minimal models to understand qualitatively this phenomenon. For instance, Schoelz \etal \cite{journalarticle:Schoelz.etal_GrapheneRipplesRealization_Phys.Rev.B15} proposed a phenomenological model where graphene is modelled as a two-dimensional Ising lattice, where each node stands for an entire ripple---they contain around a thousand carbon atoms. Every ripple is characterised by a spin $s_i$ which qualitatively mimics its curvature, either positive ($s_i>0$) or negative $(s_i<0)$. The Hamiltonian of the Schoelz model can be written as 
\begin{equation}
    \mathcal{H} = -J(M) \sum_{\langle i,j \rangle} s_i s_j - \sum_i h_i s_i,
    \label{eq:intro_schoelz_hamiltonian}
\end{equation}
so the spins interact with their nearest neighbours with a coupling $J(M)$, which depends on the total magnetisation $M=\sum_i s_i$ and, additionally, the system is coupled to a spatial-dependent external field $h_i=h(r_i)$. The external field models the influence of STM on the system that depends on the distance $r$ between the tip and the $i$-th spin. The coupling $J(M)$ is related to the elastic energy of the ripples: an antiferromagnetic interaction $J(M)<0$ destroys the order, favouring the appearance of ripples---the nearest curvatures take opposite values and effectively cancel out, whereas a ferromagnetic coupling $J(M)>0$ promotes the alignment of the spins, which provokes the buckling of the sheet---the curvatures are equal, producing a global bending. 

Another qualitative explanation of this buckling transition has been provided by the analysis of simple models made of spin-elastic lattices \cite{journalarticle:Ruiz-Garcia.etal_RipplesHexagonalLattices_J.Stat.Mech.15,journalarticle:Ruiz-Garcia.etal_STMdrivenTransitionRippled_Phys.Rev.B16,journalarticle:Ruiz-Garcia.etal_BifurcationAnalysisPhase_Phys.Rev.E17}, which motivates our approach to tackle the description of the buckling phenomenon in this thesis. Let us illustrate the model on a one-dimensional lattice with parameter---distance between nodes---$a$ in contact with a thermal bath $T$. In contrast to the Schoelz model~\cite{journalarticle:Schoelz.etal_GrapheneRipplesRealization_Phys.Rev.B15}, here each site $j$, $j=0,\ldots,N$, represents a single particle of mass $m$ that is characterised by its transversal displacement $u_j$,\footnote{We are not going to consider in-plane deformations in these spin-elastic models.} its conjugate momentum $p_j$, and a spin variable $\sigma_j=\pm 1$ that represents an electronic internal degree of freedom. The Hamiltonian in this spin-string system is given by \cite{journalarticle:Ruiz-Garcia.etal_RipplesHexagonalLattices_J.Stat.Mech.15,journalarticle:Ruiz-Garcia.etal_STMdrivenTransitionRippled_Phys.Rev.B16,journalarticle:Ruiz-Garcia.etal_BifurcationAnalysisPhase_Phys.Rev.E17}
\begin{equation}
    \hamSpinElastic = \sum_{j=0}^N \brackets{
        \frac{p_j^2}{2m} + \frac{k}{2}(u_{j+1}-u_j)^2 - h u_j\sigma_j + J \sigma_{j+1}\sigma_j},
    \label{eq:intro_spin_string_hamiltonian}
\end{equation}
where we have followed the notation $\mathbf{u}=(u_0,\ldots,u_N)$ for any set of variables. The first term on the rhs is the kinetic energy; the second one stands for the elastic contribution of the string with elastic constant $k$, which represents the internal forces that try to restore the flat/undeformed configuration of the string; whereas the rest of them involve the internal degrees of freedom. In the case of graphene, the spin variables $\sigma_j$ can be interpreted as the free out-of-plane electron. Following this idea, the coupling between the displacements and the spins, $-h u_j\sigma_j$, captures the interaction between the elastic and electronic components, modeling the electron-phonon coupling. Finally, the last term, $J \sigma_{j+1}\sigma_j$,\footnote{Here, $J>0$ indicates an antiferromagnetic coupling, in contrast to the Schoelz Hamiltonian~\eqref{eq:intro_schoelz_hamiltonian}.} represent the interaction between the nearest neighbour spins, which roughly mimics a Coulomb force between the free electrons---the antiferromagnetic case, $J>0$, would correspond to a standard Coulomb interaction, where charges with the same sign repel themselves. From a physical point of view, these contributions have a clear competition in order to stabilise the system and find the configuration that minimises the energy: all spins tend to align with the local displacement, which acts like an external field, but the antiferromagnetic coupling favours anti-aligned spins.

\begin{figure}
    \centering
    \begin{minipage}{.5\textwidth}
        \centering
        \includegraphics[width=0.85\textwidth]{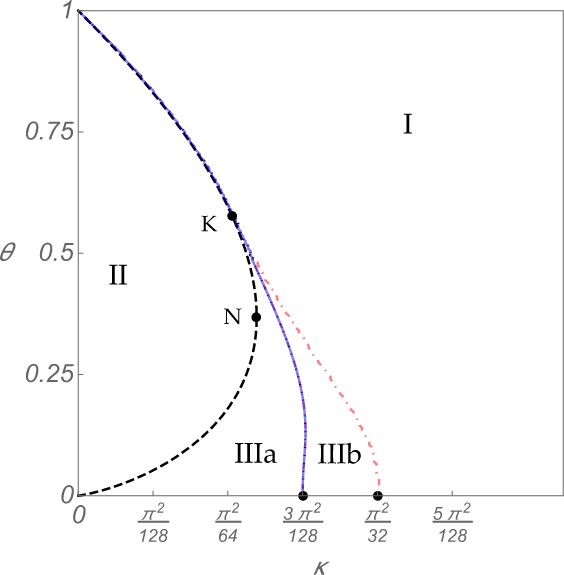}
    \end{minipage}%
    \begin{minipage}{.5\textwidth}
        \centering
        \includegraphics[width=0.95\textwidth]{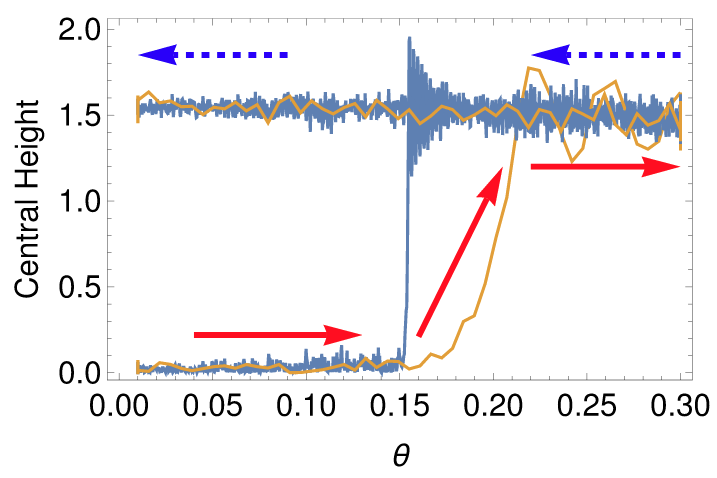}
    \end{minipage}
    \caption{Left panel: Phase diagram of the one-dimensional spin-string model in the plane $(J,T)$---($\kappa,\theta$) in the nomenclature of~\cite{journalarticle:Ruiz-Garcia.etal_BifurcationAnalysisPhase_Phys.Rev.E17}. The blue solid line corresponds to the phase transition lines of the system: a second-order one for temperatures above that of the critical point $K\equiv(J_K,T_K)$, and a first-order for temperatures below it. In region I, the only phase is the flat or rippled one, whereas a stable buckled phase appears in region II, where the rippled one becomes unstable. Region III is a coexistence zone where both phases are (locally) stable. Image taken from \cite{journalarticle:Ruiz-Garcia.etal_BifurcationAnalysisPhase_Phys.Rev.E17}. 
    Right panel: Numerical simulation of a simple model to qualitatively reproduce the Schoelz experiment. The height of the central atom is represented as a function of the temperature $\theta$, for $\kappa=0.1$. The system, prepared at a rippled phase at low temperature, is heated up until the phase transition occurs, leaving the system in an irreversible buckled state. Image taken from \cite{journalarticle:Ruiz-Garcia.etal_STMdrivenTransitionRippled_Phys.Rev.B16}.}
    \label{fig:intro_spin_elastic}
\end{figure}

The equilibrium profiles of the string are obtained by finding the displacement configuration that minimises the free energy of the system. Let us define the spatial variable $x_j = j a$, $j=0,\ldots,N$, so the string length is $L=Na$. In the continuum limit, $a\to 0^+$, the spatial variable becomes continuous $x_j\to x$, $u_j=u(x_j)\to u(x)$. Thus, a free energy functional $\mathcal{F}[u]$ can be derived \cite{journalarticle:Ruiz-Garcia.etal_BifurcationAnalysisPhase_Phys.Rev.E17}, so one can apply a variational principle to obtain an Euler-Lagrange equation for the equilibrium profiles. The Euler-Lagrange equation reads
\begin{equation}
    \cdev{2}{u}{x} = \mu(u;h/T,J/T),
    \label{eq:intro_EL_spin_string}
\end{equation}
being $\mu(u;h/T,J/T)$ the local magnetisation. The solutions of the previous equation for any pair of parameters $(J,T)$ makes it possible to characterise the phase diagram of the system---see the left panel of figure~\ref{fig:intro_spin_elastic}. In the close analogy to magnetic systems in the Landau theory of phase transitions \cite{book:Landau.Lifshitz_StatisticalPhysicsVolume_13}, a tricritical temperature $T_K$ appears, delimiting two different phase transitions. For fixed $T>T_K$, the system undergoes a second-order phase transition (blue solid line) from a rippled phase at high $J>0$ (region I), $u(x)=0$, to a buckled phase, $u(x)\neq 0$, when the antiferromagnetic coupling decreases (region II). On the contrary, at low temperature $T<T_K$, there is a range of values for $J$ where it appears a region where both phases coexist and are metastable (region III). The change of stability in this region defines a first-order transition line (blue solid line): in region IIIa, the buckled phase is stable, whereas the rippled one is metastable, and vice versa in region IIIb. The experiment reported in figure~\ref{fig:introduction_schoelz} can be qualitatively understood using this phase diagram. Let us prepare the system in a metastable rippled phase at low temperatures (region IIIa). If the temperature is increased, the system eventually crosses to region II, where the rippled phase is unstable and the system undergoes a phase transition to the stable buckled phase. This transition is irreversible, because the system remains in the stable buckled phase even if the temperature is decreased again. This magnetic hysteresis can also be computed numerically \cite{journalarticle:Ruiz-Garcia.etal_STMdrivenTransitionRippled_Phys.Rev.B16}, as shown in the right panel of figure~\ref{fig:intro_spin_elastic}.


Although the previous spin-elastic model qualitatively reproduces the buckling phenomenon observed in experiments, it is not consistent with the theory of elasticity. Let us compare the Euler-Lagrange equation of the spin-string model~\eqref{eq:intro_EL_spin_string}---the particularisation for two-dimensional model has an analogous form---with the equilibrium equation of elasticity for thin plates~\eqref{eq:elasticity_EL_general_thinplates}. If we check the left-hand sides (lhs), which correspond to elastic forces that try to keep the system flat, we observe a discrepancy: the internal forces are proportional to the Laplacian $\nabla^2$ in~\eqref{eq:intro_EL_spin_string}, instead of the Bilaplacian $\nabla^4$ in~\eqref{eq:elasticity_EL_general_thinplates}. This inconsistency stems from the elastic contribution $k(u_{j+1}-u_j)^2/2$ in the Hamiltonian~\eqref{eq:hamiltonian_spin_string_1d}, which tells us that a flat system and a system with a constant slope have different energies. In the absence of external forces, the rotational symmetry must be conserved. However, we may check how any profile with an infinitesimal transformation $u(x)\to u(x)+ c_1 x + c_2$, has a different energy for any small rotation $c_1$ and rigid translation $c_2$.

Our aim is to develop a spin-elastic model, based on the previous ones \cite{journalarticle:Ruiz-Garcia.etal_RipplesHexagonalLattices_J.Stat.Mech.15,journalarticle:Ruiz-Garcia.etal_STMdrivenTransitionRippled_Phys.Rev.B16,journalarticle:Ruiz-Garcia.etal_BifurcationAnalysisPhase_Phys.Rev.E17} that also captures the emergence of buckling phenomena, but consistent with the theory of elasticity. To do so, we propose a Hamiltonian where the elastic variables only appear in terms of the discrete curvature, \ie the second derivative of the displacement field.

\subsection{Summary of part I}

The first part of this thesis is devoted to study rotationally invariant spin-elastic models that qualitatively capture the buckling phenomenon observed in low-dimensional materials. The analysis is clearly divided into two main blocks, depending on the dimensionality of the system: one-dimensional spin-string models and two-dimensional spin-membrane models. 

Chapter~\ref{ch:materials_equilibrium} introduces our novel spin-elastic model, explaining the main difference with respect to previous proposals in the literature: its dependence on the discrete Laplacian of the displacement, \ie the discrete curvature. An Euler-Lagrange equation for the equilibrium profiles is derived in the continuum limit, which fulfils the sought rotational symmetry. Appendix~\ref{app:variational_principle_higher_order} provides a detailed derivation of the 1d Euler-Lagrange equation for functionals that depend on up to the second derivative of the field, which is the case of our spin-elastic models. The Euler-Lagrange equation becomes an algebraic equation for the curvature, that turns to be a spatially homogeneous quantity. The value of the curvature is obtained both analytically, using both bifurcation theory and Landau theory, and numerically, solving the algebraic Euler-Lagrange equation for the whole parameter space. The critical behaviour is analysed, finding a tricritical point. The phase diagram of the spin-string model is investigated in detail, characterising the different phases depending on the system parameters (temperature and spin coupling). The model qualitatively reproduces the buckling phenomenon observed in experiments with graphene sheets under local heating.

Chapter~\ref{ch:materials_2d} generalises the model to two dimensions. Different microscopic topologies for the spin-membrane model are considered, specifically square and honeycomb lattices. Both of them are analysed under the same theoretical framework, because they lead to the same continuum limit where the curvature is the crucial observable. The Euler-Lagrange equation is derived, which again predicts a spatially homogeneous curvature, and the shape of the profiles are integrated for different geometries. The possible curvatures are obtained in certain limits, specifically when either the temperature or the spin-spin interaction are small. The analysis done for low temperatures shows a way to obtain the partition function of any $d$-dimensional Ising lattice with nearest-neighbour interactions, provided that the topology does not involve triangular loops, as detailed in appendix~\ref{app:ground-state_triangular_loops}.

\section{Search processes} 
\label{sec:intro_search_processes}

As a general concept, searches are ubiquitous processes in many contexts. Essentially, these phenomena involve two ingredients: searchers and targets. Searchers are entities that explore a configurational space following certain dynamical rules. Targets are objects, placed at certain points in space, that searchers try to reach. This simple abstraction makes possible to characterise many phenomena in different interdisciplinary fields within a common framework. To name some relevant examples, we may think about: 
\begin{itemize}
    \item Biological processes like animal foraging \cite{book:Bell_SearchingBehaviourBehavioural_12,journalarticle:OBrien.etal_SearchStrategiesForaging_Am.Sci.90} or, more generally, seeking for any kind of resources, such as binding sites in biomolecular scenarios \cite{journalarticle:Reuveni.etal_RoleSubstrateUnbinding_Proc.Natl.Acad.Sci.U.S.A.14,journalarticle:Rotbart.etal_MichaelisMentenReactionScheme_Phys.Rev.E15}.
    \item Computer science applications, such as search engines on the web and related search software---including information retrieval systems in large databases \cite{journalarticle:Wu.etal_EvolutionSearchThree_ACMTrans.Manag.Inf.Syst.22,journalarticle:Schmidhuber_DeepLearningNeural_NeuralNetw.15,journalarticle:Browne.etal_SurveyMonteCarlo_IEEETrans.Comp.Intell.AIGames12}.
    \item Social sciences, ranging from economics, \eg the evolution of financial markets \cite{journalarticle:Stojkoski.etal_GeneralisedGeometricBrownian_Entropy20,journalarticle:Stojkoski.etal_GeometricBrownianMotion_Phys.Rev.E21,journalarticle:Stojkoski.etal_IncomeInequalityMobility_Philos.Trans.R.Soc.A22}, to psychology, \eg the search for appealing information in social media \cite{journalarticle:Amendola.etal_SocialSearchRetrieving_OnlineSoc.Netw.Media23}.
\end{itemize}
All preceding cases involve the archetypal problem of finding a target as fast as possible. Consequently, a crucial question in these processes is to devise search strategies that expedite reaching the target. 

\subsection{First-passage time analysis}
\label{subsec:FPT_analysis}
From a theoretical point of view, the study of search processes can be tackled by using the theory of stochastic processes \cite{book:VanKampen_StochasticProcessesPhysics_92,book:Gardiner_HandbookStochasticMethods_83,book:Redner_GuideFirstpassageProcesses_08}. Let us suppose that the searcher performs a continuous random walk in its configuration space \cite{book:Gardiner_HandbookStochasticMethods_83,book:VanKampen_StochasticProcessesPhysics_92}. The PDF of finding the object with position $x$ at time $t$ starting from $x_0$ at $t_0$ is given by $\PDF(x,t|x_0,t_0)$. The propagator $\PDF(x,t|x_0,t_0)$ evolves following~\eqref{eq:Fokker_Planck_equation_conditional}, where the initial condition is $\PDF(x,t_0|x_0,t_0) = \delta(x-x_0)$. In the absence of targets, the propagator preserves the normalisation condition $\int \d{x} \PDF(x,t|x_0,t_0) = 1$. Thus, working in an infinite domain, normalisation implies
\begin{equation}
    \lim_{x\to\pm\infty} \PDF(x,t|x_0,t_0) = 0, \quad \lim_{x\to\pm\infty} \partial_x \PDF(x,t|x_0,t_0) = 0.
\end{equation}

Mathematically, if we introduce a target at $\targetx$, the propagator must satisfy the condition $\PDF(\targetx,t|x_0,t_0)=0$. Therefore, the first-passage problems involves a boundary value problem for the Fokker-Planck equation~\eqref{eq:Fokker_Planck_equation_conditional}, with an absorbing boundary at $\targetx$. In these scenarios, the propagator does not preserve the normalisation condition because there is a non-zero probability of being sunk at that point. 

Our variable of interest is the first-passage time (FPT) $\FPT{1}$, \ie how long it takes the searcher to reach the target for the first time. As a result of the stochastic motion, $\FPT{1}$ is a random variable, which depends on the initial conditions and the target position. Thus, we can define its PDF as $\FPTpdf(\FPT{1};\targetx,x_0)$---the probability of reaching $\targetx$ in the time interval $[\FPT{1},\FPT{1}+\d{\FPT{1}}]$ is $\FPTpdf(\FPT{1};\targetx,x_0)\d{\mathcal{T}}$. Usually, a backward approach is employed to study this problem \cite{book:VanKampen_StochasticProcessesPhysics_92,book:Gardiner_HandbookStochasticMethods_83}. Let us define the survival probability as the probability that the searcher has not reached the target after a time $t$, \ie 
\begin{equation}
    \label{eq:survival_probability_def}
    \survival(t;\targetx,x_0) =  \int \d{x} \PDF(x,t|x_0,t_0) = \int_{t}^{\infty} \d{\FPT{1}} \FPTpdf(\FPT{1};\targetx,x_0),
\end{equation} 
where in the last equal we are assuming that the searcher eventually will find the target, \ie 
\begin{equation}
    \lim_{t\to\infty} \survival(t;\targetx,x_0) = 0 \quad \text{or, equivalently,} \quad \int_{0}^{\infty} \d{\FPT{1}} \FPTpdf(\FPT{1};\targetx,x_0) = 1.
\end{equation}
From~\eqref{eq:survival_probability_def}, the relation between the first-passage time distribution and the survival probability is clear,
\begin{equation}
   \label{eq:survival(t)_FPT_relation} 
    \FPTpdf(\FPT{1};\targetx,x_0) = -\partial_{\FPT{1}} \survival(\FPT{1};\targetx,x_0). 
\end{equation}  

Both $\survival(t;\targetx,x_0)$ and $\FPTpdf(\FPT{1};\targetx,x_0)$ fulfil the backward Fokker-Planck equation~\eqref{eq:backward_Fokker_Planck_equation}, which reads
\begin{equation}
    \label{eq:Backward_1d_FPT_general}
    -\partial_{\FPT{1}} \FPTpdf(\FPT{1};\targetx,x_0) = \brackets{a_1(x_0) \partial_{x_{0}} + \frac{1}{2}a_2(x_0)\partial_{x_{0}}^2}\FPTpdf(\FPT{1};\targetx,x_0),
\end{equation}
with the initial condition 
\begin{equation}
    \FPTpdf(0;\targetx,x_0)=0, \quad \forall x_0 \neq \targetx.
\end{equation}
Moreover, we have the general boundary condition
\begin{equation}
    \label{eq:Backward_1d_FPTgeneral_BCs}
    \FPTpdf(\FPT{1};\targetx,\targetx) = \delta(\FPT{1}).
\end{equation}
The other boundary conditions depend on the geometry of the problem, so we will specify them when dealing with the particular cases.

Instead of analysing the equation in the time domain, it is handy to define the Laplace transform of the FPT distribution,
\begin{equation}
    \label{eq:Laplace_transform_FPT}
    \FPTpdfLaplace(s;\targetx,x_0) = \int_0^\infty \d{\FPT{1}} e^{-s\FPT{1}} \FPTpdf(\FPT{1};\targetx,x_0).
\end{equation}
Here, $s$ is the Laplace variable. In this thesis, we are keeping the use of $\FPTpdfLaplace(s)$ to denote the Laplace transform of the FPT distributions $\FPTpdf(\FPT{1})$. However, we will also employ the notation $\laplTransformOper{A}{t}$, or $\laplTransformTilde{A}{s}$, to denote the Laplace transform of any particular function $A(t)$. This allows us to work with the ordinary differential equation (ODE) for the Laplace transform,
\begin{equation}
    \label{eq:Laplace_transform_FPT_general}
    -s \FPTpdfLaplace(s;\targetx,x_0) = \brackets{a_1(x_0) \partial_{x_0} + \frac{1}{2}a_2(x_0)\partial_{x_0}^2}\FPTpdfLaplace(s;\targetx,x_0).
\end{equation}

In order to find the best strategy to reach the target, we first must define what quantity measures the efficiency of the search. Typically, since the search time $\FPT{1}$ is a stochastic variable, we optimise search processes by using the moments of the FPT distribution $\FPTpdf(\FPT{1};\targetx,x_0)$. 
The $n$-th moment of the FPT distribution is given by
\begin{equation}
    \label{eq:FPT_moments}
    \FPT{(n)}(\targetx,x_0)=\langle \FPT{n} \rangle = \int_0^\infty \d{\FPT{1}} \FPT{n} \FPTpdf(\FPT{1};\targetx,x_0), \quad n = 0,1,2,\ldots
\end{equation}
It is common practice to use the average time to reach the target $\FPT{(1)}(\targetx,x_0)$, \ie the mean first-passage time (MFPT) as an indicator of search success. The MFPT is particularly simple to obtain because it fulfils 
\begin{equation}
    \label{eq:MFPT_def}
    \FPT{(1)}(\targetx,x_0) = \int_0^\infty \d{\FPT{1}} \FPT{1} \FPTpdf(\FPT{1};\targetx,x_0) = \int_0^\infty \d{\FPT{1}} \survival(\FPT{1};\targetx,x_0) = \lim_{s\to0} \laplTransformTilde{Q}{s;\targetx,x_0},
\end{equation}
\ie it can be computed directly from the Laplace transform of the survival probability. In fact, all the moments can be obtained through the Laplace transform of the FPT distribution~\eqref{eq:Laplace_transform_FPT}:
\begin{equation}
    \FPT{(n)}(\targetx,x_0) = (-1)^n \, \partial_s^n \FPTpdfLaplace(s;\targetx,x_0) \bigg|_{s=0}. 
    \label{eq:FPT_moments_from_Laplace}
\end{equation}
Thus, substituting them into~\eqref{eq:Laplace_transform_FPT_general}, they are the solution of the following recursive ODEs
\begin{align}
    \label{eq:FPT_moments_general}
    -n \FPT{(n-1)}(\targetx,x_0) = \brackets{a_1(x_0) \partial_{x_0} + \frac{1}{2}a_2(x_0)\partial_{x_0}^2}\FPT{(n)}(\targetx,x_0),
\end{align}
with $\FPT{(0)}(\targetx,x_0)=1$ and the general boundary conditions inherited from~\eqref{eq:Backward_1d_FPTgeneral_BCs} by the Laplace transform and the particular one that depends on the specification of boundaries in the model.

When one speaks about optimising certain combinations of moments, there are many options depending on what aspects of the search process are deemed to be more important. Most often, one is interested in shortening the average search time, thus prioritising the minimisation of $\FPT{(1)}(\targetx,x_0)$. However, one may be interested in preventing those trajectories that take too long, differing from $\FPT{(1)}(\targetx,x_0)$, and thus minimise for instance the variance $\sigma^2_{\FPT{1}}$. Expressions~\eqref{eq:FPT_moments_from_Laplace} and~\eqref{eq:FPT_moments_general} are useful to analyse these observables, since they provide a direct way to compute the moments, without needing to solve the aforementioned Fokker-Planck equations~\eqref{eq:Backward_1d_FPT_general}. Obviously, in those cases where we are able to obtain the survival probability $\survival(t;\targetx,x_0)$, we can directly compute every moment using~\eqref{eq:survival(t)_FPT_relation} and~\eqref{eq:FPT_moments}. In particular, in the following, we introduce two cases that have been considered in the literature, where it is possible to obtain the survival probability and exploit these relations: Brownian motion (BM) or free diffusion \cite{book:Redner_GuideFirstpassageProcesses_08,book:VanKampen_StochasticProcessesPhysics_92,book:Gardiner_HandbookStochasticMethods_83}, and diffusion with stochastic resetting \cite{journalarticle:Evans.Majumdar_DiffusionStochasticResetting_Phys.Rev.Lett.11,journalarticle:Evans.Majumdar_DiffusionOptimalResetting_J.Phys.A:Math.Theor.11,journalarticle:Evans.etal_StochasticResettingApplications_J.Phys.A:Math.Theor.20}.

\subsubsection*{Application to Brownian motion}
Brownian motion is a paradigmatic example of a Markovian stochastic process. It models the random motion of a particle with mass $M$, the Brownian particle, in a fluid made up of much lighter particles with mass $m$, $m \ll M$. The heavier particle undergoes random collisions with the smaller ones, which induces the stochastic motion. 

Let us consider the timescale $\text{d}t$, where the velocity $V$ of the Brownian particle changes, to be much larger than the timescale of the collisions. In that limit, collisions are considered instantaneous and independent, so that the Brownian velocity can be modelled as a Markov process, because the motion ``forgets'' previous interactions after some \emph{relaxation time}.\footnote{To be more precise, many other considerations have to be taken into account, such as there are no recollisions \cite{book:Gardiner_HandbookStochasticMethods_83,book:VanKampen_StochasticProcessesPhysics_92}.} The position of the Brownian particle $X$ is not a Markov process in the same timescale, because it is computed as $X(t)=X(t_0) + \int_{t_0}^t \d{t'} V(t')$. To be able to consider $X$ Markovian, one has to assume a much longer timescale, such that the velocity has fluctuated many times. In other words, the velocity correlations vanish instantaneously over this longer, coarse-grained, timescale. 

Let us consider the timescale over which the position is a Markov process. Here, we focus on a one-dimensional Brownian particle, which follows an isotropic movement, \ie there is no preferred direction of the motion. The FPE~\eqref{eq:Fokker_Planck_equation_conditional} that characterises this system is straightforward to obtain. Firstly, the drift coefficient must vanish because every direction is equally probable, 
\begin{equation}
    a_1(x) = 0, \quad \forall x.
\end{equation}
Secondly, because of isotropy again, the diffusion coefficient must be independent of the position, \ie 
\begin{equation}
    a_2(x) = 2D, \quad \forall x,
\end{equation}
where $D$ is usually called the \emph{diffusion coefficient}.\footnote{The value of $D$ is computed from the Fluctuation-Dissipation theorem \cite{book:VanKampen_StochasticProcessesPhysics_92,book:Gardiner_HandbookStochasticMethods_83}, which relates the diffusion coefficient with the temperature $T$ of the system and the mass $M$ of the Brownian particle.}
Therefore, the FPE that describes standard Brownian motion is analogous to the heat equation for the propagator $\PDF_0(x,t|x_0,t_0)$, 
\begin{equation}
    \label{eq:Fokker_Planck_Brownian_motion}
    \partial_t \PDF_0(x,t|x_0,t_0) = D \partial_x^2 \PDF_0(x,t|x_0,t_0). 
\end{equation}
Brownian motion is also named as \emph{free diffusion}, because~\eqref{eq:Fokker_Planck_Brownian_motion} is the diffusion equation. The subscript ``0'' refers to the solution of ``free'' Brownian motion.\footnote{That is, without stochastic resetting, which is introduced in section~\ref{subsec:standard_stochastic_resetting}.}

For an unbounded domain and not target, $x\in\mathbb{R}$, the solution of~\eqref{eq:Fokker_Planck_Brownian_motion} is
\begin{equation}
        \label{eq:Green_function_BrownianMotion}
        \PDF_0(x,t|x_0,t_0) = \frac{1}{\sqrt{4\pi D (t-t_0)}} \exp\brackets{-\frac{(x-x_0)^2}{4D(t-t_0)}}.    
\end{equation}
The propagator is thus a Gaussian distribution around the initial position $x_0$, whose mean-square displacement (MSD) or variance grows linearly with time, $\sigma_0^2(t) = \braket{x^2}-\braket{x}^2 = 2D(t-t_0)$.\footnote{Depending on the relation between the position variance and time, being generally $\sigma^2(t)\propto t^\alpha$, we can classify the stochastic process as (i) diffusive $\alpha=1$, (ii) subdiffusive $\alpha<1$, or (iii) superdiffusive $\alpha>1$.} 

Alternatively, the propagator in the presence of a target that defines an absorbing boundary is computed using the method of images \cite{book:Redner_GuideFirstpassageProcesses_08,book:VanKampen_StochasticProcessesPhysics_92},
\begin{equation}
    \label{eq:Green_function_BrownianMotion_absorbing}
    \PDF_0(x,t|x_0,t_0) = \frac{1}{\sqrt{4\pi D (t-t_0)}} \braces{\exp\brackets{-\frac{(x-x_0)^2}{4D(t-t_0)}}-{\exp\brackets{-\frac{(x+x_0-2\targetx)^2}{4D(t-t_0)}}}},
\end{equation}
which is valid for all $\targetx \in \mathbb{R}$. In this example, the searcher moves along either the semi-infinite line $x\in(-\infty,\targetx)$ or $x\in(\targetx,\infty)$, depending on whether $x_0$ is on the left or on the right side of $\targetx$, respectively. The survival probability is directly computed using its definition~\eqref{eq:survival_probability_def}, 
\begin{equation}
    \survival_0(t;\targetx,x_0) = \erf\parenthesis{\frac{|\targetx-x_0|}{\sqrt{4 D t}}},
    \label{eq:Survival_brownian_motion}
\end{equation}
and thus the FPT distribution is 
\begin{equation}
    \label{eq:FPT_Brownian_motion}
    \FPTpdf_0(\FPT{1} ; \targetx,x_0) = \frac{|\targetx-x_0|}{\sqrt{4\pi D \FPT{1}^3}} \exp\brackets{-\frac{(\targetx-x_0)^2}{4 D \FPT{1}}}.
\end{equation}
The former expression is a heavy-tailed distribution called the \emph{L\'evy distribution} \cite{journalarticle:Metzler.Klafter_RandomWalksGuide_Phys.Rep.00,journalarticle:Metzler.Klafter_RestaurantEndRandom_J.Phys.A:Math.Gen.04}, whose MFPT clearly diverges because it decays as $\FPT{1}^{-3/2}$ for large $\FPT{1}$. 

Due to the above discussion, free diffusion is not efficient as a search process. The isotropic nature of motion entails that the searcher may move away from the target for a long time, leading to an average search time that diverges due to its heavy-tailed distribution.\footnote{In fact, the searcher will eventually reach the target with probability equal to one, however it may take too long using this strategy.} 

In the next section, we introduce the stochastic resetting mechanism, which has born as a way to improve the efficiency of Brownian motion as a search process through stochastic and instantaneous resets.

\subsection{Standard stochastic resetting}
\label{subsec:standard_stochastic_resetting}

`Sometimes it is best just to give up and start all over again!' \cite{journalarticle:Evans.etal_StochasticResettingApplications_J.Phys.A:Math.Theor.20}, this simple statement summarises the key idea behind stochastic resetting (SR) \cite{journalarticle:Evans.Majumdar_DiffusionStochasticResetting_Phys.Rev.Lett.11,journalarticle:Evans.Majumdar_DiffusionOptimalResetting_J.Phys.A:Math.Theor.11,journalarticle:Evans.etal_StochasticResettingApplications_J.Phys.A:Math.Theor.20}. Let us imagine a rabbit (searcher) that is looking for carrots (target) around its burrow. If it has no information about the position of the carrots, it will wander randomly, without preferential direction, until it finds them. However, if the rabbit strays too far, it may be better to return to its burrow and start the search again. The strategy of stopping the current search, returning to the initial position and starting again the search is the core of stochastic resetting. 

Stochastic resetting is framed as one of the most elementary examples of \emph{intermittent search strategies} \cite{journalarticle:Benichou.etal_IntermittentSearchStrategies_Rev.Mod.Phys.11,journalarticle:Oshanin.etal_IntermittentRandomWalks_J.Phys.Condens.Matter07,journalarticle:Rojo.etal_IntermittentSearchStrategies_JPhysMathTheor10,journalarticle:Chupeau.etal_CoverTimesRandom_Nat.Phys.15,journalarticle:Moreau.etal_IntermittentSearchProcesses_EPL07}. These are stochastic processes where different phases of motion alternate to improve the efficiency of the search. The most paradigmatic example where one finds this intermittency is the foraging behaviour of animals \cite{journalarticle:OBrien.etal_SearchStrategiesForaging_Am.Sci.90,book:Bell_SearchingBehaviourBehavioural_12}, which can be modelled as a process in which searchers alternate between two different stages \cite{journalarticle:Benichou.etal_OptimalSearchStrategies_Phys.Rev.Lett.05,journalarticle:Benichou.etal_IntermittentSearchStrategies_Rev.Mod.Phys.11}: (i) a slower or scanning phase where the animal is able to detect the target but is almost immobile, and (ii) a faster or relocation phase where the animal moves quickly to another position but is unable to detect the target.\footnote{L\'evy flights \cite{journalarticle:Benichou.etal_IntermittentSearchStrategies_Rev.Mod.Phys.11}, run-and-tumble dynamics \cite{journalarticle:Cates_DiffusiveTransportDetailed_Rep.Prog.Phys.12}, or stochastic resetting \cite{journalarticle:Evans.Majumdar_DiffusionStochasticResetting_Phys.Rev.Lett.11,journalarticle:Evans.etal_StochasticResettingApplications_J.Phys.A:Math.Theor.20,journalarticle:Evans.Majumdar_DiffusionOptimalResetting_J.Phys.A:Math.Theor.11,journalarticle:Garcia-Valladares.etal_OptimalResettingStrategies_NewJ.Phys.23,journalarticle:Garcia-Valladares.etal_StochasticResettingRefractory_Phys.Scr.24} are particular instances of these two-stage intermittent dynamics.} Standard stochastic resetting (SSR) can be thought as a two-stage model, where the scanning phases are modelled as Brownian motion and the relocation phases are instantaneous jumps to a certain resetting position $x_r$. After each resetting event, the dynamics is renewed---starting afresh from the resetting position. In the most general framework, one may consider different initial and resetting positions, $x_0$ and $x_r$, respectively. However, it is common to take $x_r = x_0$ in the applications.

Despite the mathematical simplicity of SSR, it is powerful enough to improve the efficiency of the search. It has been successfully used as an optimisation algorithm in many applications; namely some of the aforementioned instances at the beginning of section~\ref{sec:intro_search_processes}, ranging from economics \cite{journalarticle:Stojkoski.etal_GeneralisedGeometricBrownian_Entropy20,journalarticle:Stojkoski.etal_IncomeInequalityMobility_Philos.Trans.R.Soc.A22,journalarticle:Stojkoski.etal_AutocorrelationFunctionsErgodicity_JPhysMathTheor22,journalarticle:Santra_EffectTaxDynamics_EPL22,journalarticle:Vinod.etal_TimeaveragingNonergodicityReset_Phys.Rev.E22,journalarticle:Montero.etal_ValuingDistantFuture_JPhysMathTheor22,journalarticle:Jolakoski.etal_FirstPassageResetting_ChaosSolitonFract.23} to biochemical reactions \cite{journalarticle:Reuveni.etal_RoleSubstrateUnbinding_Proc.Natl.Acad.Sci.U.S.A.14,journalarticle:Rotbart.etal_MichaelisMentenReactionScheme_Phys.Rev.E15,journalarticle:Pal.etal_ThermodynamicUncertaintyRelation_Phys.Rev.Res.21,journalarticle:Reuveni_OptimalStochasticRestart_Phys.Rev.Lett.16,journalarticle:Biswas.etal_RateEnhancementGated_J.Chem.Phys.23} or ecology \cite{journalarticle:Roldan.etal_StochasticResettingBacktrack_Phys.Rev.E16,journalarticle:Plata.etal_AsymmetricStochasticResetting_Phys.Rev.E20,journalarticle:Pal.etal_SearchHomeReturns_Phys.Rev.Res.20,journalarticle:Evans.etal_ExactlySolvablePredator_JPhysMathTheor22}---where they are mostly motivated by the beneficial effect of restart for lowering the first-passage time \cite{journalarticle:Evans.Majumdar_DiffusionStochasticResetting_Phys.Rev.Lett.11,journalarticle:Evans.Majumdar_DiffusionOptimalResetting_J.Phys.A:Math.Theor.11,journalarticle:Evans.etal_OptimalDiffusiveSearch_JPhysMathTheor13,journalarticle:Bhat.etal_StochasticSearchPoisson_JStatMechTheoryExp16,journalarticle:Ahmad.etal_FirstPassageParticle_Phys.Rev.E19,journalarticle:Bressloff_DirectedIntermittentSearch_JPhysMathTheor20,journalarticle:Faisant.etal_OptimalMeanFirstpassage_JPhysMathTheor21,journalarticle:DeBruyne.Mori_ResettingStochasticOptimal_Phys.Rev.Research23}. Moreover, it constitutes an excellent test bench for performing non-equilibrium research, providing comprehensive models to study non-equilibrium steady states (NESS) \cite{journalarticle:Majumdar.etal_DynamicalTransitionTemporal_Phys.Rev.E15,journalarticle:Mendez.Campos_CharacterizationStationaryStates_Phys.Rev.E16,journalarticle:Eule.Metzger_NonequilibriumSteadyStates_NewJ.Phys.16,journalarticle:Pal.etal_DiffusionTimedependentResetting_JPhysMathTheor16,journalarticle:Evans.Majumdar_RunTumbleParticle_JPhysMathTheor18,journalarticle:Gupta_StochasticResettingUnderdamped_JStatMechTheoryExp19,journalarticle:Basu.etal_SymmetricExclusionProcess_Phys.Rev.E19,journalarticle:Pal.etal_InvariantsMotionStochastic_NewJ.Phys.19}, stochastic thermodynamics and fluctuation theorems \cite{journalarticle:Pal.etal_ThermodynamicUncertaintyRelation_Phys.Rev.Res.21,journalarticle:Fuchs.etal_StochasticThermodynamicsResetting_EPL16,journalarticle:Pal.Rahav_IntegralFluctuationTheorems_Phys.Rev.E17,journalarticle:Busiello.etal_EntropyProductionSystems_Phys.Rev.Res.20,journalarticle:Gupta.etal_WorkFluctuationsJarzynski_Phys.Rev.Lett.20,journalarticle:Gupta.Plata_WorkFluctuationsDiffusion_NewJ.Phys.22}, large deviations \cite{journalarticle:Meylahn.etal_LargeDeviationsMarkov_Phys.Rev.E15,journalarticle:Harris.Touchette_PhaseTransitionsLarge_JPhysMathTheor17,journalarticle:Hollander.etal_PropertiesAdditiveFunctionals_JPhysMathTheor19,journalarticle:Monthus_LargeDeviationsMarkov_JStatMechTheoryExp21,journalarticle:Smith.Majumdar_CondensationTransitionLarge_J.Stat.Mech.TheoryExp.22,journalarticle:Zamparo_StatisticalFluctuationsResetting_JPhysMathTheor22}, or quantum restart \cite{journalarticle:Dhar.etal_DetectionQuantumParticle_Phys.Rev.A15,journalarticle:Rose.etal_SpectralPropertiesSimple_Phys.Rev.E18,journalarticle:Mukherjee.etal_QuantumDynamicsStochastic_Phys.Rev.B18,journalarticle:Wald.Bottcher_ClassicalQuantumWalks_Phys.Rev.E21,journalarticle:Sevilla.Valdes-Hernandez_DynamicsClosedQuantum_JPhysMathTheor23,journalarticle:Dubey.etal_QuantumResettingContinuous_JPhysMathTheor23}, to name just a few intriguing topics.

The times between resetting events is independent are the previous history of the process and independent of each other, so the position of the particle is still a Markov process. The times at which the searcher resets its position are denoted by $t_i$, $i=1,\ldots$, which depend on the time spent since the last resetting event---the subscript $i$ in $t_i$ stand for the order of occurrence. 

\begin{figure}
    \centering
    \includegraphics[width=0.7\textwidth]{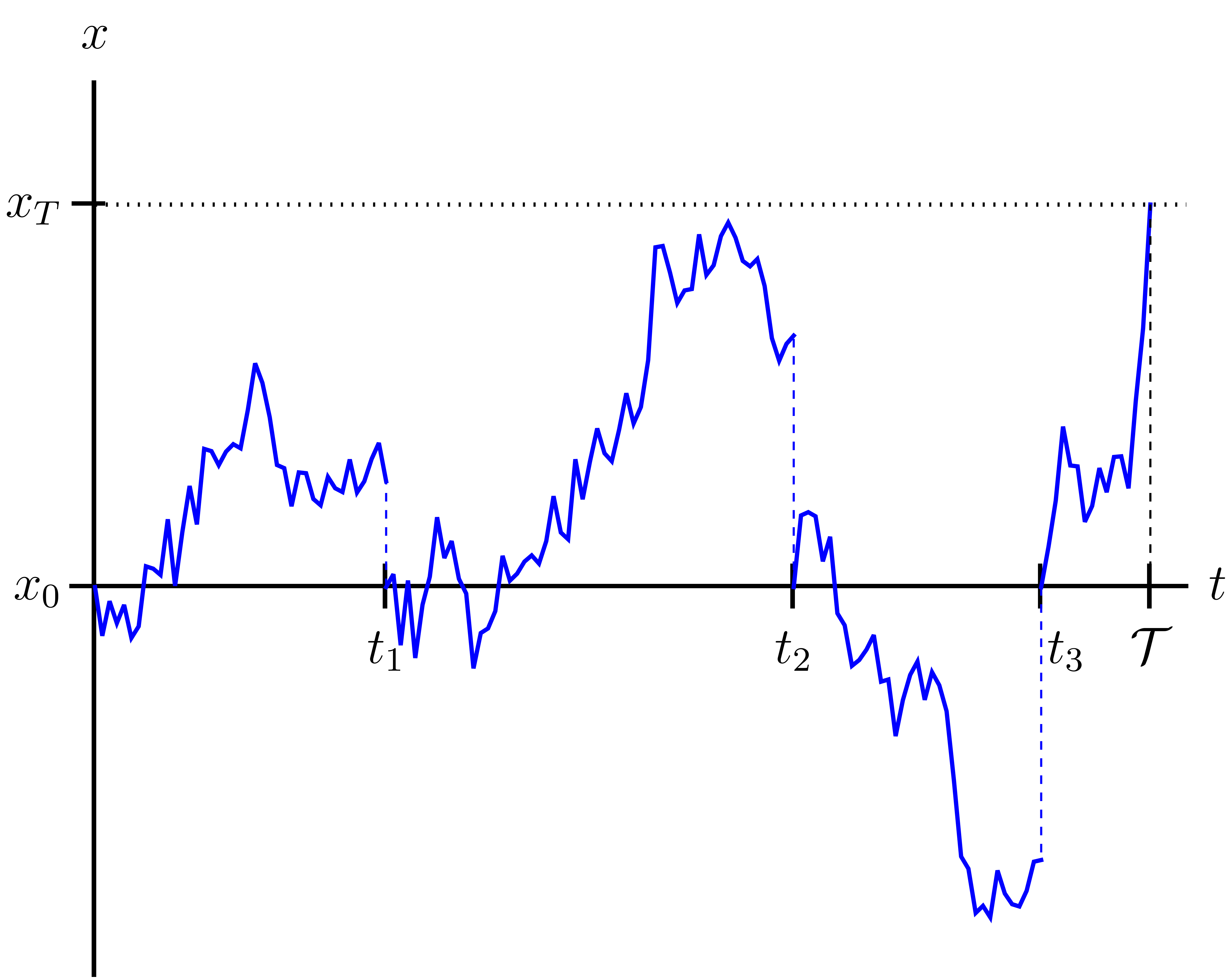}
    \caption{Illustrative trajectory of Brownian dynamics with stochastic resetting to the initial position $x_0$ with rate $r$. Dashed blue lines stand for the resetting events occurring at times $t_i$, labelled by order of occurrence. The dotted black line indicates the position of the target $\targetx$, whereas the dashed black line corresponds to the first-passage time $\mathcal{T}$ to reach $\targetx$.}
    \label{fig:Intro_SR_trajectory}
\end{figure}

Typically, the waiting time between resets, which is the time difference between two consecutive $t_i$, is drawn from an exponential distribution with rate $r$, \ie the probability $f(t) \text{d} t$ of having a reset between $[t,t+\text{d} t]$ follows an exponential distribution:
\begin{equation}
    f(t)= r e^{-rt}.
\end{equation} 
In dynamical terms, the resetting mechanism entails that the searcher's position evolves following a modification of the Langevin equation~\eqref{eq:Langevin_equation_general} for Brownian motion, specifically
 \begin{equation}
    x(t+\text{d}t) = 
    \left\{
    \begin{array}{ll}
        x(t) + \sqrt{2D \text{d}t} \mathcal{N}(0,1), & \text{with probability} \ 1 - r \text{d}t, \\
        x_0, & \text{with probability} \ r \text{d}t.
    \end{array}
    \right.
    \label{eq:Langevin_equation_standard_SR}
\end{equation}
The numbers of resets up to time $t$, $N(t)$, is a Poissonian stochastic variable. Specifically, $N(t)$ is distributed according a Poisson distribution with parameter $rt$. 
An illustrative picture of a particular trajectory of a Brownian particle under the effect of SSR is plotted in figure~\ref{fig:Intro_SR_trajectory}. Therein, we consider a target located at $\targetx$, consequently the first-passage time is given by $\FPT{1}$.

One can also obtain the ensemble approach to the stochastic process, leading to the evolution equation for the PDF $\PDF_r(x,t|x_0,t_0)$\footnote{Recall that the subscript ``r'' indicates the presence of resetting.} of finding the searcher at $(x,t)$, having started from $(x_0,t_0)$, under the effect of stochastic resetting to $x_r$,
\begin{equation}
    \label{eq:Fokker_Planck_Standard_SR}
    \partial_t \PDF_r(x,t|x_0,t_0) = D \partial_x^2 \PDF_r(x,t|x_0,t_0) - r \PDF_r(x,t|x_0,t_0) + r \delta(x-x_r).
\end{equation}
This is the modification of the Fokker-Plack equation~\eqref{eq:Fokker_Planck_Brownian_motion} with stochastic resetting. The first term on the rhs stands for the usual free diffusion~\eqref{eq:Fokker_Planck_Brownian_motion}; the second term is a sink term that expresses the loss of probability at $x$ due to resetting to $x_r$, while the last one corresponds to a source term at $x=x_r$, stemming from the probability incoming flux to $x_r$ from any point $x$.

\subsubsection*{Absence of targets. Non-equilibrium steady state}

Let us start with the problem in the absence of targets. The solution of~\eqref{eq:Fokker_Planck_Standard_SR} can be heuristically built using different renewal approaches \cite{journalarticle:Evans.Majumdar_DiffusionStochasticResetting_Phys.Rev.Lett.11,journalarticle:Evans.Majumdar_DiffusionOptimalResetting_J.Phys.A:Math.Theor.11}. On the one hand, we consider the \emph{last renewal approach}, where the solution reads
\begin{equation}
    \label{eq:last_renewal_equation_standard_SR}
    \PDF_r(x,t|x_0,t_0) = e^{-r(t-t_0)} \PDF_0(x,t|x_0,t_0) + r \int_{t_0}^t \d{t'} e^{-r(t-t')} \PDF_0(x,t|x_r,t').
\end{equation}
Here, the first term on the rhs corresponds to the contribution of those trajectories that have not experimented any reset up to time $t$, with probability 
\begin{equation}
    \label{eq:poissonian_F(t)}
    F(t-t_0) = e^{-r(t-t_0)},
\end{equation}
and thus they follow the free diffusive propagator $\PDF_0(x,t|x_0,t_0)$. The rest of contributions are obtained by integrating those trajectories that have experimented at least one reset. Let us assume that the last resetting event took place at time $t'$. From then, the dynamics naturally evolves the rest of the time $t-t'$, contributing with $\PDF_0(x,t|x_r,t')$ times the probability of a resetting event between $t'$ and $t'+\text{d}t'$, $r \text{d}t'$, times the probability of not having resets from $t'$ to $t$, $F(t-t')$. On the other hand, we can also employ the \emph{first renewal approach}, where the solution is expressed as
\begin{equation}
    \label{eq:first_renewal_equation_standard_SR}
    \PDF_r(x,t|x_0,t_0) = e^{-r(t-t_0)} \PDF_0(x,t|x_0,t_0) + r \int_{t_0}^t \d{t{''}} e^{-r(t{''}-t_0)} \PDF_r(x,t|x_r,t{''}).
\end{equation}
The difference is on the second term on the rhs, where we count those trajectories that have not been reset in the time interval $[t_0,t{''}]$, as given by the term $F(t{''}-t_0)$, experimented its first reset in the time interval $[t{''},t{''}+\text{d}t{''}]$, with probability $r\text{d}t{''}$, and evolve the rest of the time $t-t{''}$, thus contributing with $\PDF_r(x,t|x_r,t{''})$. For the sake of simplicity, we are going to consider $t_0 = 0$---corresponding to the change of variable $t-t_0\to t$, so the initial time is not explicitly written. After Laplace transformation, we can solve for $\laplTransformTilde{\PDF_r}{x,s|x_0}$, either from~\eqref{eq:last_renewal_equation_standard_SR} or~\eqref{eq:first_renewal_equation_standard_SR}, with the result  
\begin{equation}
    \label{eq:Laplace_transform_standard_SR}
    \laplTransformTilde{\PDF_r}{x,s|x_0} = \parenthesis{1 + \frac{r}{s}} \laplTransformTilde{\PDF_0}{x,s+r|x_0},
\end{equation}
where $\laplTransformTilde{\PDF_0}{x,s|x_0}$ corresponds to the Laplace transform of~\eqref{eq:Green_function_BrownianMotion}, and we have taken $x_r=x_0$ to simplify the expression.

Equation~\eqref{eq:Laplace_transform_standard_SR} gives an implicit expression of the propagator. It is possible to analytically obtain the inverse in the long-time limit $t\to\infty$. Let us assume that the limit $\lim_{s\to0} s \laplTransformTilde{\PDF_r}{x,s|x_0}$ exists; then the \emph{final value theorem} (FVT)\footnote{In general, the FVT can be applied to any function $A(t)$, provided that $A(t)$ is bounded and both limits $\lim_{t\to\infty} A(t) = \lim_{s\to 0} s \laplTransformTilde{A}{s}$ exist.} states that the long-time limit of the propagator is given by
\begin{equation}
    \label{eq:NESS_standard_SR}
    \lim_{t\to\infty} \PDF_r(x,t|x_0,t_0) = \lim_{s\to 0} s \laplTransformTilde{\PDF_r}{x,s|x_0} = \frac{1}{2} \sqrt{\frac{r}{D}} \exp\parenthesis{-\sqrt{\frac{r}{D}}|x-x_0|}.
\end{equation}
This behaviour can be directly computed from~\eqref{eq:Fokker_Planck_Standard_SR} by solving the stationary equation $\partial_t \PDF_r(x,t|x_0,t_0) = 0$. Hence, resetting makes the system reach a steady state, in particular, a non-equilibrium steady state (NESS) because there is always the flux of probability stemming from resetting. Physically, there is a competition between diffusion and resetting, which is manifested by the coefficient $(D/r)^{1/2}$, which measures the typical distance travelled by the particle between resets. This competition is notably evident if we analyse how the propagator tends to the NESS. Rewriting~\eqref{eq:last_renewal_equation_standard_SR} by using the change of variable $t' = \omega t$, it explicitly reads
\begin{equation}
    \PDF_r(x,t|x_0) =  \frac{e^{- t \Phi (1, (x-x_0)/t)}}{\sqrt{4 \pi D t}} 
                        + \frac{r t^{1/2}}{\sqrt{4 \pi D}} \int_0^1 \d{\omega} \frac{1}{\omega^{1/2}} e^{- t \Phi (\omega, (x-x_r)/t)}, 
\end{equation}
where the function $\Phi(\omega,y)$ is defined as 
\begin{equation}
    \Phi(\omega,y) = r \omega + \frac{y^2}{4 D \omega}.
\end{equation}
This previous expression is suitable for analysing the long-time behaviour by means of the Laplace or saddle-point method \cite{book:Bender.Orszag_AdvancedMathematicalMethods_99}. In section~\ref{subsec:refractory_relaxation_NESS} and appendix~\ref{app:refractory_Laplace_method}, we explore this technique in detail, though the main idea is to approximate the integral through the maximum of the integrand, because it has a very sharp peak in the limit as $t\to\infty$. Therein, for $x_r=x_0$, we obtain that the propagator behaves as
\begin{equation}
    \PDF_r(x,t|x_0) \sim e^{-t I((x-x_0)/t)}, \quad t\to\infty,
\end{equation}
where $I(y)$ is a large deviation function (LDF) \cite{journalarticle:Evans.etal_StochasticResettingApplications_J.Phys.A:Math.Theor.20,journalarticle:Touchette_LargeDeviationApproach_Phys.Rep.09}, given by
\begin{equation}
    I(y) = \left\{
    \begin{array}{ll}
        \sqrt{\frac{r}{D}} |y|, & |y| < \sqrt{ 4 D r}, \\
        r + \frac{y^2}{4 D}, & |y| > \sqrt{4 D r}.
    \end{array}
    \right.
    \label{eq:LDF_standard_SR}
\end{equation}
The physical meaning of~\eqref{eq:LDF_standard_SR} is the following: (i) for those positions close enough to $x_0$, $|x-x_0| < \sqrt{4 D r}\, t$, an internal core where the solution has already relaxed to the NESS~\eqref{eq:NESS_standard_SR} appears due to resetting; whereas (ii) for those positions outside that core, $|x-x_0| > \sqrt{4 D r}\, t$, the propagator is equivalent to the first term of~\eqref{eq:last_renewal_equation_standard_SR}, \ie the free diffusion contribution of those trajectories that have not experimented any reset up to time $t$. The internal core, where there have been enough resets to reach the functional form of NESS, grows ballistically as $\sqrt{4 D r}\, t$, so that $\PDF_r(x,t|x_0)$ tends to the stationary solution~\eqref{eq:NESS_standard_SR} when $t\to\infty$.

\subsubsection*{Presence of target. Optimal resetting rate}

Another important feature of stochastic resetting is its impact on the optimisation of search times. In presence of a target at $\targetx$, $\PDF_r(\targetx,t|x_0,t_0)=0$, the propagator is built reformulating the first renewal approach~\eqref{eq:first_renewal_equation_standard_SR}, though we have to take into account the probability of not having reached the target before $t$, \ie the survival probability, 
\begin{equation}
    \PDF_r(x,t|x_0) = e^{-r t} \PDF_0(x,t|x_0) + r \int_{0}^t \d{t{''}} e^{-r t{''}} \survival_0(t{''};\targetx,x_0) \PDF_r(x,t|x_r,t{''}). 
    \label{eq:renewal_equation_propagator_SSR}
\end{equation}
Clearly, expressions for $\PDF_0$ and $\PDF_r$ are no longer valid because of the presence of the target, since we have to regard the quantities under the absorbing boundary conditions at $\targetx$. The PDE that governs the evolution of $\survival_r(t;\targetx,x_0)$ follows an analogous procedure than the one done to obtain the general backward Fokker-Planck equation~\eqref{eq:backward_Fokker_Planck_equation_time_difference}. Considering the resetting terms from~\eqref{eq:Fokker_Planck_Standard_SR}, it reads  
\begin{equation}
    \partial_{t} \survival_r(t;\targetx,x_0) = D\partial_{x_0}^2\survival_r(t;\targetx,x_0) + r \brackets{\survival_r(t;\targetx,x_r)-\survival_r(t;\targetx,x_0)},
    \label{eq:Fokker_Planck_Backward_SSR}
\end{equation}
where they have appeared the corresponding terms to loss and gain of survival probability $- r \survival_r(t;\targetx,x_0)$ and $+ r \survival_r(t;\targetx,x_r)$, respectively. The survival probability under SSR, $\survival_r(t;\targetx,x_0)$, is obtained by integrating~\eqref{eq:renewal_equation_propagator_SSR} as defined by~\eqref{eq:survival_probability_def},
\begin{equation}
    \survival_r(t;\targetx,x_0) = e^{-r t} \survival_0(t;\targetx,x_0) + r \int_0^t \d{t{''}} e^{-r t{''}} \survival_0(t{''};\targetx,x_0) \survival_r(t-t{''};\targetx,x_r).
\end{equation}

\begin{figure}
    \centering
    \includegraphics[width=0.7\textwidth]{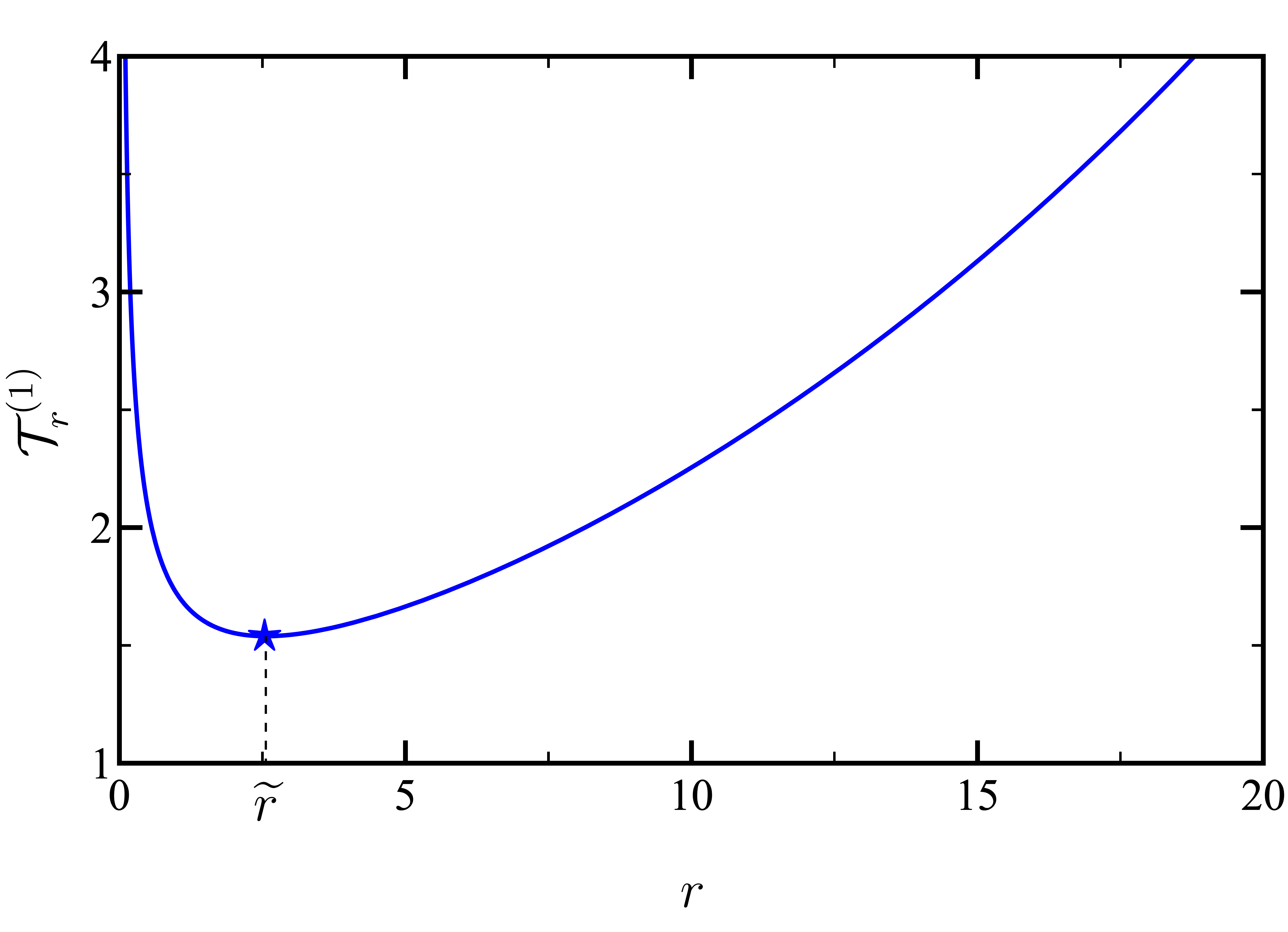}
    \caption{Mean first-passage time (MFPT) $\mathcal{T}^{(1)}_r$~\eqref{eq:MFPT_SSR} as a function of the resetting rate $r$, for a Brownian particle under SSR. There exists an optimal resetting rate $\optMFPT{r}=2.5396$ (blue star) that minimises the MFPT. Parameters have been chosen so that $\tau_D = |\targetx-x_0|^2/D = 1$.}
    \label{fig:intro_SSR_MFPT}
\end{figure}

Applying the Laplace transform, 
the mean first-passage time under SSR is computed by using~\eqref{eq:MFPT_def}, 
\begin{equation}
    \FPT{(1)}_r(r;\targetx,x_0) = \laplTransformTilde{\survival_r}{s=0;\targetx,x_0} =  \frac{1}{r}\parenthesis{e^{\sqrt{r/D}|\targetx-x_0|} - 1}.
    \label{eq:MFPT_SSR}
\end{equation} 
It is neatly observed that the MFPT is always finite for positive resetting rates, $r>0$. If $r=0$, the free diffusion case is recovered and the MFPT diverges, as expected. On the opposite limit, $r\to\infty$, the MFPT also diverges: resetting makes the searcher be stuck at $x_0$, being unable to reach the target. This behaviour indicates that there is a finite rate $\optMFPT{r}$ that minimises the MFPT and thus optimises the search, as seen in figure~\ref{fig:intro_SSR_MFPT}.\footnote{The tilde notation in parameters refers to their optimal values that minimise the MFPT.} The optimal resetting rate is computed by solving 
\begin{equation}
    \left.\partial_r \FPT{(1)}_r(r;\targetx,x_0)\right|_{r=\optMFPT{r}} = 0 \leftrightarrow \frac{\optMFPT{\alpha}_h}{2} = 1 - e^{-\optMFPT{\alpha}_h}, 
    \label{eq:optimal_rate_SSR}
\end{equation}
whose solution is $\optMFPT{\alpha}_h = (\optMFPT{r}/D)^{1/2}|\targetx-x_0| = 2 + W(-2e^{-2})\simeq 1.5936$, where $W(x)$ corresponds to the Lambert $W$ function, which is the inverse function of $x e^x$, i.e. the solution of $W(x)\exp[W(x)]=x$~\cite{journalarticle:Evans.Majumdar_DiffusionStochasticResetting_Phys.Rev.Lett.11,journalarticle:Evans.Majumdar_DiffusionOptimalResetting_J.Phys.A:Math.Theor.11}. Remarkably, resetting cuts those long excursions of Brownian motion that made the MFPT diverge, leading to a finite optimal average time to reach the target in the no-resetting case. This also translates to the FPT distribution, where stochastic resetting modifies the heavy tail of the Brownian case~\eqref{eq:FPT_Brownian_motion} to a much faster decaying distribution \cite{journalarticle:Evans.etal_StochasticResettingApplications_J.Phys.A:Math.Theor.20}, the Gumbel distribution \cite{book:Gumbel_StatisticsExtremes_19},
\begin{equation}
    \FPTpdf_r(\FPT{1};\targetx,x_0) \sim \exp\parenthesis{-r \FPT{1} e^{\sqrt{r/D}|\targetx-x_0|}}, \quad \sqrt{r/D}|\targetx-x_0|\gg 1, \; r \FPT{1}\gg 1.
\end{equation}

\subsubsection*{Limitations of searching under SSR}

SSR has been a prolific line of research due to its simplicity and its ability to optimise search problems, showing very appealing non-equilibrium properties as introduced above. Nevertheless, SSR has some issues that ask for improvement, specially for investigating more realistic scenarios of search processes. The first and most obvious drawback for SSR is the assumption of instantaneous resets. From an experimental point of view, this instantaneousness is unfeasible because it would require an infinite amount of work to bring the particle from any position in an infinitesimal amount of time. Any actual, physical, implementation of resetting must involve some finite cost. This has led to investigate more refined models, where resetting involves new phases that take a finite time to be completed before restarting the motion, so that resetting ceases to be costless. To tackle this limitation, two different minimal resetting models, based on the original one, stand out. On the one hand, a new motionless phase, called refractory period \cite{journalarticle:Evans.Majumdar_EffectsRefractoryPeriod_J.Phys.A:Math.Theor.19,journalarticle:Maso-Puigdellosas.etal_StochasticMovementSubject_J.Stat.Mech.19}, has been introduced after each resetting event, which mimics a stochastic resting time before starting the exploration again. On the other hand, resetting processes can also be followed by return phases, in which the particle takes a finite time to come back to its initial state following a certain dynamics \cite{journalarticle:Pal.etal_InvariantsMotionStochastic_NewJ.Phys.19,journalarticle:Gupta.Plata_WorkFluctuationsDiffusion_NewJ.Phys.22,journalarticle:Bodrova.etal_ScaledBrownianMotion_Phys.Rev.E19,journalarticle:Gupta.etal_StochasticResettingStochastic_J.Phys.A:Math.Theor.21,journalarticle:Gupta.etal_ResettingStochasticReturn_JStatMechTheoryExp21,journalarticle:Radice_DiffusionProcessesGammadistributed_JPhysMathTheor22,journalarticle:Olsen.etal_ThermodynamicCostFinitetime_Phys.Rev.Res.24}.

Another limitation is to consider the target fixed in space, \ie its remaining at the same position $\targetx$ for all realisations of the search process. However, in many real situations, the searcher does not know where the target is, its location is uncertain. Thus, the target position should not be a deterministic variable. To address this issue, it is relevant to analyse \emph{quenched disorder} models, \ie the location of the target varies between realisations, following a probability density function $\PDF_T(\targetx)$. The concept of quenched disorder appears in many contexts, such as foraging \cite{book:Viswanathan.etal_PhysicsForagingIntroduction_11,journalarticle:Marion.etal_UnderstandingForagingBehaviour_J.Theor.Biol.05,journalarticle:Bartumeus.etal_AnimalSearchStrategies_Ecology05,journalarticle:Boyer.Walsh_ModellingMobilityLiving_Philos.Trans.R.Soc.A10,book:Viswanathan.etal_PhysicsForagingIntroduction_11} or finding minima of a complex energy landscape, which is the main challenge in disordered systems like spin glasses \cite{journalarticle:Cavagna_SupercooledLiquidsPedestrians_Phys.Rep.09,journalarticle:Berthier.Biroli_TheoreticalPerspectiveGlass_Rev.Mod.Phys.11,journalarticle:Charbonneau.etal_FractalFreeEnergy_Nat.Commun.14,journalarticle:Folena.etal_IntroductionDynamicsDisordered_PhysStatMechAppl22,journalarticle:Ros.Fyodorov_HighdLandscapesParadigm_22}. In contrast to the usual situation in SSR, we will consider a space-dependent resetting rate $r(x)$ for the quenched disorder case, so the aim is to optimise the search process by tuning the spatial functional form of the resetting rate.

\subsection{Pathway formulation for intermittent search strategies}
\label{subsec:pathway_formulation}

Intermittent search strategies that randomly alternate between dynamics phases can be analysed by a \emph{pathway formulation}. Here, let us define a pathway as the subensemble of realisations of the stochastic process, such as the system has changed its dynamics a certain number of times up to time $t$ at specific given times. As a paradigmatic model, stochastic resetting can be studied using pathways, due to the renewal property after each resetting event. 

This pathway formulation was originally developed to systematically study the dynamics of stochastic processes with resetting. In fact, it is related to renewal theory \cite{journalarticle:Evans.Majumdar_EffectsRefractoryPeriod_J.Phys.A:Math.Theor.19,booksection:Chechkin.etal_IntroductionTheoryLevy_AnomalousTransport:FoundationsandApplications08,journalarticle:Bodrova.etal_ScaledBrownianMotion_Phys.Rev.E19,journalarticle:Wang.etal_RandomWalksComplex_Chaos21}, being inspired by similar techniques in different resetting setups \cite{journalarticle:Gupta.Plata_WorkFluctuationsDiffusion_NewJ.Phys.22,journalarticle:Chechkin.Sokolov_RandomSearchResetting_Phys.Rev.Lett.18}. The validity of this approach is proved in a very broad framework, \eg it reproduces most of the previous results of resetting with refractory \cite{journalarticle:Evans.Majumdar_EffectsRefractoryPeriod_J.Phys.A:Math.Theor.19,journalarticle:Maso-Puigdellosas.etal_StochasticMovementSubject_J.Stat.Mech.19}, and allows us to obtain general expressions for the case of stochastic resetting with refractory periods, which we deeply analyse in chapter~\ref{ch:resetting_refractory}. Nevertheless, this framework is more general and can be applied to any intermittent search strategy.

Now, we analyse in detail the general case of a two-stage dynamics, where the system alternates between two different dynamics, $A$ and $B$. At the initial time, the system starts in phase $A$ at position $x_0$, evolves up to time $t_1$, when it switches to $B$, that keeps evolving up to time $\tau_1$ when it returns to $A$. The process is repeated, switching between phases $A$ and $B$ at random times $t_i$ and $\tau_i$, respectively. The PDF of finding the system at position $x$ at time $t$ is
\begin{equation}
    \PDF(x,t|x_0) = \PDF^{(A)}(x,t|x_0) + \PDF^{(B)}(x,t|x_0), \quad 
    \PDF(x,0|x_0) = \PDF^{(A)}(x,0|x_0) = \delta(x-x_0),
\end{equation}
\ie the sum of the contributions of being in phase $A$ or $B$ at time $t$. We are interested in resetting systems, hence we assume that the system recovers the initial position after each cycle (after $B$ finishes). The waiting times between stages in each cycle  $\delta^{(A)}_i = t_i - \tau_{i-1}$, $\delta^{(B)}_i = \tau_i - t_i$, $i=1,\ldots,n$, are drawn from the joint probability distribution $h\parenthesis{\delta^{(A)}_i,\delta^{(B)}_i}=f\parenthesis{\delta^{(A)}_i}\gamma\parenthesis{\delta^{(B)}_i|\delta^{(A)}_i}$, which is independent of other cycles. Additionally, the marginal distributions are 
\begin{equation}
    f\parenthesis{\delta^{(A)}_i} = \int_0^\infty \d{\delta^{(B)}_i} h\parenthesis{\delta^{(A)}_i,\delta^{(B)}_i},
    \quad g\parenthesis{\delta^{(B)}_i} = \int_0^\infty \d{\delta^{(A)}_i} h\parenthesis{\delta^{(A)}_i,\delta^{(B)}_i},
\end{equation}
and the probabilities associated with the corresponding phases lasting less than $t$ are
\begin{subequations}
    \begin{align}
        F(t) &=1 - \int_0^t \d{\delta^{(A)}} f\parenthesis{\delta^{(A)}} = \int_t^\infty \d{\delta^{(A)}} f\parenthesis{\delta^{(A)}}, 
        \\
        G(t) &=1-\int_0^t \d{\delta^{(B)}} g\parenthesis{\delta^{(B)}}=\int_t^\infty \d{\delta^{(B)}} g\parenthesis{\delta^{(B)}}.
    \end{align}
\end{subequations}
The dynamical evolution during the $n$-th cycle for each stage $A$ and $B$ is given by the propagators $K^{(A)}(x,t|x_0,\tau_n)=K^{(A)}(x,t-\tau_n|x_0)$ and $K^{(B)}(x,t|x',t_{i+1};\delta^{(A)}_{i+1},\delta^{(B)}_{i+1})=K^{(B)}(x,t-t_{i+1}|x';\delta^{(A)}_{i+1},\delta^{(B)}_{i+1})$, respectively. The dynamics of phase $B$ depends on the time spent in phase $A$ in this cycle, $\delta^{(A)}_{i+1}$, and the needed time to return to $A$ to start the next one, $\delta^{(B)}_{i+1}$. The renewal condition implies that the propagators satisfy
\begin{equation}
    \int \d{x'} K^{(B)}(x,\tau_{n+1}|x',t_{i+1};\delta^{(A)}_{i+1},\delta^{(B)}_{i+1}) K^{(A)}(x',t_{i+1}|x_0,\tau_i) = \delta(x-x_0)
\end{equation}

\begin{figure}
    \centering
    \includegraphics[width=0.65\textwidth]{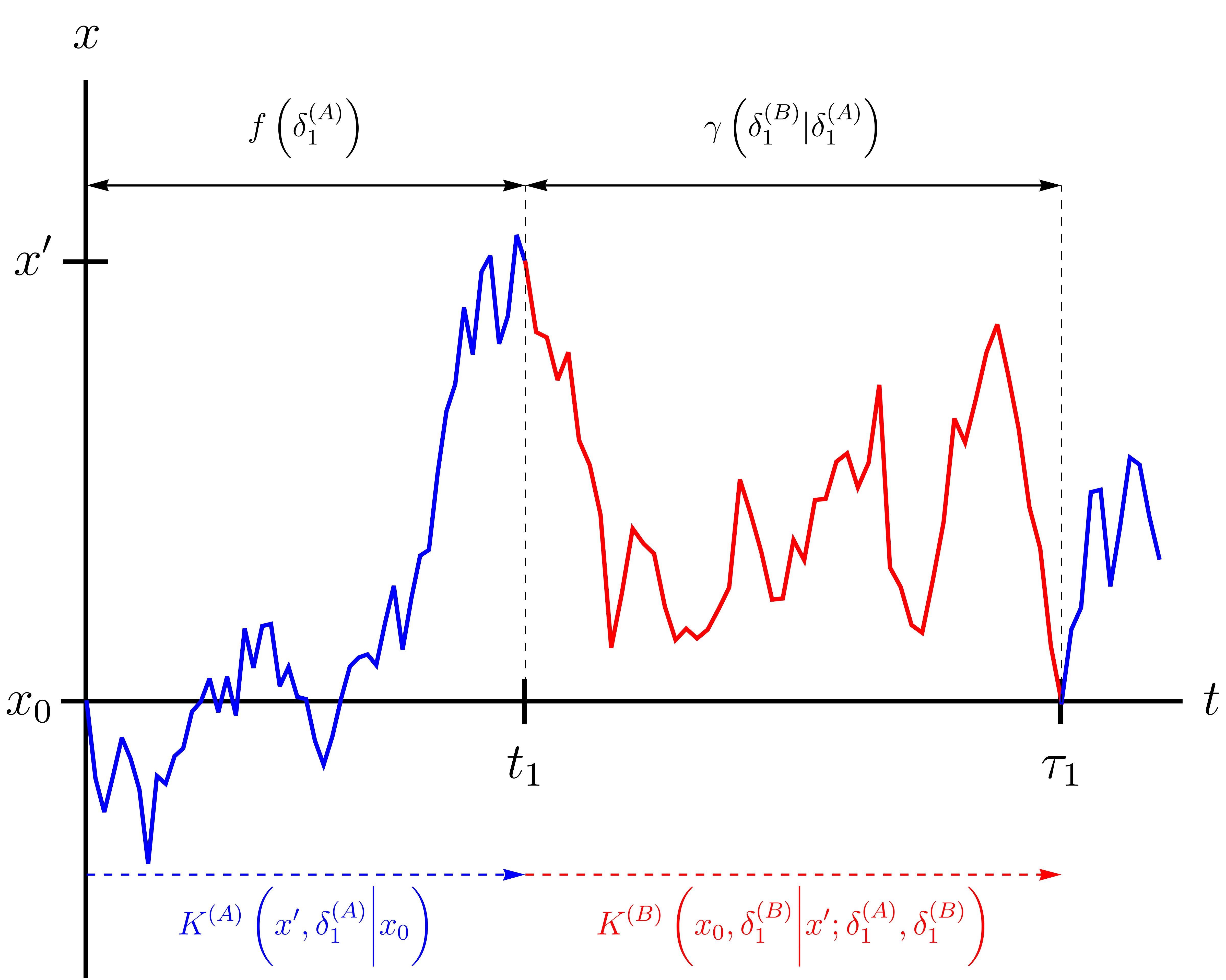}
    \caption{First cycle of a trajectory illustrating a two-stage model. The system alternates between phases $A$ (solid blue line) and $B$ (solid red line), in which it evolves following the propagators $K^{(A)}$ and $K^{(B)}$, respectively. Dashed vertical lines indicate the switching times $t_i$ ($\tau_i$), where the phase changes to $B$ ($A$). The waiting times between phases, $\delta^{(A)}_i = t_i - \tau_{i-1}$ and $\delta^{(B)}_i = \tau_i-t_i$, are drawn from distributions $f\parenthesis{\delta^{(A)}_i}$ and $\gamma\parenthesis{\delta^{(B)}_i|\delta^{(A)}_i}$, respectively. In particular, here the system always starts at $x_0$ and comes back to $x_0$, \ie $\PDF(x,0|x_0)=\delta(x-x_0)$.}
    \label{fig:Pathway_Formulation}
\end{figure}

Under the previous assumptions, we can build the propagator $\PDF(x,t|x_0)$ as a sum over all possible pathways the system may follow up to time $t$. Let us denote by $\PDF_n(x,t|x_0)$ the propagator of those trajectories that have experimented $n$ complete cycles $\{A,B\}$ $n$ times up to time $t$, so that 
\begin{equation}
    \label{eq:pathway_formulation_generalStructure_TimeDomain}
    \PDF(x,t|x_0) = \sum_{n=0}^\infty \PDF_n(x,t|x_0) = \sum_{n=0}^\infty 
    \bigg[
        \PDF^{(A)}_n(x,t|x_0) + \PDF^{(B)}_n(x,t|x_0)
        \bigg],
\end{equation}
where $\PDF^{(i)}_n(x,t|x_0)$, $i=A,B$, is the contribution of being in phase $i$ after $n$ cycles at that time. The particular case $n=0$ corresponds to the no-renewed evolution,
\begin{subequations}
    \label{eq:pathway_n0_contributions}
    \begin{align}
        \PDF_0^{(A)}(x,t|x_0)&=F(t)K^{(A)}(x,t|x_0), 
        \\
        \PDF_0^{(B)}(x,t|x_0)&= \int_{0}^{t} \d{t_1} f(t_1) \int \d{x'} K^{(A)} (x',t_1|x_0)
        \nonumber\\
        & \quad \int_{t}^{\infty} \d{\tau_1} \gamma(\tau_1-t_1|t_1)K^{(B)} (x,t|x',t_1;t_1,\tau_1 - t_1),
    \end{align}
\end{subequations}
where the system has not finished the first phase $A$ or the first cycle yet $\{A,B\}$, respectively. For generic $n\geq 0$, both $\PDF^{(A)}_n$ and $\PDF^{(B)}_n$ are built systematically,
\begin{subequations}
    \begin{align}
        \PDF^{(A)}_n(x,t|x_0) = \prod_{i=1}^n &\brackets{
            \int_{\tau_{i-1}}^t \d{t_i} \int_{t_i}^t \d{\tau_i} h(t_i - \tau_{i-1},\tau_i - t_i )} 
        \nonumber \\ & \qquad \quad \times F(t-\tau_n) K^{(A)} (x,t|x_0,\tau_n),
        \label{eq:pathway_formulation_time_1}
        \\
        \PDF^{(B)}_n(x,t|x_0) = \prod_{i=1}^n &\brackets{
            \int_{\tau_{i-1}}^t \d{t_i} \int_{t_i}^t \d{\tau_i} h(t_i - \tau_{i-1},\tau_i - t_i )} 
        \nonumber \\
            & \qquad \quad \times \int_{\tau_n}^t \d{t_{n+1}} f(t_{n+1}-\tau_{n}) 
            \int \d{x'} K^{(A)} (x',t_{n+1}|x_0,\tau_n) 
        \nonumber \\    
            & \qquad \quad \times \int_{t}^\infty \d{\tau_{n+1}} \gamma(\tau_{n+1}-t_{n+1}|t_{n+1}-\tau_{n}) 
        \nonumber\\
            & \qquad \quad \times K^{(B)} (x,t|x',t_{n+1};t_{n+1}-\tau_n,\tau_{n+1}-t_{n+1}), 
        \label{eq:pathway_formulation_time_2}
    \end{align}
\end{subequations}
where we have defined $\tau_0 = 0$ for convenience. The above construction is guided by the renewal structure of the dynamics: each term only contributes with the propagator structure during the last cycle---$K^{(A)} (x,t|x_0,\tau_n)$ or $K^{(B)} (x,t|x',t_{n+1};t_{n+1}-\tau_n,\tau_{n+1}-t_{n+1})$, weighted with the product of probabilities of having the duration of each phase. As we have to take into account all possible pathways, we integrate over all admissible switching times $t_i$ and $\tau_i$ compatible with the corresponding pathway. A schematic representation of this construction is shown in figure~\ref{fig:Pathway_Formulation}.

The main advantage of this formulation arises when one works in Laplace space, because equations~\eqref{eq:pathway_formulation_time_1} and~\eqref{eq:pathway_formulation_time_2} can be expressed as convolutions. Starting from phase $A$, 
\begin{align}
    \laplTransformTilde{\PDF^{(A)}_n}{x,s|x_0} &= \int_0^\infty \d{t} e^{-s t} \prod_{i=1}^n\brackets{
        \int_{\tau_{i-1}}^t \d{t_i} \int_{t_i}^t \d{\tau_i} h(t_i - \tau_{i-1},\tau_i - t_i )
        }
        \nonumber\\
        &\qquad \qquad \qquad \qquad \qquad \quad \times F(t-\tau_n) K^{(A)} (x,t|x_0,\tau_n)
        \nonumber\\
        &= \prod_{i=1}^n \brackets{\int_{\tau_{i-1}}^\infty \d{t_i} \int_{t_i}^\infty \d{\tau_i} h(t_i - \tau_{i-1},\tau_i - t_i )} 
        \nonumber\\
        &\qquad \qquad \qquad \times \int_{\tau_n}^\infty \d{t} e^{-s t} F(t-\tau_n) K^{(A)} (x,t-\tau_n|x_0).
\end{align}
Let us define $\widehat{K}^{(A)} (x,t-\tau_n|x_0) = F(t-\tau_n)K^{(A)} (x,t-\tau_n|x_0)$. Changing the last variable of integration to $t'=t-\tau_n$, the last integral turns to
\begin{equation}
    \int_0^\infty \d{t'} e^{-s (t'+\tau_n)} F(t') K^{(A)} (x,t'|x_0) = e^{-s\tau_n}\laplTransformTilde{\widehat{K}^{(A)}}{x,s|x_0}.
\end{equation}
Afterwards, introducing $\delta^{(B)}_n = \tau_n - t_n$ and $\delta^{(A)}_n = t_n-\tau_{n-1}$, we have that 
\begin{equation}
    \int_{\tau_{n-1}}^\infty \d{t_n} \int_{t_n}^\infty \d{\tau_n} h(t_n - \tau_{n-1},\tau_n - t_n )e^{-s\tau_n} \laplTransformTilde{\widehat{K}^{(A)}}{x,s|x_0} = e^{-s\tau_{n-1}}\laplTransformTilde{h}{s,s}\laplTransformTilde{\widehat{K}^{(A)}}{x,s|x_0}, 
\end{equation}
which can be then applied recursively to all previous integrals, leading to
\begin{equation}
    \laplTransformTilde{{\PDF}^{(A)}_n}{x,s|x_0} = \widetilde{h}^n(s,s) \laplTransformTilde{\widehat{K}^{(A)}}{x,s|x_0}.
\end{equation}
Here, we have employed the notation 
\begin{eqnarray}
    \laplTransformTilde{h}{s,m} = \int_0^\infty \d{\delta^{(A)}} e^{-s \delta^{(A)}}\int_0^\infty \d{\delta^{(B)}} e^{-m \delta^{(B)}} h\parenthesis{\delta^{(A)},\delta^{(B)}}
    \label{eq:laplace_transform_h}
\end{eqnarray}
for the bivariate Laplace transform of $h$.\footnote{Note that $\laplTransformTilde{f}{s}=\laplTransformTilde{h}{s,0}$ and $\laplTransformTilde{g}{m}=\laplTransformTilde{h}{0,m}$.} Similarly, for phase $B$, we may define 
\begin{align}
    \widehat{K}^{(B)} (x,t|x_0) &= \int_0^{t} \d{\delta^{(A)}_{n+1}} f\parenthesis{\delta^{(A)}_{n+1}} \int \d{x'}  K^{(A)} \parenthesis{x',\delta^{(A)}_{n+1}|x_0}
    \nonumber \\    
    &\qquad \times \int_{t-\delta^{(A)}_{n+1}}^\infty \d{\delta^{(B)}_{n+1}} \gamma\parenthesis{\delta^{(B)}_{n+1}|\delta^{(A)}_{n+1}} K^{(B)} \parenthesis{x,t-\delta^{(A)}_{n+1}|x';\delta^{(A)}_{n+1},\delta^{(B)}_{n+1}}.
\end{align}
Therefore, the derivation is totally analogous, yielding 
\begin{equation}
    \laplTransformTilde{\PDF^{(B)}_n}{x,s|x_0} = \laplTransformTilde{h^n}{s,s} \laplTransformTilde{\widehat{K}^{(B)}}{x,s|x_0}.
\end{equation}

The sum over all possible pathways~\eqref{eq:pathway_formulation_generalStructure_TimeDomain} can be easily performed as a geometric series in the Laplace domain, leading to 
\begin{align}
    \laplTransformTilde{\PDF}{x,s|x_0} &=
    \sum_{n=0}^\infty 
    \bigg[
        \laplTransformTilde{\PDF^{(A)}_n}{x,s|x_0} + \laplTransformTilde{\PDF^{(B)}_n}{x,s|x_0}
        \bigg]
    \nonumber \\
    &= \frac{1}{1 - \laplTransformTilde{h}{s,s}} \laplTransformTilde{\widehat{K}^{(A)}}{x,s|x_0} + \frac{1}{1 - \laplTransformTilde{h}{s,s}} 
    \laplTransformTilde{\widehat{K}^{(B)}}{x,s|x_0}.
    \label{eq:pathway_formulation_generalStructure_LaplaceDomain}
\end{align}
It only remains to invert~\eqref{eq:pathway_formulation_generalStructure_LaplaceDomain}, either numerically or analytically, to obtain the dynamical evolution of the PDF. It is worth noting that this methodology makes it possible to compute the whole propagator $\PDF(x,t|x_0)$ of any process, without needing to solve complicated differential equations. The same pathway-based reasoning also applies to any other observable of the system that can be descomposed intro contributions from each pathway.

This framework can be particularised to other scenarios, like the first-passage problem. Let us suppose the previous two-stage dynamics, where $A$ is the only scanning phase and a target is found at $\targetx$. Denoting by $\survival^{(A)}\parenthesis{\cdot}$ the probability of not having reached the target after some time interval evolving as $A$; the whole previous formulation applies by replacing 
\begin{equation}
    f\parenthesis{\cdot}\to f\parenthesis{\cdot} \survival^{(A)}\parenthesis{\cdot}, \, \text{and } F\parenthesis{\cdot}\to F\parenthesis{\cdot} \survival^{(A)}\parenthesis{\cdot}
\end{equation}
\ie the scanning phase is weighted by the survival probability to compute those pathways that have not found the target in any phase $A$. The rest of the formulation remains unchanged, so~\eqref{eq:pathway_formulation_generalStructure_LaplaceDomain} is still valid with the new definition 
\begin{equation}
    h(\delta^{(A)},\delta^{(B)}) = f(\delta^{(A)}) \survival^{(A)}(\delta^{(A)};\targetx,x_0) \gamma(\delta^{(B)}|\delta^{(A)}).
\end{equation}
Another interesting case occurs when both waiting times in each cycle are also independent. In that case, $\gamma(\delta^{(B)}|\delta^{(A)})=g(\delta^{(B)})$, and using~\eqref{eq:laplace_transform_h} we can write
\begin{equation}
   \laplTransformTilde{h}{s,s} = \laplTransformTilde{f}{s}\laplTransformTilde{g}{s}.
\end{equation}

The above general formalism developed for this thesis will be used to analyse more realistic resetting models, where a new phase is introduced before exploring again after each reset. In chapter~\ref{ch:resetting_refractory}, resetting with refractory periods is analysed as a two-stage model, where phase $A$ corresponds to free diffusion~\eqref{eq:Green_function_BrownianMotion}, $K^{(A)}(x,t|x_0,t_0) = \PDF_0(x,t|x_0,t_0)$, and phase $B$ corresponds to the refractory period, $K^{(B)}(x,t|x_0,t_0,\delta^{(A)}) = \delta(x-x_0)$, and waiting times are completely independent, $h(\delta^{(A)},\delta^{(B)}) = f(\delta^{(A)}) g(\delta^{(B)})$. 

\subsection{Summary of part II}

The second part of this thesis is devoted to studying two often disregarded aspects within the field of search strategies based on stochastic resetting. 

Regarding the cost of resetting, in chapter~\ref{ch:resetting_refractory} we analyse the effect of refractory periods as a time cost after each resetting. To study this model, a general mathematical framework is provided for analysing the time evolution of intermittent search strategies---we apply the framework elaborated in section~\ref{subsec:pathway_formulation} to this particular case. Specific properties are analysed for Poissonian stochastic waiting times, for both resetting and refractory periods. Analytical results in this particular case are obtained for the propagator, as well as for its tendency to the NESS. Furthermore, the search time is optimised by minimising the MFPT of the model. The relaxation to the NESS is thoroughly characterised resorting to the analytical calculations developed in appendix~\ref{app:refractory_Laplace_method}, which relies on the Laplace method for approximating integrals. Numerical simulations are performed, as explained in appendix~\ref{app:langevin_simulations}, which validate the approximate analytical results.

In relation to quenched disorder, the optimisation of search times in heterogeneous environments is split into two chapters. The introduction of the model and the general mathematical framework is put forward in chapter~\ref{ch:resetting_disordered}. Therein, we also discuss the simplest case, where the target position is drawn from a dichotomous distribution. The main goal is to study how the search is optimised by tuning a piecewise constant resetting rate. The first-passage time distribution is computed analytically in the long-time limit by approximately inverting its Laplace transform, using the methods detailed in appendix~\ref{app:dichotomous_Asymptotic_FPT}. The optimisation of the search is investigated in depth by minimising the average MFPT and the standard deviation of the FPT. 

A more general heterogeneous situation is analysed in chapter~\ref{ch:resetting_disordered_bulkvsboundaries}. The target position is now drawn from a general distribution. Additionally, we consider that the search is bounded within a finite domain. We introduce the concept of resetting boundaries: if the particle reaches any boundary, it is reset to its initial position. The interplay between the heterogeneous (bulk) resetting and the boundary resetting is studied in detail. The optimisation of the search times is analysed by minimising the average MFPT with respect to the functional form of the resetting rate $r(x)$ in the bulk. Analytically, we are able to obtain what conditions must be satisfied to have an optimal bulk resetting strategy different from the ``trivial case'' of no resetting. Numerically, we seek the optimal shape of $r(x)$ by employing a gradient descent algorithm, as explained in appendix~\ref{app:gradient-descent}, to minimise the functional cost defined by the average MFPT.
{\clearpage \thispagestyle{empty}}


\part{Equilibrium properties of spin-elastic models to understand the mechanical response of slender structures}
\label{part:part-materials}
\chapter{One-dimensional systems: spin-string model}
\label{ch:materials_equilibrium}
As discussed in section~\ref{sec:intro_spin_elastic_models}, the study of shape transitions and deformation profiles of slender structures are classic problems in solid mechanics. This part of the thesis follows and further develops the line of research on spin-elastic models \cite{journalarticle:Bonilla.Carpio_ModelRipplesGraphene_Phys.Rev.B12,journalarticle:Bonilla.etal_RipplesStringCoupled_Phys.Rev.E12,journalarticle:Ruiz-Garcia.etal_RipplesHexagonalLattices_J.Stat.Mech.15,journalarticle:Ruiz-Garcia.etal_STMdrivenTransitionRippled_Phys.Rev.B16,journalarticle:Ruiz-Garcia.etal_BifurcationAnalysisPhase_Phys.Rev.E17} to understand the emergence of buckling phenomena in low-dimensional systems.

Our motivation stems from the experimental observation of buckling in graphene \cite{journalarticle:Schoelz.etal_GrapheneRipplesRealization_Phys.Rev.B15}. However, although previous spin-elastic models successfully describe certain aspects of this phenomenon, they are not fully consistent with relevant physical symmetries, as explained at the end of section~\ref{subsec:intro_buckling}. Therefore, we propose a new spin-elastic model, which involves the minimal key ingredients to obtain a transition from a flat to a buckled state in the absence of external forces, while preserving rotational symmetry.

Although the model is inspired by graphene, it is not intended to be a realistic description thereof. Instead, we aim to understand the fundamental mechanisms that lead to buckling in low-dimensional systems. Thus, we consider both paradigmatic scenarios of one-dimensional and two-dimensional systems, \ie in spin-string and spin-membrane lattices, respectively. This chapter is devoted to obtain equilibrium phases of the spin-string model, characterising their stability and the corresponding phase diagrams. The two-dimensional case is addressed in chapter~\ref{ch:materials_2d}.

The rest of the chapter is organised as follows. We introduce the novel one-dimensional spin-string model in section~\ref{sec:spin-string_model}. The Euler-Lagrange equation that provides the equilibrium profiles is derived in section~\ref{sec:1d_spin-string_Euler-Lagrange}. The existence and stability of buckled states are analysed, both analytically and numerically, in section~\ref{sec:1d_spin-string_phases}. The theoretical methods used to characterise and prove the existence of buckled phases involve bifurcation theory, where we use asymptotic expansions close to the critical lines, and the low-temperature limit.

\section{Spin-string model}
\label{sec:spin-string_model}

Let us consider a string in a one-dimensional lattice, with lattice parameter $a$. Each node, indexed by $j=0,\ldots,N$, is characterised by its vertical displacement $u_j$, the conjugate momentum $p_j$, and a spin variable $\sigma_j=\pm 1$. We introduce the Hamiltonian\footnote{The Hamiltonian is slightly different from the one introduced in~\cite{journalarticle:Garcia-Valladares.etal_BucklingRotationallyInvariant_Phys.Rev.E23}, since the index runs from $j=0$ to $j=N$. This change does not affect to our results, since we are interested in the continuum limit.}
\begin{align}
    \hamSpinElastic &=
        \sum_{j=1}^{N-1} \bigg[\frac{p_j^2}{2m}+\frac{k}{2}\left(u_{j+1}-2u_j+u_{j-1}\right)^2 
        \nonumber\\
        &\quad -h \left(u_{j+1}-2u_j+u_{j-1}\right) \sigma_j\bigg] + \sum_{j=0}^{N-1} J\sigma_{j}\sigma_{j+1}. 
    \label{eq:hamiltonian_spin_string_1d}
\end{align}
We can establish a term-by-term correspondence with the physical interpretation of each contribution for the current Hamiltonian~\eqref{eq:hamiltonian_spin_string_1d} and the already introduced for previous models~\eqref{eq:intro_spin_string_hamiltonian}. Let us recapitulate it here for clarity. The first term on the rhs is the kinetic energy, the second one stands for the elastic contribution of the string with constant $k$, the third term represents the coupling between the spins and the displacements of the string tuned by $h$, and the last one accounts for the interaction between neighbouring spins. Therefore, the main and unique difference consists in how the Hamiltonian depends on the transversal displacements: any contribution that involves the displacements $u_j$ is now a function of the discrete curvature of the string, \ie $u_{j+1}-2u_j+u_{j-1}$, instead of the discrete gradient $u_{j+1}-u_j$ or the displacement itself.\footnote{Other phenomenological models have previously introduced free energies that only depend on the curvatures of the system. One of the most relevant examples is the Helfrich model, which has been extensively employed in biological studies to understand the elasticity of cell membranes \cite{journalarticle:Helfrich_ElasticPropertiesLipid_Z.Naturforsch.CBio.Sci.73,journalarticle:Wei.etal_BendingRigidityGaussian_NanoLett.13,journalarticle:Lipowsky_ConformationMembranes_Nature91,journalarticle:Kunihiro.etal_NewComputationalApproach_Nanoscale25}.} Notice that the elastic contribution is the discrete version of $F_1$ in~\eqref{eq:elasticity_free_energy_plate_functional}.

As we introduced in~\eqref{eq:intro_spin_string_hamiltonian}, spin variables are a very reductionist approach to model the electronic states of the system. In principle, they behave similarly to previous models when we analyse the equilibrium configuration: they tend to align with the curvature of the string, but the spin-spin interaction promotes anti-alignment between neighbours for $J>0$. We thus expect a phenomenology similar to that of previous spin-elastic models, in which there exists a regime where the system is frustrated, leading to the emergence of rippled and buckled stable phases. However, the novelty is the rotational invariance of the Hamiltonian~\eqref{eq:hamiltonian_spin_string_1d}: any profile $u_j$ and its rigid linear transformations $u_j'= u_j + c_1 x + c_2$ leave the energy invariant. In particular, $c_1=0$ corresponds to a rigid translation, whereas $c_1\neq 0$ corresponds to a small rotation of angle $c_1$, since $\sin(c_1)\approx c_1$ for $c_1\ll 1$. 

We have not considered specific microscopic boundary conditions because they depend on the physical situation at hand. In section~\ref{sec:1d_spin-string_Euler-Lagrange}, we will derive what conditions must fulfil the contour of the string, \ie the boundary conditions in the continuum limit, in order to ensure the minimisation of the free energy. 

Let us consider the system is in contact with a thermal bath at temperature $T$. The probability of finding the system in a certain configuration is given by the canonical distribution 
\begin{equation}
    P_{\eq}(\confSpinElastic) = \exp{-\beta \hamSpinElastic}/Z,
\end{equation} 
with $\beta = 1/(k_B T)$, where $k_B$ is the Boltzmann constant, and 
\begin{equation}
    Z = \int \d{\mathbf{u}}\int \d{\mathbf{p}} \sum_{\bm{\sigma}} \exp{-\beta \hamSpinElastic} = \prod_{j=0}^N \brackets{\int \d{u_j}\int \d{p_j} \sum_{\sigma_j=\pm 1}} \exp{-\beta \hamSpinElastic}
\end{equation}
is the partition function. We are interested in the equilibrium profiles of the string, so we would like to compute the configuration that maximises the marginal probability 
\begin{equation}
    P_{\eq}(\mathbf{u}) = \int \d{\mathbf{p}} \sum_{\bm{\sigma}} P_{\eq}(\confSpinElastic) \propto e^{-\beta \mathcal{F}(\mathbf{u})},
    \label{eq:marginal_probability_equilibrium_displacements_1d}
\end{equation}
where we have defined the free energy of the string for the profile $\mathbf{u}$ as 
\begin{equation}
    \mathcal{F}(\mathbf{u}) \equiv \sum_{j=1}^{N-1} \brackets{\frac{k}{2}(u_{j+1}-2u_j+u_{j-1})^2} - k_B T \ln Z_\sigma(\mathbf{u}).
\end{equation}
Above, $Z_\sigma(\mathbf{u})$ stands for the partition function of the spin configuration at fixed displacement field $\mathbf{u}$, \ie 
\begin{equation}
    Z_\sigma(\mathbf{u}) = \sum_{\bm{\sigma}} \exp{\beta h \sum_{j=1}^{N-1} (u_{j+1}-2u_j+u_{j-1}) \sigma_j -\beta J \sum_{j=0}^{N-1}\sigma_{j}\sigma_{j+1}}.
    \label{eq:partition_function_spins_fixed_u}
\end{equation}
Taking into account~\eqref{eq:marginal_probability_equilibrium_displacements_1d}, it is clear that the probability $P_{\eq}(\mathbf{u})$ is maximised when the free energy $\mathcal{F}(\mathbf{u})$ is minimised. 

In the continuum limit, we are able to rewrite $\mathcal{F}(\mathbf{u})$ as a functional of the continuum displacement field $u_j\to u(x = j a)$, assuming it varies slowly with $j$. Once we go to the continuum limit, our goal is to derive the Euler-Lagrange equation for the equilibrium profile $u(x)$. Nevertheless, before proceeding, it is convenient to introduce dimensionless variables. The total length of the string is $L = N a$. Let us define the discrete curvature of the profile $\chi_j$ as 
\begin{equation}
    \chi_j\equiv\frac{u_{j+1} - 2 u_j + u_{j-1}}{a^2},
\end{equation}
which becomes $\chi(x) + \mathcal{O}(a^4) = u''(x)$ in the limit $a\to 0$. The elastic contributions of the Hamiltonian can be rewritten as 
\begin{subequations}
    \begin{align}
        \frac{k}{2}(u_{j+1}-2u_j+u_{j-1})^2 &= \frac{k_0}{2}\chi_j^2, \\
        h(u_{j+1}-2u_j+u_{j-1})\sigma_j &= h_0 \chi_j \sigma_j.
    \end{align}
\end{subequations}
Here, we have introduced the scaled parameters 
\begin{equation}
    k = \frac{k_0}{a^4}, \quad h = \frac{h_0}{a^2}.
\end{equation}
These choice of the microscopic parameters ensure that the harmonic elastic interactions and the electron-phonon coupling remain finite in the continuum limit.

The dimensional analysis of the Hamiltonian~\eqref{eq:hamiltonian_spin_string_1d} tells us that 
\begin{equation}
    [\mathcal{H}] = [p]^2/m = [k][u]^2 = [h][u] = [J], 
    \label{eq:relation_dimensions_1d_spin-string}
\end{equation}
where $[X]$ stands for the chosen unit of the quantity $X=[X]X^*$, so $X^*$ is the dimensionless variable. Guided by this~\eqref{eq:relation_dimensions_1d_spin-string}, we choose the characteristic length 
\begin{equation}
    \ell_0 = [u] = [x] = [a] = \frac{k_0}{h_0},
\end{equation}
that determines the reference temperature as 
\begin{equation}
    T_0 = \frac{h_0^2}{k_B k_0}.
\end{equation}
Then, the dimensionless variables are given by 
\begin{equation}
    u^* = \frac{u}{\ell_0}, \quad x^* = \frac{x}{\ell_0}, \quad \chi^* = \ell_0 \chi, \quad p^* = \frac{p}{\sqrt{m k_B T_0}}, \quad T^* = \frac{T}{T_0}, \quad J^* = \frac{J}{k_B T_0}.
    \label{eq:dimensionless_variables_1d_spin-string}
\end{equation}
Consequently, we define $\beta^* = 1/T^*$, $L^* = L/\ell_0$. The Hamiltonian now reads 
\begin{align}
    \hamSpinElastic &=
        \sum_{j=1}^{N-1} \bigg[\frac{p_j^2}{2}+\frac{\chi_j^2}{2}-\chi_j \sigma_j \bigg] + \sum_{j=0}^{N-1} J \sigma_{j}\sigma_{j+1},
\end{align}
where we have dropped the asterisks, as also done in the following, for the sake of clarity. It is worth noting that the continuum limit does not necessarily involve a large system size limit $N\to\infty$, \ie the thermodynamic limit, but rather a ``small'' lattice constant, $a\ll \ell_0$, in contrast to previous spin-elastic models \cite{journalarticle:Ruiz-Garcia.etal_STMdrivenTransitionRippled_Phys.Rev.B16,journalarticle:Ruiz-Garcia.etal_BifurcationAnalysisPhase_Phys.Rev.E17,journalarticle:Bonilla.etal_RipplesStringCoupled_Phys.Rev.E12}.

\section{Derivation of the Euler-Lagrange equation}
\label{sec:1d_spin-string_Euler-Lagrange}

In the continuum limit, the equilibrium probability of finding the string in a certain profile $u(x)$ becomes a functional thereof,
\begin{equation}
    P_{\eq}[u] \propto e^{-\beta \mathcal{F}[u]}, \quad \mathcal{F}[u] = n \int_0^L \d{x} f(\chi),
\end{equation}
where $n = N / L = 1 / a$ is the number density, 
and $f(\chi)$ is the free energy density per unit of length. As discussed earlier, $f(\chi)$ only depends on the curvature $\chi(x) = u''(x)$. The expression for $f(\chi)$ is 
\begin{subequations}
    \label{eq:1d_spin-string_free_energy_density}
    \begin{align}
        f(\chi) &= \frac{1}{2}\chi^2 - T \ln \zeta^{(1)}(\chi/T,J/T), 
        \\
        \zeta^{(1)}(\chi/T,J/T) &= e^{-J/T} \cosh(\chi/T) + e^{J/T} \sqrt{1 + e^{-4J/T} \sinh^2(\chi/T)},
    \end{align}    
\end{subequations}
where~\eqref{eq:1d_spin-string_free_energy_density} stems from the continuum limit of~\eqref{eq:partition_function_spins_fixed_u}. Concretely, $\zeta^{(1)}(\chi/T,J/T)$ is the partition function of the one-dimensional Ising model with nearest-neighbour coupling $J$ and external field $\chi$ at temperature $T$, which can be derived using the transfer-matrix method  \cite{book:Feynman_StatisticalMechanicsSet_96}.\footnote{This result holds assuming that the displacement field $u(x)$ is smooth enough, so we can consider that $\chi(x)$ varies slowly in space to treat it as a constant external field in local subdomains.}

The equilibrium profiles $u_{\eq}(x)$ are those that minimise the free-energy functional $\mathcal{F}[u]$. Therefore, we consider the first variation thereof upon the change $u\to u + \delta u$ \cite{book:Gelfand.Fomin_CalculusVariations_00,book:Lanczos_VariationalPrinciplesMechanics_70}, 
\begin{equation}
    \delta \mathcal{F}[u] = \int_0^L \d{x} \pdev{}{f(\chi)}{\chi} \delta u''.
\end{equation}
This expression is more involved than usual, as the free energy density $f(\chi)$ depends on the second derivative of the displacement field $\chi = u''$. Making integration by parts twice, we finally obtain 
\begin{equation}
    \delta \mathcal{F}[u] = \int_0^L \d{x} \delta u 
    \cdev{2}{}{x}\parenthesis{\pdev{}{f}{\chi}}
    + 
    \brackets{\delta u' \pdev{}{f}{\chi}-\delta u \cdev{}{}{x}\parenthesis{\pdev{}{f}{\chi}}}_0^L.
    \label{eq:spin-string_variation_free_energy}
\end{equation}
The details on how the variational principle is applied to functionals with higher-order derivatives are given in appendix~\ref{app:variational_principle_higher_order}. To ensure the equilibrium profile is an extremum of the functional, both the integral and boundary terms must vanish independently for arbitrary variations $\delta u(x)$. The integral term leads to the Euler-Lagrange equation 
\begin{equation}
    \left.\cdev{2}{}{x}\parenthesis{\pdev{}{f}{\chi}}\right|_{\eq} = 0, \Rightarrow \left. \pdev{}{f}{\chi}\right|_{\eq} = C_1 x + C_0, 
    \label{eq:1d_spin-string_variational_integralterm}
\end{equation}
where $C_1$ and $C_0$ are arbitrary constants, to be determined by imposing the boundary conditions. The boundary conditions depend on the physical situation, there are different possibilities that make the boundary terms of~\eqref{eq:spin-string_variation_free_energy} vanish. Here, we are going to consider the ends of the string are fixed, so that 
\begin{equation}
    u(0) = u(L) = 0,
    \label{eq:supported_boundary_conditions_1d_spin_string}
\end{equation}
but the values of the slope at the boundaries $u'$ are free---these are the so-called supported boundary conditions \cite{book:Landau.etal_TheoryElasticityVolume_86}. With this option for the boundary conditions, employing the Euler-Lagrange equation~\eqref{eq:1d_spin-string_variational_integralterm} in~\eqref{eq:spin-string_variation_free_energy}, we have 
\begin{equation}
    \left. \pdev{}{f}{\chi}\right|_{x=0} = C_0 = 0, \quad \left. \pdev{}{f}{\chi}\right|_{x=L} = C_1 L + C_0 = 0,
\end{equation}
which entails $C_1 = C_0 = 0$. 

Thus, the equilibrium profiles $u_{\eq}(x)$ are those that satisfy $\partial f / \partial \chi|_{\eq}=0$, \ie
\begin{subequations}
    \begin{align}
        \chi_{\eq} &= \frac{e^{-2J/T}\sinh(\chi_{\eq}/T)}{\sqrt{1 + e^{-4J/T}\sinh^2(\chi_{\eq}/T)}}, \label{eq:1d_spin-string_Euler-Lagrange} 
        \\
        u(0) &= u(L) =0.
    \end{align}
\end{subequations}
Equation~\eqref{eq:1d_spin-string_Euler-Lagrange} is a transcendental equation for the equilibrium curvature $\chi_{\eq}$. Thus, the equilibrium curvature is constant, a space-independent function $\chi_{\eq}=\chi_{\eq}(J,T)$. The direct implication is that the equilibrium free energy is an extensive function $\mathcal{F}_{\eq} = n L f(\chi_{\eq}) = N f(\chi_{\eq})$, being $\chi_{\eq}$ an intensive quantity, independent of the size of the system. We restrict ourselves to study the solutions with positive curvature, since there is a clear well-defined mirror symmetry $\chi_{\eq}\leftrightarrow -\chi_{\eq}$. This is a consequence of the free-energy density~\eqref{eq:1d_spin-string_free_energy_density} being an even function of $\chi$---there is no external field breaking the symmetry.

Once we know the curvature is always constant in equilibrium, the general equilibrium string profiles are parabolas of the form 
\begin{equation}
    u_{\eq}(x;J,T) = \frac{\chi_{\eq}(J,T)}{2} x (x - L) + C_2 x + C_3.
    \label{eq:1d_spin-string_general_solution}
\end{equation}
For the supported boundary conditions~\eqref{eq:supported_boundary_conditions_1d_spin_string}, we have $C_2 = C_3 = 0$. For completely free boundary conditions, the solution is the general one~\eqref{eq:1d_spin-string_general_solution}, with $C_2$ and $C_3$ arbitrary because the boundary condition $\text{d}(\partial f/\partial \chi)/\text{d}x|_{x=0,L}=0$ is already fulfilled. For instance, the flat profile, $u_{\eq}(x)=0$, and any transversal shift plus small rotation thereof, $u_{\eq}(x) = C_2 x + C_3$, are both possible equilibrium profiles for free boundary conditions. 

Since the free-energy density only depends on the curvature $\chi$, the analysis of the buckled states that follows is valid for both supported and free boundary conditions. For this reason, we will focus here on the case of supported boundary conditions~\eqref{eq:supported_boundary_conditions_1d_spin_string}.

\section{Existence and stability of buckled states}
\label{sec:1d_spin-string_phases}

The equilibrium curvature $\chi_{\eq}$ is implicitly defined by the transcendental equation~\eqref{eq:1d_spin-string_Euler-Lagrange}. The number of solutions indicates how many equilibrium phases of the system can be found for any pair $(J,T)$. On the one hand, the trivial solution $\chi_{\eq}=0$ always exists, corresponding to the flat profile $u_{\eq}(x)=0$. On the other hand, depending on the values of the parameters $(J,T)$, buckled states with $\chi_{\eq}\neq 0$ can be found. For this reason, the curvature plays the role of the order parameter, in the sense of Landau theory, since we can describe how the phases emerge and bifurcate from the trivial solution $\chi_{\eq}=0$. The values of the curvature for the different phases are obtained either analytically, using bifurcation theory from the flat solution, or numerically, solving~\eqref{eq:1d_spin-string_Euler-Lagrange} using standard root-finding algorithms. In any case, we aim to fully characterise the existence and stability of equilibrium solutions in our string-spin model in the whole parameter space $(J,T)$. 

\begin{figure} 
    \centering
    \includegraphics[width=0.8\textwidth]{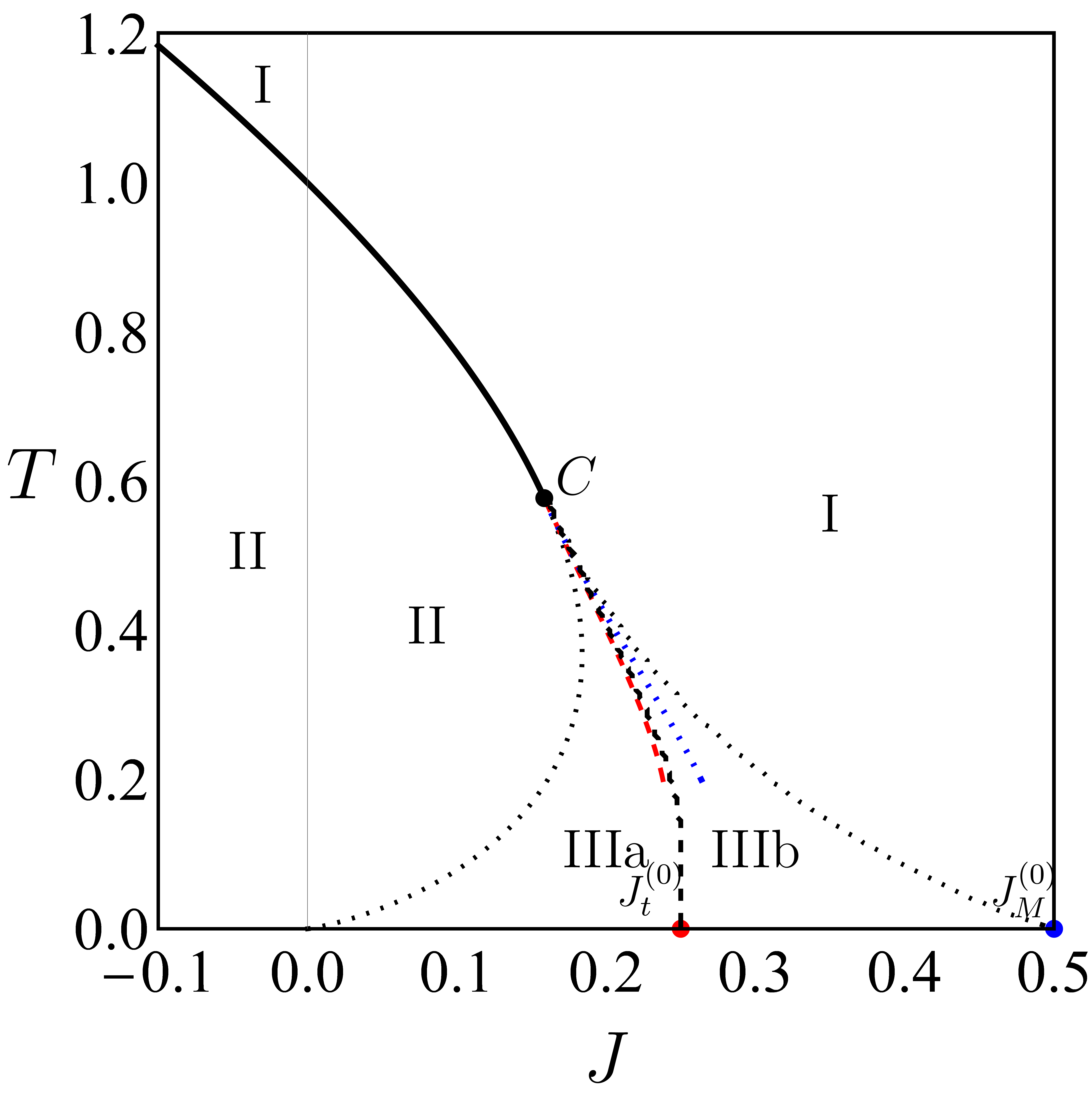}
    \caption{Phase diagram of the rotationally invariant spin-string. The plane $(J,T)$ is divided into several regions, characterised by the existing phases and their stability. In region I, only the $\zc$ phase exists. In region II, phase $\zc$ is unstable, and there appears a stable buckled phase B. In region III, three phases coexist: $\zc$, and two buckled phases, B+ and $\BuckledUnstable$; the latter being unstable, whereas $\zc$ and B+ are locally stable. Phase B+ is the most stable one in IIIa, with $\zc$ being metastable, whereas the roles are reversed in IIIb. Point $C=(J_{c},T_{c})$ is the tricritical point. Above it, the phase transition is second-order; below, it is first-order.  The curve $J_b(T)$, given by~\eqref{eq:1d_spin-string_bifurcation_curve}, is a bifurcation curve. For $T>T_c$, it defines the already mentioned second-order phase transition (black solid line). For $T<T_{c}$, it demarcates the change from region II to region III (leftmost dotted line). The curve $J_{M}(T)$ (rightmost dotted line), theoretically predicted by~\eqref{eq:limit-metastab-curve} close to the tricritical point (blue dotted line), demarcates the change from region III to region I. The curve $J_{t}(T)$ (dashed line), theoretically predicted by~\eqref{eq:first-order-curve} close to the tricritical point (red dashed line), is the first-order transition line: over it, phases B+ and $\zc$ have the same free energy. Also plotted are the theoretical low-temperature limiting values $J_{M}^{(0)}=1/4$ and $J_{t}^{(0)}=1/2$. On the one hand, the predictions for the bifurcation curve $J_b(T)$ and for the low-temperature values $J_t^{(0)}$, $J_M^{(0)}$ are exact. On the other hand, the asymptotic theoretical predictions for the curves $J_t(T)$ and $J_M(T)$ show an excellent agreement with the numerical results, within their range of validity.}
    \label{fig:1d_spin-string_phase_diagram}
\end{figure}

The complete picture of the phase diagram is presented in figure~\ref{fig:1d_spin-string_phase_diagram}. The different regions indicate the existence and stability of each phase. Both the regions and their transition lines are thoroughly explained below in the following sections. Nonetheless, we first anticipate the main results here. 

The phase diagram of figure~\ref{fig:1d_spin-string_phase_diagram} is analogous to the one plotted in figure~\ref{fig:intro_spin_elastic}. Thus, the explanation of Schoelz \emph{et al}'s experiment \cite{journalarticle:Schoelz.etal_GrapheneRipplesRealization_Phys.Rev.B15} provided in section~\ref{subsec:intro_buckling} on the basis of the previous spin-elastic models also holds for the current, rotationally invariant, one. The plane $(J,T)$ is divided into three main regions, depending on the phases---equilibrium solutions---we find. For any pair $(J,T)$, the zero-curvature solution exists (ZC phase). It is the unique solution, and thus the most stable one, in region I, for high $T$ and/or $J$. For high enough temperatures, $T>T_c$, if the spin-spin interaction decreases, we cross to region II, where a stable buckled phase B appears, whereas the $\zc$ phase becomes unstable. 
Going down along this line, it appears a tricritical point $C$ (black circle) at $(J_c,T_c)$. Below it, a third region where three phases coexist appears: two buckled phases (B+ and $\BuckledUnstable$) and the zero-curvature one. The $\BuckledUnstable$ phase is always unstable, whereas the B+ and $\zc$ phases compete to be the most stable one: (i) in region IIIa, the B+ phase is the most stable one, with $\zc$ being metastable; (ii) in region IIIb, the roles are reversed. 
More details on the physical interpretation of the curves separating the different regions are given in the caption of figure~\ref{fig:1d_spin-string_phase_diagram}. Additionally, a summary of the main characteristics of each region can be found in table~\ref{tab:1d_spin-string_phases}.

\begin{table}
    \centering
    \begin{tabular}{c c c c c c}
    \hline\hline
    Region & Definition & Phases & Stable & Unstable & Metastable \\
    \hline
    I & $J>J_b,T>T_c | J>J_M,T<T_c$ & $\zc$ & $\zc$ & None & None \\ 
    II & $J<J_b(T)$ & B+,$\zc$ & B+ & $\zc$ & None \\
    IIIa & $J_b(T)<J<J_t(T)$,$T<T_c$ & B+,$\BuckledUnstable$,$\zc$ & B+ & $\BuckledUnstable$ & $\zc$ \\
    IIIb & $J_t(T)<J<J_M(T)$,$T<T_c$ & B+,$\BuckledUnstable$,$\zc$ & $\zc$ & $\BuckledUnstable$ & B+ \\
    \hline\hline
    \end{tabular}
    \caption{Summary of each region founded in the phase diagram of figure~\ref{fig:1d_spin-string_phase_diagram}: definition, existence of the phases (zero-curvature $\zc$, stable buckled $B+$, unstable buckled $\BuckledUnstable$) and their stability.}
    \label{tab:1d_spin-string_phases}
\end{table}

\subsection{Bifurcation from the zero-curvature solution: second-order phase transition}

A theoretical analysis of the buckled states is performed by employing bifurcation theory, \ie studying how non-zero curvature solutions emerge from the flat profile as we vary $(J,T)$. 

Let us start by considering the free-energy density for the flat profile,
\begin{equation}
    f_0(J,T) = f(\chi=0,J,T) = -T \ln\brackets{2\cosh(J/T)},
\end{equation}
so we may define the difference of the free-energy density from that of flat profile as 
\begin{equation}
    \Delta f(\chi;J,T) = f(\chi;J,T) - f_0(J,T) = \frac{1}{2} \chi^2 - T\ln\brackets{\frac{\zeta^{(1)}(\chi/T,J/T)}{2\cosh(J/T)}}.
    \label{eq:free-energy_difference-1d_spin-string}
\end{equation}
The stability of the zero-curvature solution $\chi_{\eq}=0$ is determined by the sign of the second derivative of $\Delta f$, \ie
\begin{equation}
    f_2(J,T) \equiv \left.\pdev{2}{f}{\chi}\right|_{\chi=0} = 1 - \frac{e^{-2J/T}}{T}, 
\end{equation}
The change of stability takes place at the line $J_b(T)$ over which $f_2(J_b(T),T)$ vanishes, \ie 
\begin{equation}
    J_b(T) = \frac{1}{2} T \ln T,
    \label{eq:1d_spin-string_bifurcation_curve}
\end{equation}
which is called the \emph{bifurcation curve}. This line separates the plane $(J,T)$ into two regions: (i) a region with $f_{2}>0$, where the zero-curvature solution is, at least locally, stable and (ii) a region with $f_{2}<0$, where it becomes unstable and other stable solution should emerge. In figure~\ref{fig:1d_spin-string_phase_diagram}, $J_b(T)$ is represented by the black solid line which separates regions I and II above the critical point $C$, and the leftmost black dotted line between regions II and IIIa below $C$. Note that, for $J=0$, we have $f_{2}<0$ for $0<T<1$: the reference temperature $T_{0}$ is thus the transition temperature in the absence of coupling among the spins. 

Following Landau theory of phase transitions \cite{book:Landau.Lifshitz_StatisticalPhysicsVolume_13}, we expand the free-energy density in powers of $\chi$, close to the bifurcation curve, 
\begin{equation}
    \Delta f(\chi;J,T) = \frac{1}{2} f_2(J,T) \chi^2 + \frac{1}{4!} f_4(J,T) \chi^4 + \frac{1}{6!} f_6(J,T) \chi^6 + \mathcal{O}(\chi^8),
    \label{eq:free-energy_expansion_general}
\end{equation}
where $f_4$ and $f_6$ are given by the Taylor coefficients
\begin{subequations}
    \begin{align}
        f_4(J,T) &= \left.\pdev{4}{f}{\chi}\right|_{\chi=0} = \frac{e^{-6J/T}}{T^3}\parenthesis{3-e^{4J/T}}, \\
        f_6(J,T) &= \left.\pdev{6}{f}{\chi}\right|_{\chi=0} = -\frac{e^{-10J/T}}{T^5}\parenthesis{45 - 30 e^{4J/T} + e^{8J/T}}.
    \end{align}
\end{subequations}
Within this approximation, the equilibrium curvatures are the solutions of 
\begin{equation}
    \left.\pdev{}{\Delta f(\chi;J,T)}{\chi}\right|_{\eq} = f_2(J,T) \chi_{\eq} + \frac{f_4(J,T)}{3!} \chi_{\eq}^3 + \frac{f_6(J,T)}{5!} \chi_{\eq}^5 + \mathcal{O}(\chi_{\eq}^7) = 0.
\end{equation} 
In the following, this expansion is truncated at different orders, since the sign of the last retained coefficient must be positive to ensure the corresponding equilibrium distribution $P_{\eq}(\chi)$ is normalisable. 

If we truncate up to the second-order term and $f_2(J,T)>0$, $\chi_{\eq}^{\zc}=0$ is the only solution. If we want to analyse the behaviour close to the bifurcation curve, defined by $f_2 (J_b(T),T)=0$, where $f_2 (J,T)$ can be negative, we must go to the following order in the expansion, $f_4(J,T)>0$. Let us analyse how the sign of $f_2(J,T)$ varies close to $(J_b,T_b)$: we explicitly indicate the behaviour of this coefficient introducing a small parameter $0<\varepsilon\ll 1$, such that $f_2(J,T) = \varepsilon \varphi_2(J,T)$, $\varphi_2 = \mathcal{O}(1)$. This tells us that the separation of the point $(J,T)$ we are considering from the bifurcation curve is of order $\varepsilon$: it entails that $T-T_b=\mathcal{O}(\varepsilon)$ and/or $J-J_b=\mathcal{O}(\varepsilon)$. Thus, the equilibrium curvature around the bifurcation curve $(J_b,T_b)$ is obtained solving 
\begin{equation}
    0 = \varepsilon\varphi_2 \chi_{\eq} + \frac{f_{4,b}}{3!} \chi_{\eq}^3,
    \label{eq:free-energy_expansion_region_II}
\end{equation}
where 
\begin{equation}
    f_{4,b} \equiv f_4(J_b,T_b) = \frac{3 T_b^2-1}{T_b^2}.
\end{equation}
Here, $\chi_{\eq}^{\zc}=0$, which is always a solution, changes its stability from a minimum, for $f_2(J,T)>0$, to be a maximum of the free energy, for $f_2(J,T)<0$. In that case, where the zero-curvature solution becomes unstable, there appear two symmetric minima at 
\begin{equation}
    \chi_{\eq}^{\Buckled} = \pm \varepsilon^{1/2}\sqrt{-6\varphi_2/f_{4,b}}. 
    \label{eq:buckled_solutions_region_II}
\end{equation}
The superscript $B$ for these solutions correspond to buckled phases, with non-zero curvature. Hence, we have found a region II, defined by $f_2(J,T)<0$ (or $\varphi_2<0$), where the only stable phases are the buckled ones $\chi_{\eq}^{\Buckled}$, given by~\eqref{eq:buckled_solutions_region_II}, leaving the flat profile $\chi_{\eq}^{\zc}=0$ as an unstable equilibrium solution. The bifurcation curve~\eqref{eq:1d_spin-string_bifurcation_curve} that demarcates regions I and II corresponds to a second-order transition line in the sense of Landau \cite{book:Landau.Lifshitz_StatisticalPhysicsVolume_13}, because the second derivative of the free energy changes continuously in the phase transition from phase $\zc$ to $\Buckled$. In fact, we can approximately compute how the buckled solution emerges from the profile using~\eqref{eq:1d_spin-string_bifurcation_curve}. For instance, for $J=0$, we have that $T_b = 1$, $\varepsilon = T_b-T=1-T>0$, $f_{4,b}=2$, and $\varphi_2 =\lim_{T\to T_b} f_2/\varepsilon = -1$, leading to 
\begin{equation}
    \chi_{\eq}^{\Buckled} = \pm \sqrt{3 (T-1)}, \quad J=0,
    \label{eq:buckled_solutions_J=0_1d}
\end{equation}  
which corresponds to the typical universality class in Ising-like systems.

Notice that these solutions are consistent with the approximations we have introduced, since the second-order and fourth-order terms in~\eqref{eq:free-energy_expansion_region_II} are of order $\varepsilon^{3/2}$, whereas the sixth-order is negligible because it is of order $\varepsilon^{5/2}$. 

\subsection{Analysis below the tricritical point: first-order phase transition}
\label{subsec:below_tricritical_spin-string}

The analysis done in the previous section is valid as long as $f_{4}>0$, which is always satisfied for a ferromagnetic coupling $J<0$. Nonetheless, in the antiferromagnetic case $J>0$, there exists a tricritical point $C\equiv(T_c,J_c)$, 
\begin{equation}
    T_c = \frac{1}{\sqrt{3}}, \quad J_c = J_b(T_c) = \frac{\ln 3}{4\sqrt{3}}, 
\end{equation}
where $f_{4,b}$ vanishes, and thus the approximation~\eqref{eq:free-energy_expansion_region_II} breaks down. Below the tricritical point, $f_{4,b}$ becomes negative and we must retain the sixth-order term, proportional to $f_6\chi^6$ of the free-energy density.

Close to the tricritical point, a different scaling for the solutions of~\eqref{eq:free-energy_expansion_general} is needed. We still write $f_2=\varepsilon \varphi_2$: the terms involving $f_2$ and $f_6$ are of the same order if $\chi_{\eq}=\mathcal{O}(\varepsilon^{1/4})$. Then, for consistency, the three terms are comparable if $f_{4,b}=\mathcal{O}(\varepsilon^{1/2})$---this tells us the order of the distance of the point $(J_b,T_b)$ over the bifurcation curve to the tricritical point, \ie $T_b-T_c=\mathcal{O}(\varepsilon^{1/2})$ and/or $J_b-J_c=\mathcal{O}(\varepsilon^{1/2})$. Then, we can write $f_{4,b}=\varepsilon^{1/2}\varphi_4$, for $\varphi_4 = \mathcal{O}(1)$, and $f_{6,c}=f_6(T_c,J_c)=36$, so the dominant balance for the curvature equation in this region is 
\begin{equation}
    \varphi_2 \Xi + \frac{1}{6}\varphi_4 \Xi^3 + \frac{1}{120}f_{6,c}\Xi^5=0,\quad \chi_{\eq}=\varepsilon^{1/4}\Xi.
    \label{eq:free-energy_near_tricritical}
\end{equation}
We find the trivial solution $\chi_{\eq}^{\zc}=0$, and two buckled solutions 
\begin{subequations}
 \begin{align}
  \chi_{\eq}^{\BuckledStable}=&\pm\varepsilon^{1/4}\sqrt{\frac{-5\varphi_4+
    \sqrt{25\varphi_4^2-30\varphi_2f_{6,c}}}{f_{6,c}/2}}, \label{eq:chi-B+}
  \\ 
  \chi_{\eq}^{\BuckledUnstable}=&\pm\varepsilon^{1/4}\sqrt{\frac{-5\varphi_4- \sqrt{25\varphi_4^2-30\varphi_2f_{6,c}}}{f_{6,c}/2}}.
\end{align}   
\end{subequations}
The zero-curvature phase corresponds to a local minimum of the free energy as long as $f_{2}>0$, outside the bifurcation curve~\eqref{eq:1d_spin-string_bifurcation_curve}---it becomes unstable inside it, as already discussed. The curvatures $\chi_{\eq}^{\text{B}\pm}$ correspond to two buckled phases B$\pm$, the domain of existence and stability of which is discussed below.

The buckled phase B+ exists as long as $f_{4,b}<0$ (or $\varphi_{4}<0$), \ie below the tricritical point, provided that $5f_{4,b}^{2}-6f_{2}f_{6,c}>0$ (or
$5\varphi_4^2-6\varphi_2f_{6,c}>0$). The line
\begin{equation}
    \label{eq:end-metastab-v1}
    5f_{4,b}^{2}-6f_{2}f_{6,c}=0
\end{equation}
marks the end of the existence of this phase (rightmost dotted curve in figure~\ref{fig:1d_spin-string_phase_diagram}). Moreover, when traversing from right to left this line, the buckled phase pops up discontinuously with a finite curvature: specifically, from $\chi_{\eq}^{\zc}=0$ to $\chi_{\eq}^{\BuckledStable}=\pm \sqrt{-5f_{4,b}/18}\ne 0$.  It can be easily checked---by evaluating the second derivative of the free energy density with respect to $\chi$---that this buckled phase is locally stable in its domain of existence. The buckled phase $\BuckledUnstable$ also exists as long as $f_{4,b}<0$ (or $\varphi_{4,b}<0$), \ie below the tricritical point. However, not only does it need $5f_{4}^{2}-6f_{2}f_{6,c}>0$ but also $f_{2}>0$. This phase $\BuckledUnstable$ only exists between the line~\eqref{eq:end-metastab-v1} and the part of the bifurcation curve~\eqref{eq:1d_spin-string_bifurcation_curve} below the tricritical point (dotted lines in figure~\ref{fig:1d_spin-string_phase_diagram}). Phase $\BuckledUnstable$ is always unstable: the second derivative of the free energy density for $\chi=\chi_{\eq}^{\BuckledUnstable}$ is always negative.

The curve demarcating the limit of existence of the buckled phases, equation~\eqref{eq:end-metastab-v1}, can be written in a more transparent way. Let us consider, for any $T$ close to $T_{c}$, a point $(J,T)$ close to the bifurcation curve $(J_{b}(T),T)$ by taking into account that
\begin{equation}\label{eq:f4b-f2-tricrit}
  f_{4,b}\sim 6\sqrt{3} (T-T_{c}), \quad f_{2}\sim 2\sqrt{3} (J-J_{b}(T))
\end{equation}
in the approximation we are employing. Therefore, equation~\eqref{eq:end-metastab-v1} is equivalent to define the curve demarcating the end of metastability region
\begin{equation}
    \label{eq:limit-metastab-curve}
    J_{M}(T)\equiv J_{b}(T)+\frac{5}{12}\sqrt{3} (T-T_{c})^{2}.
\end{equation}
This approximation is represented by the blue dotted line in figure~\ref{fig:1d_spin-string_phase_diagram}, showing a good agreement with the numerical results within its range of validity.

The analysis above entails the emergence of the region III in the plane $(J,T)$, also depicted in figure~\ref{fig:1d_spin-string_phase_diagram}. This region extends over the zone of the plane $(J,T)$ below the tricritical point $C$, with its left border at the bifurcation line~\eqref{eq:1d_spin-string_bifurcation_curve} and its right border at the line~\eqref{eq:limit-metastab-curve}. Inside region III, the three phases $\zc$, B+, and $\BuckledUnstable$ coexist. Phases B+ and $\zc$ correspond to local minima of the free energy, whereas $\BuckledUnstable$ always corresponds to a local maximum. The relative stability of phases B+ and $\zc$ is elucidated in the following.

In region III, the phases $\zc$ and B+ are both locally stable: one of them corresponds to the deepest minimum, being the most favourable thermodynamic state, whereas the other one corresponds to a metastable state. This means that the transition changes from second-order above the tricritical point to
first-order below it, since the curvature (order parameter) of the most stable phase changes abruptly below the tricritical point. The change of stability takes place at the first-order transition line, determined by the condition $\Delta \mathcal{F}(\BuckledStable)=\Delta\mathcal{F}(\zc)=0$. Close to the tricritical point, this is equivalent to
\begin{equation}
  f_{2}+\frac{1}{12}f_{4,b}(\chi_{\eq}^{\BuckledStable})^{2}+
\frac{1}{360}f_{6,c}(\chi_{\eq}^{\BuckledStable})^{4}=0.
\end{equation}
Bringing to bear equations~\eqref{eq:free-energy_near_tricritical} and~\eqref{eq:chi-B+}, over the first-order line one has $8f_{2}f_{6,c}=5f_{4,b}^{2}$, which translates into the definition of the transition curve over which both states are equally stable:
\begin{equation}
    \label{eq:first-order-curve}
    J_{t}(T)\equiv J_{b}(T)+\frac{5}{16}\sqrt{3}(T-T_{c})^{2},
\end{equation}
making use of~\eqref{eq:f4b-f2-tricrit}. The analytical curve $J_{t}(T)$ is shown in figure~\ref{fig:1d_spin-string_phase_diagram} as a red dashed line, providing an excellent approximation to the numerical results near the tricritical point. 

The buckled phase B+ is the most stable one in region IIIa, see figure~\ref{fig:1d_spin-string_phase_diagram},
\ie between the branch of the bifurcation line below the tricritical point (leftmost dotted line) and the first-order line~\eqref{eq:first-order-curve} (dashed line), with the $\zc$ phase being metastable. The situation is just reversed in region IIIb between the first-order line~\eqref{eq:first-order-curve} and the curve demarcating the limit of existence of  phases B$\pm$, equation~\eqref{eq:limit-metastab-curve} (rightmost dotted line): therein, the $\zc$ phase is the most stable and B+ is metastable.

The emergence of region III is very important from a physical point of view. As we anticipated, it stems from the competition between the elastic and electronic contributions in the Hamiltonian~\eqref{eq:hamiltonian_spin_string_1d}: the electron-phonon coupling makes the spins have the same sign as the curvature in the buckled phase, but the antiferromagnetic coupling of the spins tries to maintain neighbouring spins antiparallel---destroying the buckled states and making the system recover a zero-curvature profile.

\subsection{Low-temperature limit}

Analytical exact results can be obtained in the low-temperature limit $T\ll1$. Let us define the energy per site in the ground state of the one-dimensional Ising system with external field $\chi$ and antiferromagnetic coupling $J$ as
\begin{equation}
    e^{(1)}_{\gs} = \lim_{T\to0^+} -T \ln \zeta^{(1)} = 
    - (|\chi|-2J)\Theta(|\chi|-2J)-J,
    \label{eq:ground-state_energy_1d_spin-string}
\end{equation}
where $\Theta(\cdot)$ stands for the Heaviside step function.\footnote{Here, $\gs$ subscript marks we are working with the ground state in the thermal sense.}
Therefore, the free-energy density~\eqref{eq:1d_spin-string_free_energy_density} becomes 
\begin{equation}
    f(\chi;J,T) \sim \frac{\chi^2}{2} - (|\chi|-2J)\Theta(|\chi|-2J)-J, 
\end{equation}
while the Euler-Langrage equation turns out to be 
\begin{equation}
    |\chi_{\eq}| = \Theta(|\chi_{\eq}|-2J).
    \label{eq:euler-lagrange_spin-string_lowT}
\end{equation}

The discussion in section~\ref{subsec:below_tricritical_spin-string} shows us the most interesting case is the antiferromagnetic one, $J>0$, where the system exhibits a rich behaviour due to the emergence of a coexistence region.  In~\eqref{eq:ground-state_energy_1d_spin-string}, for very small curvature, $|\chi|\ll 1 $, the antiferromagnetic coupling wins and the ground state corresponds to the antiferromagnetic ordering, $f(\chi;J,T) \sim e^{(1)}_{\gs} = -J$, whereas for large $\chi$, the interactions between the spins and string dominate, it corresponds to the case of all spins aligned with the curvature, $e^{(1)}_{\gs}=-|\chi|+J$. Thus, we expect to find the following behaviour in the low-temperature limit: (i) for large enough $J$, the antiferromagnetic coupling destroy the buckled state and then the unique stable solution is the zero-curvature one, whereas (ii) for small $J$, stable buckled states can exist.

\begin{figure}
    \centering
    \includegraphics[width=0.8\textwidth]{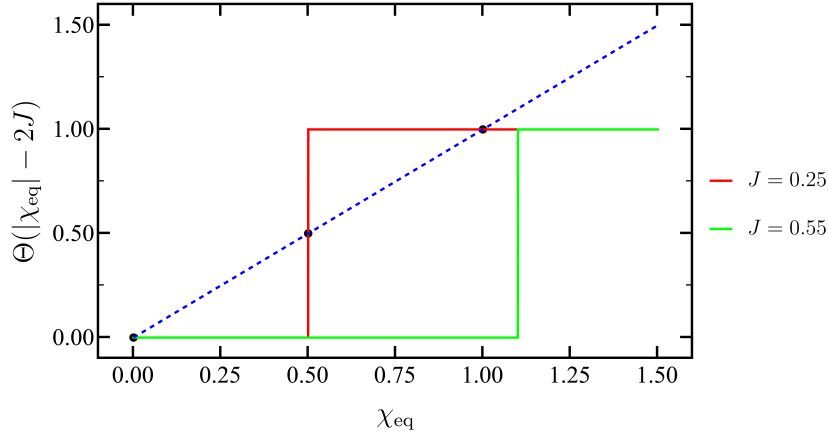}
    \caption{Graphical method for solving~\eqref{eq:euler-lagrange_spin-string_lowT}. For each $J$, the intersection points of the functions $\Theta(|\chi_{\eq}|-2J)$ (solid lines) and $\chi_{\eq}$ (dashed blue line) give the equilibrium curvature  in the low temperature limit. For $J > J_{M}^{(0)}=1/2$, for instance $J=0.55$ (green solid line), the only solution is $\chi_{\eq}=0$, \ie  the $\zc$ phase. For $J\leq J_{M}^{(0)}$, for instance $J=0.25$ (red solid line), there appear three solutions: in addition to the ZC phase corresponding to $\chi_{\eq}=0$, we have two buckled phases corresponding to $\chi_{\eq}^{\BuckledUnstable}=2J$ (unstable) and $\chi_{\eq}^{\BuckledStable}=1$ (stable).}
    \label{fig:spin-string_lowT_solutions}
\end{figure}

The equilibrium curvatures are the solutions of~\eqref{eq:euler-lagrange_spin-string_lowT}, where Heaviside's step function must be understood in a physical way, as a limit of the rhs of~\eqref{eq:1d_spin-string_Euler-Lagrange}. Therefore, $\Theta(x)$ comprises three strokes: two flat strokes, equal to zero and unity for $x<0$ and $x>0$, respectively, and a vertical stroke at $x=0$, which joins the previous two. In figure~\ref{fig:spin-string_lowT_solutions}, we show how the Euler-Lagrange equation~\eqref{eq:euler-lagrange_spin-string_lowT} is graphically solved by finding the intersection points of the curvature and the Heaviside function. It is clearly seen that $\chi_{\eq}^{\zc}=0$ is always a solution, for all $J$: the zero-curvature profile survives in the low-temperature limit. Additionally, there appear two buckled states for $2J \leq 1$, \ie $J\leq J_M^{(0)}=1/2$, which are, specifically, $\chi_{\eq}=2J$ and $\chi_{\eq}=1$. They coalesce at $J_M^{(0)}=1/2$ that marks the end of coexistence and the line separating regions IIIb and I---see also figure~\ref{fig:1d_spin-string_phase_diagram}.

Let us identify these two buckled phases by calculating their free-energy density. The free-energy density of the $\zc$ phase is given by $f_0(J,T)=-J$, for $T\ll 1$, so that
\begin{subequations}
    \begin{align}
        \Delta f (\chi_{\eq}=2J;J,T\to0) &\sim 2J^2, \\
        \Delta f (\chi_{\eq}=1;J,T\to0) &\sim -\frac{1}{2}+2J.
    \end{align}
\end{subequations}
On the one hand, the phase with $\chi_{\eq}=2J$ has always a larger free energy than the $\zc$ phase, so it corresponds to the low-temperature limit of phase $\BuckledUnstable$. On the other hand, the phase with $\chi_{\eq}=1$ changes the sign at $J=J_t^{(0)}=1/4$: it is the low-temperature limit of phase B+, being the most stable phase for $0<J<J_t^{(0)}$, and metastable for $J_t^{(0)}<J<J_M^{(0)}$. 

\subsection{Numerical analysis}
The numerical analysis of the phase diagram is performed by solving~\eqref{eq:1d_spin-string_Euler-Lagrange}. We construct a $200\times 200$ mesh in the $(J,T)$ plane and find numerically the solutions, using standard root-finding algorithms. 
\begin{figure}
  \centering
  \includegraphics[width=0.66\textwidth]{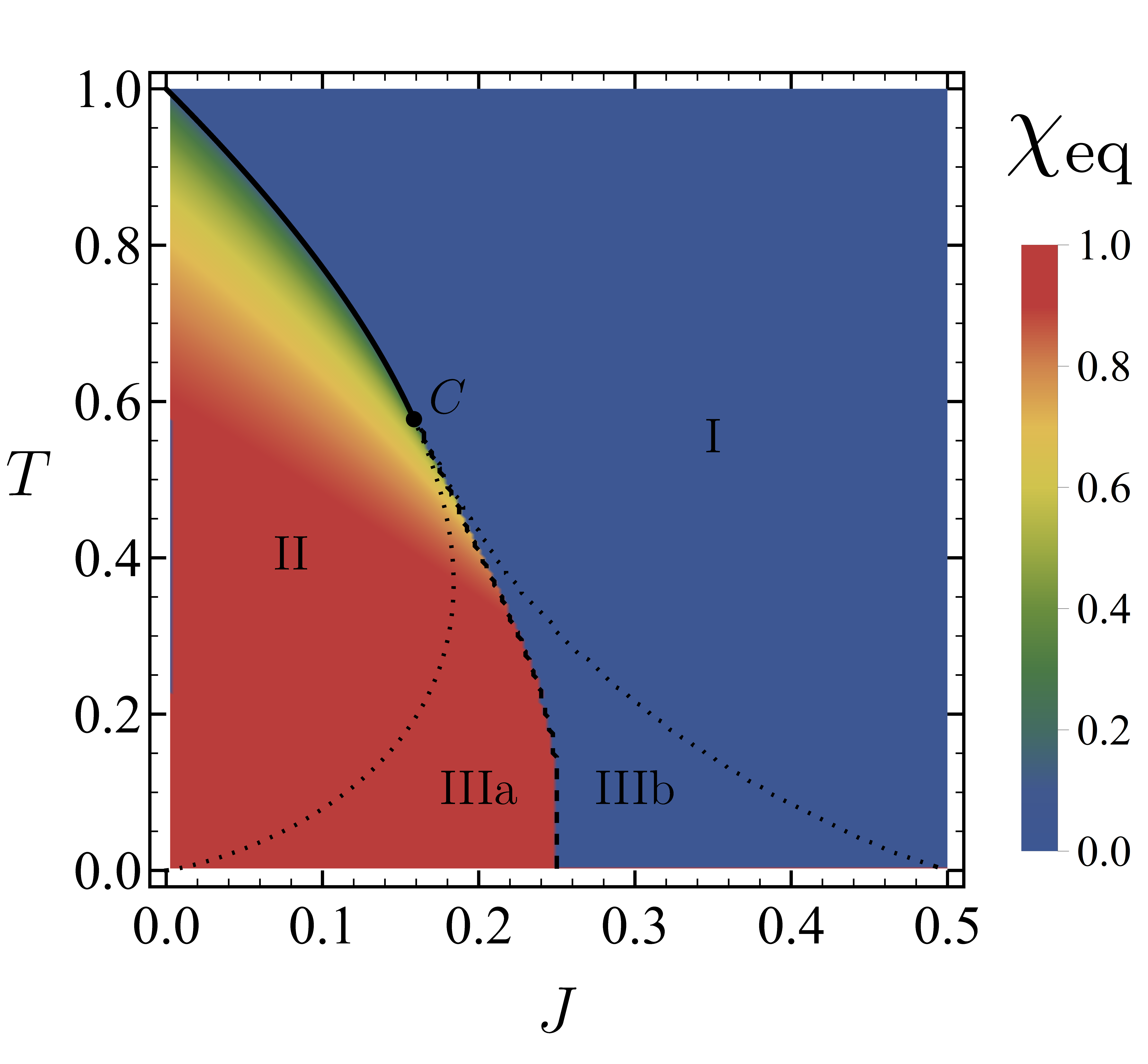}
  \caption{Curvature of the most stable phase. The labelling of the different regions and the code of the different lines is the same as in figure~\ref{fig:1d_spin-string_phase_diagram}. The change of order of the transition, from second-order (solid line) above $C$ to first-order (dashed line) below, is clearly observed.}
  \label{fig:chieq-num}
\end{figure}
\begin{figure}
    \centering  
    \includegraphics[width=0.9\textwidth]{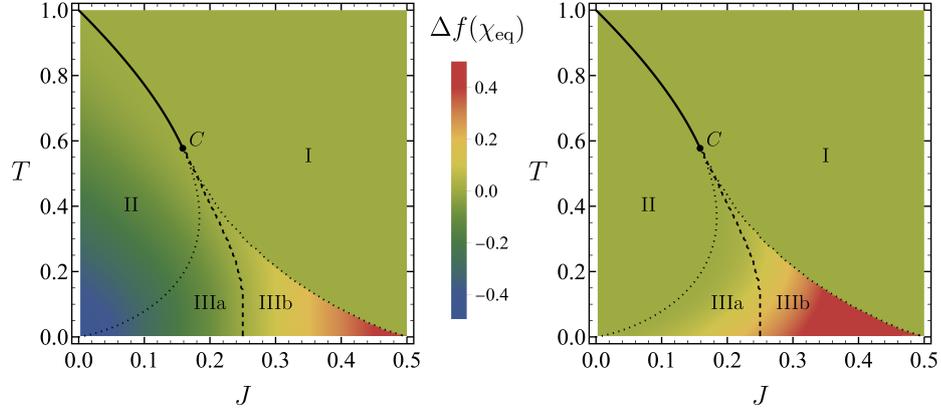} 
    \caption{Free energy density difference $\Delta f(\chi_{\eq};J,T)$ of the buckled states with respect to the ZC phase (numerically obtained). The labelling of the different regions and the code of the different lines is the same as in figures~\ref{fig:1d_spin-string_phase_diagram} and~\ref{fig:chieq-num}. On the left panel, we show a density plot for the (at least locally) stable buckled phase: B in region II and B+ in region III. On the right panel, we show the density plot for the unstable buckled phase $\BuckledUnstable$, which only exists in region III. In regions where there is no buckled phase, a constant value zero is plotted.}
    \label{fig:phase-diagram_num_200x200}
\end{figure} 

Figure~\ref{fig:chieq-num} shows the values of $\chi_{\eq}$ for the thermodynamically stable phase---the most stable one when there is coexistence. Therein, it is clearly observed that the transition changes from second-order to first-order at the tricritical point. Above the tricritical point $C$, $\chi_{\eq}$ changes continuously between zero and non-zero values at the bifurcation curve $J_{b}(T)$ (solid line). Below the tricritical point, $\chi_{\eq}$ changes abruptly between zero and non-zero values at the first-order line $J_{t}(T)$ (dashed line). Recall that the curvature plays the role of the order parameter in our model. In fact, $\chi_{\eq}$ equals the magnetisation of the spin, which is given by the rhs of~\eqref{eq:1d_spin-string_Euler-Lagrange}. 

Once we have obtained the equilibrium curvature, we can compute the free-energy density~\eqref{eq:free-energy_difference-1d_spin-string} over the considered mesh. In figure~\ref{fig:phase-diagram_num_200x200}, we plot the difference of the free energy density with respect to the ZC in two scenarios, obtaining the same behaviour as we anticipated in the theoretical phase diagram shown in figure~\ref{fig:1d_spin-string_phase_diagram}. In the left panel, the value for the stable buckled phase, wherever it exists, is shown. Therefore, phase B is shown in region II, inside the bifurcation curve~\eqref{eq:1d_spin-string_bifurcation_curve} (solid line above the tricritical point $C$, leftmost dotted line below it), and phase B+ is plotted in region III. Region III is demarcated by the bifurcation curve and the curve $J_{M}(T)$ that marks the limit of existence of the phases B$\pm$ (rightmost dotted line).  In the right panel, $\Delta f(\chi_{\eq};J,T)$ is plotted for the unstable phase $\BuckledUnstable$, which only exists in region III. The three phases, ZC, B+, and \BuckledUnstable coexist in region III, which is divided into two subregions by the first-order line $J_{t}(T)$ (dashed line): IIIa, where $\Delta f(\chi_{\eq}^{\BuckledStable};J,T)<0$ and the most stable phase is B+, being ZC metastable; and IIIb, where the roles are reversed, the most stable phase is ZC, being B+ metastable.
%

\chapter{Two-dimensional systems: spin-membrane model}
\label{ch:materials_2d}
Here, we extend the analysis done for the spin-string model in chapter~\ref{ch:materials_equilibrium} to the two-dimensional case, proposing a spin-membrane model. The physical idea is the same as before: we have a two-dimensional elastic lattice, where there is a particle at each node characterised by its vertical displacement, transversal to the plane of the membrane, and an internal degree of freedom represented by a spin variable. The contributions to the Hamiltonian depend again on the discrete curvature of the membrane, so we must take into account what kind of topology the lattice has. 

The rest of the chapter is organised as follows. First, we introduce the two-dimensional spin-membrane model in section~\ref{sec:spin-membrane_model}, where we present two different geometries: the honeycomb and square lattices. We prove that the continuum limit allows us to study both of them at the same time in dimensionless variables. The Euler-Lagrange equation that provides the equilibrium profiles is derived in section~\ref{sec:2d_spin-membrane_equilibrium_profiles}. The existence and stability of buckled states are analysed in section~\ref{sec:2d_spin-membrane_limits}, where we consider two limit situations: the absence of spin-spin interaction, $J=0$, and the low-temperature limit, $T=0$. 

\section{Spin-membrane model}
\label{sec:spin-membrane_model}

Let us now consider the two-dimensional version of the spin-elastic model, \ie a spin-membrane system. Now, the label of a node needs two indices to move along a two-dimensional lattice. For each node, we associate a vertical displacement $u_{i,j}$, its momentum $p_{i,j}$, and the internal degree of freedom $\sigma_{i,j}=\pm 1$. They are no longer vectors, \ie $\mathbf{u}$, but matrices, $\hat{u}$. We must take into account the specific arrangement of the nodes in the two-dimensional lattice, \ie the crystal structure of the membrane. In particular, we consider two kinds of structures: (i) the honeycomb lattice, in resemblance of graphene sheets, and (ii) the square lattice. However, as we will see, any structure can be treated similarly, as long as we correctly define the discrete energetic contributions.

\subsection{Honeycomb lattice}
\label{subsec:spin-membrane_honeycomb_lattice}

\begin{figure}
    \centering
    \includegraphics[width=0.55\textwidth]{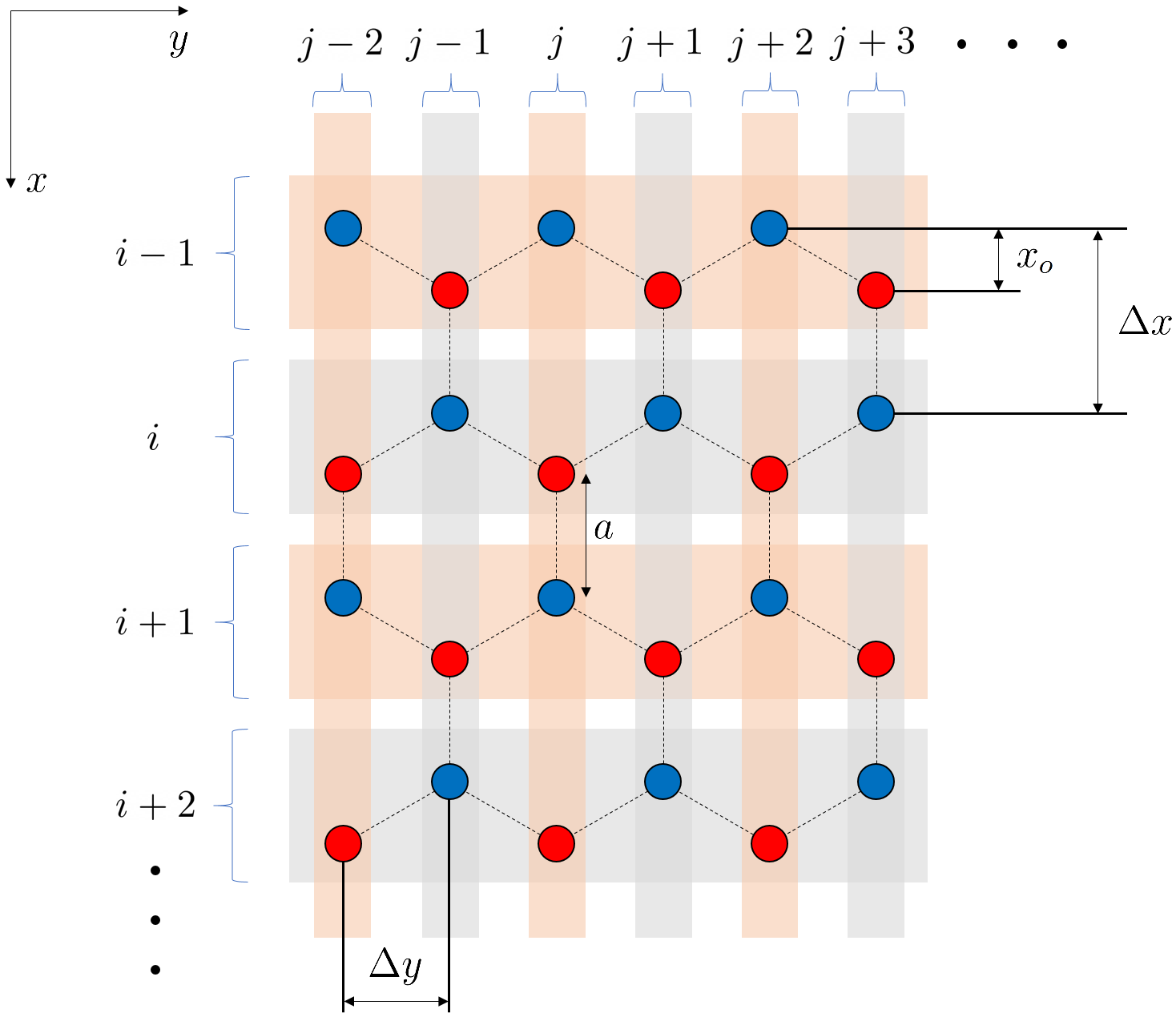}
    \caption{Sketch of the honeycomb lattice with parameter $a$. The lattice is bipartite and can be thus split into two sublattices: e-sites (blue), two nearest neighbours below and one above, and o-sites (red), two nearest neighbours above and one below. The distance between sites of the same kind corresponding to consecutive indexes are $\Delta x=3a/2$, $\Delta y=\sqrt{3}a/2$. The spacing between e- and o-sites with the same value of the index $i$ is $x_o=a/2$.}
    \label{fig:honeycomb_lattice}
\end{figure}

Let us start with a honeycomb lattice. A sketch of this structure is shown in figure~\ref{fig:honeycomb_lattice}, characterised by the lattice parameter $a$. Indices $i$ and $j$ are employed for rows and columns, respectively. There are two types of sites, depending on the parity of $i+j$: (i) e-sites (blue circles), where $i+j$ is even, and (ii) o-sites (red circles), where $i+j$ is odd. Each site has three nearest neighbours, one above and two below for e-sites, and vice versa for o-sites. Note that the two-dimensional domain $\Omega$ can have an arbitrary shape, so we have not introduced anything about boundary sites so far. As we did in the one-dimensional case, the appropriate boundary conditions over the contour $\partial\Omega$ are computed later on using a variational principle. 

The Hamiltonian of the spin-membrane for a honeycomb lattice is
\begin{align} 
    \mathcal{H}(\hat{u},\hat{p},\hat{\sigma})&= \sum_{|i-j|=\text{even}} \bigg[\frac{p_{i,j}^2}{2m} +
    \frac{k}{2}\left(u_{i-1,j}+u_{i,j-1}+u_{i,j+1}-3u_{i,j}\right)^2 
    \nonumber \\
    & \qquad  -h
    \sigma_{i,j}\left(u_{i-1,j}+u_{i,j-1}+u_{i,j+1}-3u_{i,j}\right)+\frac{J}{2} \sigma_{i,j} \left(\sigma_{i-1,j}+\sigma_{i,j-1}+
    \sigma_{i,j+1}\right) \bigg]
    \nonumber\\
    &\quad +\sum_{i-j=\text{odd}}
    \bigg[ \frac{p_{i,j}^2}{2m} + \frac{k}{2}\left(u_{i+1,j}+u_{i,j-1}+u_{i,j+1}-3u_{i,j}\right)^2  
    \nonumber \\
    & \qquad -h
    \sigma_{i,j}\left(u_{i+1,j}+u_{i,j-1}+u_{i,j+1}-3u_{i,j}\right) + \frac{J}{2} \sigma_{i,j} \left(\sigma_{i+1,j}+\sigma_{i,j-1}+
    \sigma_{i,j+1}\right) \bigg],
\end{align}
for any configuration $(\hat{u},\hat{p},\hat{\sigma})$. For the sake of clarity, the Hamiltonian is split into two sums, for e-sides and o-sites, respectively. The physical interpretation of the different terms is completely analogous to the one-dimensional case~\eqref{eq:hamiltonian_spin_string_1d}. The elastic terms depend on the discrete curvature, whose form is specific for this topology, but it can be generalised to any structure. It is nothing but the discrete Laplace operator acting on the vertical displacements $u_{i,j}$, which can be defined as the sum of differences over the nearest neighbours of each site.  

The procedure to obtain the equilibrium profiles is the same as before, we integrate the canonical distribution $P_{\eq}(\hat{u},\hat{p},\hat{\sigma}) \propto \exp(-\beta\mathcal{H})$ over the momenta and spins, and look for the profile maximising the marginal probability $P_{\eq}(\hat{u})\propto e^{-\beta \mathcal{F}}$. The free energy $\mathcal{F}$ is computed in the continuum limit, $a\to 0^+$, as done in the one-dimensional case.

The continuum limit here is more involved than that for the spin-string model. Let us introduce the continuous spatial variables $x$ and $y$ depicted in figure~\ref{fig:honeycomb_lattice}, $u_{i,j}=u(x,y)$, so that the position of any site is univocally determined by 
\begin{subequations}
    \begin{align}
        x = i\Delta x, \ y=j\Delta y, \quad \text{for e-sites sublattice}, \\
        y = x_o + i \Delta x, \ y = j\Delta y, \quad \text{for o-sites sublattice},
    \end{align}
\end{subequations}
where we have defined $x_o=a/2$, $\Delta x = 3a/2$, and $\Delta y = \sqrt{3}a/2$. All nearest neighbours of o-sites are e-sites, and vice versa, \ie the honeycomb lattice is bipartite \cite{book:Newman_NetworksIntroduction_10}. Therefore, if $(i,j)$ corresponds to an o-site, the discrete terms of elastic contributions to the Hamiltonian can be written as 
\begin{subequations}
    \begin{align}
    u_{i+1,j}&=u(x+a,y) \simeq u(x,y)+a \pdev{}{u}{x}+\frac{a^2}{2}\pdev{2}{u}{x}, 
    \\
    u_{i,j-1}&=u\parenthesis{x-x_o,y-\Delta y} \simeq u(x,y)-\frac{a}{2} \pdev{}{u}{x}
    -\frac{\sqrt{3}a}{2}\pdev{}{u}{y}
    \nonumber\\
    &\qquad \qquad \qquad \qquad \qquad \quad
    +\frac{a^2}{8} \pdev{2}{u}{x}+\frac{3a^2}{8}\pdev{2}{u}{y}
    +\frac{\sqrt{3}a^2}{4}\frac{\partial^2 u}{\partial x\partial y}, 
    \\
    u_{i,j+1}&=u\parenthesis{x-x_o,y+\Delta y} \simeq u(x,y)-\frac{a}{2} \pdev{}{u}{x}
    +\frac{\sqrt{3}a}{2}\pdev{}{u}{y}
    \nonumber\\
    &\qquad \qquad \qquad \qquad \qquad \quad
    +\frac{a^2}{8} \pdev{2}{u}{x}+\frac{3a^2}{8}\pdev{2}{u}{y}
    -\frac{\sqrt{3}a^2}{4}\frac{\partial^2 u}{\partial x\partial y},
    \end{align}
\end{subequations}
so that 
\begin{equation}
    u_{i+1,j}+u_{i,j-1}+u_{i,j+1}-3u_{i,j} \simeq \frac{3a^2}{4}\left(\pdev{2}{u}{x}+\pdev{2}{u}{y}\right).
\end{equation}
As expected, the elastic terms correspond to the two-dimensional curvature $\chi_{i,j}\to\chi(x,y) \equiv \nabla^2 u(x,y)$, where $u_{i+1,j}+u_{i,j-1}+u_{i,j+1}-3u_{i,j} = (\sqrt{3}a/2)^2\chi_{i,j}$. The same result holds for e-sites, $u_{i-1,j}+u_{i,j-1}+u_{i,j+1}-3u_{i,j} = (\sqrt{3}a/2)^2\chi_{i,j}$. Then, the contributions to the Hamiltonian read
\begin{subequations}
    \label{eq:2d_spin-membrane_elastic_contributions_honeycomb}
    \begin{align}
        \frac{k}{2}(u_{i+1,j}+u_{i,j-1}+u_{i,j+1}-3u_{i,j})^2 &= \frac{k_0}{2}\chi_{i,j}^2,
        \\
        h(u_{i+1,j}+u_{i,j-1}+u_{i,j+1}-3u_{i,j})\sigma_{i,j} &= h_0 \chi_{i,j} \sigma_{i,j},
    \end{align}
\end{subequations}
in which we have defined 
\begin{equation}
    \label{eq:scaled_parameters_honeycomb}
    k_0 = \parenthesis{\frac{\sqrt{3}a}{2}}^4 k, \quad h_0 = \parenthesis{\frac{\sqrt{3}a}{2}}^2 h.
\end{equation}

\subsection{Square lattice}
\label{subsec:spin-membrane_square_lattice}

\begin{figure}
    \centering
    \includegraphics[width=0.50\textwidth]{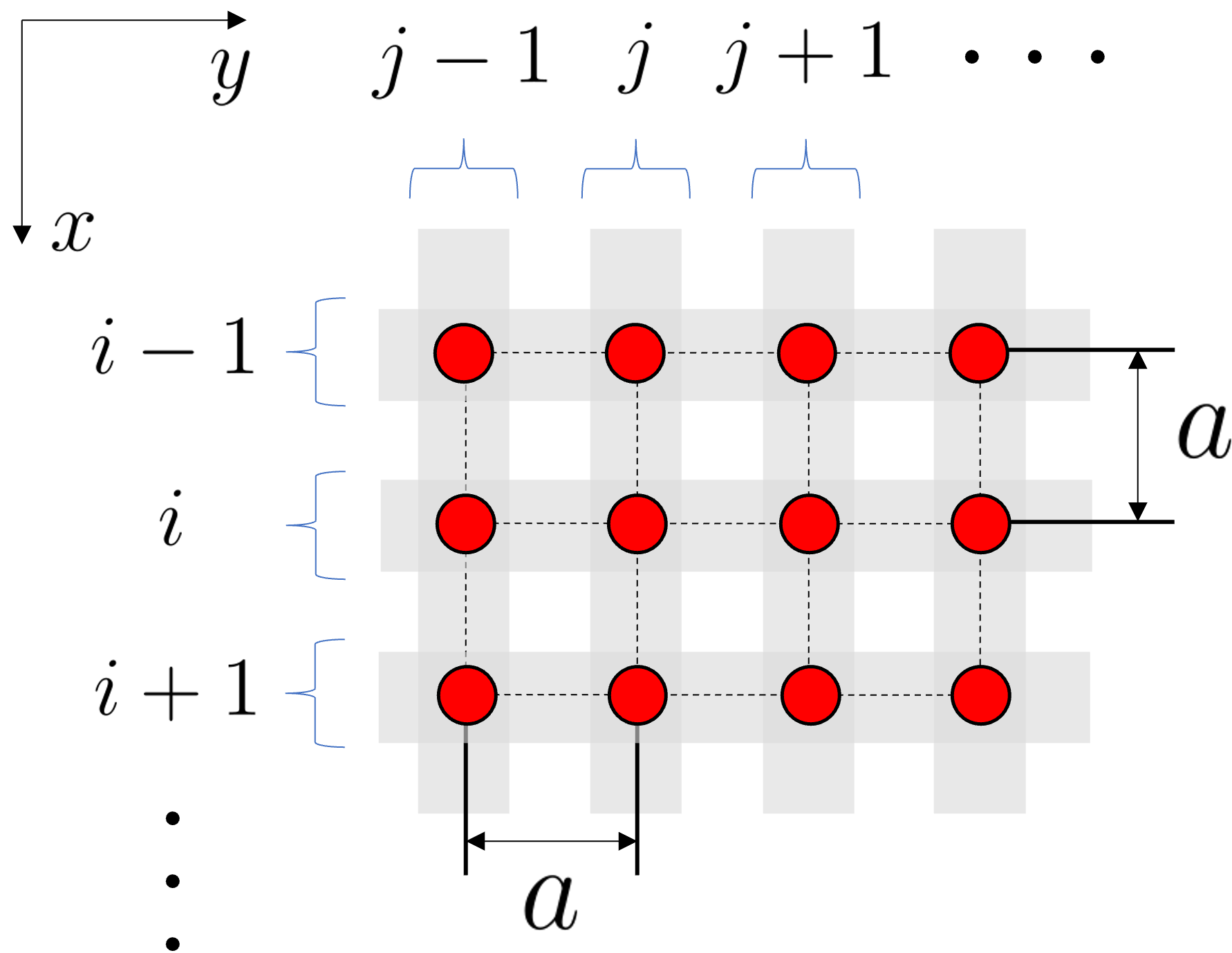}
    \caption{Sketch of the square lattice with parameter $a$. The distance between two consecutive rows or columns is always $\Delta x = \Delta y = a$.}
    \label{fig:square_lattice}
\end{figure}

Let us move to the case of a square lattice. The definitions of the variables for each node are the same as before. The structure can be observed in figure~\ref{fig:square_lattice}. The Hamiltonian of the spin-membrane for a square lattice is
\begin{align}
    \mathcal{H}(\hat{u},\hat{p},\hat{\sigma})&=
    \sum_{i,j} \bigg[\frac{p_{i,j}^2}{2m}
    +\frac{k}{2}\left(u_{i+1,j}+u_{i-1,j}+u_{i,j+1}+u_{i,j-1}-4u_{i,j}\right)^2
    \nonumber\\
    & \qquad h \sigma_{i,j}\left(u_{i+1,j}+u_{i-1,j}+u_{i,j+1}+u_{i,j-1}-4u_{i,j}\right)
    \nonumber\\
    & \qquad + \frac{J}{2}\sigma_{i,j}(\sigma_{i+1,j}+\sigma_{i-1,j}+\sigma_{i,j+1}+\sigma_{i,j-1})\bigg]
\end{align}
for any configuration $(\hat{u},\hat{p},\hat{\sigma})$. Notice that the discrete curvature is again written as the discrete Laplace operator applied on $u_{i,j}$, but now each site has four nearest neighbours.

In the continuum limit, the equilibrium profiles depend on the spatial variables $x = i a$ and $y = j a$, \ie $\Delta x = \Delta y = a $, as shown in figure~\ref{fig:square_lattice}. The displacement at each node is $u_{i,j}=u(x,y)$, so the contributions to the Hamiltonian read
\begin{subequations}
    \begin{align}
    u_{i+\pm 1,j}&=u(x\pm a,y) \simeq u(x,y)\pm a \pdev{}{u}{x}+\frac{a^2}{2}\pdev{2}{u}{x}, 
    \\
    u_{i,j\pm 1}&= u(x,y\pm a)\simeq u(x,y)\pm a \pdev{}{u}{y}+\frac{a^2}{2}\pdev{2}{u}{y},
    \end{align}
\end{subequations}
and 
\begin{equation}
    u_{i+1,j}+u_{i-1,j}+u_{i,j+1}+u_{i,j-1}-4u_{i,j} \simeq a^2\left(\pdev{2}{u}{x}+\pdev{2}{u}{y}\right).
\end{equation}
Again, we identify the two-dimensional curvature $\chi_{i,j}\to\chi(x,y) \equiv \nabla^2 u(x,y)$, where $u_{i+1,j}+u_{i-1,j}+u_{i,j+1}+u_{i,j-1}-4u_{i,j} = a^2 \chi_{i,j}$. This result allows us to write the contributions to the Hamiltonian in a similar way as in~\eqref{eq:2d_spin-membrane_elastic_contributions_honeycomb}, \ie 
\begin{subequations}
    \begin{align}
        \frac{k}{2}(u_{i+1,j}+u_{i-1,j}+u_{i,j+1}+u_{i,j-1}-4u_{i,j})^2 &= \frac{k_0}{2}\chi_{i,j}^2,
        \\
        h (u_{i+1,j}+u_{i-1,j}+u_{i,j+1}+u_{i,j-1}-4u_{i,j})\sigma_{i,j} &= h_0 \chi_{i,j} \sigma_{i,j}.
    \end{align}
\end{subequations}
The difference resides in the definition of the parameters: (i) for a square lattice, they are given by $k_0 = a^4 k$ and $h_0 = a^2 h$, whereas for a honeycomb lattice (ii), they are defined as in~\eqref{eq:scaled_parameters_honeycomb}.

This similarity allows us to treat both geometries---or any other if the Hamiltonian depends on the discrete Laplacian---at the same time, just taking into account the right definitions of $k_0$ and $h_0$ for each case when the continuum limit is taken.

\section{Derivation of the Euler-Lagrange equation}
\label{sec:2d_spin-membrane_equilibrium_profiles}

Before proceeding with the derivation of the Euler-Lagrange equation for the model, we introduce dimensionless variables for the two-dimensional lattices. Dimensionless variables are defined using the same criteria as in the one-dimensional case. Let us define the characteristic length $\ell_0=k_0/h_0$, where $k_0$ and $h_0$ depend on the topology, so dimensionless variables follow the same structure as in~\eqref{eq:dimensionless_variables_1d_spin-string}, except for the addition of the new spatial variable $y^*=y/\ell_0$.

The two-dimensional free energy functional is 
\begin{equation}
    \mathcal{F}[u]=n \int_{\Omega} \d{x}\d{y} f(\chi;J,T),
\end{equation}
where $n = (\Delta x\Delta y)^{-1}$ is the surface density of nodes and the free-energy density reads
\begin{equation}
    f(\chi;J,T) = \frac{1}{2}\chi^2 - T \ln \zeta^{(2)}(\chi/T,J/T).
    \label{eq:free-energy_density_2d_spin-membrane}
\end{equation}
Following the same notation as in the one-dimensional case, $\chi(x,y)=\nabla^2 u(x,y)$ is the local curvature of the membrane, and $\zeta^{(2)}(\chi/T,J/T)$ is the partition function of the two-dimensional Ising model with coupling $J$ and external field $\chi$ at temperature $T$. Unfortunately, in contrast with the one-dimensional case, there is no analytical expression for $\zeta^{(2)}(\chi/T,J/T)$ when $J\neq0$ \cite{journalarticle:Onsager_StatisticalHydrodynamics_NuovoCimento49}. 

Equilibrium profiles minimises $\mathcal{F}[u]$, so they are determined by the condition $\delta \mathcal{F}=0$. Since the free-energy density only depends on the curvature $\chi$, it is convenient to define the field 
\begin{equation}
    \Phi(x,y) \equiv \frac{\partial f}{\partial \chi} = \chi - \pdev{}{\ln \zeta^{(2)}(\chi/T,J/T)}{\chi}.
\end{equation}
Then, the variational principle can be written as 
\begin{align}
    \delta F[u] &= \int_\Omega \d{x} \d{y} \Phi \,\nabla^2 (\delta u) \nonumber \\
    &= \int_\Omega \d{x} \d{y} \nabla\!\cdot\!\brackets{ \Phi \, \nabla(\delta u)-\nabla \Phi \delta u}
   + \! \int_\Omega \d{x} \d{y} \nabla^2\Phi \delta u \nonumber \\
   &= \int_{\partial\Omega} \d{s} \bm{n}\!\cdot\!\parenthesis{ \Phi \, \nabla(\delta u)-\nabla \Phi \,\delta u}
   + \int_\Omega \d{x} \d{y}  \nabla^2\Phi \,\delta u,
   \label{eq:deltaF-2d}
\end{align}
which is the two-dimensional version of~\eqref{eq:spin-string_variation_free_energy}. In the contour integral along the boundary $\partial \Omega$, $\text{d}s$ is the length element and $\bm{n}$ is the outward pointing unit normal vector. The surface integral provides us the Euler-Lagrange equation, which results to be the Laplace equation for the field $\Phi(x,y)$,
\begin{equation}
    \nabla^2 \Phi = 0,
    \label{eq:2d_spin-membrane_LaplaceEquation}
\end{equation}
which is the two-dimensional version of~\eqref{eq:1d_spin-string_variational_integralterm}. The boundary conditions are obtained from the contour integral, imposing that the two terms therein vanish separately, 
\begin{equation}
    \Phi \frac{\partial}{\partial n}\delta u =0, \quad \delta u \frac{\partial}{\partial n}\Phi=0,
\end{equation}
where $\partial/\partial n = \bm{n}\cdot\nabla$. For consistency with the analysis done in the spin-string model, let us assume supported boundary conditions: the membrane displacement vanishes at the contour, 
\begin{equation}
    u(x,y)=0, \quad (x,y)\in\partial\Omega,
    \label{eq:supported_boundary_conditions_membrane_1}
\end{equation}
but the value of the normal derivative $\partial u /\partial n$ is free. Therefore, the field must also vanish at the boundaries, 
\begin{equation}
    \Phi(x,y) = 0, \quad (x,y)\in\partial\Omega.
    \label{eq:supported_boundary_conditions_membrane_2}
\end{equation}
Interestingly, the combination of the Euler-Lagrange equation~\eqref{eq:2d_spin-membrane_LaplaceEquation} with the homogeneous Dirichlet boundary condition~\eqref{eq:supported_boundary_conditions_membrane_2} tells us that the unique possible solution for $\Phi$ is the trivial one, \ie the field $\Phi(x,y)$ vanishes everywhere in the domain $\Omega$, or equivalently, the curvature $\chi(x,y)$ is a solution of the implicit algebraic equation
\begin{equation}
    \chi - \pdev{}{\ln \zeta^{(2)}(\chi/T,J/T)}{\chi} = 0.
    \label{eq:2d_spin-membrane_Euler-Lagrange}
\end{equation}
Although we do not have an explicit expression for $\zeta^{(2)}(\chi/T,J/T)$, equation~\eqref{eq:2d_spin-membrane_Euler-Lagrange} tells us the equilibrium curvature is spatially homogeneous, an independent function of the coordinates $\chi(x,y)=\chi_{\eq}$, as in the one-dimensional case. Thus, the equilibrium free energy is again an extensive quantity 
\begin{equation}
    \mathcal{F}_{\eq}(J,T) \equiv N f(\chi_{\eq};J,T).
\end{equation}

Let us consider a certain equilibrium profile $u_{\eq}(x,y)$ with curvature $\chi_{\eq}$, which is a solution of~\eqref{eq:2d_spin-membrane_Euler-Lagrange}. This profile and any other rigid transformation of it, \ie $u_{\eq}'(x,y)=u_{\eq}(x,y)+Ax+By+C$,\footnote{The family of transformations defined by $Ax+By+C$ contains any small two-dimensional rigid rotation.} have associated the same free energy, so our two-dimensional proposal keeps the rotational invariance of the spin-string model. The functional form of $u_{\eq}(x,y)$ is obtained by solving Poisson's equation with Dirichlet boundary conditions~\eqref{eq:supported_boundary_conditions_membrane_1},
\begin{equation}
    \nabla^2 u_{\eq} = \chi_{\eq}, \quad (x,y)\in\Omega; \quad u_{\eq}(x,y)=0, \quad (x,y)\in\partial\Omega.
    \label{eq:2d_spin-membrane_PoissonEquation}
\end{equation}
The shape of the profile depends on the geometry of the domain $\Omega$. The shape and the domain are related to the contour of the system---circular, rectangular or whatever it is. Note that it is important to distinguish the shape of the domain from the structure of the lattice---honeycomb, square, hexagonal, etc.

\begin{figure}
    \centering
    \includegraphics[width=0.6\textwidth]{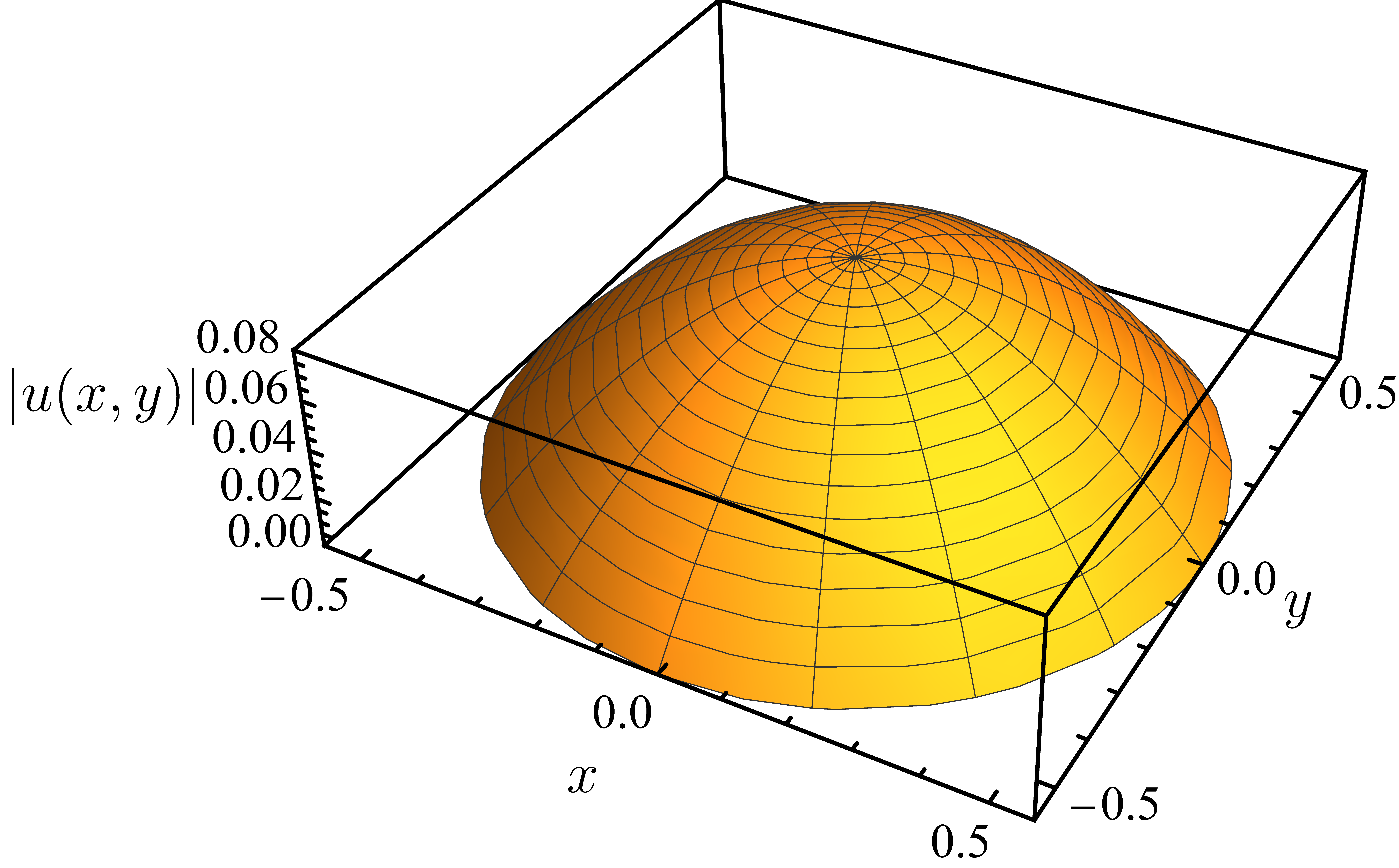}
    \caption{Low-temperature stable buckled configuration for the spin-membrane model. The plotted surface corresponds to the absolute value of~\eqref{eq:spin-membrane_profile_circular_domain} for a unit area circle. 
    }
    \label{fig:spin-membrane_circular_domain}
\end{figure}

The simplest situation appears for a circular membrane of radius $R$; therein, the solution only depends on the distance $r$ to the centre of the membrane and a straightforward integration of the Poisson equation provides
\begin{equation}
    u_{\eq}(r) = \frac{\chi_{\eq}}{4}(r^2-R^2), \quad r\in[0,R].
    \label{eq:spin-membrane_profile_circular_domain}
\end{equation}
The profile for a circular membrane of unit area is shown in figure~\ref{fig:spin-membrane_circular_domain}. Therein, the curvature equals unity, which corresponds to the stable buckled phase $\BuckledStable$ in the low-temperature limit, as we prove later in~\ref{subsec:2d_spin-membrane_low-temperature_limit}. 

\begin{figure}
    \centering
    \includegraphics[width=0.6\textwidth]{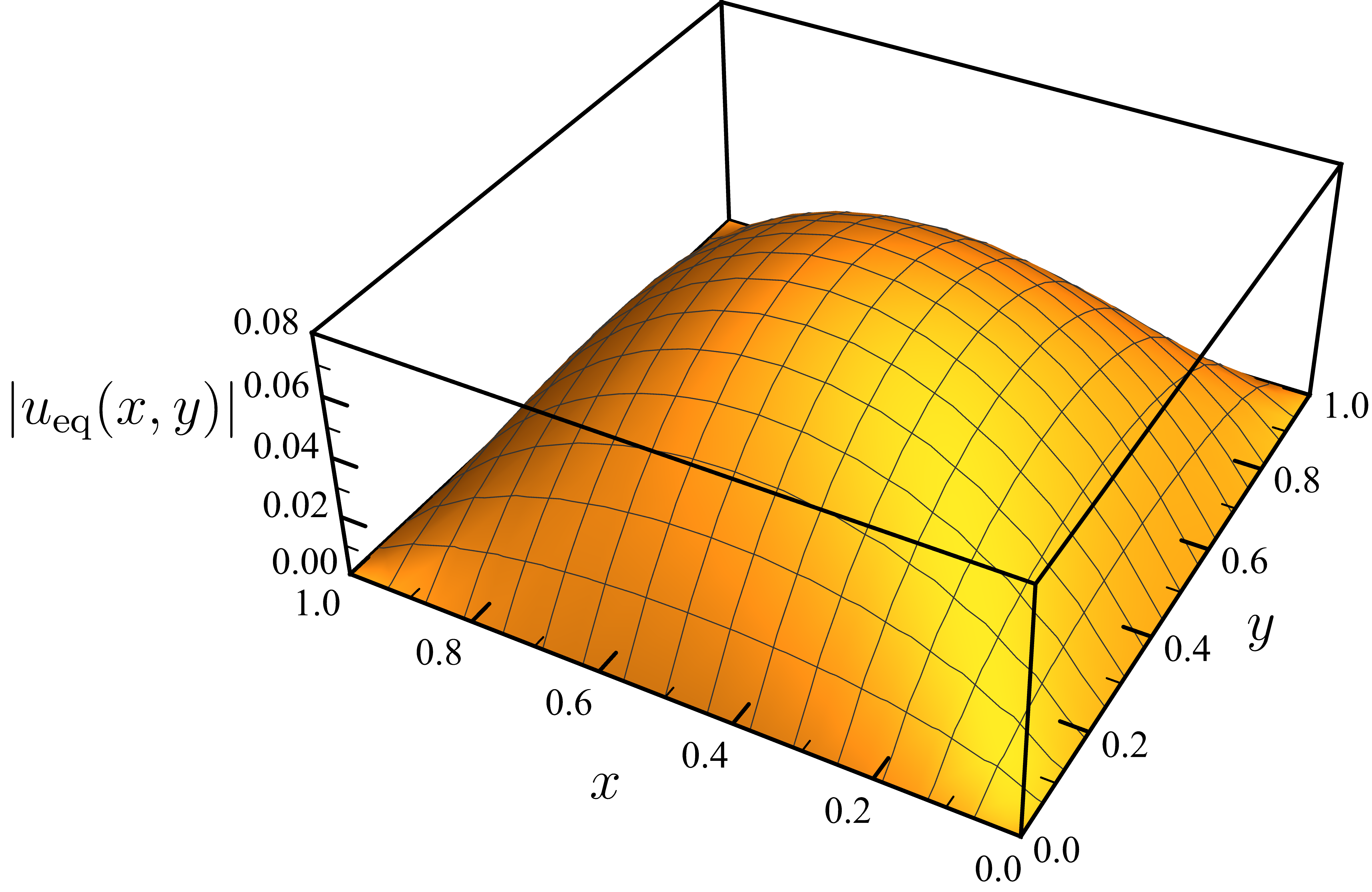}
    \caption{Absolute value of the stable buckled state on a square domain.  The graph corresponds to the low-temperature limit, in which $\chi_\eq=1$, for a unit area square with supported boundary conditions.}
    \label{fig:spin-membrane_rectangular_domain}
\end{figure}

Another possibility is to consider a rectangular domain of sides $L_x$ and $L_y$. We propose the solution
\begin{equation}
    u_{\eq}(x,y)=\chi_{\eq}\brackets{\frac{x(x-L_x)}{2}+w(x,y)+z(x,y)}.
    \label{eq:buckled_profile_rectangular_domain}
\end{equation}
The first term on the rhs verifies the Poisson's equation and the homogeneous Dirichlet boundary conditions at the sides $x=0$ and $x=L_x$, but not at the sides $y=0$ and $y=L_y$. The other functions $w(x,y)$ and $z(x,y)$ must satisfy Laplace equation and are introduced to ensure the boundary conditions at $y=0$ and $y=L_y$, specifically
\begin{subequations}
    \begin{align}
        w(0,y) &= w(L_x,y) = w(x,0) = 0, 
        \\
        w(x,L_y) &= -x(x-L_x)/2,
        \\
        z(0,y) &= z(L_x,y) = z(x,L_y) = 0,
        \\
        z(x,0) &= -x(x-L_x)/2.
    \end{align}
\end{subequations}
Applying separation of variables, we obtain the following Fourier series:
\begin{subequations}
    \label{eq:buckled_profile_rectangular_domain_wz}
    \begin{align}
        w(x,y) &= \frac{4L_x^2}{\pi^3}\sum_{n\, \text{odd}}\frac{\sinh(n\pi y/L_x)}{\sinh(n\pi L_y/L_x)}\frac{\sin(n\pi x/L_x)}{n^3},
        \\
        z(x,y) &= \frac{4L_x^2}{\pi^3}\sum_{n\, \text{odd}}\frac{\sinh(n\pi(L_y-y)/L_x)}{\sinh(n\pi L_y/L_x)}\frac{\sin(n\pi x/L_x)}{n^3}.
    \end{align}
\end{subequations}
Figure~\ref{fig:spin-membrane_rectangular_domain} illustrates the solution~\eqref{eq:buckled_profile_rectangular_domain}, where the functions in~\eqref{eq:buckled_profile_rectangular_domain_wz} have been calculated with a cutoff at $n=11$.

\section{Existence of buckled states: analytical insight for certain limits}
\label{sec:2d_spin-membrane_limits}

Even though we are not able to compute the exact expression for $\zeta^{(2)}(\chi/T,J/T)$, we can analytically work it out in two limits: (i) absence of spin-spin coupling, $(J=0,T\neq0)$, and (ii) low temperatures $(J\neq0,T=0)$. 


\subsection{Absence of spin-spin interaction: second-order phase transition}
\label{subsec:2d_spin-membrane_no_spin-spin_interaction}

Let us consider the case $J=0$ and $T\neq0$. The partition function $\zeta^{(2)}(\chi/T,0)$ of the non-interacting two-dimensional Ising model with external field $\chi$ can be obtained \cite{book:Landau.Lifshitz_StatisticalPhysicsVolume_13} and the free-energy density in that case reads 
\begin{equation}
    f(\chi;0,T) = \frac{1}{2}\chi^2 - T \ln\brackets{2\cosh\parenthesis{\frac{\chi}{T}}},
\end{equation}
and the Euler-Lagrange for the curvature reduces to 
\begin{equation}
    \chi_{\eq} (J=0,T) = \tanh\parenthesis{\frac{\chi_{\eq}(J=0,T)}{T}}.
\end{equation}
This coincides with the one-dimensional equation~\eqref{eq:1d_spin-string_Euler-Lagrange} for $J=0$---this is logical, in the absence of spin-spin interactions, the partition function is obtained as $N$ times the partition function of a single spin, being $N$ the number of spins, since it does depend neither on the dimension nor on the geometry.

For $T>1$, the only solution is $\chi_{\eq}^{\zc}=0$, which corresponds to an absolute minimum of the free energy. On the other hand, for $T<1$, one buckled phase with non-zero curvature $\pm\chi_{\eq}^{\Buckled}$ appears, which is the most stable one, whereas $\chi_{\eq}^{\zc}=0$ becomes unstable. At $T=1$, a critical behaviour emerges, with the stable buckled phase $B$ continuously bifurcating from the $\zc$ phase, namely 
\begin{equation}
    \chi_{\eq}^{\Buckled} \sim \pm\sqrt{3(1-T)}, \quad \text{for}\ T\to1^{-},
\end{equation}
which exactly coincides with the one-dimensional result~\eqref{eq:buckled_solutions_J=0_1d}.

\subsection{Low-temperature limit: first-order phase transition}
\label{subsec:2d_spin-membrane_low-temperature_limit}

Let us now consider the low-temperature limit $T\ll1$, with $J\neq0$, in the case of the honeycomb lattice---the case of the square lattice is completely analogous. Similarly to the one-dimensional case, we define the ground state contribution of the spin variables to the free-energy density as
\begin{equation}
    e_{\gs}^{(2)} = \lim_{T\to0^+} -T \ln \zeta^{(2)}(\chi/T,J/T) = - (|\chi|-3J)\Theta(|\chi|-3J)-\frac{3}{2}J.
    \label{eq:ground-state_energy_2d_spin-membrane}
\end{equation}
Analogously with the one-dimensional case, this expression has a clear physical interpretation. For small curvature, $|\chi|<3J$, the spin-spin coupling dominates and the system is ordered according to $J$, $e_{\gs}^{(2)}=-3J/2$. In the opposite limit, $|\chi|>3J$, the spins are aligned with the curvature, so $e_{\gs}^{(2)}=-|\chi|+3J/2$. Notice that this result does depend on the topology we are analysing: once the spin-spin coupling is taking into account, the coordination number, \ie the number of nearest neighbours, plays a crucial role. We have found that this behaviour of the ground state always appears in any dimension as long as the lattice does not have triangular loops---see appendix~\ref{app:ground-state_triangular_loops} for more details.

The free-energy density is then
\begin{equation}
    f(\chi;J,T) \sim \frac{1}{2}\chi^2 - (|\chi|-3J)\Theta(|\chi|-3J)-\frac{3}{2}J, \quad T\ll1,
\end{equation}
and the equilibrium curvature is the solution of 
\begin{equation}
    |\chi_{\eq}| = \Theta(|\chi_{\eq}|-3J), \quad T\ll1.
    \label{eq:2d_spin-membrane_Euler-Lagrange_lowT}
\end{equation}
The analysis of the low-temperature limit profiles follows along the same lines as in the one-dimensional case, so we only indicate the quantitive results that are obtained. The $\zc$ phase is always a solution~\eqref{eq:2d_spin-membrane_Euler-Lagrange_lowT}, being locally stable for all $J$. Additionally, two buckled phases emerge for $3J\leq1$, \ie $J\leq J_{M}^{(0)}=1/3$. For $J>J_{M}^{(0)}$, the antiferromagnetic coupling is strong enough to destroy the buckled states, so the unique stable phase is $\zc$. 

The curvatures of the buckled states are $\chi_{\eq}=3J$ and $\chi_{\eq}=1$. Their stability can be elucidated by evaluating their free-energy density difference with respect to the $\zc$ phase, $f_0(J,T)=-3J/2$. Therefore,
\begin{subequations}
    \begin{align}
        \Delta f (\chi_{\eq}=3J;J,T\to0) &\sim \frac{9}{2}J^2, \\
        \Delta f (\chi_{\eq}=1;J,T\to0) &\sim -\frac{1}{2}+3J.
    \end{align}
\end{subequations}
The phase with $\chi_{\eq}=3J$ has always a larger free energy than the $\zc$ phase, $\Delta f\geq0$, so it corresponds to the low-temperature limit of the unstable phase $\BuckledUnstable$. The free energy difference of the phase with $\chi_{\eq}=1$ changes sign at $J_{t}^{(0)}=1/6$: it is the low-temperature limit of phase B+, being the most stable one for $0<J<J_{t}^{(0)}$, and metastable for $J_{t}^{(0)}<J<J_{M}^{(0)}$. The transition from the $\zc$ phase to the buckled phase B+ is first-order, since there is an abrupt change of the curvature at $J_{t}^{(0)}$.

The above identification of the phases is physically sensible, since in the limit $J\to0$, the B+ phase for $T\ll 1$ converges to the buckled phase B for $J=0$ in the limit $T\to0^+$.

{\clearpage \thispagestyle{empty}}
\part{Realistic implementations of stochastic resetting in search processes}
\label{part:resetting}

\chapter{Stochastic resetting with refractory periods}
\label{ch:resetting_refractory}
In this chapter, we investigate a particular two-stage search strategy: the stochastic resetting with refractory periods. In general, a refractory period is envisioned as a recovery time paid after performing some action, \eg it may follow a resetting event \cite{journalarticle:Evans.Majumdar_DiffusionStochasticResetting_Phys.Rev.Lett.11,journalarticle:Evans.Majumdar_DiffusionOptimalResetting_J.Phys.A:Math.Theor.11,journalarticle:Evans.etal_StochasticResettingApplications_J.Phys.A:Math.Theor.20}. Concretely, we are focusing on an intermittent dynamical model where a particle is spreading following a conventional Brownian motion (phase $A$), which experiences instantaneous resets to its initial position at the end of the exploration phase. Then, the particle remains motionless (phase $B$), the refractory period, before resuming the exploration again. Even though it still involves instantaneous resets, these class of models is a physically sound approach to certain chemical and biological reactions. A paradigmatic example concerns the nervous system, where neurons fire electric signals that are followed by a quiescent state, \ie an ineffective time to any stimulus \cite{journalarticle:Fetz.Gustafsson_RelationShapesPostsynaptic_J.Physiol.83,booksection:Maida_Chapter2Cognitive_CognitiveComputing:TheoryandApplications16}.
Furthermore, stochastic resetting with refractory periods has been shown to be useful in the context of enzymatic reactions following the Michaelis-Menten scheme \cite{journalarticle:Reuveni.etal_RoleSubstrateUnbinding_Proc.Natl.Acad.Sci.U.S.A.14,journalarticle:Rotbart.etal_MichaelisMentenReactionScheme_Phys.Rev.E15,journalarticle:Pal.etal_SearchHomeReturns_Phys.Rev.Res.20,journalarticle:Reuveni_OptimalStochasticRestart_Phys.Rev.Lett.16}. Therein, an enzyme interacts with a substrate in a reversible binding-unbinding reaction, which, in a second step from the bound state, releases a certain product. The enzymatic dynamics can be enhanced by the unbinding step, which can be interpreted as a refractory period (waiting time before the release) after the resetting event (binding). 

Understanding this model as an intermittent search process, our general analysis of the observables of interest relies on the pathway formulation introduced in section~\ref{subsec:pathway_formulation}. Specifically, exact results for the case of Poissonian resets with Poissonian refractory periods are derived---the evolution of the PDF of a resetting Brownian particle in an infinite domain with refractory periods is explicitly worked out. Moreover, we obtain the MFPT as a function of the rates governing the exponential distributions for the duration of both phases. 

The chapter is organised as follows. The fundamental ingredients of the model are described in section~\ref{sec:refractory_model}. Section~\ref{sec:refractory_evolution} is devoted to the detailed analysis of the PDF of the stochastic process through our pathway formulation. We explicitly obtain the whole evolution of the PDF, which reaches a NESS in the long-time limit. Section~\ref{sec:refractory_MFPT} deals with the MFPT. In addition to obtaining a general expression for the MFPT with refractory periods within our framework, an explicit formula is derived for the case of both Poissonian resets and refractory periods. Finally, the optimal resetting rate is obtained as a function of the rate that governs the duration of the refractory periods.

\section{Ingredients of the model}
\label{sec:refractory_model}

Let us consider a two-stage stochastic resetting process, where we are able to apply the analysis done in section~\ref{subsec:pathway_formulation}. Herein, the system is a particle that, in the absence of resetting, stochastically evolves following a given propagator $\PDF_0$.\footnote{Following the notation used in section~\ref{subsec:pathway_formulation}, the phase $A$ is given by the free diffusive propagator~\eqref{eq:Green_function_BrownianMotion}, $K^{(A)}(x,t|x_0,t_0)=\PDF_0(x,t|x_0,t_0)$.} On top of this natural dynamics, random resets to the initial position $x_r=x_0=x(t_0 = 0)$ occur. Time events at which the particle instantaneously suffers resetting events to $x_0$ are again denoted by $t_i$, where the subscript $i=1,2,\ldots$, stands for the order of occurrence. Using the same notation as section~\ref{subsec:pathway_formulation}, the PDF that a resetting event occurs after an exploration time $\deltaPhase{\free}$ is $f\parenthesis{\deltaPhase{\free}}$; whereas $F\parenthesis{\deltaPhase{\free}}=\int_{\deltaPhase{\free}}^\infty \d{t} f(t)$ is the probability that no resetting events have occurred up to time $\deltaPhase{\free}$. In other words, $F(\deltaPhase{\free})$ is the probability of having an uninterrupted exploration phase lasting at least $\deltaPhase{\free}$. 

\begin{figure}
    \centering
    \includegraphics[width=0.7\textwidth]{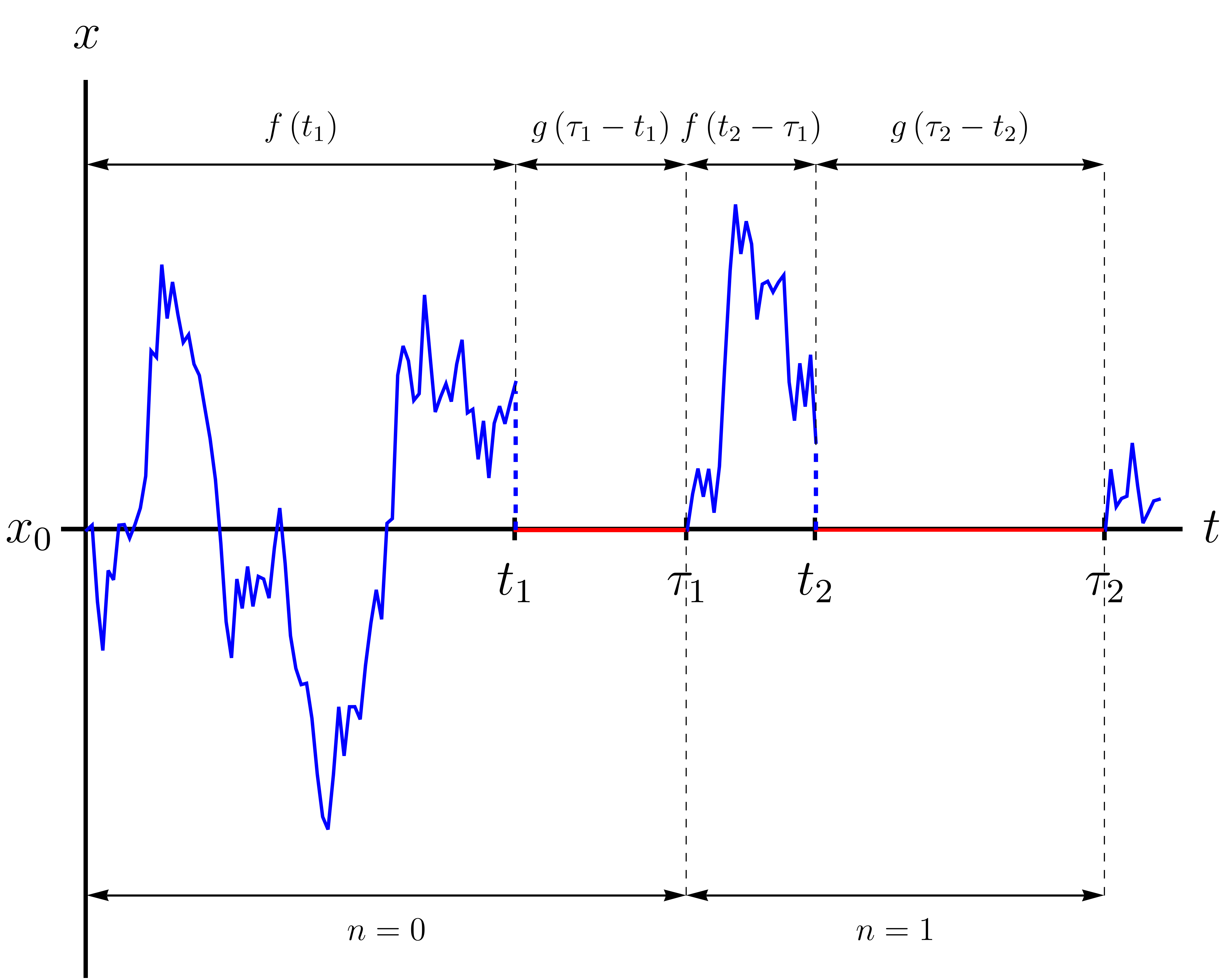}
    \caption{Single trajectory for stochastic resetting with refractory periods. The labels $t_i$, $i=1,2,\ldots$, mark a new reset to $x_0$ where the exploration phase (blue line) ends with an instantaneous reset (dashed blue line) and the refractory period begins (red line). Analogously, $\tau_i$ marks the end of the refractory phase. Herein, $n$ stands for the number of times the system is renewed, for $\tau_n < t < \tau_{n+1}$. The duration $\delta$ of each exploration (refractory) phase comes from the PDF $f(\delta)$ ($g(\delta)$).}
    \label{fig:refractory_model}
\end{figure}

As previously discussed, instantaneous resets are difficult to motivate within a realistic dynamics, since they involve no payment---neither energetic nor temporal. In contrast to SSR, here we consider refractory periods that entail a temporal cost after each event \cite{journalarticle:Evans.Majumdar_EffectsRefractoryPeriod_J.Phys.A:Math.Theor.19,journalarticle:Maso-Puigdellosas.etal_StochasticMovementSubject_J.Stat.Mech.19}. Specifically, the particle is assumed to be motionless at $x_0$ after the $i$-th resetting event up to time $\tau_i$, $i=1,2\ldots$,\footnote{For consistency, $\tau_0=0$.} for an independent refractory time $\deltaPhase{\refr}_i=\tau_i-t_i$.\footnote{Bringing back the notation in section~\ref{subsec:pathway_formulation}, the propagator would be $K^{(B)}(x,t|x_0,t_0;\delta^{(A)},\delta^{(B)})=\delta(x-x_0)$, and the waiting time distribution does not depend on the previous time, so $g(\deltaPhase{\refr}|\deltaPhase{\free}) = g(\deltaPhase{\refr})$.} 
This time interval is a stochastic variable drawn from the PDF $g(\tau_i-t_i)=g\parenthesis{\deltaPhase{\refr}}$, and its integral $G\parenthesis{\deltaPhase{\refr}}=\int_{\deltaPhase{\refr}}^\infty \d{t} g(t)$ is the probability of not having finished the refractory period after a time interval $\deltaPhase{\refr}$. An illustrative depiction of the whole dynamics for a one-dimensional model is shown in figure~\ref{fig:refractory_model}, where blue and red lines stand for the exploration and refractory phases, respectively.

The renewal structure of the dynamics can be exploited, despite the inclusion of refractory periods. In fact, let us consider the first resetting event being at $t_1$ and the first renewal being at $\tau_1$; then the PDF $\PDF_r(x,t|x_0)$ of finding the particle at position $x$, starting from $x_0$ at $t=0$, after a time evolution of duration $t$ is given by
\begin{align}
    \PDF_r(x,t|x_0) &= F(t)\PDF_0(x,t|x_0)
    +\int_0^t \d{t_1} f(t_1) G(t-t_1) \delta (x-x_0) \nonumber\\
    &\quad +\int_0^t \d{t_1} f(t_1) \int_{t_1}^t \d{\tau_1} g(\tau_1-t_1) \PDF_r(x,t-\tau_1|x_0),
    \label{eq:refractory_renewal_structure}
\end{align}
Note that we are writing $\PDF_r(x,t|x_0)$ using the first renewal approach, as in~\eqref{eq:first_renewal_equation_standard_SR}. Similarly to that case, the first term on the right-hand side of~\eqref{eq:refractory_renewal_structure} is the contribution of the pathways where there has been no resetting up to time $t$, thus weighted by the probability $F(t)$. The second term stems from pathways for which there has been a resetting event in the interval $(0,t)$ and the subsequent refractory phase has not ended at time $t$; therefore, it contributes with $\delta(x-x_0)$. The last term comes from trajectories in which the particle has been reset at $t_1\in(0,t)$, has had a subsequent refractory period ending at $\tau_1\in(t_1,t)$, and has reached $x$ at time $t$ following its renewed dynamics in the interval $(\tau_1,t)$.

\section{Evolution of the probability density function} 
\label{sec:refractory_evolution}
The aim of our work is to find out an explicit expression for $\PDF_r(x,t|x_0)$ at any time, thus going beyond the solution for the PDF that can be found in the literature \cite{journalarticle:Evans.Majumdar_EffectsRefractoryPeriod_J.Phys.A:Math.Theor.19,journalarticle:Maso-Puigdellosas.etal_StochasticMovementSubject_J.Stat.Mech.19}. 

The density probability function $\PDF_r(x,t|x_0)$ can be computed using the pathway formulation introduced in section~\ref{subsec:pathway_formulation}, based on how many renewals (cycles) of the dynamics have occurred up to time $t$, 
\begin{eqnarray}
    \PDF_r(x,t|x_0) = \sum_{n=0}^\infty \bigg[
        \PDF_n^{\free}(x,t|x_0) + \PDF_n^{\refr}(x,t|x_0).
    \bigg]
    \label{eq:refractory_pathway_expansion}
\end{eqnarray} 
On the right-hand side we have distinguished between the contributions of each phase---exploration $\free$ and refractory $\refr$, depending on the final state of the evolution at time $t$.

The general expressions for these contributions are computed by substituting the corresponding propagators---$K^{(A)}(x,t|x_0,t_0)$ for free diffusion~\eqref{eq:Green_function_BrownianMotion} and $K^{(B)}(x,t|x_0,t_0;\allowbreak \delta^{(A)},\delta^{(B)})=\delta(x-x_0)$ for the refractory period, and waiting time distributions in~\eqref{eq:pathway_formulation_time_1} and~\eqref{eq:pathway_formulation_time_2},
\begin{subequations}
    \begin{align}
        \PDF^{\free}_n(x,t|x_0, t_0 = 0) = \prod_{i=1}^n &\brackets{
            \int_{\tau_{i-1}}^t \d{t_i} f(t_i - \tau_{i-1}) \int_{t_i}^t \d{\tau_i} g(\tau_i - t_i)} 
        \nonumber \\ & \qquad \times F(t-\tau_n) \PDF_0 (x,t|x_0,\tau_n),
        \label{eq:refractory_pathway_contributions_free}
        \\
        \PDF^{\refr}_n(x,t|x_0,t_0 = 0) = \prod_{i=1}^n &\brackets{
            \int_{\tau_{i-1}}^t \d{t_i} f(t_i - \tau_{i-1}) \int_{t_i}^t \d{\tau_i} g(\tau_i - t_i)} 
        \nonumber \\
            & \qquad \times \int_{\tau_n}^t \d{t_{n+1}} f(t_{n+1}-\tau_{n})    
            G(t-t_{n+1})\delta(x-x_0),
        \label{eq:refractory_pathway_contributions_refr}
    \end{align}
\end{subequations}
where we recall that $\tau_0=0$ for the sake of a compact notation.\footnote{Note that~\eqref{eq:refractory_pathway_contributions_free} and~\eqref{eq:refractory_pathway_contributions_refr} admit a recurrence relation between two consecutive renewals, specifically $\PDF_n^{(s)}(x,t|x_0)=\int_{0}^{t_1}\d{t_1} f(t_1)\int_{t_1}^{t} \d{\tau_1} g(\tau_1-t_1) \PDF_{n-1}^{(s)}(x,t|x_0,\tau_1)$.} Since this intermittent process is completely memoryless, we have directly written $h\parenthesis{\deltaPhase{\free},\deltaPhase{\refr}}=f\parenthesis{\deltaPhase{\free}}g\parenthesis{\deltaPhase{\refr}}$.

After applying the Laplace transform and summing over all the contributions, we obtain
\begin{equation}
    \label{eq:propagator_each_phase_LaplaceDomain_refractory}
    \laplTransformTilde{\PDF^{\free}}{x,s|x_0} = \frac{1}{1-\laplTransformTilde{f}{s}\laplTransformTilde{g}{s}} \laplTransformTilde{F \PDF_0}{x,s|x_0}, 
    \quad
    \laplTransformTilde{\PDF^{\refr}}{x,s|x_0} = \frac{\laplTransformTilde{f}{s}\laplTransformTilde{G}{s}}{1-\laplTransformTilde{f}{s}\laplTransformTilde{G}{s}} \delta(x-x_0),
\end{equation}
which is a particularisation of~\eqref{eq:pathway_formulation_generalStructure_LaplaceDomain} to our situation. This general expression---valid for any $\PDF_0$, $f$ and $g$---coincides with the one obtained in \cite{journalarticle:Maso-Puigdellosas.etal_StochasticMovementSubject_J.Stat.Mech.19}, which reduces to equation (4) in \cite{journalarticle:Evans.Majumdar_EffectsRefractoryPeriod_J.Phys.A:Math.Theor.19} when diffusive propagation and Poissonian resetting are assumed. 

\subsection{Poissonian resetting and refractory period}
\label{subsec:refractory_Poissonian_dynamics}
Let us consider now that the resetting events and the refractory periods both follow exponential distributions, but with different rates:
\begin{subequations}
    \begin{align}
        f(t)&=r_1 e^{-r_1 t} \implies F(t)=e^{-r_1 t},\label{eq:refractory_DistributionResetting} \\
        g(t)&=r_2 e^{-r_2 t} \implies G(t)=e^{-r_2 t}. \label{eq:refractory_DistributionRefractory}
    \end{align}
\end{subequations}
The Laplace transform of $f(t)$ and its integral are, respectively, $\laplTransformTilde{f}{s}=r_1/(r_1+s)$ and $\laplTransformTilde{F}{s}=1/(r_1+s)$, with analogous expressions for $\laplTransformTilde{g}{s}$ and $\laplTransformTilde{G}{s}$ with the exchange $r_1\leftrightarrow r_2$.
Then, the Laplace transforms in~\eqref{eq:propagator_each_phase_LaplaceDomain_refractory} turn out to be
\begin{subequations}
    \begin{align}
        \laplTransformTilde{\PDF^{\free}}{x,s|x_0} &= \laplTransformTilde{\PDF_0}{x,s+r_1|x_0} + \frac{r_1 r_2}{r_1+r_2}\parenthesis{\frac{1}{s}-\frac{1}{s+r_1+r_2}}\laplTransformTilde{\PDF_0}{x,s+r_1|x_0}, 
        \\
        \laplTransformTilde{\PDF^{\refr}}{x,s|x_0} &= \frac{r_1}{r_1+r_2}\parenthesis{\frac{1}{s}-\frac{1}{s+r_1+r_2}}\delta(x-x_0),
    \end{align}
\end{subequations}
which can be readily inverted,
\begin{subequations}
    \begin{align}
        \PDF^{\free} (x,t|x_0)&=
        e^{-r_1 t}\PDF_0(x,t|x_0)
        +\frac{r_2}{r_1+r_2}r_1 e^{-r_1 t}\int_0^t \,\d{\tau}
        \parenthesis{e^{r_1 \tau}-e^{-r_2 \tau}}\PDF_0(x,t-\tau|x_0),
        \label{eq:refractory_pprop_timedomain_exponential}
        \\
        \PDF^{\refr} (x,t|x_0)&=
        \frac{r_1}{r_1+r_2}\left(1-e^{-(r_1+r_2)t}\right)\delta(x-x_0).
        \label{eq:refractory_prefr_timedomain_exponential}
    \end{align}
\end{subequations}

The former expressions become especially simple if both distributions follow the same Poissonian statistics, \ie the exponential rates are equal, $r_1=r_2=r$,
\begin{subequations}
    \begin{align}
        \PDF^{\free} (x,t|x_0)&=
        e^{-r t}\PDF_0(x,t|x_0)+ r e^{-r t}\int_0^t \d \tau
        \sinh\left(r\tau\right)\PDF_0(x,t-\tau|x_0), 
        \\
        \PDF^{\refr} (x,t|x_0)&=
        e^{-rt}\sinh\left(r t\right)\delta(x-x_0),
    \end{align} 
\end{subequations} 
which can also be directly computed in the time domain from the pathway formulation~\eqref{eq:refractory_pathway_contributions_free} and~\eqref{eq:refractory_pathway_contributions_refr}. Plugging the exponential distributions therein, they become\footnote{Note that because both rates are equal, the integrals only depend on the last time variable. For that reason, the rest of them are simplified using $\int_0^t \d{t_1} \int_{t_1}^t \d{t_2} \ldots \int_{t_{n-1}}^t \d{t_n} = \parenthesis{\int_0^t d{t'}}^n/ n! =t^n/n!$. Additionally, $\sinh(x)=\sum_{n=0}^\infty x^{2n+1}/(2n+1)!$.}
\begin{align}
    \PDF_0^{\free} (x,t|x_0)
    &= e^{-rt} \PDF_0(x,t|x_0)
    \\ 
    \PDF_n^{\free} (x,t|x_0)
    &=r^{2n}e^{-rt}\int_0^t \d{t_1} \int_{t_1}^t \d{\tau_1} \int_{\tau_1}^t \d{t_2} \ldots
    \int_{t_n}^t \d{\tau_n}\, \PDF_0(x,t-\tau_n|x_0), 
    \nonumber\\
    &=r^{2n}e^{-rt}\int_0^t \d{\tau_n} \frac{\tau_n^{2n-1}}{(2n-1)!}\PDF_0(x,t-\tau_n|x_0),  \quad n \geq 1,
    \\
    \PDF_n^{\refr}(x,t|x_0) &= e^{-rt} \frac{(rt)^{2n+1}}{(2n+1)!}\delta(x-x_0).
\end{align}

\subsection{Relaxation to the non-equilibrium steady state}
\label{subsec:refractory_relaxation_NESS}
The long-time limit of the PDF $\PDF_r(x,t|x_0)$ is easily computed once the expressions in the Laplace domain are known. Specifically, using the final value theorem, it follows
\begin{equation}
    \PDF_r^{st}(x|x_0)=\lim_{t\to\infty} \PDF_r(x,t|x_0) 
    = \lim_{s\to 0} s \laplTransformTilde{\PDF_r}{x,s|x_0} 
    = \lim_{s\to 0} s \brackets{\laplTransformTilde{\PDF^{\free}}{x,s|x_0} + \laplTransformTilde{\PDF^{\refr}}{x,s|x_0}}.
\end{equation}

Considering the particular case of Poissonian resetting and refractory periods, the long-time behaviour for arbitrary rates $(r_1,r_2)$ is derived,
\begin{subequations}
    \begin{align}
        \lim_{t\to\infty} \PDF^{\free}(x,t|x_0)&=
        \frac{1}{2}\frac{r_2}{r_1+r_2}\sqrt{\frac{r_1}{D}}\exp\left[-\sqrt{\frac{r_1}{D}}|x-x_0|\right], \label{eq:refractory_NESS_pprop}
        \\
        \lim_{t\to\infty} \PDF^{\refr}(x,t|x_0) &=
        \frac{r_1}{r_1+r_2}\delta(x-x_0). \label{eq:refractory_NESS_prefr}
    \end{align}
\end{subequations}
Of course, these results are consistent with those obtained by taking the long-time limit in~\eqref{eq:refractory_pprop_timedomain_exponential} and~\eqref{eq:refractory_prefr_timedomain_exponential}, as well as with the results found in \cite{journalarticle:Evans.Majumdar_EffectsRefractoryPeriod_J.Phys.A:Math.Theor.19,journalarticle:Maso-Puigdellosas.etal_StochasticMovementSubject_J.Stat.Mech.19}. Herein, we have found that the resetting systems with refractory periods also reach a non-equilibrium steady state for exponential distributions. One can easily check that the normalisation of~\eqref{eq:refractory_NESS_pprop} and~\eqref{eq:refractory_NESS_prefr} is coherent with the fraction of time spent in each phase, \ie the probability of finding the system in each phase is given by 
\begin{eqnarray}
    \prob^{\free} = \frac{r_2}{r_1+r_2}, \quad \prob^{\refr} = \frac{r_1}{r_1+r_2}.
\end{eqnarray}
In the limit $r_2\to 0$, the system cannot escape from the refractory phase after the first resetting event, so $\prob^{\refr}\to 1$. On the other hand, when $r_2\to\infty$, the refractory period is negligible, and we recover the SSR result~\eqref{eq:NESS_standard_SR}.

\begin{figure}
    \centering
    \includegraphics[width=0.7\textwidth]{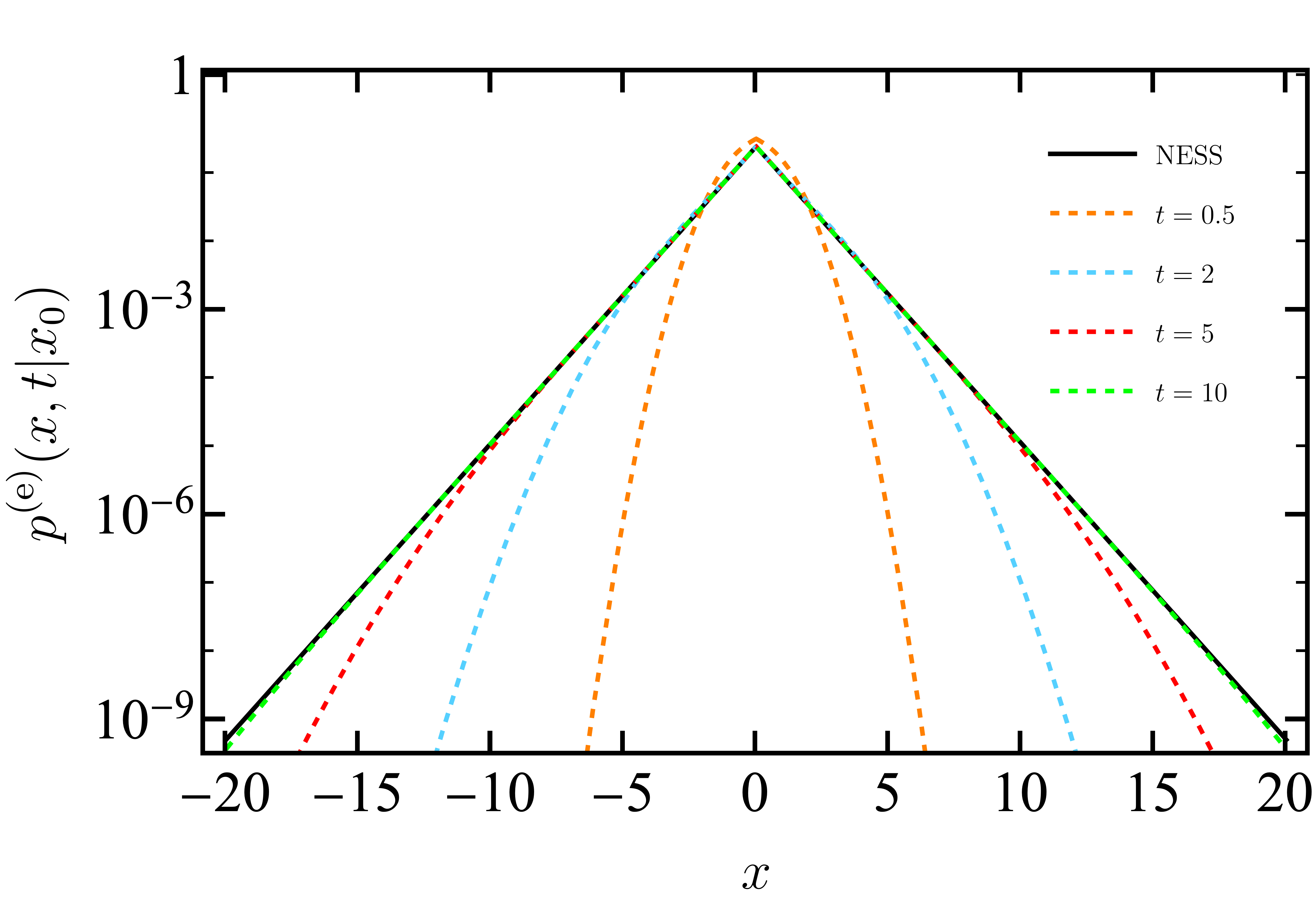} 
   \caption{PDF of the exploration phase with exponential waiting times for resetting and refractory periods. Numerical integration of $\PDF^{\free}(x,t|x_0)$ (colourful dashed lines), given by~\eqref{eq:refractory_pprop_timedomain_exponential}, at different times and the infinite time NESS~\eqref{eq:refractory_NESS_pprop} (solid black line) are shown. All the results correspond to $x_0=0$, $D=1$, $r_1=r_2=1$ as parameters.}
    \label{fig:refractory_Ness_NumericalIntegration}
\end{figure}

A representative picture of the evolution and convergence to the NESS is shown in figure~\ref{fig:refractory_Ness_NumericalIntegration}. Therein, it is remarkable that the convergence of the PDF is pretty similar to the one for SSR and other setups \cite{journalarticle:Majumdar.etal_DynamicalTransitionTemporal_Phys.Rev.E15,journalarticle:Gupta_StochasticResettingUnderdamped_JStatMechTheoryExp19}. As in section~\ref{subsec:standard_stochastic_resetting}, an inner core around $x_0$ that has almost reached the non-equilibrium steady state (solid black line) is reported. This region grows as $t$ increases, whereas the outer region is still relaxing. Thus, we can establish a dynamical separation between $|x-x_0| < \chi (t)$, where the system has relaxed, and a transient layer for $|x-x_0| > \chi (t)$.

This long-time behaviour can be analytically characterised. Let us rewrite~\eqref{eq:refractory_pprop_timedomain_exponential} as 
\begin{align}
    \PDF^{\free} (x,t|x_0)&=
    \frac{1}{\sqrt{4 \pi D t}} \exp \brackets{-t  \, \Phi_1 \parenthesis{1;\frac{x-x_0}{t} }}
    \nonumber\\
    &\quad +\frac{r_1 r_2}{r_1+r_2} \sqrt{\frac{t}{4\pi D }}
    \int_0^1 \,\frac{\d \omega}{\sqrt{\omega}}\,\exp \brackets{-t \, \Phi_1\left(\omega;\frac{x-x_0}{t}\right) }
    \nonumber \\
    &\quad -\frac{r_1 r_2}{r_1+r_2} \sqrt{\frac{t}{4\pi D }}
    e^{-(r_1+r_2)t} \int_0^1 \,\frac{\d \omega}{\sqrt{\omega}}\,\exp \brackets{ t \, \Phi_2\left(\omega;\frac{x-x_0}{t}\right) }  \label{eq:refractory_Laplace_prop},
\end{align}
where we have performed the change of variable $\omega = 1 - \tau/t$, and defined the functions
\begin{equation}
    \Phi_1(\omega;y)\equiv r_1 \omega + \frac{y^2}{4 D \omega}, \quad \Phi_2(\omega;y) \equiv r_2 \omega - \frac{y^2}{4 D \omega}.
\end{equation}
This integral can be approximated for long times $t\gg 1$ using Laplace's method \cite{book:Bender.Orszag_AdvancedMathematicalMethods_99}. The method consists of estimating the integral by computing the dominant contribution of the exponent, which is given by the minimum of the function $\Phi_1(\omega;y)$, or the maximum of $\Phi_2(\omega;y)$. Herein, we tackle the simplified analysis of the leading contribution of~\eqref{eq:refractory_Laplace_prop} for long times $t\gg 1$, emphasising intuitive ideas and focusing on the qualitative behaviour. A rigorous derivation of the complex formula for every regime is relegated to appendix~\ref{app:refractory_Laplace_method}, where all the terms and subtleties of Laplace's method are considered in detail.

On the one hand, $\Phi_2$ is a monotonically increasing function of $\omega$ and its maximum is always at the upper limit of the integration. The corresponding contribution is always subdominant against the non-integral term in~\eqref{eq:refractory_Laplace_prop}, \ie the one involving trajectories that have not suffered any resetting up to time $t$, as we prove at the end of appendix~\ref{app:refractory_Laplace_method}. On the other hand, $\Phi_1(\omega;y)$ presents a single absolute minimum at $\widetilde{\omega}=|y|/\sqrt{4 D r_1}$, since $r_1$ and $D$ are strictly positive. Depending on the position of $\widetilde{\omega}$ with respect to the integration interval $(0,1)$, we can distinguish two different regimes. If the minimum is inside the interval, $\widetilde{\omega}\in(0,1)$, the contribution becomes a Gaussian integral,
\begin{equation}
    \frac{r_1 r_2}{r_1+r_2} \sqrt{\frac{t}{4\pi D }}
    \int_0^1 \,\frac{\d \omega}{\sqrt{\omega}}\,\exp \brackets{-t \, \Phi_1\left(\omega;\frac{x-x_0}{t}\right) } \sim \frac{1}{2}\frac{r_2}{r_1+r_2}\sqrt{\frac{r_1}{D}}\exp\left[-\sqrt{\frac{r_1}{D}}|x-x_0|\right], 
\end{equation}
which corresponds to the NESS~\eqref{eq:refractory_NESS_pprop}. If the minimum is outside the interval, $\widetilde{\omega}>1$, the minimum within the integration interval is at the upper limit $\omega=1$. In this case, this contribution is also subdominant against the non-resetting term. 

Therefore, the PDF of the exploration phase is roughly estimated as 
\begin{equation}
    \PDF^{\free} (x,t|x_0)  \sim
    \left\{
    \begin{aligned}        
        \frac{1}{2}\frac{r_2}{r_1+r_2}\sqrt{\frac{r_1}{D}}\exp\left[-\sqrt{\frac{r_1}{D}}|x-x_0|\right], \quad |x-x_0| < \sqrt{4 D r_1}t,\\
        \frac{1}{\sqrt{4 \pi D t}}\exp\left[-r_1 t -\frac{(x-x_0)^2}{4 D t}\right], \quad |x-x_0| > \sqrt{4 D r_1}t.
    \end{aligned} \right. \label{eq:Approximation_NESS}
\end{equation}
As expected, we have found there are two clearly differentiated regimes in the dynamical evolution of $\PDF^{\free}(x,t|x_0)$, as already happened in SSR. On the one hand, an inner region around $x_0$ that has almost reached the non-equilibrium steady state~\eqref{eq:refractory_NESS_pprop} whose width grows linearly with time, $|x-x_0| / t < $$\sqrt{4 D r_1}$. On the other hand, outside that region, the system is still relaxing---dominated by the non-resetting contribution $F(t)\PDF_0(x,t|x_0)$, so a transient behaviour is observed. Interestingly, the change of behaviour does not depend on the inclusion of refractory periods, \ie the matching point $y=\sqrt{4 D r_1}$ is the same as for SSR~\eqref{eq:LDF_standard_SR}, regardless of $r_2$.

\begin{figure}
     \centering
     \begin{subfigure}[b]{0.48\textwidth}
         \centering
         \includegraphics[width=\textwidth]{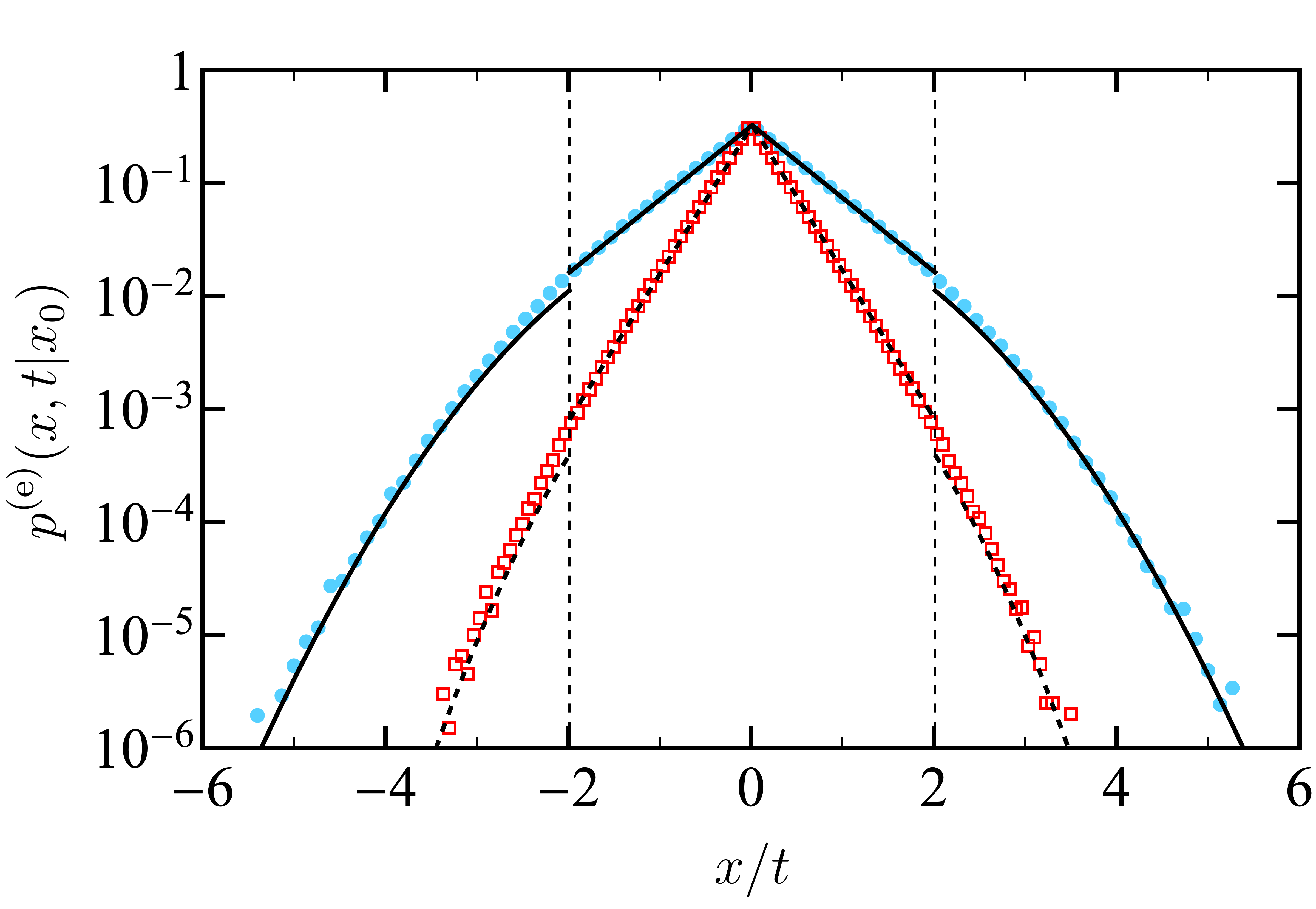}
         \caption*{}
     \end{subfigure}
     \begin{subfigure}[b]{0.48\textwidth}
         \centering
         \includegraphics[width=\textwidth]{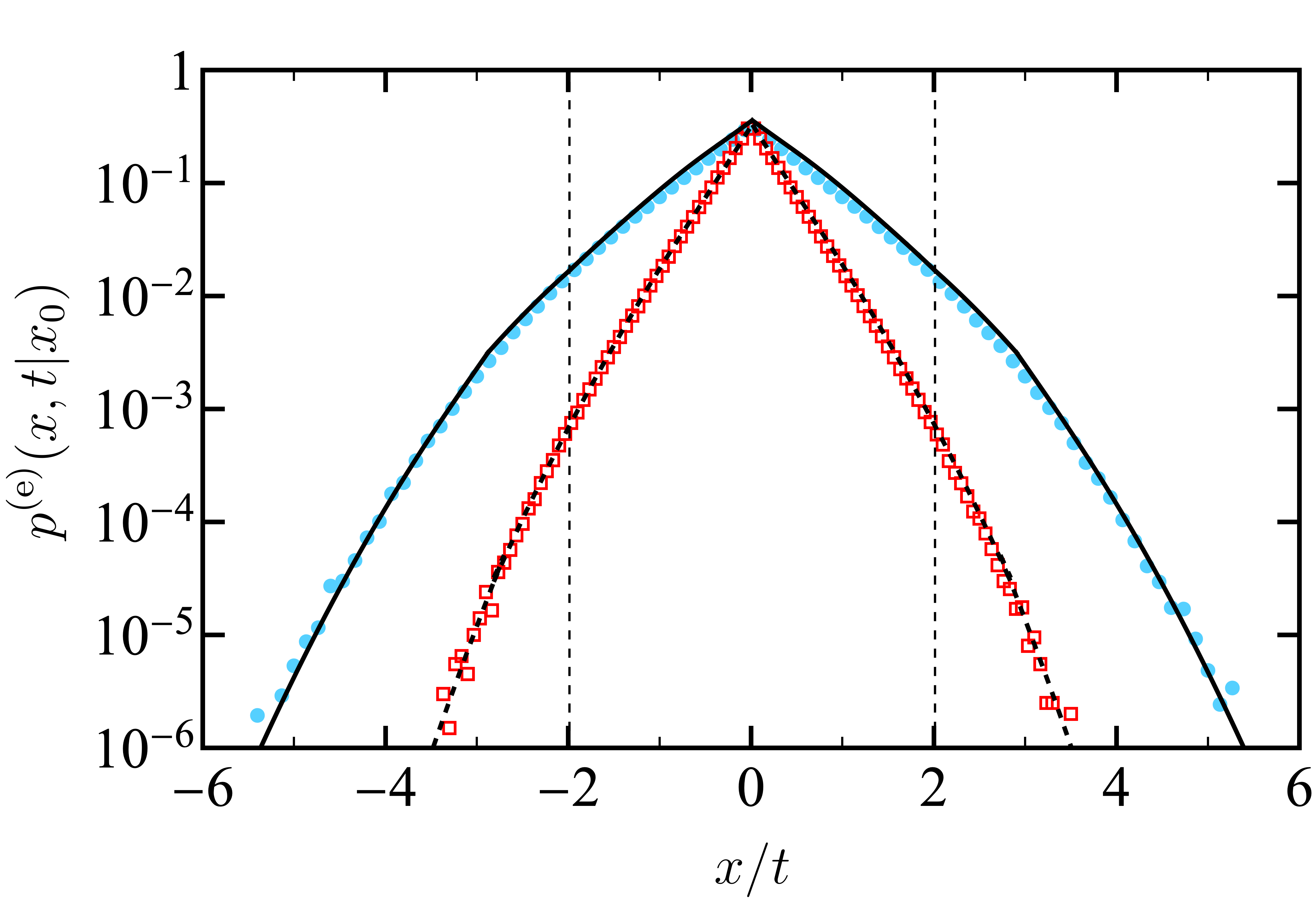}
         \caption*{}
     \end{subfigure}
        \caption{Comparison of the numerical and analytical PDFs of the exploration phase. Parameter values are $D=1$, $r_1=1$, $r_2=2$ and $x_0=0$. Symbols stand for numerical simulations for $t=1.5$ (light blue circles) and $t=3$ (red squares), while black lines stand for the analytical approximation obtained in~\eqref{eq:Approximation_NESS} (left panel) or~\eqref{eq:refractory_approximation_tendencyNESS} (right panel). Vertical dashed lines at $|x|/t= \sqrt{4 D r_1}$ indicate the separation between the inner region, where the NESS has already been reached, and the outer region, where the transient behaviour is still observed.}
    \label{fig:refractory_RelaxationNESS}
\end{figure}

The agreement between the approximation~\eqref{eq:Approximation_NESS} and simulations is illustrated in figure~\ref{fig:refractory_RelaxationNESS}.\footnote{Simulations of a resetting Brownian particle with refractory periods have been obtained using the algorithm explained in appendix~\ref{app:langevin_simulations}.} Therein, we observe that the rough approximation on the left panel, where we are using only leading contributions, is accurate enough to describe the relaxation to the NESS. The more precise expressions derived in appendix~\ref{app:refractory_Laplace_method} are shown in the right panel.


\section{First-passage time with refractory periods} 
\label{sec:refractory_MFPT}
Now we focus on the first-passage time problem in the presence of resetting with refractory periods. Considering a target at position $\targetx$, we already know there appears a finite optimal resetting rate $\optMFPT{r}_1$ that minimises the MFPT. On a physical basis, with the inclusion of refractory periods, it is clear that MFPT will increase as $r_2$ increases, due to the resting times after each reset. Still, since the probability distributions of resetting events and refractory periods are independent, one might naively think that the optimal resetting rate $\optMFPT{r}_1$ would remain unaffected by $r_2$. The analysis below shows that this expectation does not hold: in fact, the optimal resetting rate $\optMFPT{r}_1$ presents a non-trivial dependence on $r_2$. 

\subsection{General formulation}
The previous expressions so far in this chapter are no longer valid when a target is present, since normalisation is no longer preserved. We have to redefine the time evolution taking into account the possibility of reaching the target during the exploration phase. Let $\survival_{r}(t;\targetx, x_0)$ denote the probability of not having reached the target up to time $t$ under the presence of stochastic resetting and refractory periods, starting from $x_0$ and having an absorbing boundary at $\targetx$. Keep in mind that all functions depend on the target position from now on, though it is not explicit in our notation---to avoid clutter.

The PDF for the particle to be found at position $x$ at time $t$ fulfils the renewal equation
\begin{align}
    \PDF_r(x,t|x_0) &= F(t) \survival_0(t;\targetx,x_0)\PDF_0(x,t|x_0) \nonumber\\
    &\quad +\int_0^t \d{t_1}  f(t_1)\survival_0(t_1;\targetx,x_0) G(t-t_1) \delta(x-x_0) \nonumber\\
    &\quad +\int_0^t \d{t_1} f(t_1)\survival_0(t_1;\targetx,x_0) \int_{t_1}^t \d{\tau_1} g(\tau_1-t_1) \,\PDF_r(x,t-\tau_1|x_0).
\end{align}
And by definition~\eqref{eq:survival_probability_def}, the survival probability is the integral of $\PDF_r$ over all space,
\begin{align}
    \survival_{r}(t;\targetx,x_0) &= F(t) \survival_0(t;\targetx,x_0) 
    \nonumber\\
    &\quad +\int_0^t \d{t_1}  f(t_1)\survival_0(t_1;\targetx,x_0) G(t-t_1)
    \nonumber\\
    &\quad +\int_0^t \d{t_1} f(t_1)\survival_0(t_1;\targetx,x_0) \int_{t_1}^t \d{\tau_1} g(\tau_1-t_1) \,\survival_{r}(t-\tau_1;\targetx,x_0)
    .
\end{align}
The dynamics of the first-passage problem is again computed using the pathway formulation that was introduced in section~\ref{subsec:pathway_formulation}, bringing to bear the changes $f(t)\to f(t)\survival_0 (t;\targetx,x_0)$ and $F(t)\to F(t)\survival_0 (t;\targetx,x_0)$. Expanding the PDF in terms of the number of renewals $n$, as in~\eqref{eq:refractory_pathway_expansion}, we have 
\begin{subequations}
    \begin{align}
        \PDF^{\free}_n(x,t|x_0) &= \prod_{i=1}^n \brackets{
            \int_{\tau_{i-1}}^t \d{t_i} \survival_0(t_i-\tau_{i-1};\targetx,x_0)f(t_i - \tau_{i-1}) \int_{t_i}^t \d{\tau_i} g(\tau_i - t_i)} 
        \nonumber \\ 
        &\quad \survival_0(t-\tau_{n};\targetx,x_0) F(t-\tau_n) \PDF_0 (x,t|x_0,\tau_n)
            ,
        \\
        \PDF^{\refr}_n(x,t|x_0) &= \prod_{i=1}^n \brackets{
            \int_{\tau_{i-1}}^t \d{t_i} \survival_0(t_i-\tau_{i-1};\targetx,x_0)f(t_i - \tau_{i-1}) \int_{t_i}^t \d{\tau_i} g(\tau_i - t_i)} 
        .
    \end{align}
\end{subequations}
With this approach, we obtain 
\begin{equation}
    \laplTransformTilde{\survival_r}{s;\targetx,x_0} = \frac{1}{1-\laplTransformTilde{f \survival_0}{s;\targetx,x_0}\laplTransformTilde{g}{s}} \bigg[ 
        \laplTransformTilde{F \survival_0}{s;\targetx,x_0} + \laplTransformTilde{f \survival_0}{s;\targetx,x_0}\laplTransformTilde{G}{s},
    \bigg]
    \label{eq:refractory_Laplace_survival_general}
\end{equation}
which is consistent with previous results in the literature \cite{journalarticle:Evans.Majumdar_EffectsRefractoryPeriod_J.Phys.A:Math.Theor.19,journalarticle:Maso-Puigdellosas.etal_StochasticMovementSubject_J.Stat.Mech.19}. 

\subsection{Poissonian resetting and refractory period}
\label{subsec:refractory_Poissonian_MFPT}
Let us consider the same model as in section~\ref{subsec:refractory_Poissonian_dynamics}, the distributions of resetting and refractory periods are given by~\eqref{eq:refractory_DistributionResetting} and~\eqref{eq:refractory_DistributionRefractory}, respectively. 
It is handy to introduce the characteristic diffusion time between the initial position and the target as 
\begin{equation}
    \tau_D = \frac{(\targetx -x_0)^2}{D}\geq 0.
    \label{eq:refractory_characteristic_time}
\end{equation}
Substituting all the expressions into~\eqref{eq:refractory_Laplace_survival_general}, we notice that the survival probability only spatially depends on the distance $|\targetx-x_0|$, 
\begin{equation}
    \laplTransformTilde{\survival_r}{s;\tau_D} = (s+r_1+r_2)\frac{1-e^{-\sqrt{\tau_D(s+r_1)}}}{s(s+r_1+r_2)+r_1r_2e^{-\sqrt{\tau_D(s+r_1)}}}. 
\end{equation}

Deriving the MFPT is straightforward from~\eqref{eq:MFPT_def}, yielding
\begin{equation}
    \FPT{(1)}(r_1,r_2;\tau_D) = \lim_{s\to 0} \laplTransformTilde{\survival_r}{s;\tau_D} = \parenthesis{e^{\sqrt{\tau_D r_1}}-1}\parenthesis{\frac{1}{r_1}+\frac{1}{r_2}}.
    \label{eq:refractory_MFPT_Poissonian}
\end{equation}
Since we are interested in the dependence of the MFPT on the parameters controlling the typical duration of the reset events, $r_1$, and the refractory periods, $r_2$, we have introduced them explicitly in the notation. As expected on a physical basis, two contributions appear in the MFPT---coming from the two summands in the second set of parentheses. The first one, which depends exclusively on $r_1$, corresponds to the instantaneous resetting without refractory period \cite{journalarticle:Evans.Majumdar_DiffusionStochasticResetting_Phys.Rev.Lett.11,journalarticle:Evans.Majumdar_DiffusionOptimalResetting_J.Phys.A:Math.Theor.11}. The second one stems from the refractory period that we have introduced after each resetting event, and depends on both $r_1$ and $r_2$. 

\begin{figure}
    \centering
    \includegraphics[width=0.7\textwidth]{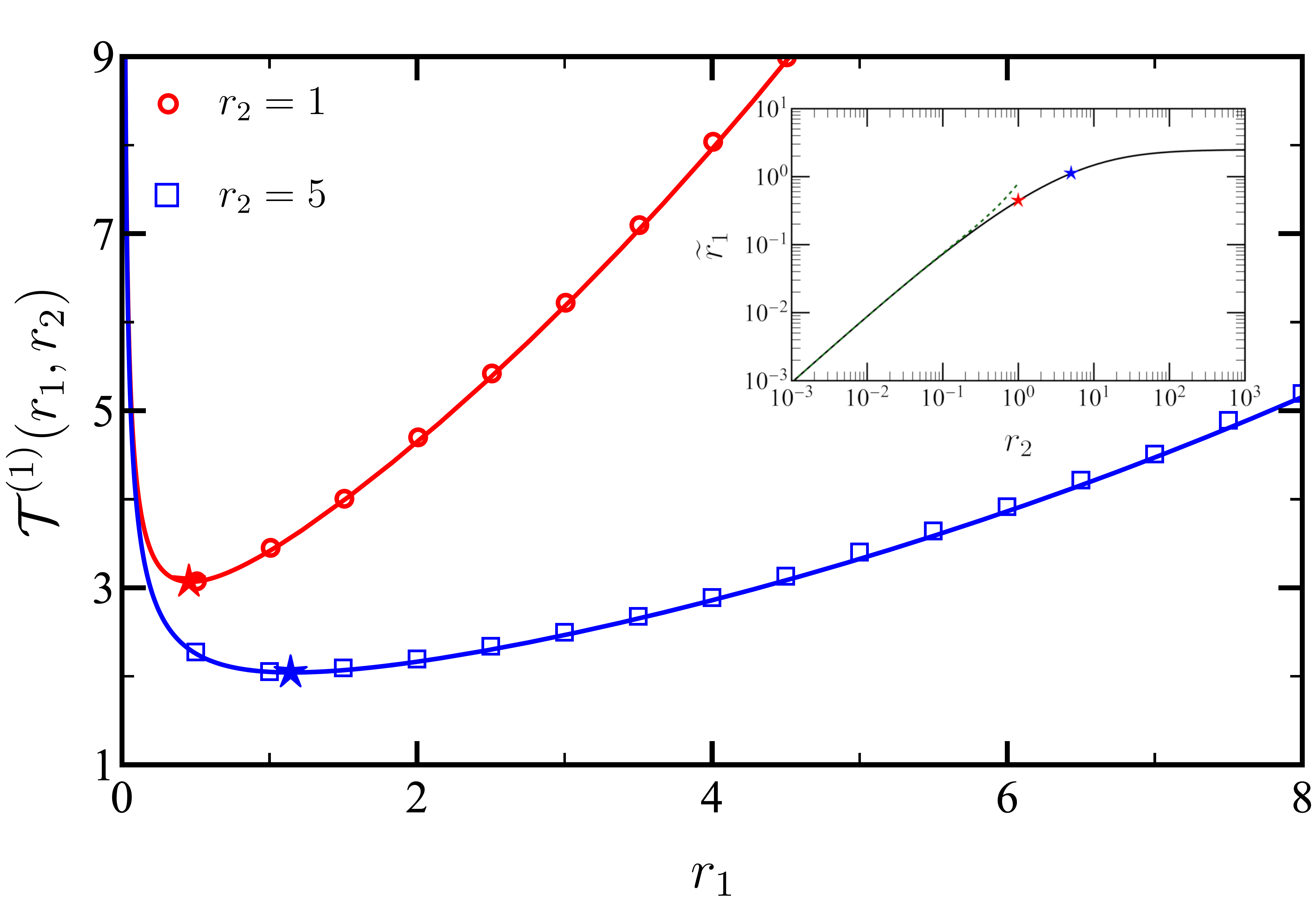}
    \caption{Mean first-passage time $\mathcal{T}^{(1)}(r_1,r_2)$ as a function of the resetting rate $r_1$ for fixed refractory rate $r_2$. An excellent agreement between simulations (symbols) and theory (solid lines) is found. The optimal resetting rate $\optMFPT{r}_1(r_2)$, represented by five-pointed stars, monotonically increases with $r_2$---in the limit $r_2\to \infty$, the value $\optMFPT{r}=\left[2-W(-2e^{-2})\right]^2= 2.5396$ is reached. Inset: Optimal resetting rate as a function of the refractory period rate. We show the numerical solution of the implicit equation for $\optMFPT{r}_1$, as given by~\eqref{eq:DerivativeMFPT} (solid line), and the analytical approximation for small $r_2$, as given by~\eqref{eq:expansion_r1(r2)} (dashed line).}
    \label{fig:Tr1r2}
\end{figure}

Despite its simple functional form, equation~\eqref{eq:refractory_MFPT_Poissonian} exhibits atypical and rich behaviour. We are interested in how the MFPT changes as a function of $r_1$ and $r_2$---clearly, it monotonically increases with $\tau_D$ in~\eqref{eq:refractory_characteristic_time}. With this knowledge, we are able to write~\eqref{eq:refractory_MFPT_Poissonian} as a function that only depends on $r_1$ and $r_2$, by introducing dimensionless parameters 
\begin{equation}
    r_1^* = r_1 \tau_d, \quad r_2^* = r_2 \tau_d, \quad \FPT{(1)*} = \frac{\FPT{(1)}}{\tau_d}.
\end{equation}
Therefore, the MFPT is
\begin{equation}
    \FPT{(1)}(r_1,r_2)=\left(e^{\sqrt{r_1}}-1\right)\left(\frac{1}{r_1}+\frac{1}{r_2}\right) \label{eq:refractory_MFPT_dimensionless},
\end{equation}
where we have dropped the asterisk in order not to clutter the formulae. This result was also obtained in an apparently different context~\cite{journalarticle:Mercado-Vasquez.etal_IntermittentResettingPotentials_JStatMechTheoryExp20}, when considering and intermittent V-shaped potential that was randomly switched on and off with different rates. Reasonably, in the limit of the stiffness of the potential going to infinity, the stochastic resetting with refractory periods is recovered, since the particle during the return stage is pinned at the centre of the trap.

Let us compute the minimum of $\FPT{(1)}(r_1,r_2)$ as a function of $r_2$, for fixed $r_1$. We clearly get that $r_2\to \infty$ is the best choice, reobtaining the MFPT in the absence of refractory period~\eqref{eq:MFPT_SSR},
\begin{equation}
\lim_{r_2 \to \infty} \FPT{(1)}(r_1,r_2)=\frac{e^{\sqrt{r_1}}-1}{r_1}.
\end{equation}

Now we focus on looking for the minimum of $\FPT{(1)}(r_1,r_2)$ as a function of $r_1$, for fixed $r_2$. In figure~\ref{fig:Tr1r2}, we show a couple of instances of $\FPT{(1)}(r_1,r_2)$ as a function of $r_1$ for $r_2 \in \{1,5\}$, finding an excellent agreement between theory and simulations. The figure illustrates the existence of a certain optimal curve $\optMFPT{r}_1=\optMFPT{r}_1(r_2)$, which is obtained by solving
\begin{equation}
    0=\left.\frac{\partial \FPT{(1)}(r_1,r_2)}{\partial r_1}\right|_{\optMFPT{r}_1}
    \implies
    2 r_2\left(1- e^{\sqrt{\optMFPT{r}_1}}\right)
    +\sqrt{\optMFPT{r}_1}e^{\sqrt{\optMFPT{r}_1}}\left(\optMFPT{r}_1+r_2\right)=0.
    \label{eq:DerivativeMFPT}
\end{equation}

The limiting behaviour of $\optMFPT{r}_1$ for both $r_2\to\infty$ (no refractory period) and $r_2\to 0$ (infinite refractory period) can be exactly solved. On the one hand, for $r_2 \to \infty$, we find that the limiting value $\lim_{r_2 \to \infty } \optMFPT{r}_1(r_2)$ is determined by
\begin{equation}
   \lim_{r_2 \to \infty} 2 \left[\optMFPT{r}_1(r_2) \right]^{-1/2}\left[ 1-e^{-\sqrt{ \optMFPT{r}_1(r_2)}}\right]= 1,
\end{equation}
whose solution is the expected one $\optMFPT{r}_1(r_2\to\infty)=\brackets{2+W(-2e^{-2})}^2$~\eqref{eq:MFPT_SSR}, since it is equivalent to minimise~\eqref{eq:refractory_MFPT_dimensionless}. On the other hand, it is clear from~\eqref{eq:DerivativeMFPT} that $\lim_{r_2\to 0}\optMFPT{r}_1(r_2)=0$, because $\optMFPT{r}_1^{3/2}e^{\sqrt{\optMFPT{r}_1}}=0$, which is logical from a physical point of view: if the particle has to remain motionless for ever, the best strategy for minimising the MFPT is to avoid resetting. 

A dominant balance argument shows that $\optMFPT{r}_1\sim r_2$ in the limit of long refractory periods (small $r_2$). To compute that dependence, it is handy to expand $\optMFPT{r}_1$ in a power series of $\sqrt{r_2}$. Let us define $u=\sqrt{\optMFPT{r}_1}$ and expand it around $v=\sqrt{r_2}=0$,
\begin{subequations}
    \begin{align}
    u=u(v)\simeq& u(0)+u'(0)v+\frac{u''(0)}{2}v^2 + \mathcal{O}(v^3), \\
    f(u(v))\simeq& f(u(0))
    +\left.\pdv{f(u)}{u}\pdv{u(v)}{v}\right|_{v=0}v
    \nonumber\\
    &+\left[\pdv[2]{f(u)}{u}\left(\pdv{u(v)}{v}\right)^2
    +\pdv{f(u)}{u}\pdv[2]{u(v)}{v}\right]_{v=0}\frac{v^2}{2}.
    \end{align}
\end{subequations}
Substituting this expansion into~\eqref{eq:DerivativeMFPT}, after a bit of algebra and imposing the positivity of $u$ one gets the following coefficients
\begin{subequations}
    \begin{align}
    v^0&: \quad u(0)^3 e^{u(0)}=0 \Rightarrow u(0)=0,
    \\
    v^3&: \quad u'(0)\left(-1+u'(0)\right)\left(1+u'(0)\right)=0 \Rightarrow u'(0)=1,
    \\
    v^4&: \quad \left(1+u''(0)\right)=0 \Rightarrow u''(0)=-1.
    \\
    v^5&: \quad \frac{1}{12}\left(-7+4u^{(3)}(0)\right)=0 \Rightarrow u^{(3)}(0)=\frac{7}{4}.
    \end{align}
\end{subequations}
Thus, the optimal curve $\optMFPT{r}_1$ in the limit as $r_2\to 0$ has the form
\begin{equation}
    \optMFPT{r}_1=r_2-r_2^{3/2}+\frac{5}{6}r_2^2+\mathcal{O}\left(r_2^{5/2}\right).
   \label{eq:expansion_r1(r2)}
\end{equation}
Interestingly, if we go back to dimensional variables, $\optMFPT{r}_1$ does not depend on $\tau_d$ until the second, non-linear, order contribution proportional to $r_2^{3/2}$. A comparison between the expansion~\eqref{eq:expansion_r1(r2)} and the numerical estimate for $\optMFPT{r}_1$ is shown in the inset of figure~\ref{fig:Tr1r2}.

Our result shows that there appears a ``resonance'' phenomenon, which optimises the MFPT---making it minimum---for a resetting rate that is linked with the refractory period rate. When the resetting point $x_0$ and the target $\targetx$ are close, in the sense that $r_2 \tau_d\ll 1$, $\optMFPT{r}_1 \simeq r_2$. 
As $r_2$ grows, $\optMFPT{r}_1$ consequently increases, but it asymptotically saturates for large enough values of $r_{2}$: for $r_2 \tau_d\gg 1$, the MFPT tends to~\eqref{eq:MFPT_SSR}, \ie corresponding to stochastic resetting without refractory periods.

\chapter{Resetting in disordered environments: dichotomous distributions}
\label{ch:resetting_disordered}
As introduced in section~\ref{subsec:standard_stochastic_resetting}, resetting is a powerful strategy to expedite search times in certain stochastic processes like free diffusion. Often, in search processes, the position of the target is well-defined, \ie there is a single target that remains at the same deterministic position for all realisations of the process. However, in real-world scenarios, the position of the target is unknown and may vary during the whole search process. In fact, if we knew with certainty where the target is, the best strategy would be to go directly to it, without the need for a search process. This chapter focuses on the effect of randomness in the target position, as a way to mimic the unawareness of searchers, for resetting processes. In analogy to other physical systems like disordered glasses \cite{journalarticle:Cavagna_SupercooledLiquidsPedestrians_Phys.Rep.09,journalarticle:Berthier.Biroli_TheoreticalPerspectiveGlass_Rev.Mod.Phys.11,journalarticle:Charbonneau.etal_FractalFreeEnergy_Nat.Commun.14,journalarticle:Folena.etal_IntroductionDynamicsDisordered_PhysStatMechAppl22,journalarticle:Ros.Fyodorov_HighdLandscapesParadigm_22}, this feature is going to be referred as \emph{disorder}. 

On the one hand, the position of the target may vary from one realisation to another, but it is fixed for each individual trajectory. We call this feature \emph{quenched disorder}. Quenched disorder naturally emerges in contexts where the target position is uncertain and fluctuates slowly compared to the dynamics of the searcher. One of this most exemplifying scenarios appears in the context of foraging \cite{book:Viswanathan.etal_PhysicsForagingIntroduction_11,journalarticle:Marion.etal_UnderstandingForagingBehaviour_J.Theor.Biol.05,journalarticle:Bartumeus.etal_AnimalSearchStrategies_Ecology05,journalarticle:Boyer.Walsh_ModellingMobilityLiving_Philos.Trans.R.Soc.A10,journalarticle:Viswanathan.etal_OptimizingSuccessRandom_Nature99,journalarticle:Berger-Tal.Bar-David_RecursiveMovementPatterns_Ecosphere15,journalarticle:Pal.etal_SearchHomeReturns_Phys.Rev.Res.20}, where a certain animal looks for nutrients or resources throughout an environment that is heterogeneous. Here, we may imagine a predator (searcher) looking for prey (target) at night, when it is almost motionless until daybreak.

On the other hand, a different situation arises when one considers the target to be mobile, following its own dynamics. According to the terminology employed in disordered systems, this is known as \emph{annealed or dynamical disorder} \cite{journalarticle:Zwanzig_RateProcessesDynamical_Acc.Chem.Res.90}, and it must be contemplated when the timescale of the target and searcher dynamics are comparable. In the foraging context, we would have that both predator and prey are mobile 
\cite{journalarticle:Diz-Pita.Otero-Espinar_PredatorPreyModels_Mathematics21,journalarticle:Anderson_ModelCalculationsCooperative_J.Chem.Phys.70,journalarticle:Alpern.etal_StochasticGameModel_J.R.Soc.Interface19,journalarticle:Mercado-Vasquez.Boyer_LotkaVolterraSystems_JPhysMathTheor18,journalarticle:Evans.etal_ExactlySolvablePredator_JPhysMathTheor22,journalarticle:Toledo-Marin.etal_PredatorpreyDynamicsChasing_19}. Intermittent targets, which have been recently studied in the context of stochastic resetting \cite{journalarticle:Biswas.etal_RateEnhancementGated_J.Chem.Phys.23,journalarticle:Mercado-Vasquez.Boyer_FirstHittingTimes_Phys.Rev.Lett.19}, are a particular case of annealed disorder where targets fluctuate between reactive and non-reactive states. This intermittency is often applied to model biological processes within the cell, from the accessibility of DNA-binding sites to the opening and closing of ion channels \cite{journalarticle:McAdams.Arkin_StochasticMechanismsGene_Proc.Natl.Acad.Sci.U.S.A.97,journalarticle:Tian.Burrage_StochasticModelsRegulatory_Proc.Natl.Acad.Sci.U.S.A.06}.  


In this thesis, we focus on search processes in which the target is subject to a quenched disorder scheme. We have a searcher under the effect of resetting events trying to reach a unique target, whose position varies from one trajectory to another. Our goal is to find the optimal space-dependent resetting rate $\optMFPT{r}(x)$ that minimises the search time for any distribution of targets $\PDF_T(\targetx)$. Despite this intuitive and simple idea, there have been few previous works in this direction in the field of stochastic resetting. A first attempt to include heterogeneity involved the optimisation of the distribution of the resetting position in the presence of a non-resetting window around the resetting point for exponentially distributed targets~\cite{journalarticle:Evans.Majumdar_DiffusionOptimalResetting_J.Phys.A:Math.Theor.11}. Another work studied the optimal problem for the MFPT when resetting events only take place beyond a threshold value, which is a first approximation to non-homogeneous resetting~\cite{journalarticle:Plata.etal_AsymmetricStochasticResetting_Phys.Rev.E20,journalarticle:DeBruyne.etal_OptimizationFirstPassageResetting_Phys.Rev.Lett.20}. Regarding other interesting works outside the MFPT problem, a general framework has been developed to obtain the stationary distribution in the presence of heterogeneous resetting~\cite{journalarticle:Roldan.Gupta_PathintegralFormalismStochastic_Phys.Rev.E17}. In the context of systems with discrete states, a random walker on complex networks with node-dependent resetting rate has been considered~\cite{journalarticle:Ye.Chen_RandomWalksComplex_JStatMechTheoryExp22}. Heterogeneity has also been investigated in several resetting systems introducing spatially-dependent diffusion coefficients~\cite{journalarticle:Wang.etal_TimeAveragingEmerging_Phys.Rev.E21,journalarticle:Lenzi.etal_TransientAnomalousDiffusion_PhysStatMechAppl22,journalarticle:Sandev.etal_HeterogeneousDiffusionStochastic_JPhysMathTheor22}. 
Rigorous results have been derived for the MFPT of a spatially-dependent resetting problem with a given distribution of the target position, assuming that explicit solutions for the MFPT equation are known~\cite{journalarticle:Pinsky_OptimizingDriftDiffusive_Electron.J.Probab.19}. Recently, a very close and complementary analysis to ours has been performed, where the minimisation of the MFPT is done by finding the quenched target distribution that makes optimal a particular waiting time distribution between resetting events, which can be considered non-Poissonian~\cite{journalarticle:Evans.Ray_StochasticResettingPrevails_Phys.Rev.Lett.25}.

Computing the optimal resetting profile $\optMFPT{r}(x)$ for any possible target distribution is certainly a mathematical challenge. For that reason, the analysis of quenched disorder in resetting processes has been split into two parts within this thesis. First, the current chapter introduces the concepts and generic mathematical framework to study the effect of quenched disorder in resetting processes. Afterwards, we apply this formalism to a simplified setup, where the target and the resetting profiles have particularly simple forms. Second, chapter~\ref{ch:resetting_disordered_bulkvsboundaries} explores more general distributions of targets, with an arbitrary shape for the resetting profile.

This chapter is organised as follows. The model is described in section~\ref{sec:dichotomous_model}. Section~\ref{sec:dichotomous_FPT} is devoted to the detailed analysis of the first-passage problem, computing analytical expressions for the FPT distributions and its moments. The optimisation of those moments is investigated in section~\ref{sec:dichotomous_Optimisation}, working with both the minimisation of the average MFPT and the average variance of the FPT. Analytical implicit expressions for the optimal quantities are obtained and solved, either numerically or approximately, in certain limits. 

\section{Description of the disordered resetting model}
\label{sec:dichotomous_model}

Let us consider the one-dimensional movement of an overdamped Brownian particle, under the effect of stochastic resetting to the initial position $x_0$. From now on, we consider a space-dependent resetting rate, $r(x)$. The propagator $\PDF_r(x,t|x_0,\targetx)$ indicates the PDF of finding the searcher at position $x$ at time $t$, having an absorbing boundary at $\targetx$ if it started at $x_0$.\footnote{We are using the same notation as~\eqref{eq:Fokker_Planck_Standard_SR}, though the dependence on the target is considered, and we always choose $t_0=0$ (equivalent to $t-t_0\to t$). Furthermore, the dependence on the heterogeneous resetting $r(x)$ is not explicitly written.} It obeys the following forward Fokker-Planck equation
\begin{align}
    \partial_t \PDF_r(x,t|x_0,\targetx) &= D \partial_{x}^2 \PDF_r(x,t|x_0,\targetx) - r(x) \PDF_r(x,t|x_0,\targetx) 
    \nonumber\\
    &\quad + \delta(x-x_r) \int \d{y} r(y)\, \PDF_r(y,t|x_0,\targetx), 
    \label{eq:Fokker_Planck_Heterogeneous_SR}
\end{align}
where $D$ is the diffusion coefficient as usual. The interpretation of contributions found in the rhs is totally analogous to the standard case~\eqref{eq:Fokker_Planck_Standard_SR}, except for our taking into account a heterogeneous resetting rate $r(x)$. Concretely, the gain term from any point to $x_0$ has changed due to the heterogeneity, $r\delta(x-x_r)\to \delta(x-x_r) \int \d{y} r(y)\, \PDF_r(y,t|x_0,\targetx)$. Clearly, we recover the SSR case when the constant case $r(x)=r$ is considered. 

\begin{figure}
    \centering
    \includegraphics[width=0.7\textwidth]{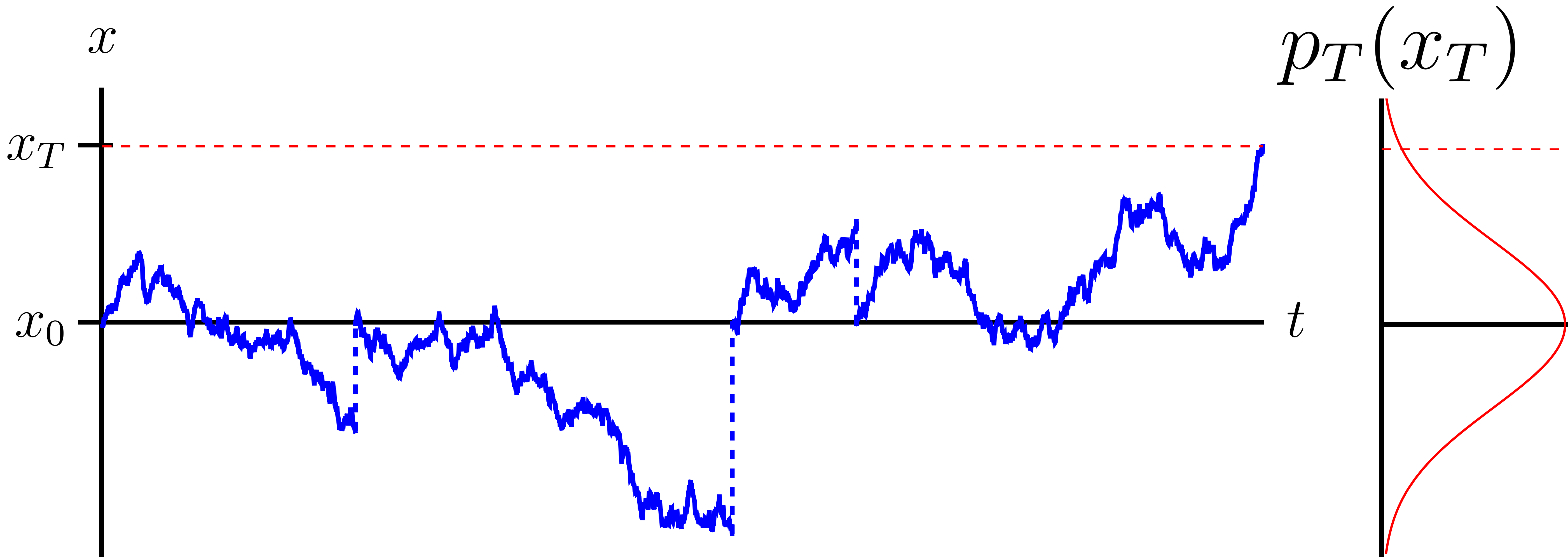}
    \caption{Single trajectory for stochastic resetting with quenched disorder. The target position $\targetx$ is drawn from a given distribution $\PDF_T(\targetx)$, and it remains fixed for each realisation of the dynamics. The searcher starts at $x_0$ and resets to $x_r=x_0$ depending on a heterogeneous rate $r(x)$. In this particular trajectory, the searcher can only move within the semi-infinite domain $(-\infty,\targetx)$, because the target is at $\targetx>x_0$. We have chosen a Gaussian distribution for the target position as an example.}
    \label{fig:disordered_general_model}
\end{figure}

The initial condition is always $\PDF_r(x,0|x_0,\targetx) = \delta(x-x_0)$, whereas the boundary conditions depend on the physical situation. For an unbounded domain, \ie when there is no target at all, the probability current vanishes at $x\to\pm\infty$,
\begin{equation}
    \lim_{x\to \pm \infty} \partial_x \PDF_r(x,t|x_0,\targetx) = 0,
\end{equation}
whereas if there is an absorbing boundary at $\targetx$, the propagator must satisfy
\begin{equation}
    \PDF_r(\targetx,t|x_0,\targetx)=0.
\end{equation}
If the searcher can move freely, the last condition implies that the particle may only look for in the semi-infinite domain defined by $(-\infty,\targetx)$ or $(\targetx,\infty)$, depending on the position of $x_0$ with respect to $\targetx$. A single realisation of the dynamics is sketched in figure~\ref{fig:disordered_general_model}. Therein, we clearly identify resetting events that are beneficial (detrimental) when the particle is at the opposite (same) side of the target with respect to $x_0$.

Disorder is incorporated by considering that the position of the target $\targetx$ is drawn from a certain distribution $\PDF_T(\targetx)$. The propagator with disorder is obtained by integrating over $\targetx$, 
\begin{equation}
    \overline{\PDF}_r(x,t|x_0) = \int \d{\targetx} \PDF_T(\targetx) \PDF_r(x,t|x_0,\targetx).
    \label{eq:averaging_propagator_disorder}
\end{equation}
This provides a clear interpretation of the quenched disorder problem, since the position of the target is fixed for each realisation of the dynamics. This construction is also inherited by other quantities, such as the survival probability, the FPT distribution, or its moments. 

\subsection{Dimensionless variables}
\label{subsec:dimensionless_variables}
For the analysis in these chapters on stochastic resetting with quenched disorder, it is handy to introduce dimensionless variables---in order to identify the natural units of our problem. Let $\ell$ be the radius of the target distribution, \ie the target distribution has a compact support $\targetx\in[x_r-\ell,x_r+\ell]$. Then, we may take the dimensionless space and time variables by 
\begin{equation}
    x^* = (x-x_r)/\ell, \quad t^* = t/(\ell^2/D), \quad r^*(x^*) = (\ell^2/D) r(x).
\end{equation}
Then, if we drop the asterisk not to clutter our formulae,~\eqref{eq:Fokker_Planck_Heterogeneous_SR} becomes 
\begin{align}
    \partial_t \PDF_r(x,t|x_0,\targetx) = \partial_{x}^2 \PDF_r(x,t|x_0,\targetx) - r(x) \PDF_r(x,t|x_0,\targetx) + \delta(x) \int \d{y} r(y)\, \PDF_r(y,t|x_0,\targetx), 
    \label{eq:Fokker_Planck_Heterogeneous_SR_dimensionless}
\end{align}
with initial and boundary condition 
\begin{equation}
    \PDF_r(x,0|x_0,\targetx)=\delta(x-x_0), \quad \PDF_r(\targetx,t|x_0,\targetx)=0.
\end{equation}

\subsection{Dichotomous disorder and piecewise resetting rates}
Along this chapter, we focus on the particular simple case of a dichotomous target distribution, where the target can be only at two possible locations,
\begin{equation}
    \PDF_T(\targetx) = \disorder \delta(x_T-1) + (1-\disorder) \delta(x_T+1), 
    \label{eq:averaging_dichotomous}
\end{equation}
with $0<\disorder<1$. 
The distance from the resetting position to the target is always unity in dimensionless variables, and the probability of the target being to the right and left of the origin is $\disorder$ and $1-\disorder$, respectively.
In addition to the choice for our target distribution, we restrict ourselves to the simplest non-trivial family of resetting functions, \ie the piecewise constant function
\begin{equation}
    r(x) = r_+ \Theta(x) + r_- \Theta(-x) = \begin{cases}
        r_+, & x>0, \\
        r_-, & x<0,
    \end{cases}
\end{equation}
where $\Theta(x)$ is again the Heaviside step function and $r_\pm \geq 0$. 

The rest of quantities, which are obtained after averaging over disorder, read 
\begin{equation}
    \overline{z}(x,t) = \disorder \, \positiveSol{z}(x,t) + (1-\disorder) \negativeSol{z}(x,t),
    \label{eq:averaging_dichotomous_dimensionless}
\end{equation}
\ie the linear combination depending on $\disorder$. Here, we have conveniently defined 
\begin{equation}
    \positiveSol{z}(x,t|x_0) \equiv z(x,t|x_0,+1), \quad \negativeSol{z}(x,t|x_0) \equiv z(x,t|x_0,-1).
\end{equation}
Note that the subscripts of the piecewise resetting function, $r_\pm$ (or later, $\alpha_\pm$), do not refer to the target position, but to the spatial regions $x>0$ and $x<0$. For instance, the averaged propagator is
\begin{equation}
    \overline{\PDF}_r(x,t|x_0) = \disorder \, \positiveSol{\PDF}(x,t|x_0) + (1-\disorder) \negativeSol{\PDF}(x,t|x_0).
\end{equation}
Now, the optimal problem for this particular choice for the target distribution and the resetting function is finding the pair $(\optMFPT{r}_+,\optMFPT{r}_-)$ that optimises any observable of interest~\eqref{eq:averaging_dichotomous_dimensionless}.

Before the first-passage analysis, we first derive how the dynamics behaves in the long-time limit in the absence of target. In this context, the propagator
\begin{equation}
    \PDF_r^{st}(x)\equiv \lim_{t\to\infty}\PDF_r(x,t|x_0)
\end{equation} 
is the solution of 
\begin{subequations}
    \begin{align}
        \partial_x^2 \PDF_r^{st}(x) + r_+ \PDF_r^{st}(x) = 0, & \quad x>0, \\
        \partial_x^2 \PDF_r^{st}(x) + r_- \PDF_r^{st}(x) = 0, & \quad x<0,
    \end{align} 
\end{subequations}
with (i) the boundary conditions $\lim_{x\to\pm\infty} \partial_x \PDF_r^{st}(x) = 0$ and (ii) the continuity of $\PDF_r^{st}(x)$ and its derivative at $x=0$. Thus, the non-equilibrium stationary distribution reads
\begin{equation}
    \PDF_r^{st}(x) = \frac{\alpha_{+}\alpha_{-}}{\alpha_{+}+\alpha_{-}} 
    \brackets{e^{-\alpha_{+}x} + e^{\alpha_{-}x}},
\end{equation}
where the competition between diffusion and resetting defines the length scales\footnote{The length scales are defined as $\alpha_{\pm}^{-1}=\sqrt{D/r_{\pm}}$ in the original variables.}
\begin{equation}
    \alpha_{\pm} = \sqrt{r_{\pm}}.
\end{equation}
Hereinafter, we use $\alpha_\pm$ as the resetting function to optimise instead of $r_\pm$, since expressions are more compact.

\section{First-passage analysis}
\label{sec:dichotomous_FPT}

Our aim is to solve the first-passage problem for the quenched disordered model described in section~\ref{sec:dichotomous_model}. As stated below~\eqref{eq:averaging_dichotomous_dimensionless}, every observable of interest after averaging over disorder follows the linear combination~\eqref{eq:averaging_dichotomous_dimensionless}, so we are able to use the backward approach from section~\ref{subsec:standard_stochastic_resetting} for both $\PDF_+$ and $\PDF_-$. We start by defining the survival probabilities 
\begin{equation}
    \survival_\pm(t;x_0) =  \int \d{x} \PDF_\pm(x,t|x_0),
\end{equation}
\ie the probability of not having reached the target at $\targetx=\pm 1$ at time $t$, starting from $x_0$. Their corresponding FPT distributions are 
\begin{equation}
    \FPTpdf_\pm(\FPT{1};x_0) = -\partial_{\FPT{1}} \survival_\pm(\FPT{1};x_0).
\end{equation}

The backward differential equation fulfilled by $\PDF_\pm(x,t|x_0)$, $\survival_\pm(t;x_0)$, and $\FPTpdf_\pm(\FPT{1};x_0)$ is~\eqref{eq:Fokker_Planck_Backward_SSR}. For the sake of concreteness, since we are focusing on the FPT distribution, it holds 
\begin{equation}
    \partial_{\FPT{1}} \FPTpdf_\pm(\FPT{1};x_0) = \partial_{x_0}^2 \FPTpdf_\pm(\FPT{1};x_0) -\alpha^2(x_0) \bigg[ \FPTpdf_\pm(\FPT{1};x_0) - \FPTpdf_\pm(\FPT{1};0)\bigg].
\end{equation}
It must be complemented by the boundary conditions
\begin{equation}
    \FPTpdf_\pm(\FPT{1};\pm 1) = \delta(t), \quad \lim_{x_0\to\mp\infty} \partial_{x_0} \FPTpdf_\pm(\FPT{1};x_0) = 0, 
\end{equation}
and the initial condition $\FPTpdf_\pm (0;x_0\neq \pm 1)=0$. Notice that the distribution is well normalised because eventually the searcher always reaches the target, 
\begin{equation}
    \int_0^\infty \d{\FPT{1}} \FPTpdf_\pm(\FPT{1};x_0) = \survival_\pm(0;x_0) -\lim_{t\to\infty} \survival_\pm(t;x_0) = 1.
\end{equation}

The property~\eqref{eq:averaging_dichotomous_dimensionless} is inherited by the FPT distribution,
\begin{equation}
    \overline{\FPTpdf}(\FPT{1};x_0) = \disorder \, \positiveSol{\FPTpdf}(\FPT{1};x_0) + (1-\disorder) \negativeSol{\FPTpdf}(\FPT{1};x_0),
\end{equation}
and all its moments and any linear combination thereof. Hereinafter, we just focus on solutions where the target is on the positive side, $\positiveSol{\FPTpdf}(\FPT{1};x_0)$, due to the left-right symmetry of our system: $\negativeSol{\FPTpdf}(\FPT{1};-x_0)$ is obtained by exchanging $\alpha_+\leftrightarrow \alpha_-$ in $\positiveSol{\FPTpdf}(\FPT{1};x_0)$. This invariance stems from a mirror symmetry when one simultaneously exchanges $\targetx\leftrightarrow -\targetx$ and $r_+\leftrightarrow r_-$, the system is just the mirror image of the original one with respect to the resetting point $x_r=0$.\footnote{This symmetry holds while the searcher does not have any preferred direction, \eg our diffusive process.} Following the same scheme as in section~\ref{subsec:FPT_analysis}, the Laplace transform of the FPT distribution $\positiveSol{\FPTpdfLaplace}(s;x_0)$ fulfils the ODE in~\eqref{eq:Fokker_Planck_Backward_SSR},
\begin{equation}
    - s\positiveSol{\FPTpdfLaplace}(s;x_0)  = \partial_{x_0}^2 \positiveSol{\FPTpdfLaplace}(s;x_0) -\alpha^2(x_0) \bigg[ \positiveSol{\FPTpdfLaplace}(s;x_0) - \positiveSol{\FPTpdfLaplace}(s;0)\bigg].
    \label{eq:disordered_FPT_distribution_Laplace}
\end{equation}
whereas for the moments we obtain the hierarchy of equations
\begin{align}
    \label{eq:disordered_FPT_moments}
    -n \positiveSol{\FPT{(n-1)}}(x_0) = \partial_{x_0}^2 \positiveSol{\FPT{(n)}}(x_0) - \alpha(x_0)^2 \bigg[
        \positiveSol{\FPT{(n)}}(x_0) - \positiveSol{\FPT{(n)}}(0)
    \bigg],
\end{align}
with 
\begin{equation}
    \positiveSol{\FPT{(n)}}(x_0) = (-1)^n \partial_s^n \positiveSol{\FPTpdfLaplace}(s;x_0) \bigg|_{s=0}, \quad \positiveSol{\FPT{(0)}}(x_0)=1. 
\end{equation}

Instead of solving the whole hierarchy of moments, we focus on~\eqref{eq:disordered_FPT_distribution_Laplace} to obtain the full distribution of FPT. 
Since resetting takes different values on each side of $x_r=0$, we must solve~\eqref{eq:disordered_FPT_distribution_Laplace} in the two semi-infinite domains $(-\infty,0)$ and $(0,\infty)$, 
\begin{equation}
    \positiveSol{\FPTpdfLaplace}(s;x_0) = \begin{cases}
        \positiveSol{\FPTpdfLaplace}^{(R)}(s;x_0), & x_0 > 0, \\
        \positiveSol{\FPTpdfLaplace}^{(L)}(s;x_0), & x_0 < 0,
    \end{cases}
\end{equation}
and afterwards match both solutions, enforcing consistency. On each side, an inhomogeneous second-order ODE with constant coefficients holds. The general solutions are 
\begin{subequations}
    \begin{align}
        \positiveSol{\FPTpdfLaplace}^{(R)}(s;x_0) &=  A e^{\lambda_+ x_0} + B e^{-\lambda_+ x_0} + \frac{\alpha_+^2}{\lambda_+^2} \positiveSol{\FPTpdfLaplace}(s;0), \\
        \positiveSol{\FPTpdfLaplace}^{(L)}(s;x_0) &=  C e^{\lambda_- x_0} + D e^{-\lambda_- x_0} + \frac{\alpha_-^2}{\lambda_-^2} \positiveSol{\FPTpdfLaplace}(s;0),
    \end{align} 
\end{subequations}
where we have defined $\lambda_\pm = \sqrt{s+\alpha_\pm^2}$, and $A,B,C,D$ are integration constants to be determined. The set $\{A,B,C,D,\positiveSol{\FPTpdfLaplace}(s;0)\}$ is obtained by imposing: (i)-(ii) the boundary conditions 
\begin{equation}
    \positiveSol{\FPTpdfLaplace}^{(R)}(s;1) = 0, \quad \lim_{x_0\to -\infty} \partial_{x_0} \positiveSol{\FPTpdfLaplace}^{(L)}(s;x_0) = 0,
\end{equation}
(iii)-(iv) the continuity of $\positiveSol{\FPTpdfLaplace}(s;x_0)$ at $x_0=0$, $\positiveSol{\FPTpdfLaplace}^{(R)}(s;0)=\positiveSol{\FPTpdfLaplace}^{(L)}(s;0)$, plus the consistency with the value $\FPTpdfLaplace_+(s;0)$, and (v) the continuity of the first derivative at $x_0=0$, $\partial_{x_0}\positiveSol{\FPTpdfLaplace}^{(R)}(s;0)=\partial_{x_0}\positiveSol{\FPTpdfLaplace}^{(L)}(s;0)$. Solving this linear system of equations, we obtain
\begin{align}
    \positiveSol{\FPTpdfLaplace}^{(R)}(s;x_0)  &= \positiveSol{\FPTpdfLaplace}(s;0) + s\frac{\lambda_-\brackets{\cosh\parenthesis{\lambda_+ x_0}-1}+\lambda_+\sinh\parenthesis{\lambda_+ x_0}}{\lambda_-\parenthesis{\alpha_+^2+s\cosh\lambda_+}+s\lambda_+\sinh\lambda_+}, 
    \\
    \positiveSol{\FPTpdfLaplace}^{(L)}(s;x_0) &= \positiveSol{\FPTpdfLaplace}(s;0) + s\lambda_+^2\frac{e^{\lambda_- x_0}-1}{\lambda_-^2\parenthesis{\alpha_+^2+s\cosh\lambda_+}+s\lambda_+\lambda_- \sinh\lambda_+},
    \\
    \positiveSol{\FPTpdfLaplace}(s;0) &= \frac{\lambda_+^2 \lambda_-}{\lambda_-\parenthesis{\alpha_+^2+s\cosh\lambda_+}+s\lambda_+ \sinh\lambda_+}.
\end{align}

\begin{figure}
    \centering
    \includegraphics[width=\textwidth]{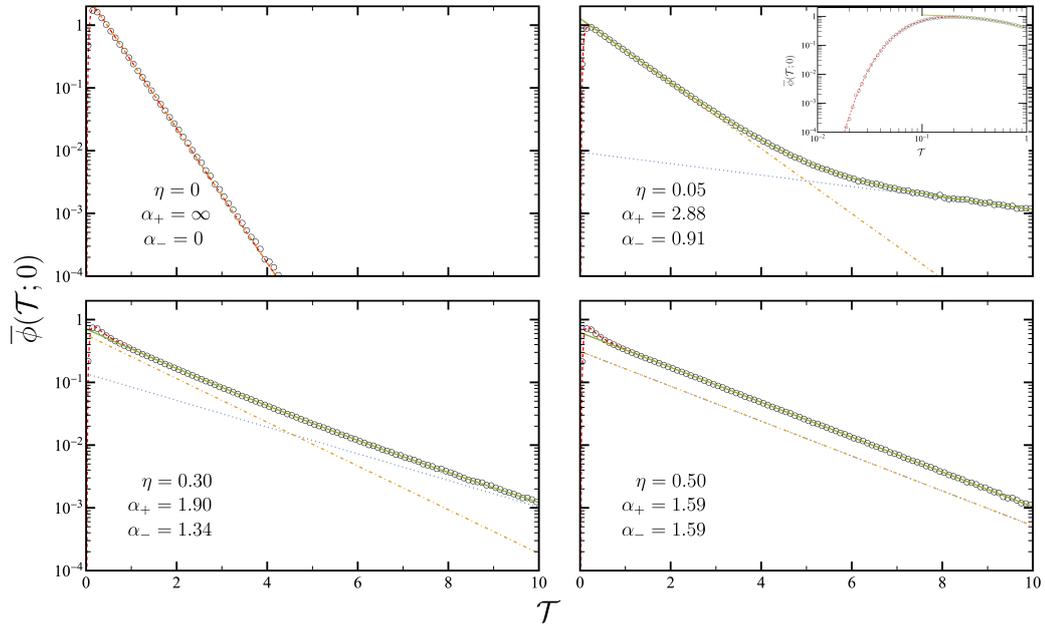}
    \caption[First-passage time distribution for the disordered resetting system. Specifically, we plot results from the numerical simulations (circles), the numerical Laplace inversion of the total Laplace transform using~\eqref{eq:disordered_FPT_distribution_Laplace_solution} (dashed red line), the asymptotic contribution from $\positiveSol{\FPTpdf}(\mathcal{T};0)$ given by~\eqref{eq:asymp+1} (dotted blue line), the asymptotic contribution from $\negativeSol{\FPTpdf}(\mathcal{T};0)$ given by~\eqref{eq:asymp-1} (dot-dashed orange line), and the sum of the two latter asymptotic contributions (solid green line). In the inset of the right top panel, we check how the numerical Laplace inversion and simulations agree at short times and how both lines overlap with the asymptotic estimation from $t=2\cdot 10^{-1}$.]{First-passage time distribution for the disordered resetting system. Specifically, we plot results from the numerical simulations (circles), the numerical Laplace inversion of the total Laplace transform using~\eqref{eq:disordered_FPT_distribution_Laplace_solution} (dashed red line), the asymptotic contribution from $\positiveSol{\FPTpdf}(\mathcal{T};0)$ given by~\eqref{eq:asymp+1} (dotted blue line), the asymptotic contribution from $\negativeSol{\FPTpdf}(\mathcal{T};0)$ given by~\eqref{eq:asymp-1} (dot-dashed orange line), and the sum of the two latter asymptotic contributions (solid green line). In the inset of the right top panel, we check how the numerical Laplace inversion and simulations agree at short times and how both lines overlap with the asymptotic estimation from $t=2\cdot 10^{-1}$ onwards.} 
    \label{fig:dichotomous_FPT_distribution}
\end{figure}

We are interested in the scenario where the searcher starts at the resetting position, $x_0=0$, \ie the midpoint of the two target positions. After substituting $\lambda_\pm = \sqrt{s+\alpha_\pm^2}$, the Laplace transform of the FPT distribution reads
\begin{equation}
    \positiveSol{\FPTpdfLaplace}(s;x_0 = 0) = \brackets{
        \frac{\alpha_+^2+s\cosh\parenthesis{\sqrt{s+\alpha_+^2}}}{s+\alpha_+^2}
        + \frac{s \sinh\parenthesis{\sqrt{s+\alpha_+^2}}}{\sqrt{s+\alpha_+^2}\sqrt{s+\alpha_-^2}}}^{-1},
    \label{eq:disordered_FPT_distribution_Laplace_solution}
\end{equation}
Analytical inverting~\eqref{eq:disordered_FPT_distribution_Laplace_solution} for general parameters $(\alpha_+,\alpha_-)$, however one can compute its long-time behaviour, which provides information about the tail of the distribution. The asymptotic complex analysis of the moment generating function indicates that the FPT distribution decays exponentially as---see appendix~\ref{app:dichotomous_Asymptotic_FPT} for the detailed derivation,
\begin{equation}
    \positiveSol{\FPTpdf}(\FPT{1};0) \sim \left.\frac{(s+\alpha_+^2)(s+\alpha_-^2)}{\partial_s\zeta(s,\alpha_+,\alpha_-)} e^{s\FPT{1}}\right|_{s=s_+^*}, \quad \FPT{1}\to\infty,
    \label{eq:asymp+1}
\end{equation}
where 
\begin{equation}
    \zeta(s,\alpha_+,\alpha_-) \equiv \frac{(s+\alpha_+^2)(s+\alpha_-^2)}{\positiveSol{\FPTpdfLaplace}(s;0)},
\end{equation}
and $s_+^*$ is the largest real zero of $\zeta(s,\alpha_+,\alpha_-)$,\footnote{Equivalently, the largest real pole of~\eqref{eq:disordered_FPT_distribution_Laplace_solution}.} which is always negative. As already mentioned, $\negativeSol{\FPTpdfLaplace}(s;0)$ is immediately obtained from $\positiveSol{\FPTpdfLaplace}(s;0)$ by exchanging $\alpha_+\leftrightarrow \alpha_-$,
\begin{equation}
    \negativeSol{\FPTpdf}(\FPT{1};0) \sim \left.\frac{(s+\alpha_+^2)(s+\alpha_-^2)}{\partial_s\zeta(s,\alpha_-,\alpha_+)} e^{s\FPT{1}}\right|_{s=s_-^*}, \quad \FPT{1}\to\infty,
    \label{eq:asymp-1}
\end{equation}
with $s_-^*$ the largest real zero of $\zeta(s,\alpha_-,\alpha_+)$.
Therefore, we have proved that the average FPT distribution is a weighted sum of two exponential functions for long times, 
\begin{equation}
    \overline{\FPTpdf}(\FPT{1};0) \sim \disorder \left.\frac{(s+\alpha_+^2)(s+\alpha_-^2)}{\partial_s\zeta(s,\alpha_+,\alpha_-)} e^{s\FPT{1}}\right|_{s=s_+^*} 
    +
    (1-\disorder) \left.\frac{(s+\alpha_+^2)(s+\alpha_-^2)}{\partial_s\zeta(s,\alpha_-,\alpha_+)} e^{s\FPT{1}}\right|_{s=s_-^*}.
    \label{eq:disordered_global_asymptotic_FPT}
\end{equation}
This result has been checked numerically and using simulations, as shown in figure~\ref{fig:dichotomous_FPT_distribution}. In each panel, a different value of the disorder $\disorder$ is displayed; the rest of parameters, \ie $(\alpha_+,\alpha_-)$, are those minimising the average MFPT for that value of $\disorder$---derived later in section~\ref{subsec:optimal_MFPT_dichotomous}. The simulation data (circles) have been computed by building a histogram of $N=10^7$ trajectories for the parameters considered in each panel, integrated with a time step $\Delta t= 10^{-5}$ up to the first-passage time---using the method explained in appendix~\ref{app:langevin_simulations}. They show an excellent agreement with our theoretical predictions,  both asymptotic (solid green) and numerical Laplace inversion (dashed red). Although the asymptotic prediction is expected to work only in the limit of long times, it fails just at very short times. Therein, numerical inversion of the exact result still perfectly matches the simulations---as shown in the inset of the second panel.

An especially interesting feature of figure~\ref{fig:dichotomous_FPT_distribution} is the crossover between the two exponential decays corresponding to the long-time behaviours of~\eqref{eq:disordered_global_asymptotic_FPT}, provided that $\disorder\ne \{0,1\}$. For $\disorder=0$ (first panel), \ie no disorder, only $\negativeSol{\FPTpdf}(\FPT{1};0)$ contributes, and therefore the long-time behaviour is exponential with $s_-^*$. For $0<\disorder<1/2$ (rest of panels), the largest pole (smallest in absolute value) is $s_+^*$ for our choice of $(\alpha_+,\alpha_-)$; whose value decreases as $\disorder$ becomes smaller---it vanishes in the limit $\disorder \to 0$. The crossover appears because, for any $0<\disorder<1/2$, the long-term behaviour is first dominated by $s_-^*$ (since the contribution of the $+$ mode is weighted with $\disorder\ll 1$) but then turns to be dominated by $s_+^*$. The crossover is not present for $\disorder=1/2$, because $s^*_+=s^*_-$ for our choice of parameters. The results for $1/2<\disorder<1$ are not shown due to the mirror symmetry of the system under the exchange $\alpha_+\leftrightarrow\alpha_-$. Note that each panel has the same scale, this decision has been made to stress the emergence of a longer time exponential tail. 

\begin{figure}
    \centering
    \includegraphics[width=0.7\textwidth]{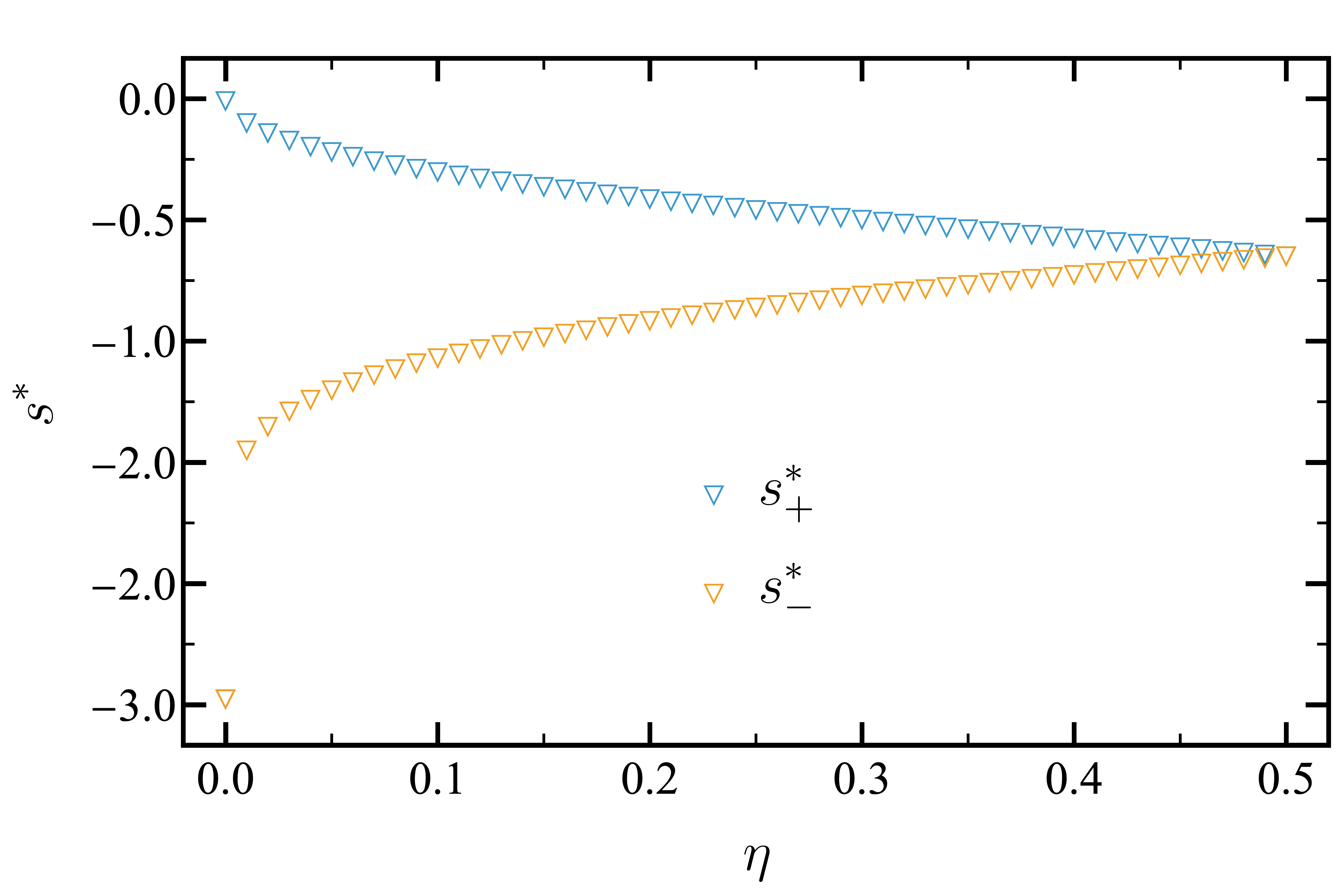}
    \caption[Poles $s_{\pm}^*$ as a function of the disorder $\disorder$. Resetting rates $(\alpha_+,\alpha_-)$ have been chosen as those minimising $\overline{M_1}$---consistently with figure~\ref{fig:dichotomous_FPT_distribution}.]{Poles $s_{\pm}^*$ as a function of the disorder $\disorder$. Resetting rates $(\alpha_+,\alpha_-)$ have been chosen as those minimising the average MFPT---consistently with figure~\ref{fig:dichotomous_FPT_distribution}.}
    \label{fig:dichotomous_FPT_poles_vs_p}
\end{figure}

This asymptotic exponential behaviour, given by the poles $(s_+^*,s_-^*)$, depends completely on the set of parameters $(\alpha_+,\alpha_-,\disorder)$. The relation between the poles and the disorder $\disorder$ is shown in figure~\ref{fig:dichotomous_FPT_poles_vs_p} for the optimal parameters that minimise the average MFPT $(\optMFPT{\alpha}_+,\optMFPT{\alpha}_-)$, which are derived in section~\ref{subsec:optimal_MFPT_dichotomous}. For $0<\disorder< 1/2$, it is always $s_+^*>s_-^*$ and thus the dominant contribution for long times comes from $s_+^*$. Both poles become closer as $\disorder$ increases and merge for $\disorder=1/2$, consistently with the behaviour shown in the last panel in figure~\ref{fig:dichotomous_FPT_distribution}. In the limit as $\disorder\to 0^+$, see the second panel, $s_+^*$ tends to zero: $\positiveSol{\FPTpdf}(\FPT{1};0)$ decays slower as $\disorder$ becomes smaller. Comparing the first two panels, we check that the behaviour of $\positiveSol{\FPTpdf}(\FPT{1};0)$ for $\disorder=0$ is very different from that for $\disorder=0^+$, since $\negativeSol{\FPTpdf}(\FPT{1};0)$ is the only survivor despite $\positiveSol{\FPTpdf}(\FPT{1};0)$ would have the largest pole. This has a clear connection with the optimal choice of the parameters, since we prove that the optimal resetting rates at $\disorder=0$ (the target is certainly to the left of the resetting point) are $\alpha_+ \to \infty$, $\alpha_-=0$. This means that the Laplace transform reduces to $\FPTpdfLaplace(s;0)=\negativeSol{\FPTpdfLaplace}(s;0)=\sech(\sqrt{s})$, whose pole with the largest real part is $s_{-}^*=-2.47$. 

\clearpage

\section{Optimisation of observables of interest}
\label{sec:dichotomous_Optimisation}

The two main observables we are analysing are the mean and the standard deviation of the FPT distribution. From the hierarchy of moments~\eqref{eq:disordered_FPT_moments}, we can obtain 
\begin{subequations}
    \begin{align}
    \positiveSol{\FPT{(1)}}(0) &= \alpha_+^{-2} \parenthesis{\cosh\alpha_+ + \frac{\alpha_+}{\alpha_-}\sinh\alpha_+ -1}\equiv M_1(\alpha_+,\alpha_-), 
    \label{eq:disordered_MFPT}
    \\
    \positiveSol{\FPT{(2)}}(0) &= \frac{1}{\alpha_-^3\alpha_+^4}
    \bigg\{
        \big(\alpha_-^2-\alpha_+^2\big)\alpha_- 
        + \big[\alpha_+^2-\alpha_-^2(\alpha_-+3)\big]\alpha_+\sinh\alpha_+
    \nonumber\\ &\quad 
        - \big(\alpha_+^2+2\alpha_- -4\alpha_+\sinh\alpha_+\big)\alpha_-^2\cosh\alpha_+
    \nonumber\\ &\quad
        + \big(\alpha_-^2+\alpha_+^2\big)\alpha_-\cosh(2\alpha_+)
        \bigg\}\equiv M_2(\alpha_+,\alpha_-),
    \label{eq:disordered_secondMoment}
    \end{align}
\end{subequations}
where we have defined $M_n(\alpha_+,\alpha_-)$ to highlight the dependence on the resetting parameters. We recall that we are always considering $x_0=x_r=0$.  

Recalling that the moments for $\targetx=-1$ are obtained by exchanging $\alpha_+\leftrightarrow \alpha_-$, the average first and second moment can be written as the linear combination 
\begin{align}
    \overline{M_n} = \disorder M_n(\alpha_+,\alpha_-) + (1-\disorder) M_n(\alpha_-,\alpha_+), \quad n=1,2,
    \label{eq:disordered_totalMFPT}
\end{align}
whereas the standard deviation is defined as 
\begin{equation}
    \overline{\sigma_{\FPT{1}}}(\alpha_+,\alpha_-) = \sqrt{\overline{M_2}-\overline{M_1}^2}.
    \label{eq:disordered_std}
\end{equation}
The explicit expressions are not very illuminating, so we do not write them. For a homogeneous resetting profile, $\alpha_+=\alpha_-=\alpha_h$, the expressions simplify to 
\begin{equation}
    \overline{M_1}(\alpha_h,\alpha_h)  = \frac{e^{\alpha_h}-1}{\alpha_h^2}, \quad
    \overline{\sigma_{\FPT{1}}} = \frac{\sqrt{(2\sinh\alpha_h-\alpha_h)e^{\alpha_h}}}{\alpha_h^2},
\end{equation}
which is a robust benchmark, since we recover the same expression as the standard stochastic resetting with a fixed target~\eqref{eq:MFPT_SSR}. Obviously, the former expressions do not depend on $\disorder$.

\begin{figure}
    \centering
    \includegraphics[width=0.49\textwidth]{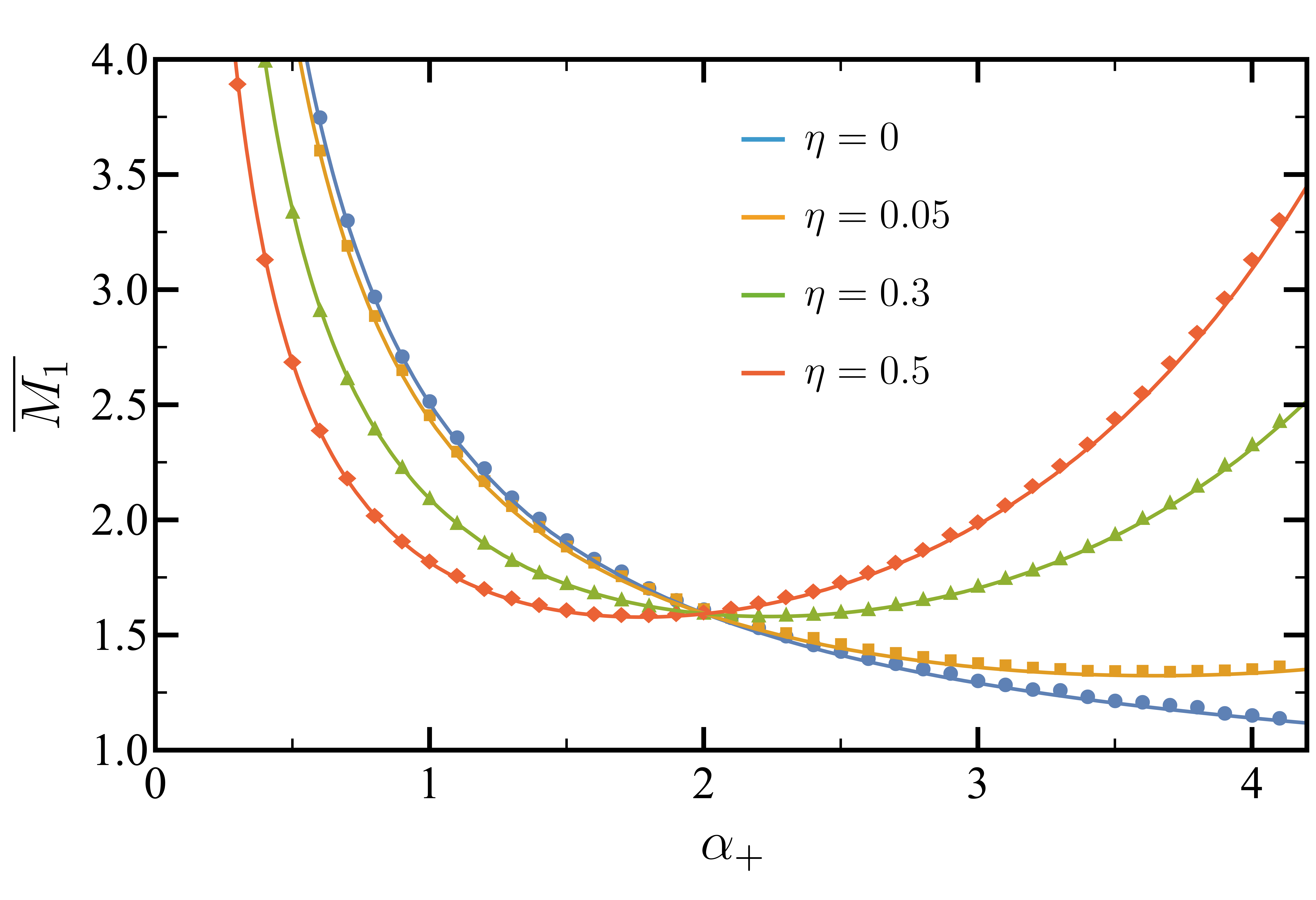}
    \includegraphics[width=0.49\textwidth]{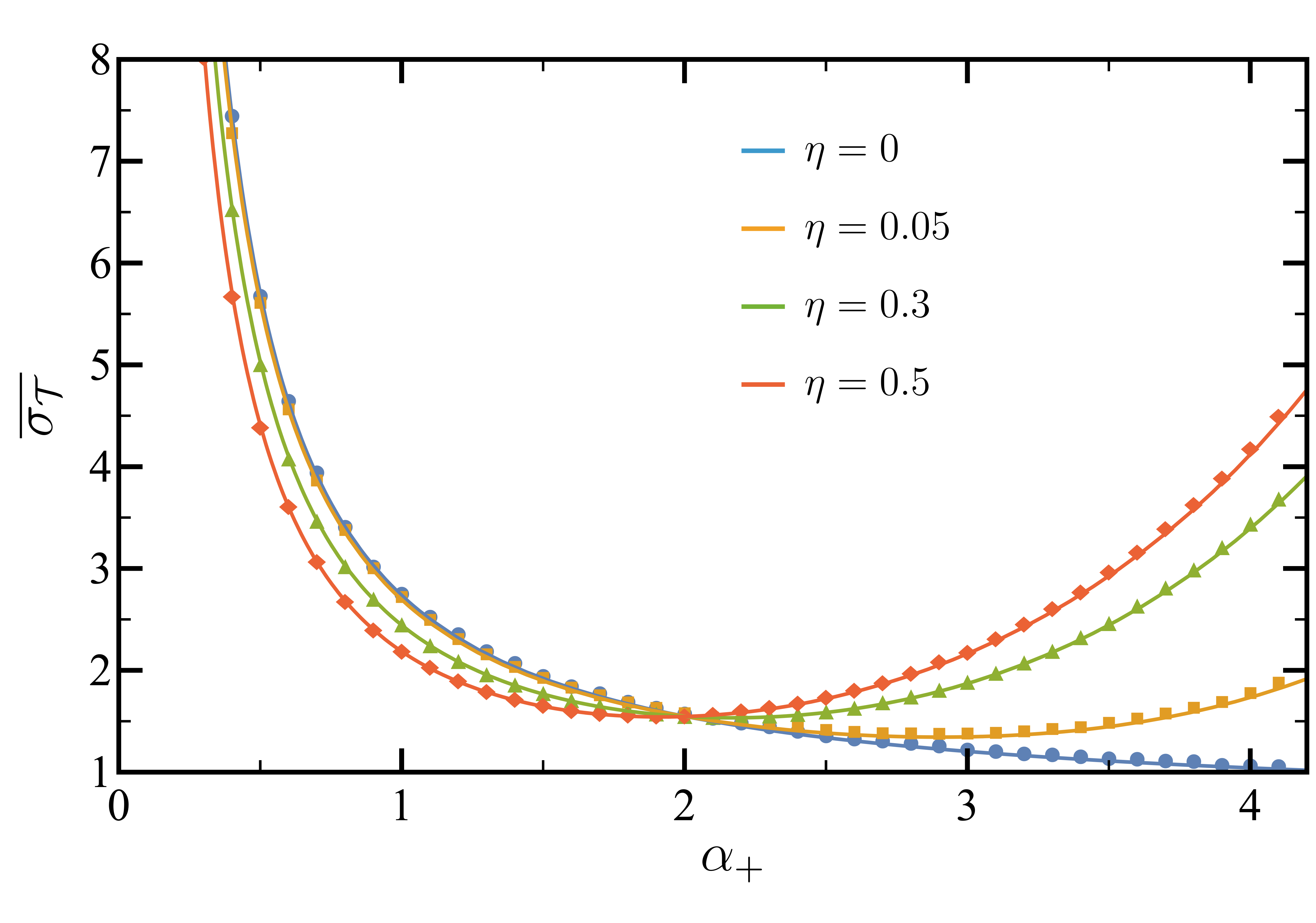}
    \caption{Average mean first-passage time (left panel) and its standard deviation (right panel) as functions of $\alpha_+$, at fixed $\alpha_-=2$. Simulation data (symbols) are compared with the exact analytical results (solid lines), showing a perfect agreement. Different values of the disorder $\disorder$ have been used to produce the different curves, as detailed in the legend.}
    \label{fig:disordered_obversables_analytical_simulations}
\end{figure}

\begin{figure}
    \centering
    \includegraphics[width=0.49\textwidth]{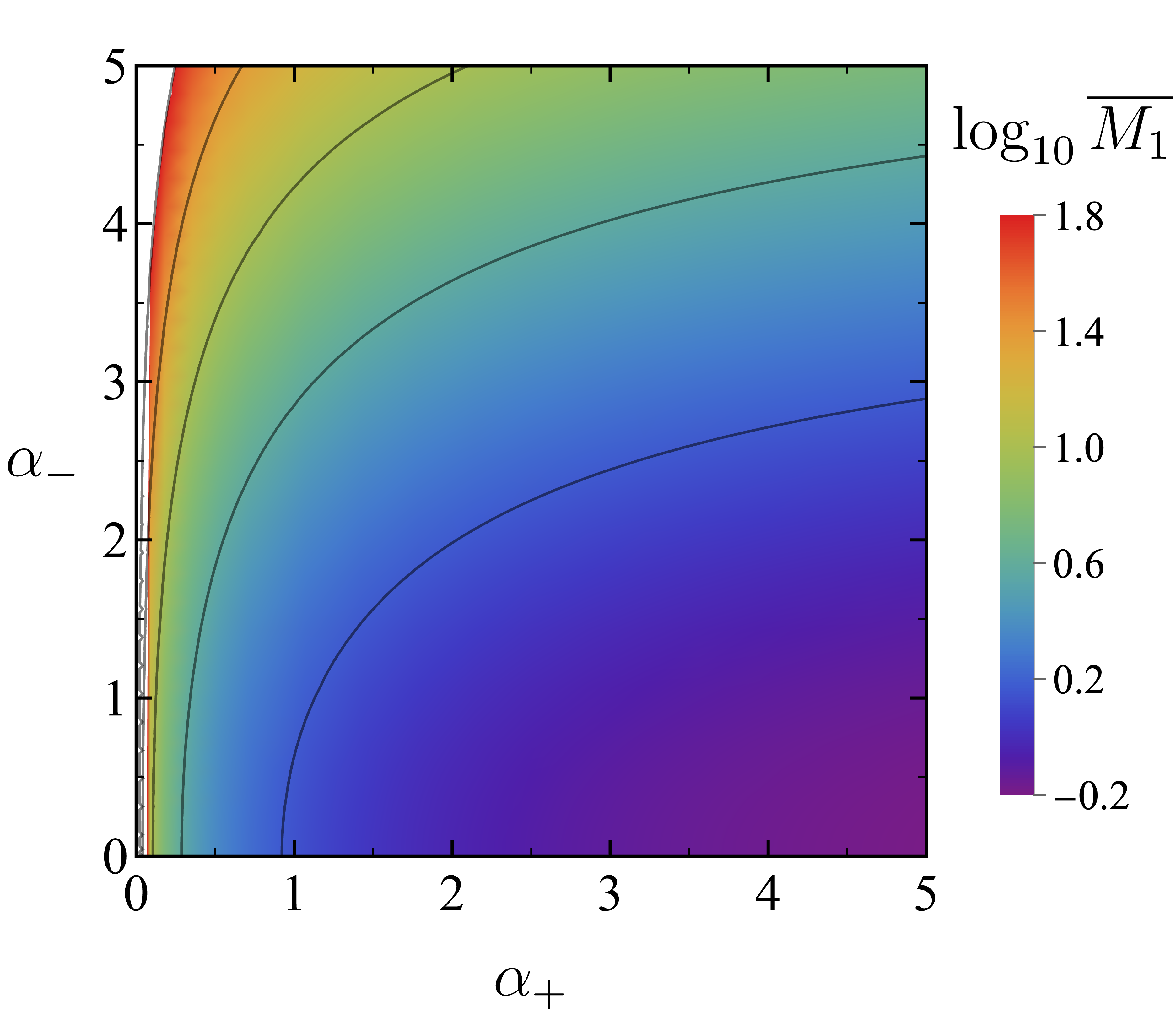}
    \includegraphics[width=0.49\textwidth]{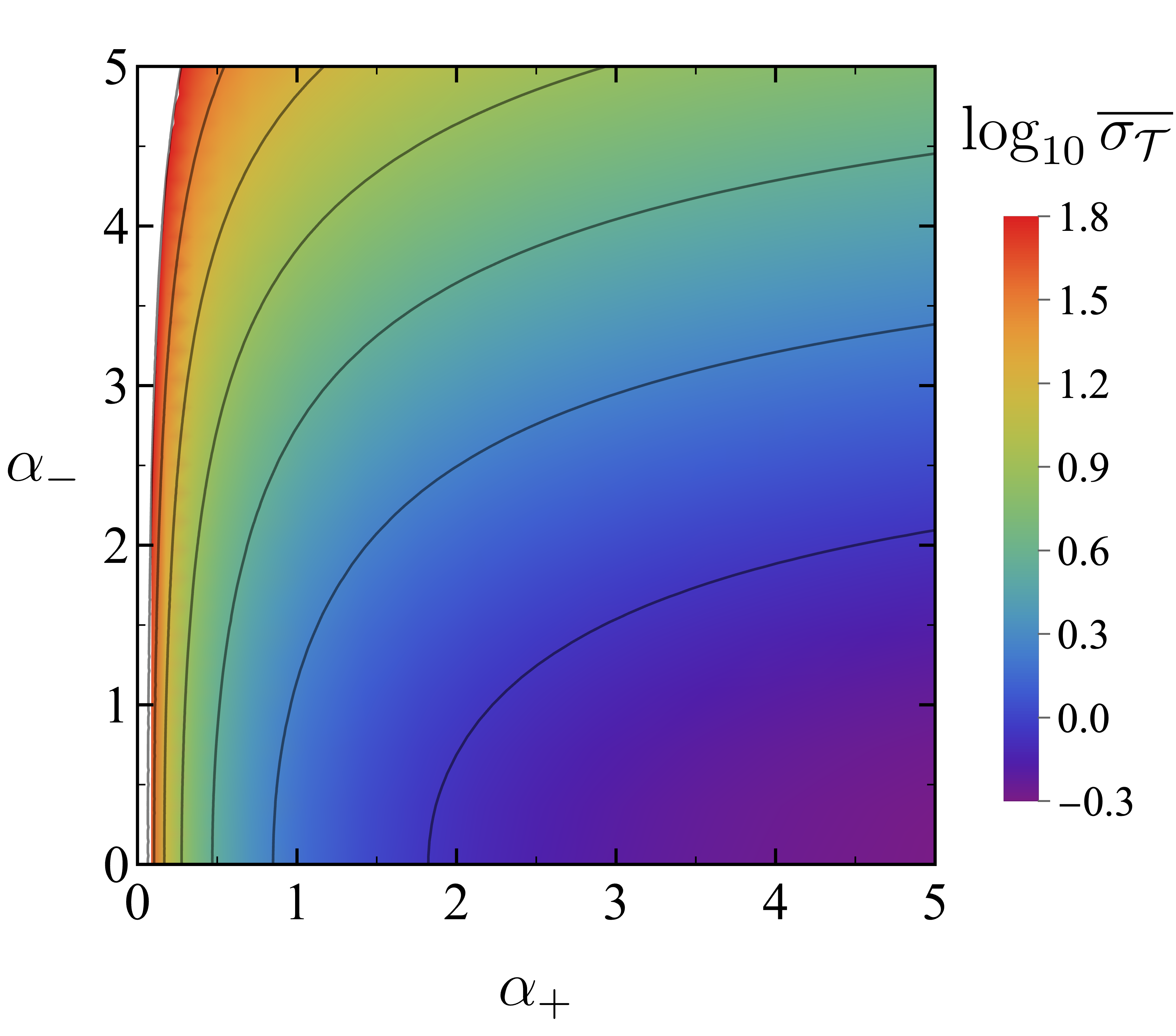}
    \includegraphics[width=0.49\textwidth]{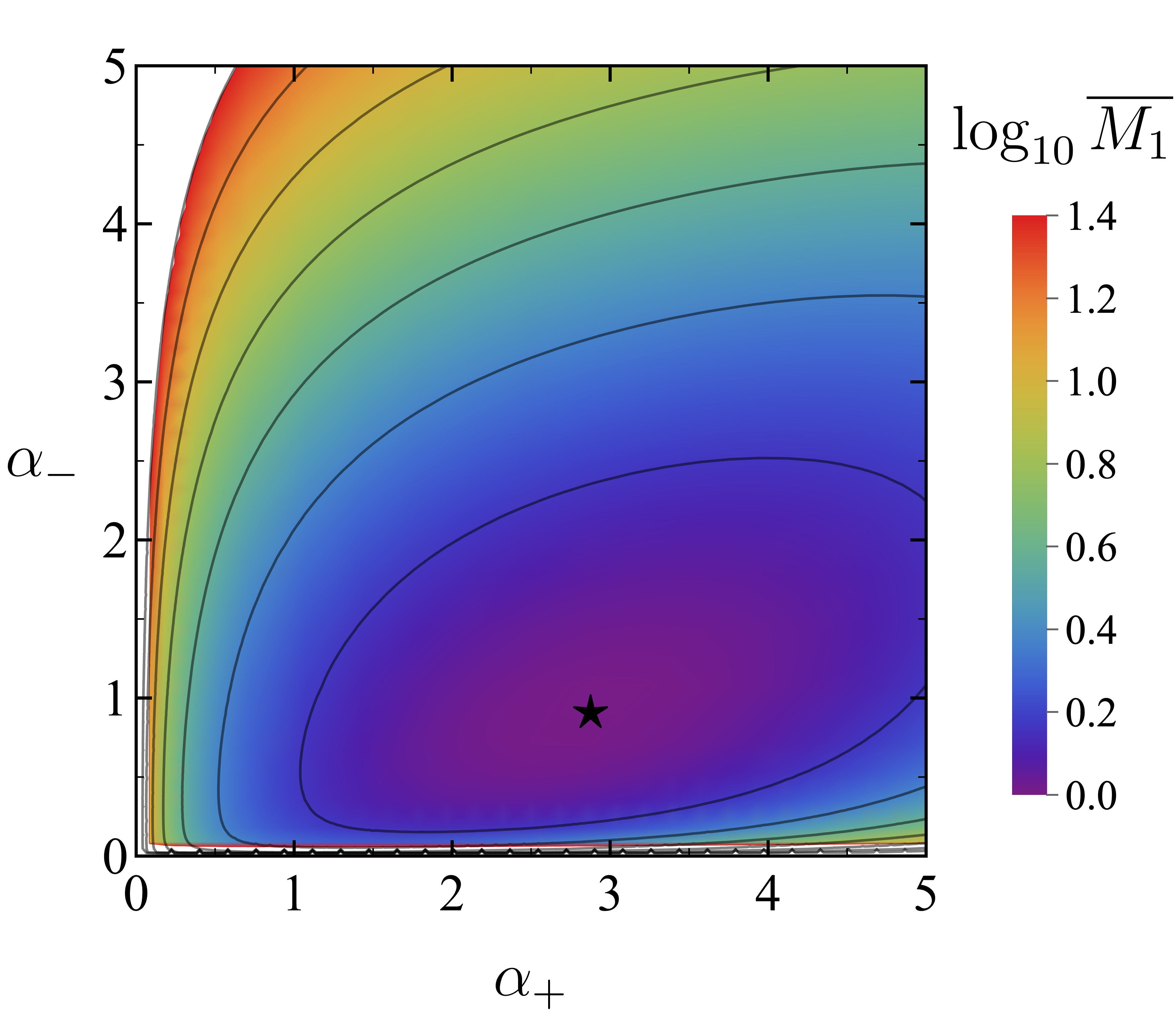}
    \includegraphics[width=0.49\textwidth]{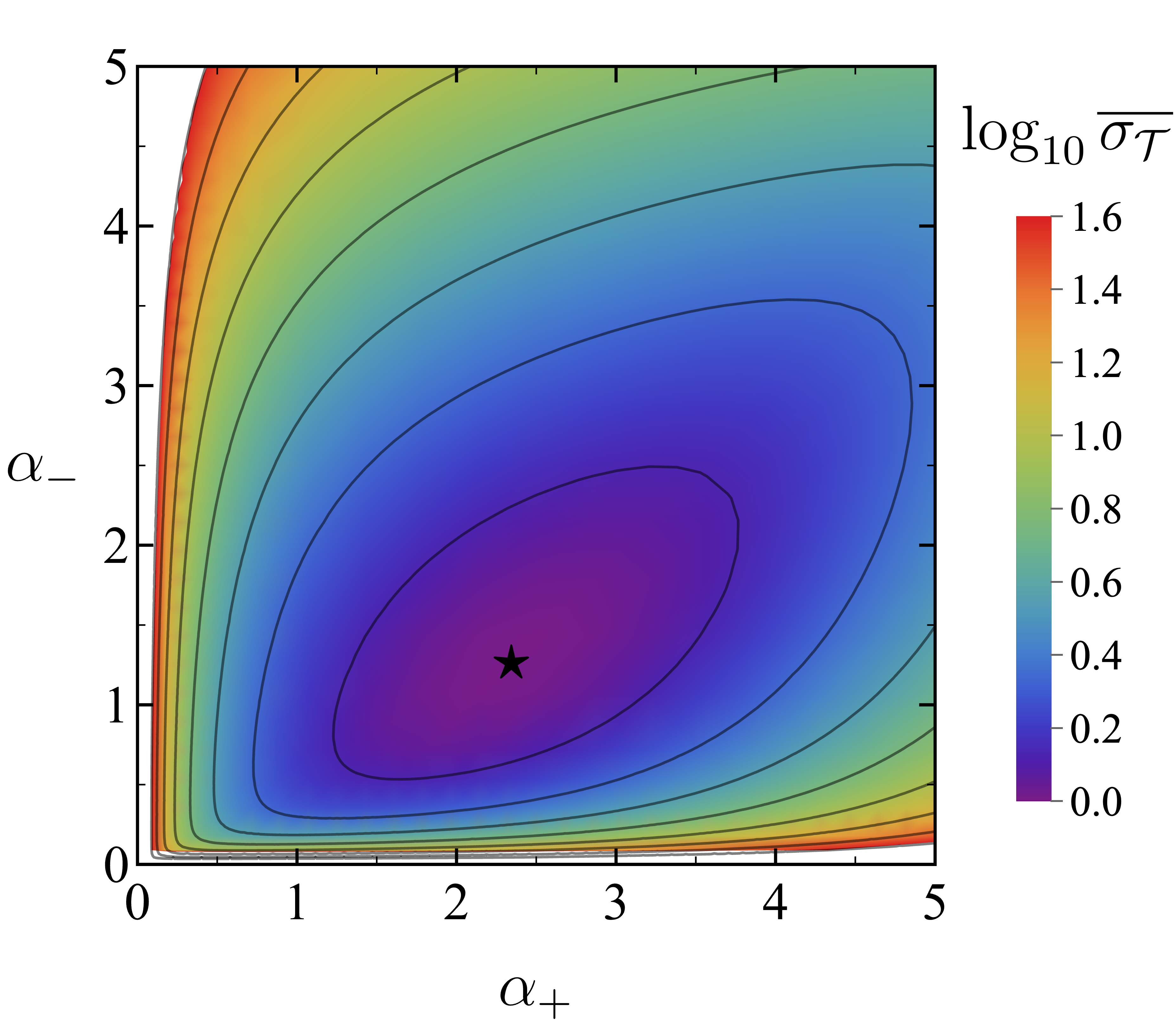}
    \includegraphics[width=0.49\textwidth]{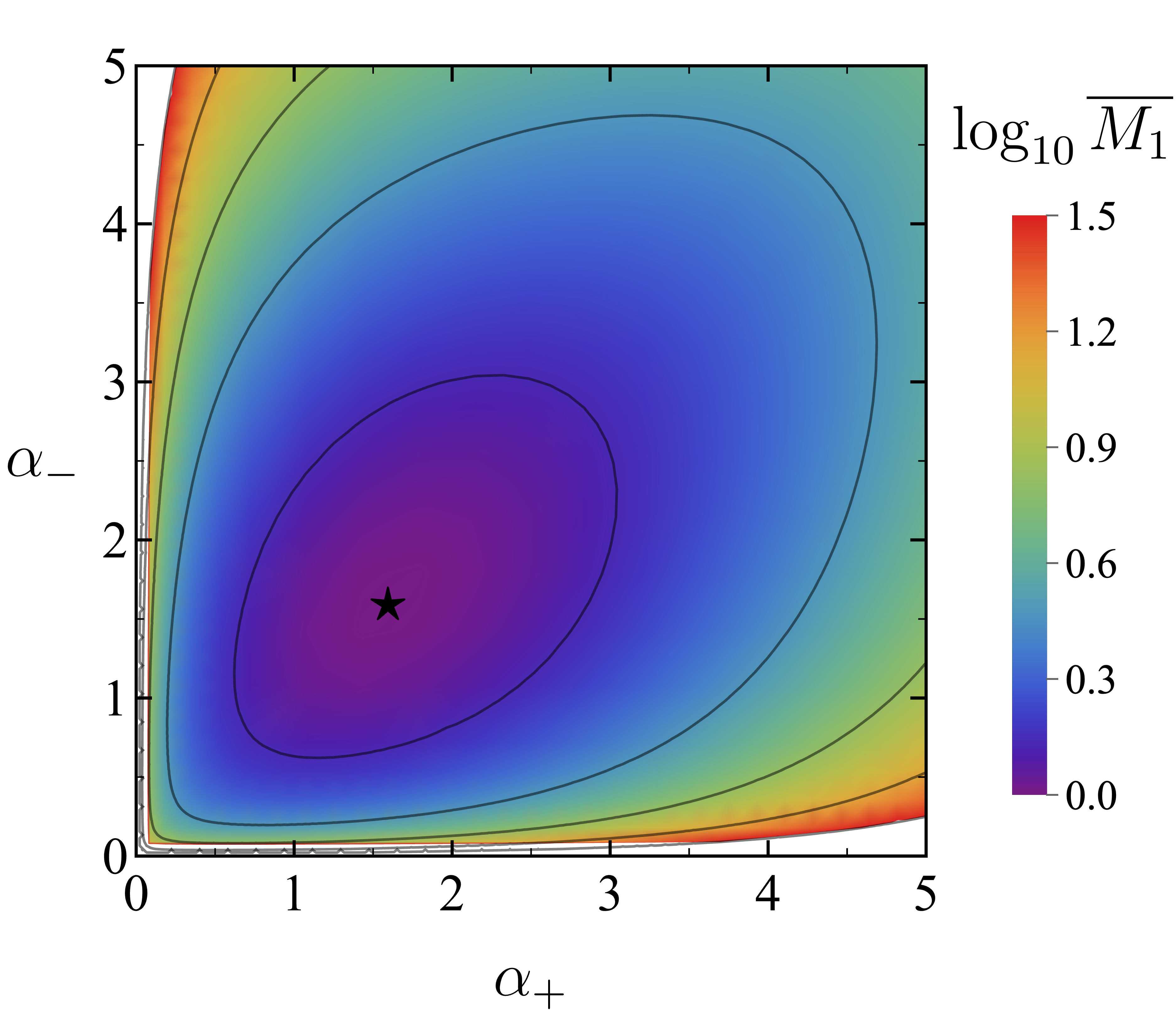}
    \includegraphics[width=0.49\textwidth]{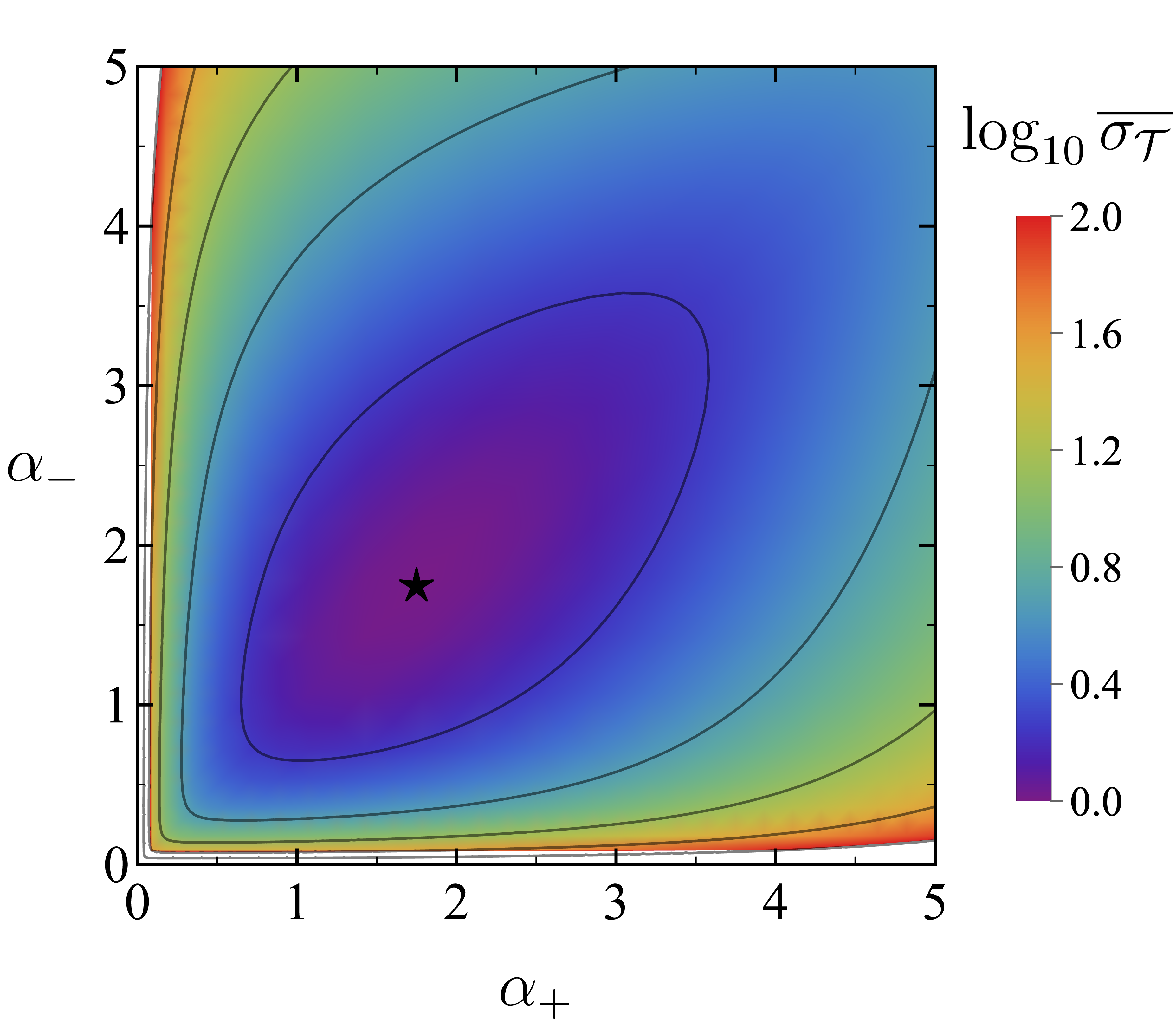}
    \caption{Density plots of the average mean first-passage time (left panels) and its standard deviation (right panels) on the $(\alpha_+,\alpha_-)$ plane.  They are obtained by evaluating equations~\eqref{eq:disordered_totalMFPT} and~\eqref{eq:disordered_std}. Different rows stand for different values of the disorder $\disorder$, namely $\disorder=\{0, 0.05,0.5\}$ from top to bottom. In each panel, white is used to indicate values larger than the maximum one in the corresponding colour legend, whereas the filled black star indicates the minimum---for $\disorder=0$, it is reached at $\alpha_+\to \infty$ and $\alpha_-=0$.}
    \label{fig:disordered-densityplot-full}
\end{figure}

A comparison between theoretical predictions and numerical simulations of these quantities is shown in figure~\ref{fig:disordered_obversables_analytical_simulations}; we find an excellent agreement between them. The simulations have been performed as explained in appendix~\ref{app:langevin_simulations}, averaging over $N=10^7$ trajectories and using a time step $\Delta t = 10^{-5}$. Specifically, fixing $\alpha_-=2$, we have plotted the average MFPT and its standard deviation as functions of $\alpha_+$ for different values of the disorder $\disorder$. One can clearly observe that there exist optimal pairs that minimise both observables, which depend on the value of $\disorder$. A more complete picture of these observables is presented in figure~\ref{fig:disordered-densityplot-full}, where the full dependence on both resetting rates $(\alpha_+,\alpha_-)$ is illustrated, for several values of $\disorder$---from top to bottom, $\disorder=\{0,0.05,0.5\}$. Some interesting features must be highlighted here. First, in the non-disordered case, $\disorder=0$, the optimal values are reached when $\alpha_+\to\infty$ and $\alpha_-\to0^+$, \ie the searcher cannot explore on the right side, and it is free to diffuse on the left side $(-1,0)$---this is equivalent to include a reflecting boundary at $x_0=x_r=0$. The other extreme case, $\disorder=1/2$, behaves as expected, since the reflection symmetry with respect to the line $\alpha_+=\alpha_-$ stands out. In intermediate cases, $0<\disorder<1/2$, both the MFPT and the standard deviation have a non-trivial dependence on $(\alpha_+,\alpha_-)$, and the optimal values are finite. Specifically, it is worth noting that both quantities diverge in the limits $\alpha_+\to 0^+$ and/or $\alpha_-\to 0^+$, even for $\disorder\to0^+$, remarking the impact of any non-vanishing disorder.

\subsection{Optimal mean first-passage time}
\label{subsec:optimal_MFPT_dichotomous}

In this section, we focus on finding the optimal pair $(\optMFPT{\alpha}_+,\optMFPT{\alpha}_-)$ that minimises the average MFPT~\eqref{eq:disordered_totalMFPT}, \ie we have to solve 
\begin{equation}
    \left.\pdev{}{\overline{M_1}}{\alpha_+}\right|_{\raisebox{-0.0ex}{$\stackrel{\scriptstyle \alpha_+=\optMFPT{\alpha}_+}{\scriptstyle \alpha_-=\optMFPT{\alpha}_-}$}}
    =\left.\pdev{}{\overline{M_1}}{\alpha_-}\right|_{\raisebox{-0.0ex}{$\stackrel{\scriptstyle \alpha_+=\optMFPT{\alpha}_+}{\scriptstyle \alpha_-=\optMFPT{\alpha}_-}$}} = 0. 
\end{equation}
This pair $(\optMFPT{\alpha}_+,\optMFPT{\alpha}_-)$ is clearly a function of $\disorder$, as we notice in figure~\ref{fig:disordered-densityplot-full}. For $\disorder=1$ ($\disorder=0$), we expect the optimal resetting to forbid the exploration of the region on the opposite side of the fixed target with $\alpha_-\to\infty$ ($\alpha_+\to\infty$). This physical intuition is numerically confirmed in the first panel of figure~\ref{fig:disordered-densityplot-full}. Mathematically, it can be proved by the fact that (i) $M_1$ in~\eqref{eq:disordered_MFPT} is a monotonically increasing (decreasing) function with respect to its first (second) argument, and (ii) just one $M_1$ survives in the average~\eqref{eq:disordered_totalMFPT} for extreme cases. For any $\disorder\in[0,1]$, there just exists one solution $(\optMFPT{\alpha}_+,\optMFPT{\alpha}_-)$. This can be obtained numerically for any $\disorder$, as we check in figure~\ref{fig:disordered-densityplot-full}. Nevertheless, we can also derive explicit theoretical approximations in some limits. Recall that we focus on analysing the case $\disorder\in[0,1/2]$, due to the mirror symmetry.

\subsubsection*{Almost symmetric target location}
When the dichotomous target distribution is almost symmetric, \ie $\disorder = 1/2 +\delta \disorder$, with $|\delta \disorder| \ll 1$, it is handy to rewrite~\eqref{eq:disordered_totalMFPT} as 
\begin{equation}
    \overline{M_1}(\alpha_+,\alpha_-) = M_s(\alpha_+,\alpha_-) + \delta \disorder M_a(\alpha_+,\alpha_-),
\end{equation}
where we have defined the symmetric and antisymmetric functions 
\begin{subequations}
    \begin{align}
        M_s(\alpha_+,\alpha_-) &= \frac{M_1(\alpha_+,\alpha_-) + M_1(\alpha_-,\alpha_+)}{2}, \\
        M_a(\alpha_+,\alpha_-) &= M_1(\alpha_+,\alpha_-) - M_1(\alpha_-,\alpha_+).
    \end{align}
\end{subequations}
From homogeneous resetting, the symmetric part is optimised at $\alpha_+=\alpha_-=\optMFPT{\alpha}_h$, whose value is obtained by 
\begin{equation}
    \optMFPT{\alpha}_h = 2\parenthesis{1-e^{-\optMFPT{\alpha}_h}} \implies \optMFPT{\alpha}_h = 2 + W(-2e^{-2}),
\end{equation}
which is the same reported in the case of SSR, see~\eqref{eq:optimal_rate_SSR}.

\begin{figure}[!ht]
    \centering
    \includegraphics[width=0.7\textwidth]{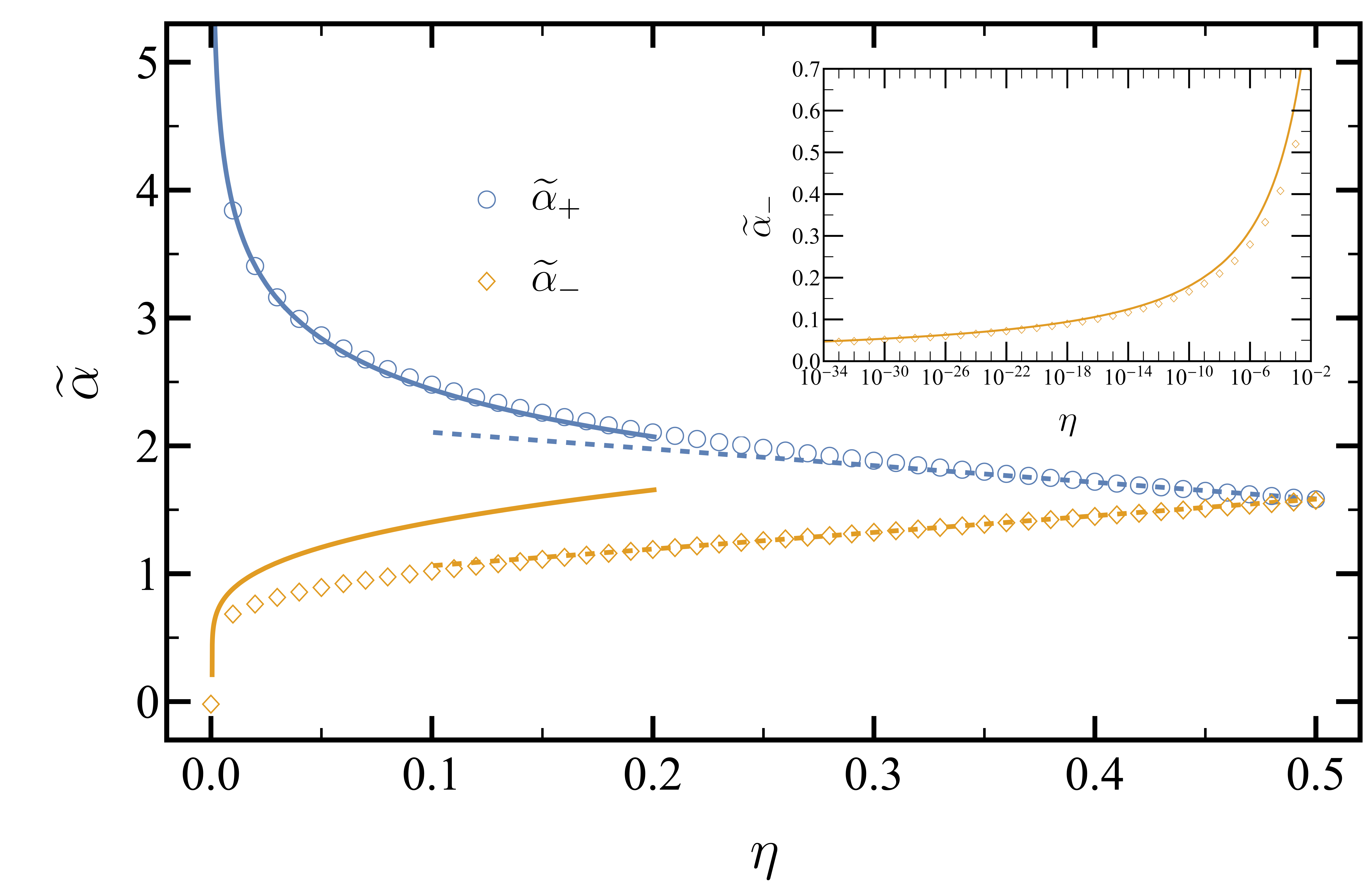}
    \includegraphics[width=0.7\textwidth]{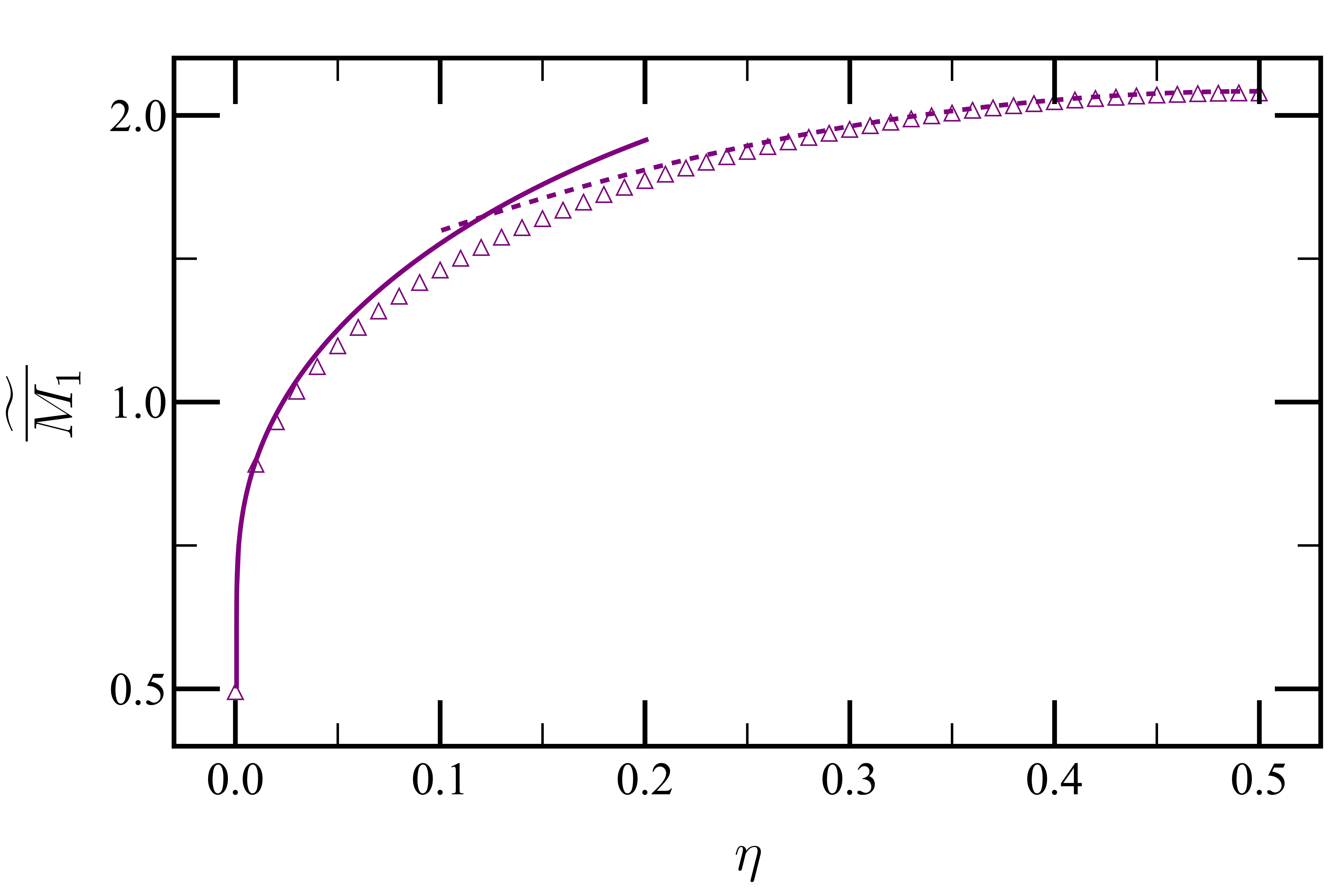}
    \caption{Minimum of the average mean first-passage time as a function of the disorder. Empty symbols stand for the numerical values obtained from numerically minimising the average MFPT, in~\eqref{eq:disordered_totalMFPT} with $n=1$, whereas the lines correspond to asymptotic expansions in the limits $\disorder \to 0$ (solid line) and $\disorder \to 1/2$ (dashed line). The top panel shows the optimal resetting rates $(\optMFPT{\alpha}_+,\optMFPT{\alpha}_-)$ as a function of $\disorder$ ($\optMFPT{\alpha}_+$: blue, $\optMFPT{\alpha}_-$: orange). The inset is a zoom of the region close to $\disorder=0$---note the logarithmic scale on the horizontal axis. The bottom panel shows the associated average MFPT $\optMFPT{\overline{M_1}}$, again as a function of $\disorder$.}
    \label{fig:MFPT_optimal}
\end{figure}

For $|\delta \disorder|\ll 1$, we write the average MFPT using a perturbative method around the symmetrical scenario $\disorder=1/2$. Using $(\alpha_+,\alpha_-)= (\optMFPT{\alpha}_h,\optMFPT{\alpha}_h) + (c_+,c_-) \delta \disorder$, it reads
\begin{equation}
    \overline{M_1} \simeq M_s + {\bf c}^T \cdot {\nabla M_s} \, \delta \disorder + \frac{1}{2} {\bf c}^T \cdot \mathbb{H}_s \cdot {\bf c} \, (\delta \disorder)^2 + M_a \delta \disorder + {\bf c}^T \cdot {\nabla M_a} \, (\delta \disorder)^2
\end{equation}
where ${\bf c}^T\equiv(c_+,c_-)$, $\nabla = (\partial_{\alpha_+},\partial_{\alpha_-})^T$, and $\mathbb{H}_s$ is the Hessian matrix of $M_s$. All the coefficients are computed at $\alpha_+=\alpha_-=\optMFPT{\alpha}_h$, so $\nabla M_s=0$ is due to optimality and $M_a=0$ is due to antisymmetry. The optimal coefficients $\optMFPT{\bf c}$ that minimises $\overline{M_1}$ are the solution of 
\begin{equation}
    \label{eq:c_opt}
    \optMFPT{\bf c} = - \mathbb{H}_s^{-1} \cdot \nabla M_a.
\end{equation} 
Then the minimum of the average MFPT is
\begin{equation}
    \overline{M_1}(\optMFPT{\alpha}_+,\optMFPT{\alpha}_-) \simeq M_s - \frac{1}{2} \left({\nabla M_a}\right)^T \cdot \mathbb{H}^{-1}_s \cdot \nabla M_a (\delta \disorder)^2 \leq M_s, 
    \label{eq:disordered_MFPT_approx_p=1/2}
\end{equation}
where
\begin{subequations}
    \begin{align}
    \left( \nabla M_a \right)^T &= \frac{4-\optMFPT{\alpha}_h}{2\optMFPT{\alpha}_h^2(2-\optMFPT{\alpha}_h)} (1,-1),
    \\
   \mathbb{H}_s &= \frac1{4\optMFPT{\alpha}_h^3(2-\optMFPT{\alpha}_h)} 
    \left(\begin{array}{cc}
        \optMFPT{\alpha}_{s}(1+\optMFPT{\alpha}_{s}) & -4+\optMFPT{\alpha}_{s}(3-\optMFPT{\alpha}_{s})\\
        -4+\optMFPT{\alpha}_{s}(3-\optMFPT{\alpha}_{s}) & \optMFPT{\alpha}_{s}(1+\optMFPT{\alpha}_{s})
    \end{array}\right),
    \\
    \frac{1}{2} \left({\nabla M_a}\right)^T \cdot \mathbb{H}_s^{-1} \cdot \nabla M_a &= \frac{(4-\optMFPT{\alpha}_h)^2}{2\optMFPT{\alpha}_h (2-\optMFPT{\alpha}_h)(2+\optMFPT{\alpha}_h(\optMFPT{\alpha}_h-1))} 
    .
    \label{eq:disordered_asymptotic_MFPT_nearSymmetric}
    \end{align}
\end{subequations}
Thus, the optimal rates are $\optMFPT{\alpha}_\pm = \optMFPT{\alpha}_h + \optMFPT{c}_\pm$, with coefficients 
\begin{equation}
    \optMFPT{\bf c}^T = -\frac{\optMFPT{\alpha}_h(4-\optMFPT{\alpha}_h)}{2+\optMFPT{\alpha}_h(\optMFPT{\alpha}_h-1)} (1,-1).
\end{equation}
These approximations are represented as solids lines in figure~\ref{fig:MFPT_optimal}, providing good agreement with the numerical results---even for values of $\disorder$ not so close to $\disorder=1/2$.

\subsubsection*{Highly asymmetric target location}

In the opposite limit, $\disorder=\delta \disorder \ll 1$, it is almost sure that the target lies on the half-line to the left of the resetting point. As physically expected, for $\disorder=0$, the optimal rates are $\optMFPT{\alpha}_+\to\infty$ and $\optMFPT{\alpha}_-\to0^+$. Therefore, one expects that to have $\optMFPT{\alpha}_-\ll 1$ and $\optMFPT{\alpha}_+\gg 1$ as $\disorder\to 0^+$. In these limits, the average MFPT~\eqref{eq:disordered_totalMFPT} can be approximated as
\begin{equation}
    \overline{M_1} = M_1(\alpha_-,\alpha_+) + \delta \disorder \brackets{M_1(\alpha_+,\alpha_-)-M_1(\alpha_-,\alpha_+)}, 
\end{equation}
where the leading terms of each function become
\begin{subequations}
    \begin{align}
        M_1(\alpha_+,\alpha_-) &\sim \frac{e^{\alpha_+}}{2 \alpha_+\alpha_-}
        \\
        M_1(\alpha_-,\alpha_+) &\sim \frac{1}{2} + \frac{\alpha_-^2}{24}+\frac{1}{\alpha_+}.
    \end{align}
\end{subequations}
This implies that the derivative of the average MFPT follows 
\begin{subequations}
    \begin{align}
        \pdev{}{\overline{M_1}}{\alpha_+} &\sim -\frac{1}{\alpha_+^2} + \frac{e^{\alpha_+}}{2 \alpha_+\alpha_-}\delta p, 
        \\
        \pdev{}{\overline{M_1}}{\alpha_-} &\sim \frac{\alpha_-}{12} - \frac{e^{\alpha_+}}{2 \alpha_+ \alpha_-^2}\delta p,
    \end{align}
\end{subequations}
so the optimal rates yield
\begin{equation}
    \optMFPT{\alpha}_+ \sim 2 W\parenthesis{\frac{3^{1/4}}{\sqrt{\delta p}}}, \quad 
    \optMFPT{\alpha}_- \sim \frac{2\sqrt{3}}{\optMFPT{\alpha}_+}.
    \label{eq:disordered_MFPT_asymptotic_optimalRates_nearZero}
\end{equation}

The optimal values of the resetting rates $(\optMFPT{\alpha}_+,\optMFPT{\alpha}_-)$ and the resulting average MFPT $\optMFPT{\overline{M_1}}$ are shown in figure~\ref{fig:MFPT_optimal}. The asymptotic analysis for $(\optMFPT{\alpha}_+,\optMFPT{\alpha}_-)$ provides good estimations in both limits. It is just when $\disorder\to0^+$ that the approximation for $\optMFPT{\alpha}_-\to 0$ seems to fail, due to the quite rapid variation of $\optMFPT{\alpha}_-$ in a narrow interval of $\disorder$---this abrupt variation arises from the emergence of the disorder when changing from $\disorder=0$ (no disorder) to $\disorder\to0^+$ (weak disorder). However, the inset in the top panel proves that the approximation~\eqref{eq:disordered_MFPT_asymptotic_optimalRates_nearZero} and the numerical results are reconciled when plotted in a semi-log scale.  

\subsection{Optimal standard deviation of the first-passage time}
\label{subsec:optimal_std_dichotomous}

The optimisation of the standard deviation of the average FPT distribution involves looking for the solution $(\optVar{\alpha}_+,\optVar{\alpha}_-)$ of the system of equations 
\begin{equation}
    \left.\pdev{}{\overline{\sigma_{\FPT{1}}}}{\alpha_+}\right|_{\raisebox{-0.0ex}{$\stackrel{\scriptstyle \alpha_+=\optVar{\alpha}_+}{\scriptstyle \alpha_-=\optVar{\alpha}_-}$}}
    =\left.\pdev{}{\overline{\sigma_{\FPT{1}}}}{\alpha_-}\right|_{\raisebox{-0.0ex}{$\stackrel{\scriptstyle \alpha_+=\optVar{\alpha}_+}{\scriptstyle \alpha_-=\optVar{\alpha}_-}$}} = 0. 
\end{equation}
We use the hat notation to refer to the optimisation of the standard deviation, in contrast to the tilde used for the minimisation of the average MFPT. Since the analytical expression is quite cumbersome, only the numerical results are presented in the following. 

\begin{figure}
    \centering
    \includegraphics[width=0.49\textwidth]{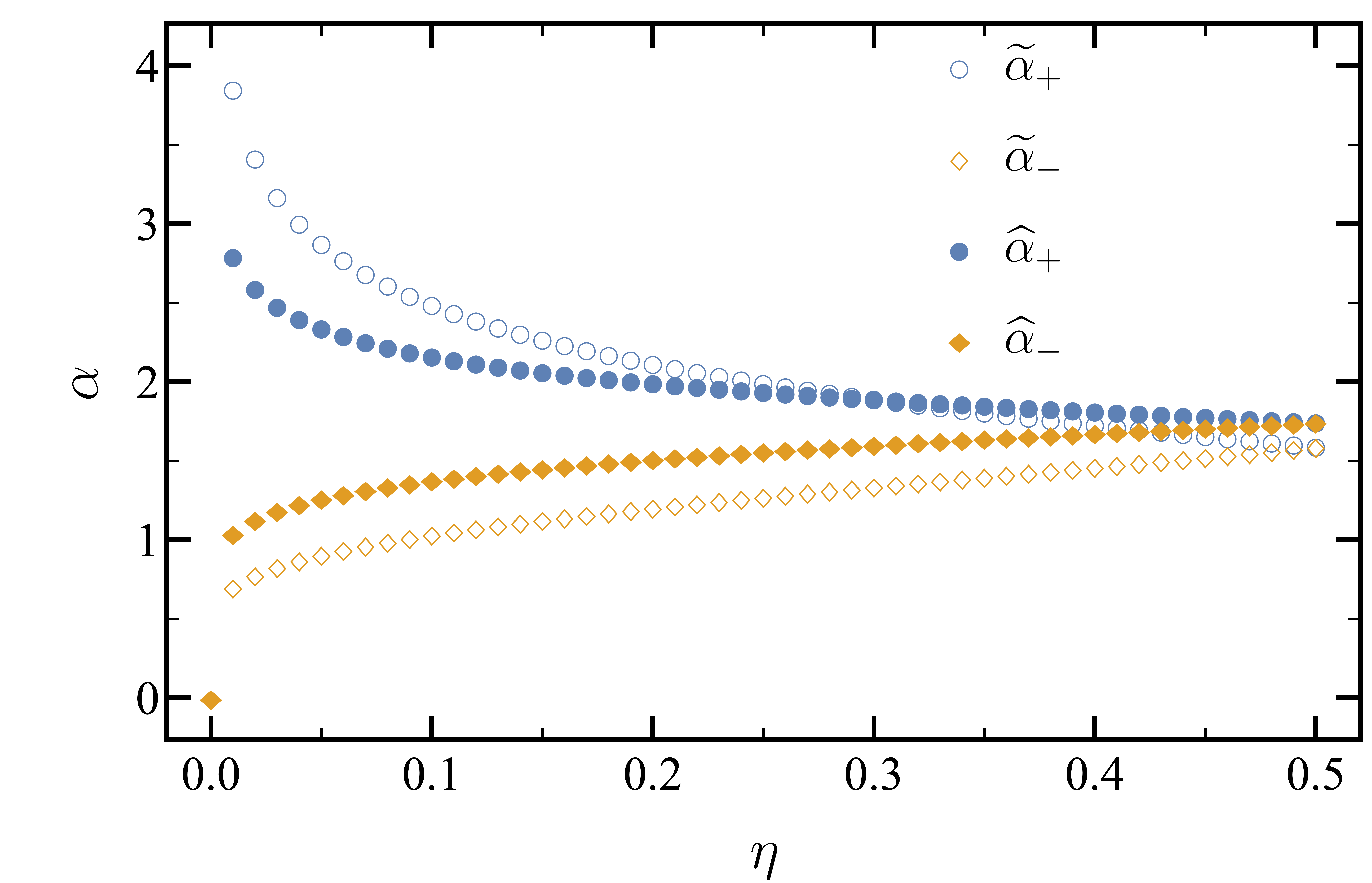}
    \includegraphics[width=0.49\textwidth]{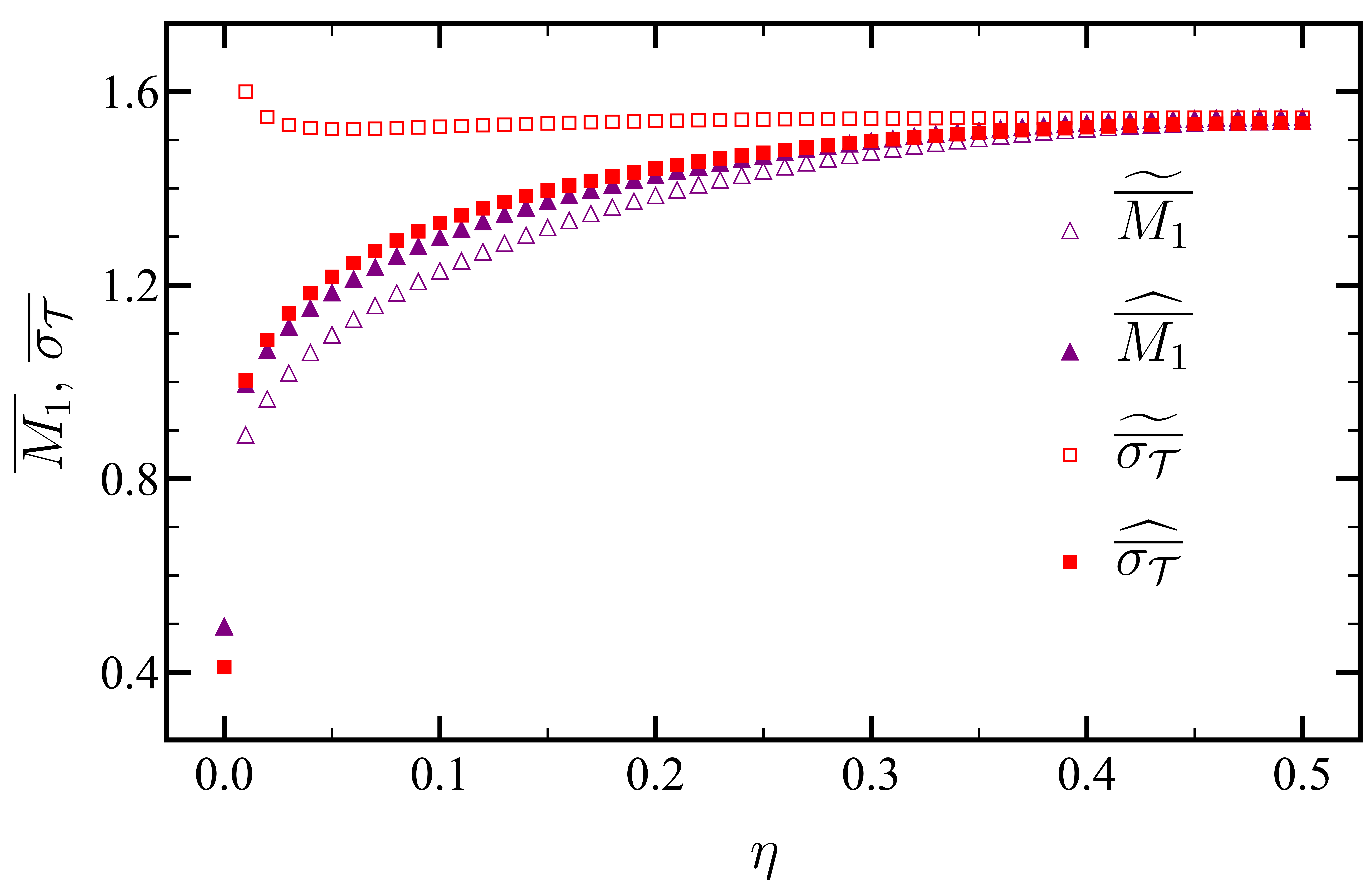}
    \caption{Comparison between the optimal solutions for the average mean first-passage time and the standard deviation, all of them obtained numerically. On the left panel, we show the optimal resetting rates $(\alpha_+,\alpha_-)$ (blue circles and orange diamonds, respectively) when either the average MFPT or the standard deviation are minimised. On the other hand, on the right panel, we compare all the possible cases of the average MFPT and standard deviation (purple triangles and red rectangles, respectively) when one of them is optimal. Filled (empty) symbols correspond to the optimisation of the standard deviation (average MFPT).
    }
    \label{fig:disordered_comp_optimalvalues}
\end{figure}

The optimal values $(\optVar{\alpha}_+,\optVar{\alpha}_-)$ and the resulting minimum $\optVar{\overline{\sigma}}$ are shown in figure~\ref{fig:disordered_comp_optimalvalues}. In the non-disordered scenario, $\disorder=0$, the expected optimal rates $\optVar{\alpha}_+\to\infty$ and $\optVar{\alpha}_-=0$ are recovered. From a qualitative point of view, the behaviour of $(\optVar{\alpha}_+,\optVar{\alpha}_-)$ seems to be quite similar to the optimal MFPT values $(\optMFPT{\alpha}_+,\optMFPT{\alpha}_-)$, but they show a slower dependence of $\disorder$. In the right panel of figure~\ref{fig:disordered_comp_optimalvalues}, we compare the minimum average MFPT $\optMFPT{\overline{M_1}}$ with $\optVar{\overline{M_1}}$, calculated at the rates that minimise the standard deviation; and also the minimum standard deviation $\optVar{\overline{\sigma_{\FPT{1}}}}$ with the standard deviation $\optMFPT{\overline{\sigma_{\FPT{1}}}}$ calculated at the rates that minimise the average MFPT. It is important to note that the standard deviation explodes in the vicinity of $\disorder=0$ when the average MFPT is minimised, \ie $\lim_{\disorder\to0^+}\optMFPT{\overline{\sigma_{\FPT{1}}}}=\infty$, even though it is finite for exactly $\disorder=0$. The divergence of the standard deviation stems from the contribution to the second moment of the less probable target: the interplay between the low value of $\disorder$, which means that trajectories with $\targetx=1$ are very rare, and the divergent waiting times over such rare trajectories. The different behaviour of $\optVar{\alpha}_\pm(\disorder)$ regularises the standard deviation $\optVar{\overline{\sigma_{\FPT{1}}}}$; although the contribution of the second moment stemming from the less probable target still diverges, it remains finite when multiplied by the probability of finding the target $\disorder$.

\begin{figure}[t]
    \centering
    \includegraphics[width=0.6\textwidth]{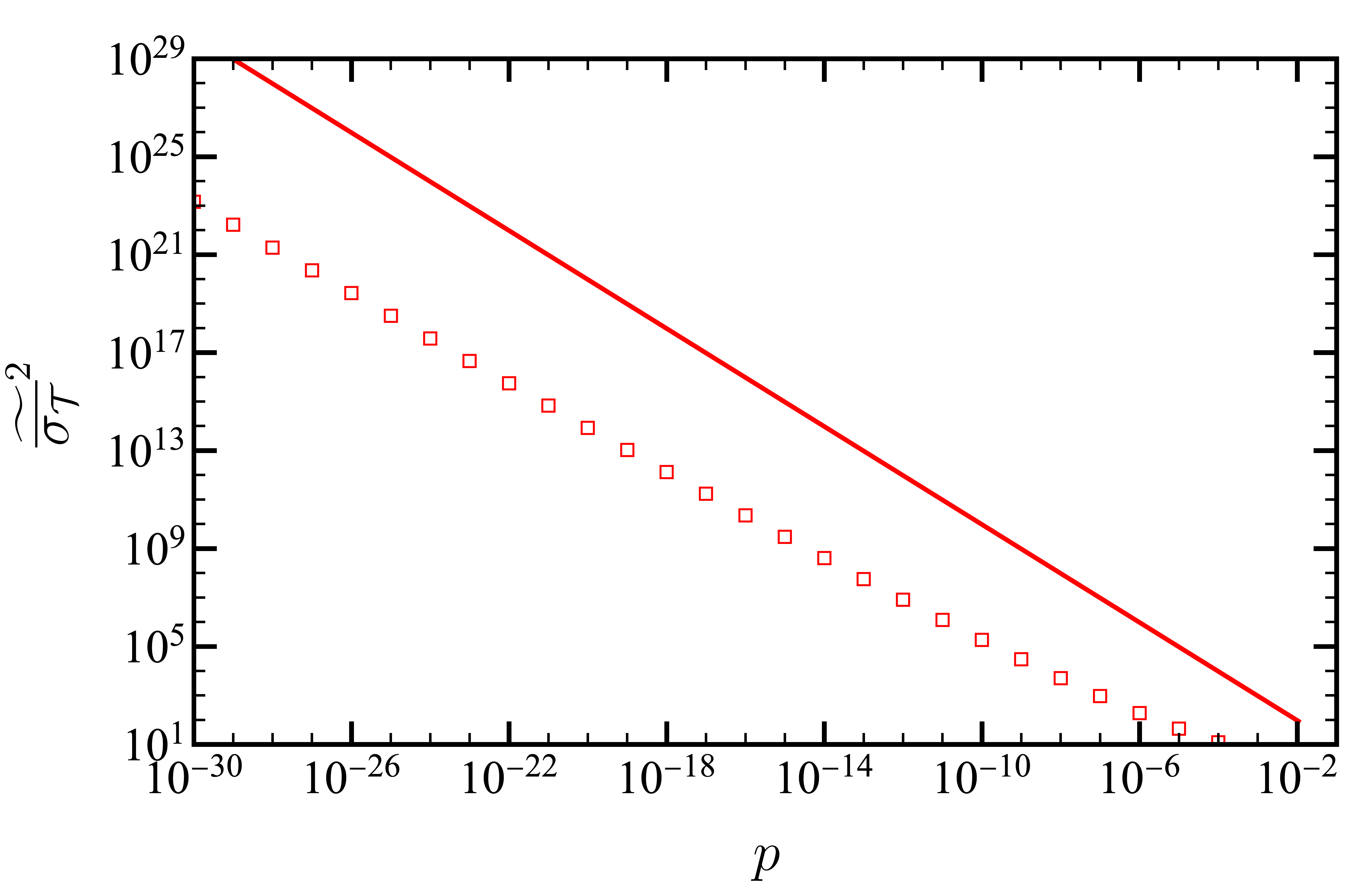}
    \caption{Divergence of the variance at minimum average MFPT in the limit as $\disorder \to 0^+$. The numerical solutions (symbols) are compared with the leading order of the asymptotic prediction $\disorder^{-1}$ (solid line).
    }
    \label{fig:disordered_divergence_std}
\end{figure}

This divergence can be quantified using the previous estimation for $\disorder=\delta \disorder \ll 1$. In that limit, we may also estimate the value of the pole $s_+^*$ by looking at~\eqref{eq:asymp+1}. A dominant balance argument leads to the asymptotic behaviour 
\begin{equation}
    s_+^*\sim -4\sqrt{3}e^{-\alpha_+}.
\end{equation} 
Therefore, substituting the asymptotic values~\eqref{eq:disordered_MFPT_asymptotic_optimalRates_nearZero}, we obtain 
\begin{equation}
    \optMFPT{s_+^*}\sim -4\sqrt{3}\disorder(\ln \disorder)^2.
\end{equation} 
This entails that the contribution of this pole to the average MFPT is proportional to $\disorder/s_+^*$ and therefore vanishes. On the other hand, the contribution to the second moment is proportional to $\disorder/(s_+^*)^2$,\footnote{Recall that the moments are related to the poles by~\eqref{eq:FPT_moments_from_Laplace}.} which diverges as $\disorder^{-1}$ with slowly varying logarithmic corrections, as shown in figure~\ref{fig:disordered_divergence_std}.

{\clearpage \thispagestyle{empty}}
\chapter{Resetting in disordered environments: general approach}
\label{ch:resetting_disordered_bulkvsboundaries}
We have already studied in chapter~\ref{ch:resetting_disordered} a first, simple, scenario to understand how heterogeneous stochastic resetting can expedite search times in media with quenched disorder. However, we would like to go further by looking into the general situation: how can heterogenous stochastic resetting $r(x)$ be introduced to optimise the search, given an arbitrary quenched target distribution $\PDF_T(\targetx)$? 

As already mentioned in chapter~\ref{ch:resetting_disordered}, this is a very difficult question to tackle mathematically. In order to simplify the mathematical problem, we focus on stochastic processes defined within a bounded domain, \ie the searcher's position is restricted to a finite interval $[x_r-\ell,x_r+\ell]$. Obviously, the target position is also restricted to this interval, which means that the target distribution $\PDF_T(\targetx)$ has a compact support $\targetx\in[x_r-\ell,x_r+\ell]$. This feature is also interesting because stochastic resetting has predominantly been studied in infinite domains, due to its capacity to cut the heavy tails of the first-passage time distributions. For finite domains, the literature is not so vast \cite{journalarticle:DeBruyne.etal_OptimizationFirstPassageResetting_Phys.Rev.Lett.20,journalarticle:Christou.Schadschneider_DiffusionResettingBounded_JPhysMathTheor15,journalarticle:Pal.Prasad_FirstPassageStochastic_Phys.Rev.E19,journalarticle:Durang.etal_FirstpassageStatisticsStochastic_JPhysMathTheor19,journalarticle:Chen.Huang_FirstPassageDiffusing_Phys.Rev.E22}, and stochastic resetting has been observed to be detrimental in certain situations \cite{journalarticle:Christou.Schadschneider_DiffusionResettingBounded_JPhysMathTheor15,journalarticle:Pal.Prasad_FirstPassageStochastic_Phys.Rev.E19,journalarticle:Durang.etal_FirstpassageStatisticsStochastic_JPhysMathTheor19}. In those cases, an optimal non-vanishing resetting rate was found only when the resetting position is close enough to the target position. Inspired by the results obtained in chapter~\ref{ch:resetting_disordered}, we introduce here a new kind of boundary conditions, which we term \emph{resetting boundaries}. 

The chapter is organised as follows. Section~\ref{sec:resetting_boundaries} introduces the model, as well as the observables and equations we analyse. In section~\ref{sec:resetting_boundaries_homogeneous_bulk}, the average MFPT is optimised over a homogeneous bulk resetting rate. The analytical optimisation is done considering a heterogeneous strategy in section~\ref{sec:resetting_boundaries_heterogeneous_bulk}. Therein, we elucidate when the best strategy is to reset only at the boundaries. Finally, exact numerical profiles of bulk resetting are provided for different target distributions in section~\ref{sec:resetting_boundaries_numerical}. 

\section{Brownian search with bulk resetting and resetting boundaries}
\label{sec:resetting_boundaries}

Let us consider that our one-dimensional Brownian searcher is confined in the finite interval $[x_r-\ell,x_r+\ell]$. If the searcher has the certainty that the target is not in one side of the domain, from a physical point of view, it is clearly optimal to immediately reset if it crosses to that side, as illustrated for example if the first panel of figure~\ref{fig:disordered-densityplot-full}. Similarly, if the system is within a compact support and the searcher reaches either wall without having found the target, the optimal strategy is to restart the search from its initial position, to explore the other side of the box. Resetting boundaries are born as an implementation of this natural strategy, we argue that resetting is always beneficial at the boundaries. An analogous idea was introduced in processes with multiple searchers \cite{journalarticle:Biswas.etal_TargetSearchOptimization_25}.

\begin{figure}          
    \centering
    \includegraphics[width=0.85\textwidth]{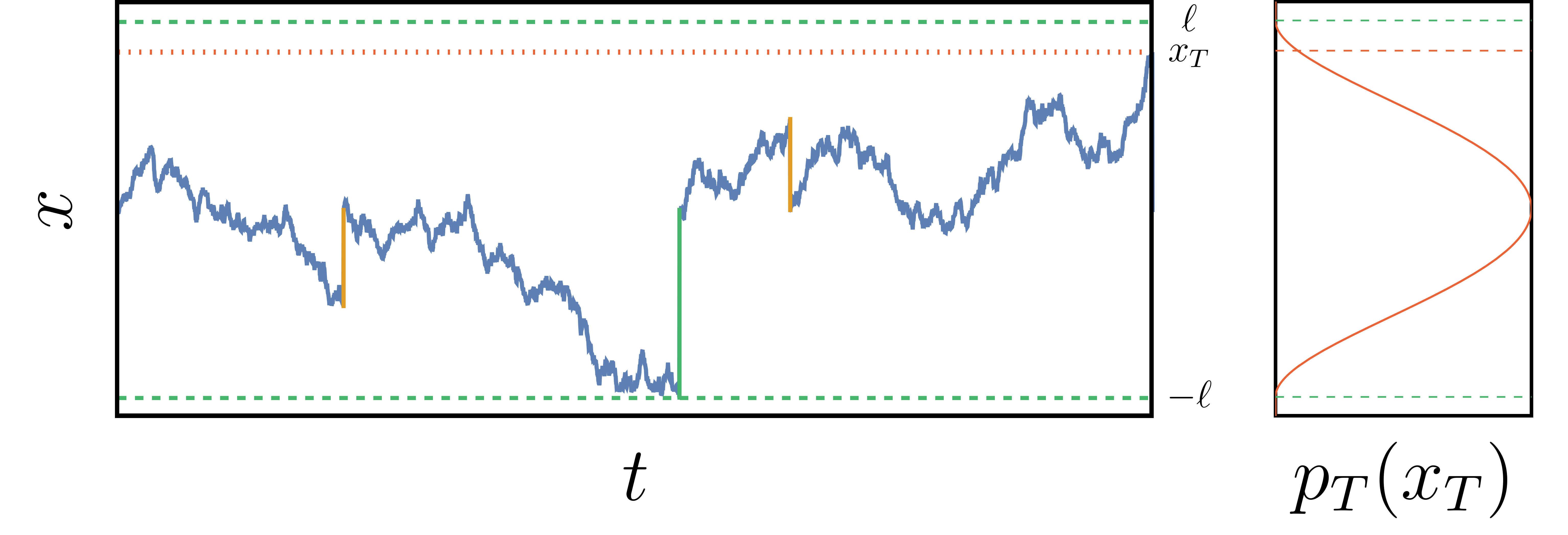}
    \caption{Sketch of a single trajectory for the Brownian search of a spatially distributed target. Left: One trajectory starting from $x_0=x_r=0$. The blue solid trace stands for diffusive motion, with instances of a boundary reset (vertical green stroke) and bulk resets (vertical yellow strokes). The target position $\targetx$ (horizontal dotted red line) has been drawn from its distribution $\PDF_T(\targetx)$. Domain boundaries are at $\pm \ell$ (horizontal dashed green lines). Right: Spatial distribution of the target.}
    \label{fig:resetting-boundaries-scheme}
\end{figure}

Therefore, we are going to analyse the search problem of a Brownian particle inside a one-dimensional box with the effect of two kinds of resetting processes: 
\begin{itemize}
    \item A stochastic \emph{bulk resetting} with a space-dependent rate $r(x)$, which is the extension of the typical SSR, \ie the searcher at position $x$ has a probability per unit of time $r(x)$ to instantaneously jump to $x_r$.
    \item A deterministic \emph{boundary resetting} at the edges of the box, which makes the searcher restart the search from $x_r$ whenever it reaches any boundary at $x_r\pm\ell$.
\end{itemize}
On the one hand, by definition, resetting boundaries clearly expedite the search in a one-dimensional domain---as compared with the usual reflecting boundaries. On the other hand, we expect bulk resetting to be beneficial only in certain situations. In a given trajectory of the stochastic process, it would be better to reset in the bulk when the searcher's position is further from the target than the resetting position. An illustrating scheme of a particular realisation of the process is shown in figure~\ref{fig:resetting-boundaries-scheme}: therein, the system experiments a beneficial bulk reset (first vertical orange stroke), followed by another favorable reset at the boundary (green vertical stroke), and finally a detrimental bulk reset (second vertical orange stroke). In fact, figure~\ref{fig:resetting-boundaries-scheme} is an analogous realisation as the one shown in figure~\ref{fig:disordered_general_model}, though now the system is confined in a finite domain with resetting boundaries. Two natural questions arise from this model:
\begin{enumerate}
    \item When is bulk resetting advantageous in the presence of resetting boundaries for a given distribution $\PDF_T(\targetx)$?
    \item What is the optimal bulk resetting strategy $r(x)$ for a given $\PDF_T(\targetx)$?
\end{enumerate}

Let us introduce the same dimensionless variables as in section~\ref{subsec:dimensionless_variables}. Thus, the resetting position is always $x_r=0$, and the domain of the system is $[-1,1]$. In the quenched disorder framework, we have already introduced that every observable can be computed as an average over the target distribution $\PDF_T(\targetx)$, \eg the average propagator in~\eqref{eq:averaging_propagator_disorder}. On this occasion, the equation governing the forward evolution of the propagator with a target at $\targetx$ is 
\begin{align}
    \partial_t \PDF_r(x,t|x_0,\targetx) &= \partial_{x}^2 \PDF_r(x,t|x_0,\targetx) - r(x) \PDF_r(x,t|x_0,\targetx) + \delta(x) \int \d{y} r(y)\, \PDF_r(y,t|x_0,\targetx) 
    \nonumber \\
    & \quad +\delta (x) \left[ \partial_x \PDF_r(x,t|x_0,\targetx) \right]_{x=1}^{x=-1},
\end{align}
which is equivalent to~\eqref{eq:Fokker_Planck_Heterogeneous_SR_dimensionless} with the addition of the last term, which accounts for the flux of probability because of resetting boundaries. Additionally, we must include the initial condition $\PDF_r(x,0|x_0,\targetx)=\delta(x-x_0)$ and boundary conditions
\begin{equation}
    \PDF_r(\targetx,t|x_0,\targetx) = 0, \quad \PDF_r(\pm 1,t|x_0,\targetx) = 0.
\end{equation}
The first boundary condition is the absorbing condition at the target position as usual, whereas the second ones represent the resetting boundaries---the particle is never found at the boundaries due to it is instantaneously reset. Mathematically, since the search is terminated when the searcher hits the target, $x=\targetx$, the only active boundary in the first-passage analysis is the one opposite to the target, \ie if $\targetx>0$, the active boundary is $x=-1$, and vice versa.

In this chapter, the optimisation of the search time is always performed by minimising the average MFPT, 
\begin{equation}
    \label{eq:functional_average_MFPT}
    \overline{\FPT{(1)}}[r]= \int \d{\targetx} \PDF_T(\targetx) \FPT{(1)}(x_0=0,\targetx).
\end{equation}
This quantity is a functional of the bulk resetting rate $r(x)$ through the MFPT for fixed target position $\FPT{(1)}(x_0,\targetx)$. The ODE for $\FPT{(1)}(x_0,\targetx)$ corresponds to the case $n=1$ of~\eqref{eq:disordered_FPT_moments}, \ie 
\begin{subequations}
    \begin{align}
        & -1 = \partial_{x_0}^2 \FPT{(1)}(x_0,\targetx)- r(x_0)^2 \bigg[
        \FPT{(1)}(x_0,\targetx) - \FPT{(1)}(0,\targetx)\bigg],
        \\
        & \FPT{(1)}(\targetx,\targetx) = 0, \quad \FPT{(1)}(\pm 1,\targetx) = \FPT{(1)}(0,\targetx).
        \label{eq:resetting_boundaries_FPT_BCs}
    \end{align}
\end{subequations}
The conditions for the resetting boundaries, \ie the second equality in~\eqref{eq:resetting_boundaries_FPT_BCs}, physically indicate that the MFPT is the same if one starts from a resetting boundary or from the resetting position. We are particularly interested in the case where the searcher starts at the resetting position, $x_0=x_r=0$, so we will denote $\FPT{(1)}(\targetx) \equiv \FPT{(1)}(0,\targetx)$ for simplicity. Performing the change of variable
\begin{equation}
    \chi(x_0) = \FPT{(1)}(x_0,\targetx) - \FPT{(1)}(\targetx),
\end{equation}  
the ODE for $\chi(x_0)$ simplifies to 
\begin{subequations}
    \label{eq:resetting_boundaries_FPT_ODE}
    \begin{align}
        -1 &= \partial_{x_0}^2 \chi(x_0)- r(x_0)^2 \chi(x_0),
        \\
        \chi(0) &= 0, \quad \chi(\pm 1) = 0.
    \end{align}
\end{subequations}
Since we are interested in the MFPT starting from the origin, we only need to evaluate the solution at $x_0=\targetx$, \ie $\FPT{(1)}(\targetx) = -\chi(\targetx)$.

The main challenge of minimising $\overline{\FPT{(1)}}[r]$ is its functional dependence on the bulk resetting profile $r(x)$. The MFPT with fixed target is given by the solution that solves the second-order differential equation~\eqref{eq:resetting_boundaries_FPT_ODE}, \ie $\FPT{(1)}(\targetx)$, whose form is unknown for a general $r(x)$. Next sections are devoted to analytically tackle this minimisation problem for different choices of $r(x)$.

\section{Optimal homogeneous bulk resetting}
\label{sec:resetting_boundaries_homogeneous_bulk}

Let us start by considering a homogeneous bulk resetting rate, $r(x)=r$. The average MFPT~\eqref{eq:functional_average_MFPT} is then no longer a functional, but a standard function of $r$, so we denote it as $\overline{\FPT{(1)}}(r)$. The solution of the MFPT is directly obtained from~\eqref{eq:resetting_boundaries_FPT_ODE}, yielding
\begin{align}
    \FPT{(1)}(\targetx;r)=\frac{1}{r}\brackets{\frac{\cosh[(|\targetx|+\frac{1}{2})\sqrt{r}]}{\cosh[\sqrt{r}/2]}-1}.
\label{eq:MFPT_homogeneous_bulk_resetting}
\end{align}
In fact, for homogeneous bulk resetting, the MFPT can also be obtained for other boundary conditions like reflecting boundaries,\footnote{The reflecting boundaries are obtained by substituting $\chi(\pm 1) = 0$ for $\partial_{x_0} \chi(x_0)|_{x_0=\pm 1}=0$.} whose result is
\begin{equation}
    \mathcal{T}^{(1),\text{(refl)}}(\targetx;r)=\frac{1}{r}\left[\frac{\cosh[(|\targetx|+1)\sqrt{r}]}{\cosh[\sqrt{r}]}-1\right].
    \label{eq:MFPT_homogeneous_bulk_resetting_reflecting}
\end{equation}
Herein, we prove that resetting boundaries always outperform reflecting ones for fixed homogeneous resetting rate, \ie $\FPT{(1)}(\targetx;r)<\mathcal{T}^{(1),\text{(refl)}}(\targetx;r)$, $\forall(\targetx;r)$, which also holds for the average MFPT, $\overline{\FPT{(1)}}(r)<\overline{\FPT{(1)}}^{\text{(refl)}}(r)$---due to the linearity of the integral. 

Let $\optMFPT{r}$ be the optimal homogeneous bulk resetting rate that minimises the average MFPT $\overline{\FPT{(1)}}(r)$. We observe that $\FPT{(1)}(\targetx;r)$ is a convex function of $r$, \ie $\partial_r^2 \FPT{(1)}(\targetx;r)>0$. Convexity is inherited by $\overline{\FPT{(1)}}(r)$ for arbitrary $\PDF_T(\targetx)$, which means any extremum $\optMFPT{r}$ is a global minimum of the average MFPT, provided that $\optMFPT{r}\geq0$. Note that the global minimum may be reached at $r=0$. Precisely, due to the global convexity, if the derivative is positive at that point, $\partial_r \overline{\FPT{(1)}}(r)|_{r=0}>0$, the best strategy is no resetting in the bulk, $\optMFPT{r}=0$. Instead, for negative derivative at $r=0$, the global minimum of $\overline{\FPT{(1)}}(r)$ is attained at some finite value $\optMFPT{r}\neq0$. As happened in SSR, the average MFPT diverges for $r\to\infty$ because the particle becomes trapped at the resetting point.

\begin{figure}
    \centering
    \includegraphics[width = 0.7\textwidth]{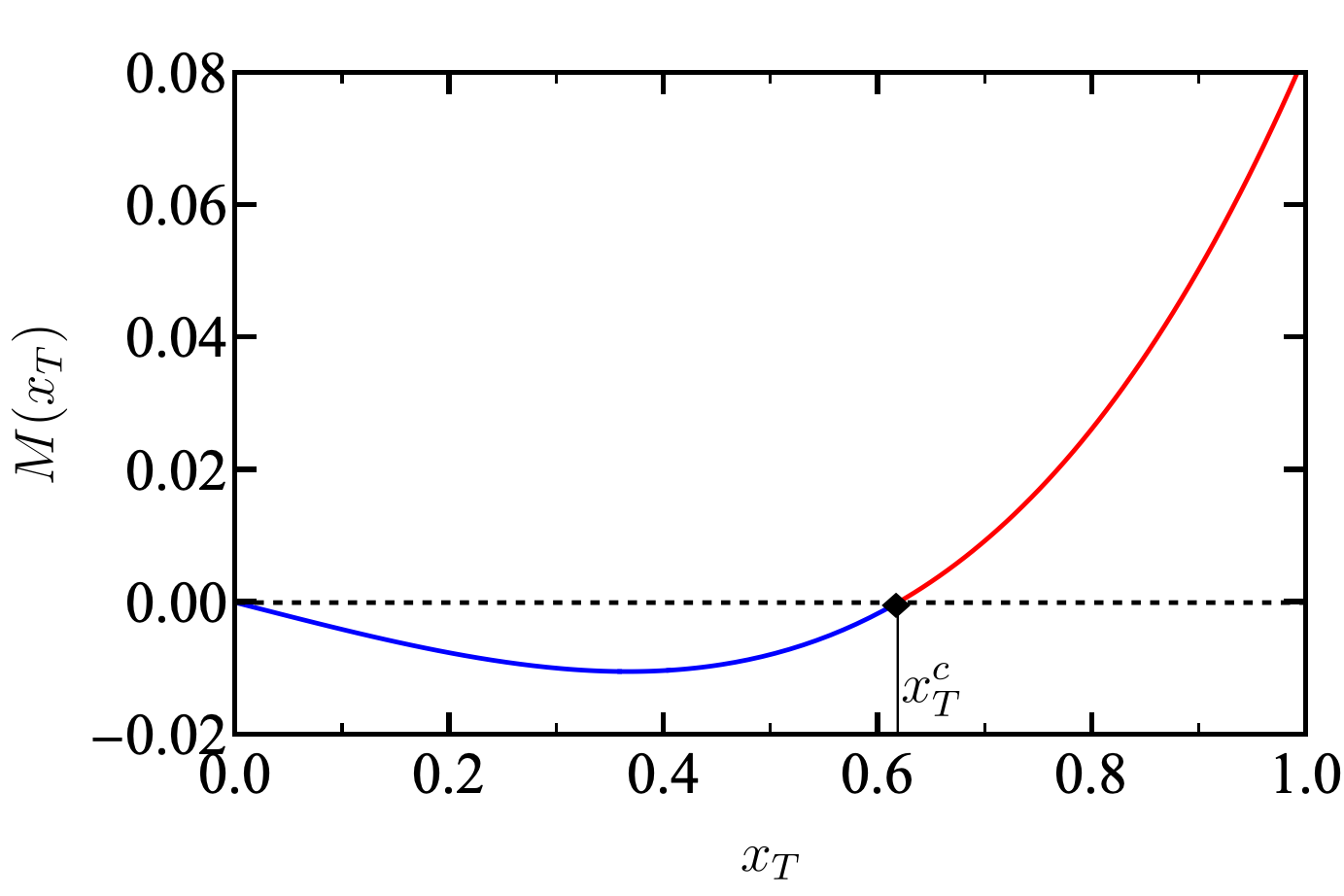}
    \caption{Derivative controlling the stability of the non-resetting strategy for homogeneous resetting, \ie $M(\targetx)$ defined in~\eqref{eq:Mxt_homogeneous}. Positive (negative) values thereof, represented as warm (cold) colors, contribute to zero (non-zero) bulk resetting as the best search strategy. The change of stability takes place at $\targetx^c=(\sqrt{5}-1)/2$ (black diamond).}
    \label{fig:Mxt_homogeneous}
\end{figure}

The average MFPT is computed by integrating~\eqref{eq:MFPT_homogeneous_bulk_resetting} over the target distribution $\PDF_T(\targetx)$. We are interested in studying the stability of the non-resetting strategy, $\optMFPT{r}=0$, \ie in checking if resetting boundaries are enough to optimise the average MFPT for a given target distribution. In this chapter, we use the term stability in a mathematical sense, meaning that we are interested in discriminating if a fixed point is either  a local minimum or maximum, \ie if it is stable or unstable, respectively. Expanding the average MFPT around $r=0$, we obtain
\begin{equation}
    \overline{\FPT{(1)}}(r)=\frac{\left\langle \targetx^2 + \left|\targetx\right|\right\rangle_T}{2}+ \frac{\left\langle \targetx^4 - \left|\targetx\right| \left( 1-2\targetx^2\right) \right\rangle_T}{24} r + \mathcal{O}(r^2),
    \label{eq:Expansion_homogeneous_FPT}
\end{equation}
where $\langle\cdot\rangle_T$ stands for the average over the target distribution $\PDF_T(\targetx)$. The key quantity is the linear coefficient, which we denote in the following as  
\begin{subequations}
    \begin{align}
        m &= \langle M(\targetx)\rangle_T \equiv \left.\pdev{}{\overline{\FPT{(1)}}}{r}\right|_{r=0},
        \label{eq:slope_MFPT_homogeneous_bulk_resetting}
        \\
        M(\targetx) &\equiv \left.\pdev{}{\FPT{(1)}(\targetx;r)}{r}\right|_{r=0} = \frac{1}{24}\brackets{\targetx^4 - |\targetx|\parenthesis{1-2\targetx^2}}.
        \label{eq:Mxt_homogeneous}
    \end{align}
\end{subequations}
The sign of $m$ controls the optimality of non-resetting in the bulk. Specifically, $m>0$ ($m<0$) implies that the minimum is reached at $\optMFPT{r}=0$ ($\optMFPT{r}>0$). This idea of analysing the slope of the MFPT has been successfully used in the past for optimising different resetting configurations \cite{journalarticle:Ahmad.etal_FirstPassageParticle_Phys.Rev.E19,journalarticle:Christou.Schadschneider_DiffusionResettingBounded_JPhysMathTheor15,journalarticle:Ray.etal_PecletNumberGoverns_J.Phys.A:Math.Theor.19,journalarticle:Pal.Prasad_LandaulikeExpansionPhase_Phys.Rev.Res.19}. For fixed target, the sign of $M(\targetx)$ changes at $\targetx^c\equiv (\sqrt{5}-1)/2$, \ie $M(\targetx)<0$ for $0<|\targetx|<\targetx^c$ and $M(\targetx)>0$ for $\targetx^c<|\targetx|<1$, as figure~\ref{fig:Mxt_homogeneous} illustrates. The physical interpretation of this quantity is clear. On the one hand, if the target is close enough to the centre of the domain, resetting in the bulk is beneficial, since it prevents the searcher from moving too far from the target. On the other hand, if the target is far from the centre, resetting in the bulk is detrimental because the searcher must explore the whole domain to find the target. 

Although we are interested in the behaviour of $m$, the sign of $M(\targetx)$ gives us a good intuition of the optimal strategy. Concretely, it is worth noting some general results for different families of distributions. First, $m<0$ for any target distribution with finite support in the subinterval $0\leq |\targetx|<\targetx^c$, because $M(\targetx)<0$ where $\PDF_T(\targetx)\neq0$---thus, homogeneous resetting is always beneficial and $\optMFPT{r}>0$. Second, $m>0$ for any target distribution with finite support in the subinterval $\targetx^c<|\targetx|\leq 1$, because $M(\targetx)>0$ where $\PDF_T(\targetx)\neq0$---thus, homogeneous resetting is always detrimental and $\optMFPT{r}=0$. Thus, for a continuous uniform distribution, we have that $m>0$, since there is too much probability concentrated on the boundaries, so homogeneous resetting does not give any advantage for a homogeneously distributed target.

\begin{figure}
    \includegraphics[width=0.49\textwidth]{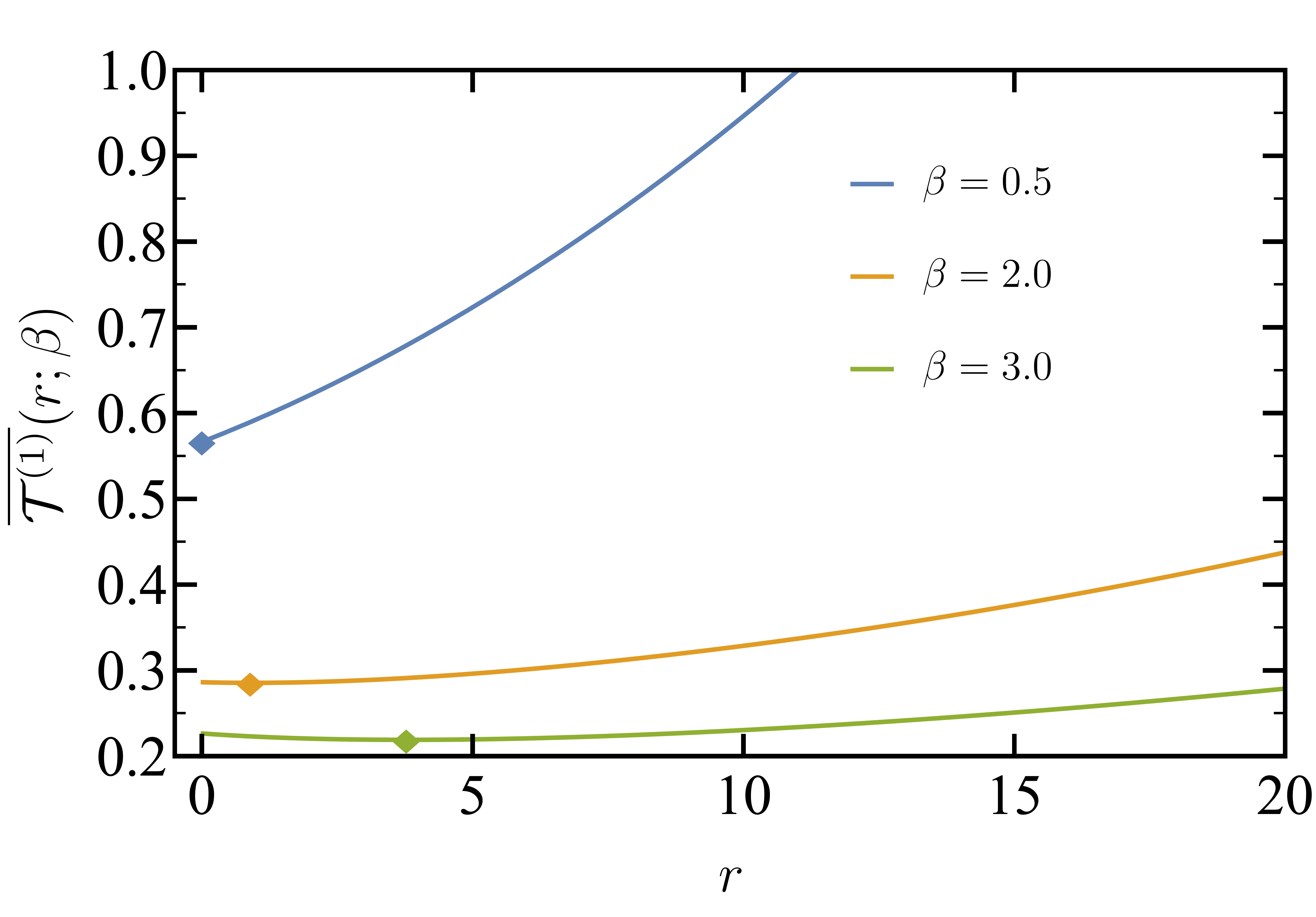}
    \includegraphics[width=0.49\textwidth]{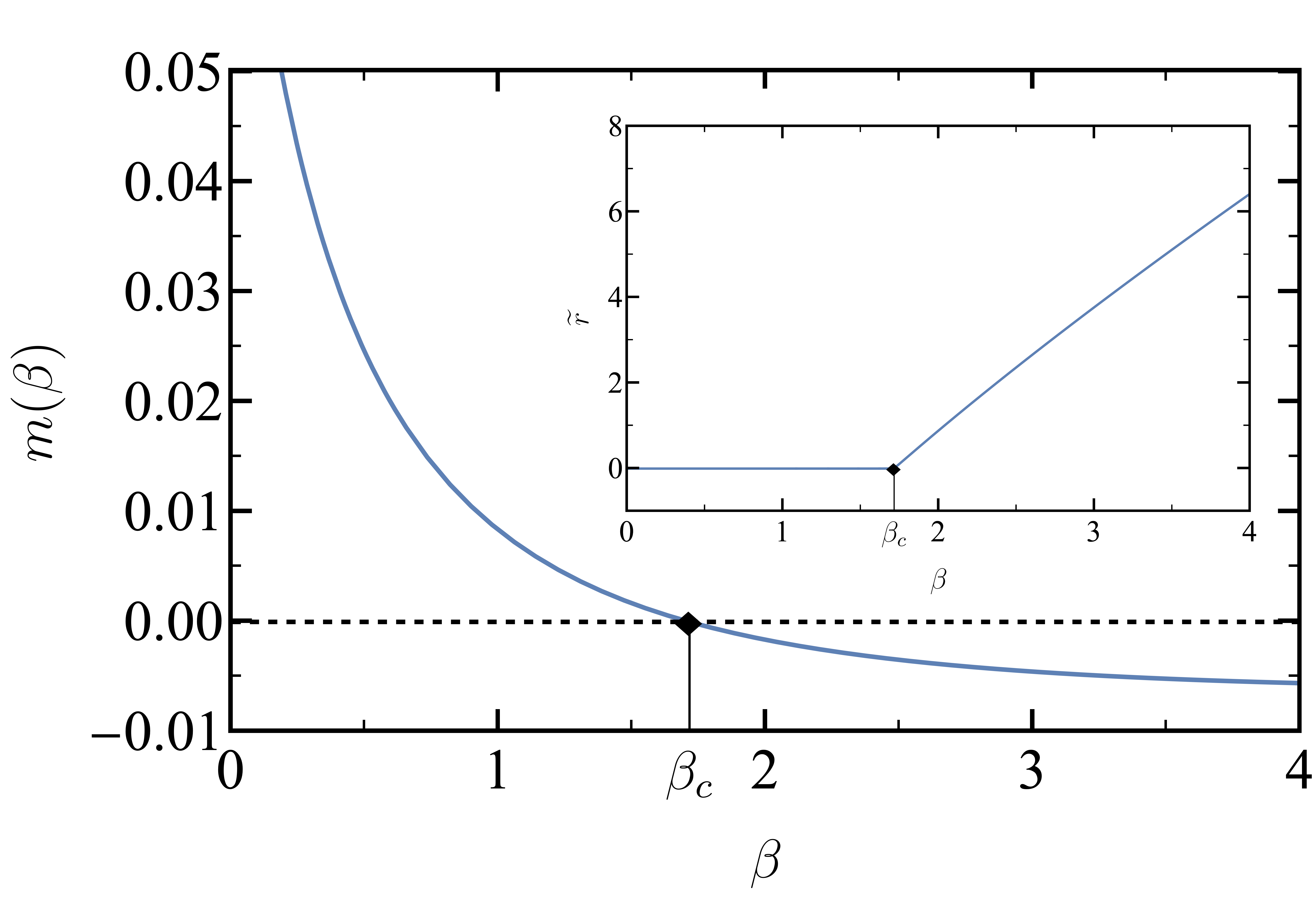}
    \caption{Left: Average MFPT for different $\beta$ distributions, as given by~\eqref{eq:average_MFPT_homogeneous_beta_distribution}. For $\beta=0.5$, the minimum is attained at $\optMFPT{r}=0$ (blue diamond), while for $\beta=2,\, 3$ is attained at $\optMFPT{r}>0$ (orange and green diamonds, respectively). The acceleration factor for the optimal value with respect a vanishing resetting strategy in the bulk is around 0.2\% and 3\% for $\beta=2$ and $\beta=3$, respectively. Right: Slope of the average MFPT at $r=0$, as given by~\eqref{eq:m-hom-beta} of the main text. It crosses zero at $\beta=\beta_c\simeq 1.71$ (black diamond). In the inset, the continuous phase transition for the optimal homogeneous resetting rate $\optMFPT{r}$ is shown.}
    \label{fig:homogeneous_bulk_resetting_beta}
\end{figure}

To further illustrate our discussion, we introduce a specific monoparametric family of symmetric target distributions. In particular, we choose a variant of the beta distribution,  
\begin{equation}
    \label{eq:PTbeta}
    \PDF_T(\targetx;\beta)=\frac{\Gamma\left(\beta+\frac{1}{2}\right)}{\sqrt{\pi} \Gamma(\beta)}(1-\targetx^2)^{\beta-1},
\end{equation}
defined for $-1 < \targetx < 1$, with $\beta >0$.  This $\beta$-family interpolates really different behaviours as one varies the parameter $\beta$: (i) the probability is accumulated at the centre and vanishes at the boundaries for $\beta > 1$, tending to a single delta peak for $\beta \to +\infty$; (ii) the probability is uniformly distributed for $\beta=1$; and (iii) the probability is concentrated at the boundaries for $0<\beta<1$, tending to two equally weighted delta peaks at the boundaries for $\beta \to 0$---it would be equivalent to the dichotomous distribution studied in chapter~\ref{ch:resetting_disordered} for $\eta=1/2$. Since its support is the whole domain $[-1,1]$, the general observations made above for especially supported distributions in a subinterval do not apply, because there is a competition between the positive and negative contributions to $m$ from $M(\targetx)$. Substituting the solution for homogeneous bulk resetting~\eqref{eq:MFPT_homogeneous_bulk_resetting} and the distribution~\eqref{eq:PTbeta} into~\eqref{eq:functional_average_MFPT}, the average MFPT becomes
\begin{align}
    \overline{\FPT{(1)}}(r;\beta) = &r^{-1} \Bigg[ {}_0F_1 \left(; \beta+\frac{1}{2} ; \frac{r}{4}\right)  -1 +2^{\beta-\frac{1}{2}} r^{\frac{1}{4}-\frac{\beta}{2}}
    \Gamma\left( \beta +\frac{1}{2}\right) 
    L_{\beta-\frac{1}{2}} \left(\sqrt{r}\right) \tanh \left( \frac{\sqrt{r}}{2}\right) \Bigg],
    \label{eq:average_MFPT_homogeneous_beta_distribution}
\end{align}
being ${}_0F_1(;a;z)$ the confluent hypergeometric function and $L_n(z)$ the modified Struve function \cite{journalarticle:Abramowitz.etal_HandbookMathematicalFunctions_Am.J.Phys.88}. The linear coefficient $m$ can be obtained directly from~\eqref{eq:slope_MFPT_homogeneous_bulk_resetting} for the $\beta$-family~\eqref{eq:PTbeta}, or differentiating the average MFPT~\eqref{eq:average_MFPT_homogeneous_beta_distribution} at $r=0$. Anyway, the slope becomes a function of $\beta$:
\begin{align}
    m(\beta)=\frac{1}{8[3+4\beta(\beta+2)]}-\frac{(\beta-1)\Gamma(\beta +\frac{1}{2})}{24\sqrt{\pi} \Gamma(\beta+2)},
    \label{eq:m-hom-beta}
\end{align}
whose only root is $\beta_c \simeq 1.71$. For $\beta<\beta_c$, $m(\beta)>0$, so the global minimum of $\overline{\FPT{(1)}}(r;\beta)$ is reached at $\optMFPT{r}=0$. Instead, for $\beta>\beta_c$, $m(\beta)<0$, so $\overline{\FPT{(1)}}(0^+;\beta)<\overline{\FPT{(1)}}(0;\beta)$ and the optimal rate is attained at a certain value $\optMFPT{r}>0$. This behaviour is reminiscent of a continuous phase transition, where $\optMFPT{r}$ plays the role of the order parameter. In figure~\ref{fig:homogeneous_bulk_resetting_beta}, we capture these results for homogeneous bulk resetting and the $\beta$-family of target distributions. In the left panel, the average MFPT is shown as a function of $r$ for several values of $\beta$. The slope $m(\beta)$ is reported in the right panel, as well as the optimal resetting rate $\optMFPT{r}(\beta)$. 

Summarising, for homogeneous resetting, non-resetting in the bulk is the optimal strategy for certain target distributions. Specifically, those for which regions with positive $M(\targetx)$ are more relevant than those with negative $M(\targetx)$. Physically, this means that non-resetting in the bulk is advantageous if $\PDF_T(\targetx)$ is concentrated enough at the boundaries of the domain. For the monoparametric $\beta$-family~\eqref{eq:PTbeta}, this corresponds to $\beta<\beta_c$. 

\section{Optimisation of heterogeneous bulk resetting}
\label{sec:resetting_boundaries_heterogeneous_bulk}

Let $r(x)$ be a general resetting rate in the bulk. Similarly to the homogeneous case, the stability of the non-resetting strategy $r(x)=0$ is determined by the sign of the derivative of the average MFPT. Nevertheless, since we are working with a functional, the expansion differs from the standard Taylor expansion used in~\eqref{eq:Expansion_homogeneous_FPT},
\begin{align}
    \overline{\FPT{(1)}}[r]= \frac{\langle \targetx^2 + |\targetx| \rangle_T}{2} + \int dx \left\langle \left.  \frac{\delta\FPT{(1)}(\targetx)}{\delta r(x)}\right|_{r=0}\right\rangle_T r(x) + \mathcal{O}((r(x))^2).
\end{align} 
Then, our analysis involves now the sign of
\begin{equation}
    \label{eq:mu(x)_heterogeneous_resetting}
    \mu(x) \equiv \left. \frac{\delta \overline{\FPT{(1)}}}{\delta r(x)}\right|_{r(x)\equiv 0} =\braket{M(\targetx,x)}_T,
\end{equation}
where the functional derivative $M(\targetx,x)$ at fixed target is defined by
\begin{align}
    M(\targetx,x) \equiv \left. \frac{\delta\FPT{(1)}(\targetx)}{\delta r(x)}\right|_{r(x)\equiv 0} =\lim_{\varepsilon \to 0^+} \frac{\left.\FPT{(1)}(\targetx)\right|_{r(x_0)=\varepsilon\delta(x_0-x)}-\left.\FPT{(1)}(\targetx)\right|_{r(x_0)=0}}{\varepsilon}.
    \label{eq:M(xt,x)_heterogeneous_resetting}
\end{align}
The interpretation of $M(\targetx,x)$ is clear: it measures how the MFPT varies when the non-resetting strategy is locally perturbed at any point $x$.

Let us analyse the sign of $\mu(x)$. On the one hand, positivity of $\mu(x)$, $\forall x$, entails that the average MFPT attains, at least, a local minimum for $\optMFPT{r}(x)=0$, since any perturbation from that flat profile would lead to a longer search time. On the other hand, the emergence of any subinterval inside which $\mu(x)<0$ implies the decrease of the average MFPT for a perturbation such that $r(x)\neq0$ in that subinterval. In that case, the profile $r(x)=0$ is no longer a local minimum. Notice that we cannot guarantee the uniqueness of the minimum of the functional, since we have not been able to prove that the average MFPT is a convex functional of $r(x)$. However, physical intuition tells us that convexity is hold for heterogeneous resetting, because of the numerical results we display in section~\ref{sec:resetting_boundaries_numerical}.

For computing $\mu(x)$, we have first to solve the equation for MFPT~\eqref{eq:resetting_boundaries_FPT_ODE} for a delta-like perturbation of the resetting rate, $r(x_0) = \varepsilon\delta(x_0-x)$. The solution is obtained by solving the equation in the two subdomains $[-1,x)$ and $(x,1]$ with $r(x)=0$, and then matching the solutions at $x_0=x$ by imposing (i) the continuity condition $\chi(x^-)=\chi(x^+)$, and (ii) the delta kick condition $\chi'(x^+)-\chi'(x^-)=\varepsilon \chi(x)$. After a careful calculation, we get
\begin{align}
    \delta \FPT{(1)}(\targetx) &\equiv \left.\FPT{(1)}(\targetx)\right|_{r(x_0)=\varepsilon\delta(x_0-x)}-\left.\FPT{(1)}(\targetx)\right|_{r(x_0)=0}
    = \dfrac{\varepsilon}{2} \dfrac{| x\, \targetx | (1-| x|)^2}{1+\varepsilon | x | (1-| x |)}\Theta\left(-\targetx \, x\right) 
    \nonumber\\ &\quad +  \dfrac{\varepsilon}{2} |x|(|\targetx|-|x|)(1+|x|)
    \Theta\left(\targetx \, x\right)\Theta\left(|\targetx|-|x|\right),
\end{align}
so~\eqref{eq:M(xt,x)_heterogeneous_resetting} becomes
\begin{align}
    M(\targetx,x) =
    \dfrac{ x \targetx (1-| x|)^2}{2}\Theta\left(-\targetx x\right) 
    +  \frac{x(\targetx-x)(1+|x|)}{2}\Theta\left(\targetx x\right)\Theta\left(|\targetx|-|x|\right)
    \label{eq:M(x_t,x)_explicit_heterogeneous}
\end{align}
The two contributions of $M(\targetx,x)$ have opposite signs: the first one is always negative, whereas the second is always positive. Thus, any single perturbation on $x$ provokes that bulk resetting is 
\begin{enumerate}
    \item Beneficial if $\targetx \, x<0$ (target and particle positions at opposite sides of the box);
    \item Detrimental for $\targetx \, x>0$ and $|\targetx|>|x|$ (particle between the centre and the target);
    \item Irrelevant if $\targetx \, x>0$ and  $|\targetx|<|x|$, since $M(\targetx,x)=0$ in that case.
\end{enumerate}
These results are physically intuitive. Resetting is advantageous to avoid exploring the side of the box opposite to the target position, whereas it is detrimental if the resetting event moves the particle away from the target. Finally, it does not affect the MFPT if the position is beyond the target, since the searcher is unable to reach that point before hitting the target. 

\begin{figure}
    \centering
    \includegraphics[width = 0.8\textwidth]{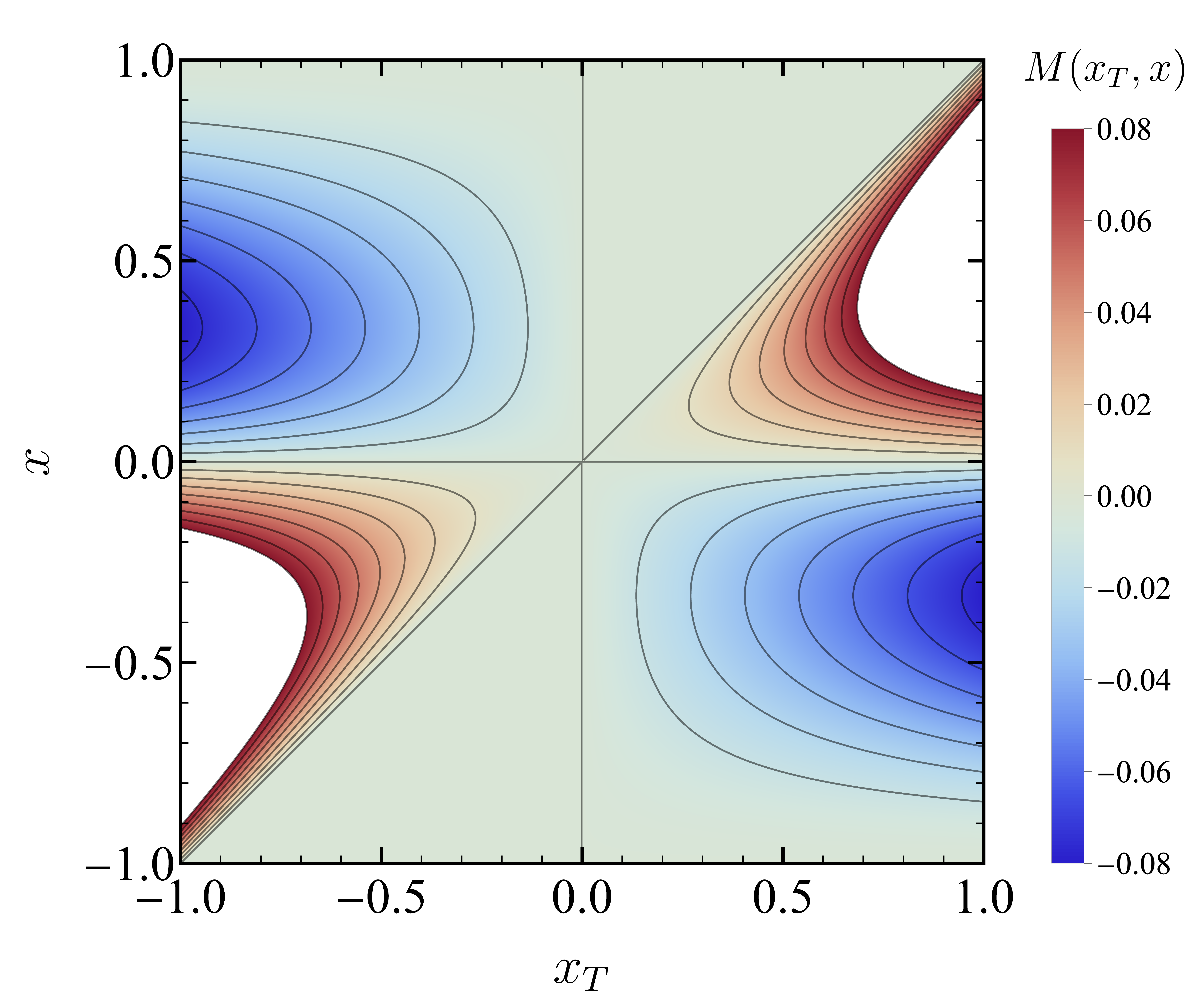}
    \includegraphics[width = 0.8\textwidth]{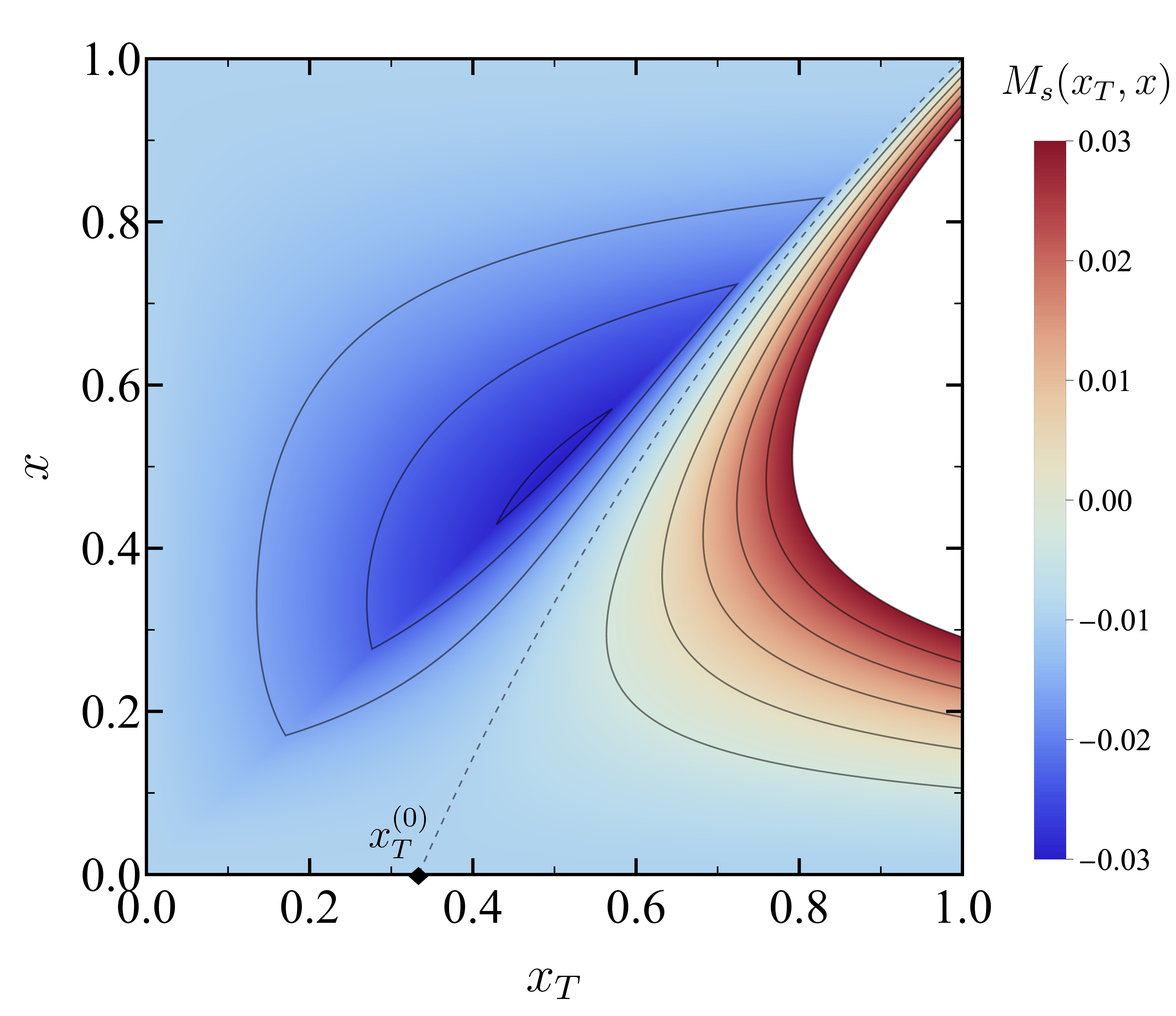}
    \caption{Derivative controlling the optimality of the non-resetting strategy for heterogeneous resetting. Positive (negative) values thereof, represented as warm (cold) colors, contribute to zero (non-zero) bulk resetting as the best search strategy. White color indicates saturated values. Top: Functional derivative $M(\targetx,x)$ defined in~\eqref{eq:M(x_t,x)_explicit_heterogeneous}. Bottom: Symmetrised functional derivative $M_s(\targetx,x)$ defined in~\eqref{eq:Ms(x_t,x)_explicit_heterogeneous} for the heterogeneous case. The contour line at which $M_s=0$ is marked (dashed line), as well as the value $\targetx^{(0)}$ (black diamond) below which $M_s(\targetx,x)<0$, $\forall x$.}
    \label{fig:stability-non-resetting-bulk}
\end{figure}

In the top panel of figure~\ref{fig:stability-non-resetting-bulk}, equation~\eqref{eq:M(x_t,x)_explicit_heterogeneous} is displayed. For $M(\targetx,x) \neq 0$,  the sign of $M(\targetx,x)$ is given by the sign of the product $\targetx x$. Hence, some general statements on the stability of the non-resetting strategy in the bulk can be made without further knowledge of the target distribution. Specifically, $r(x)=0$ does not minimise the MFPT for any target distribution with support in one half of the box, \ie either $\targetx\in (-1,0)$ or $\targetx\in (0,1)$. This can be physically understood, if we know the sign of the target position, the optimal strategy is clearly to forbid exploration of the side of the box opposite to the target's position, \ie such that $x\, \targetx<0$, sending the resetting rate to infinity therein, similarly with our finding for $\eta=0$ or $\eta=1$ in chapter~\ref{ch:resetting_disordered}.

\subsection{Symmetric target distributions}

Let us focus on symmetric target distributions, $\PDF_T(\targetx)=\PDF_T(-\targetx)$. In that case, the optimal resetting strategy inherits the symmetry of the problem, so it suffices to analyse the even function 
\begin{equation}
    M_s(\targetx,x) \equiv \frac{M(\targetx,x)+M(-\targetx,x)}{2},
\end{equation}
and its average over $\PDF_T(\targetx)$ in the positive side $[0,1]$. Taking into account~\eqref{eq:M(x_t,x)_explicit_heterogeneous}, the symmetrised function reads 
\begin{equation}
    \label{eq:Ms(x_t,x)_explicit_heterogeneous}
    M_s(\targetx,x)=
    \begin{cases}
     -\dfrac{1}{4}x^2[1+x+(x-3)\targetx], & 0 \leq x \leq \targetx ,\\ \, \\
     -\dfrac{1}{4} (1-x)^2x\, \targetx, &  \targetx \leq x \leq 1,
    \end{cases} 
\end{equation}
which is displayed in the bottom panel of figure~\ref{fig:stability-non-resetting-bulk}. Interestingly, there exists an interval of ``small'' target positions, $0\leq \targetx \leq \targetx^{(0)}$, where $M_s(\targetx,x)<0$, $\forall x$. This implies that $\mu(x)<0$, $\forall x$, for any target distribution with finite support in that subinterval; thus, the optimal strategy involves a non-vanishing bulk resetting rate, regardless of further details of $\PDF_T(\targetx)$. The dashed line that indicates the change of sign of $M_s(\targetx,x)$ is given by 
\begin{equation}
    \left.x(\targetx)\right|_{M_s=0}=\frac{3\targetx-1}{1+\targetx},
\end{equation}
so the critical value we have mentioned is $\targetx^{(0)}=1/3$. Conversely, there is no interval of ``large'' target positions above which $M_s$ is positive for all $x$ and, thus, we cannot guarantee that a finite-support distribution leads, without further knowledge of its details, to suppression of resetting in the bulk as the optimal strategy. However, looking at the figure, it is clear that target positions that contribute to $r(x)=0$ as the optimal strategy are those close to the boundaries, i.e., $|\targetx|$ close to unity---as already discussed in the simpler case of homogeneous resetting in section~\ref{sec:resetting_boundaries_homogeneous_bulk}.

\begin{figure}
    \centering
    \includegraphics[width = 0.8\textwidth]{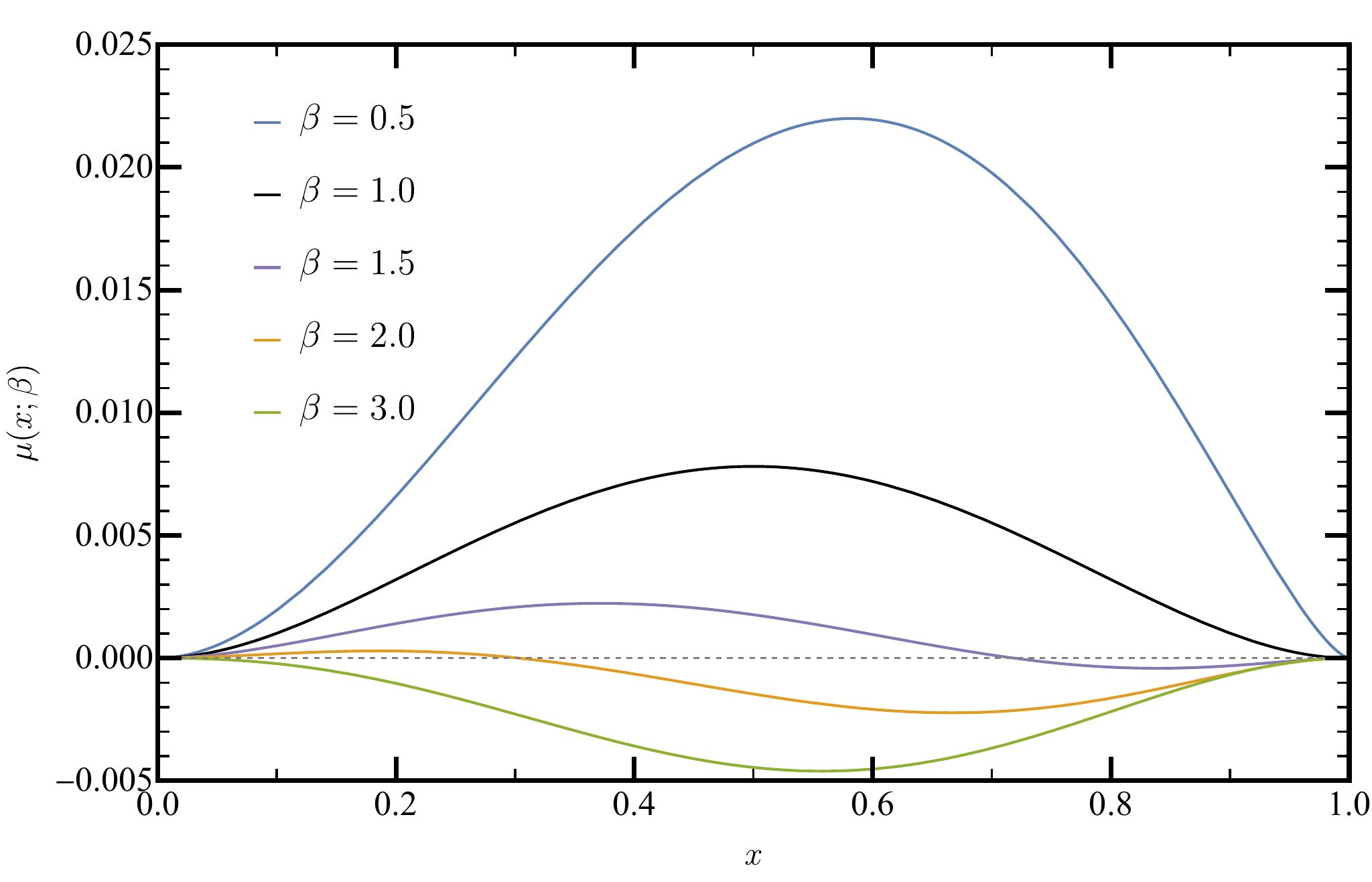}
    \caption{Functional derivative $\mu$ controlling the optimality of the non-resetting strategy in the bulk. Specifically, we show $\mu$ as a function of $x$, given by~\eqref{eq:mu(x)_beta_distribution}, for the $\beta$-family of target distributions introduced in~\eqref{eq:PTbeta}. The critical value $\beta_c=1$ (black solid line) marks the change to a not-always-positive $\mu(x;\beta)$, which entails an optimal non-zero resetting strategy in the bulk.}
    \label{fig:heterogeneous_bulk_resetting_beta_mu(x)}
\end{figure}

Mathematically, we are able to provide a rigorous criterion for the stability of the non-resetting strategy in the bulk for symmetric target distributions. The global derivative is defined as~\eqref{eq:mu(x)_heterogeneous_resetting}, so  
\begin{equation}
    \mu(x) = 2\int_0^1 d\targetx \PDF_T(\targetx) M_s(\targetx,x).
\end{equation}
Equation~\eqref{eq:Ms(x_t,x)_explicit_heterogeneous} directly shows that $\mu(x=0)=\mu(x=\pm 1)=0$, \ie the average MFPT remains unchanged for perturbations at the boundaries or at the centre of the domain. Physically, this behaviour makes sense, because we have resetting boundaries and the centre of the box is the resetting point. In fact, we can also check that $\mu'(x=0)=\mu'(x=\pm 1)=0$. The behaviour nearly the boundaries is thus given by the second derivative, which reads
\begin{subequations}
    \label{eq:mu''(x)_heterogeneous_resetting}
    \begin{align}
        \left. \mu''(x) \right|_{x=0} &= \frac{3}{2} \left\langle | \targetx | \right\rangle_T -\frac{1}{2},
        \\ 
        \left. \mu''(x) \right|_{x=1} &= - \frac{1}{2}\left\langle | \targetx | \right\rangle_T +  \PDF_T(1).
    \end{align}
\end{subequations}
In order to ensure $r(x)=0$ constitutes a stable minimum, convexity at $x=0,1$ is a necessary condition for optimality---recall that $\mu(x)>0$ to guarantee a local minimum. Therefore, we impose $\mu''(x=0,1)>0$, which leads to 
\begin{equation}
    1/3 < \braket{|\targetx|}_T < 2 \PDF_T(1).
\end{equation}
This inequality tells us that bulk resetting should be avoided unless the target is close enough to the origin, or the target distribution decays significantly at the boundaries. We must emphasise that this condition is only necessary, but not sufficient, for having a local minimum MFPT $r(x)=0$, because $\mu(x)$ may become negative in some subinterval of $(0,1)$.

Particularising for the same $\beta$-family of target distributions defined in~\eqref{eq:PTbeta}, our final result for the functional derivative is 
\begin{align}
    \mu(x;\beta)=\frac{1}{4} x \Bigg\{ &-x (1 + x) + 
    \frac{\Gamma \left(\frac{1}{2} + \beta \right)}{\sqrt{\pi} \Gamma(1 + \beta)} 
     \times\Bigg[-(1 - x)^2 + (1 - x)^\beta (1 + x)^{1 + \beta} 
    \nonumber \\ 
    &+2 \beta x^2 (1 + x) \, _2F_1\left(\frac{1}{2}, 1 - \beta; \frac{3}{2}, x^2\right) \Bigg] \Bigg\} .
    \label{eq:mu(x)_beta_distribution}
\end{align}
In figure~\ref{fig:heterogeneous_bulk_resetting_beta_mu(x)}, we show $\mu(x;\beta)$ as a function of $x$ for several values of $\beta$. If $\mu(x;\beta)<0$, $\forall x$, the optimal bulk resetting rate is non-zero. The black line stands for the critical value $\beta_c = 1$, \ie the flat distribution, which indicates the change of stability: for $\beta < \beta_c$, the non-resetting strategy is the optimal one because of $\mu(x,\beta)>0$, $\forall x\in(0,1)$. For $\beta>1$, $\mu(x;\beta)$ becomes negative close to the wall $x=1$ (boundary instability), signaling that $r(x)=0$ is no longer the best strategy. Notice that the critical value $\beta_c=1$ is consistently smaller than the one found in the homogeneous case, $\beta_c\simeq 1.71$, as shown in section~\ref{sec:resetting_boundaries_homogeneous_bulk}. 

\section{Optimal heterogeneous resetting: numerical results}
\label{sec:resetting_boundaries_numerical}

In order to find the exact bulk resetting profile that minimises the average MFPT, we resort to numerical methods because of the mathematical complexity of the problem---stemming from the generality. The numerical computation of the average MFPT has two main steps: (i) solving the ODE for the MFPT~\eqref{eq:resetting_boundaries_FPT_ODE} for a given $r(x)$ and $\targetx$; and (ii) minimising the average MFPT to obtain for the optimal profile $\optMFPT{r}(x)$.

The numerical scheme is based on a gradient descent algorithm to minimise the functional $\overline{\FPT{(1)}}[r]$, whose whole details are provided in appendix~\ref{app:gradient-descent}. Here, a brief summary follows. The spatial coordinate $x$ is discretised into a mesh with $N$ nodes. Initially, some resetting profile $r_0(x)$ is considered and the average MFPT is computed for such an initial strategy. Then the following steps are executed iteratively:
\begin{enumerate}
    \item For the current search strategy, the functional derivative $\mu$ is computed in the mesh. 
    \item The new potential search strategy is given by $r(x) \to \max[r(x)-\lambda \,\mu,0]$, where $\lambda$ is an adaptive factor with a given value $\lambda_0$ in the first trial.
    \item The average MFPT of the new strategy is computed. If it is lower or equal than the previous one, the strategy is updated and we go to step 1. Otherwise, $\lambda$ is reduced, $\lambda \to \lambda/2$, and we go to step 2.
\end{enumerate}
As usual, gradient descent is expected to converge to the local minimum whose attraction basin includes the initial condition. Since we have not been able to prove the convexity of our functional, it may have multiple local minima. We mitigate the risk of not finding the global minimum by varying the initial resetting profile, $r_0(x)$. Numerical parameters typically adopted are $N\in\{501,1001,2001\}$, $\lambda_0\in\{10^5,10^6,10^7\}$. 

The following sections are devoted to present the numerical results for several target distributions. We only report results for the model with resetting boundaries: as we will show in section~\ref{sec:resetting_boundaries_numerical}, the optimal non-zero bulk resetting profile diverges at the boundaries, which corresponds to an effective resetting boundary.

\subsection{\texorpdfstring{$\beta$-family of target distributions}{beta-family of target distributions}}

\begin{figure}
    \centering
    \includegraphics[width = 0.7\textwidth]{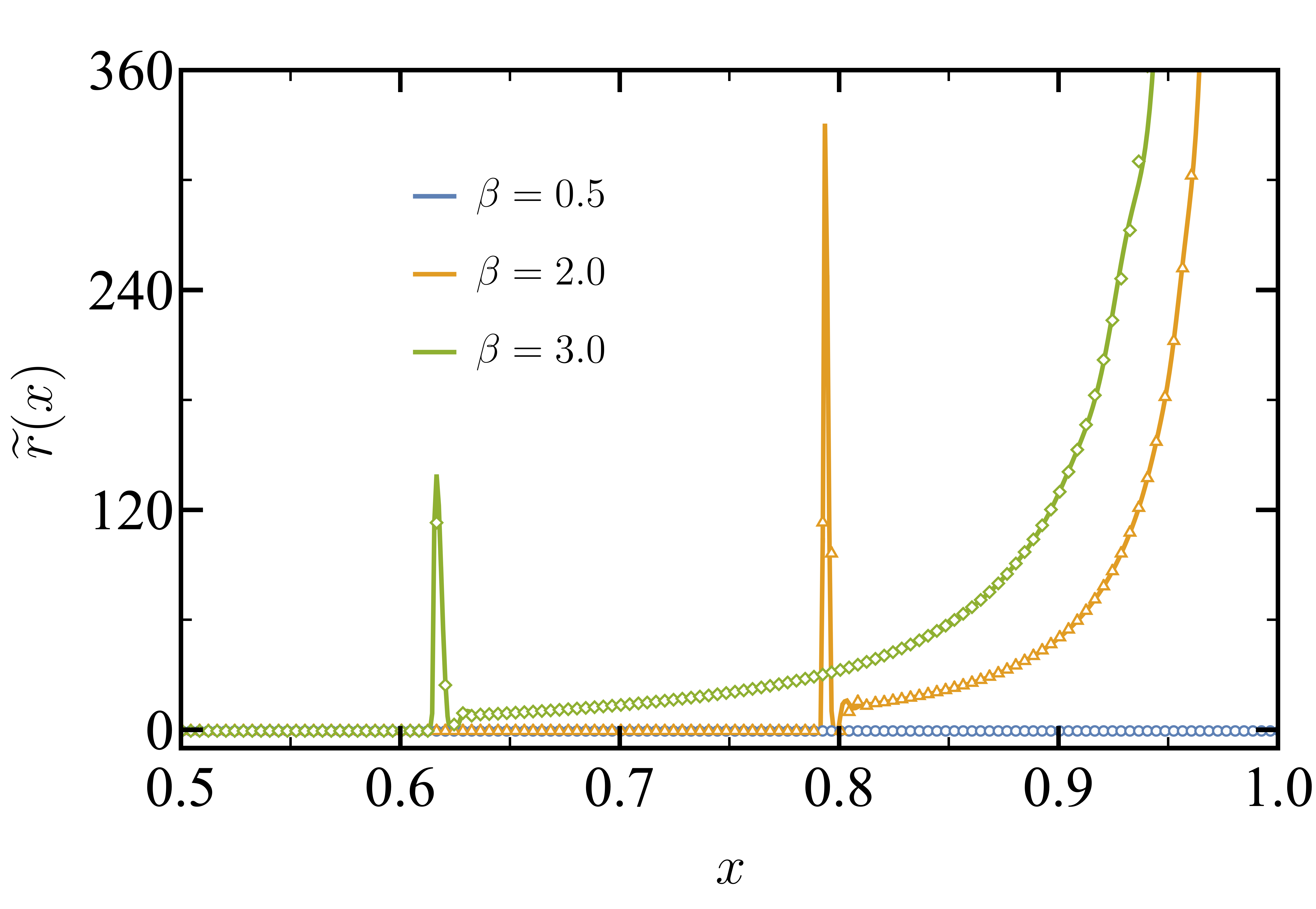}
    \caption{Optimal heterogeneous resetting profiles in the bulk. Specifically, they have been obtained numerically for the target distributions of the family~\eqref{eq:PTbeta} corresponding to $\beta=0.5,2,3$. For $\beta\leq1$, the optimal strategy consistently involves no resetting in the bulk. For $\beta>1$, for which non-zero bulk resetting is the optimal strategy, the features displayed---central region with no resetting for $|x|<x_J$, strongly heterogeneous behaviour close to the wall---are robust.  The acceleration factor with respect a vanishing resetting bulk is around 2\% and 7\% for $\beta=2$ and $\beta=3$, respectively. Symbols and lines stand for meshes with $N=501$ and $N=2001$ nodes, respectively. 
    }
    \label{fig:beta_optimal_r}
\end{figure}

We start by analysing the $\beta$-family of target distributions defined in~\eqref{eq:PTbeta}, the optimal profiles of which are displayed in figure~\ref{fig:beta_optimal_r}. For $\beta\leq1$, we obtain the same analytical prediction as in the previous section, \ie the optimal bulk resetting rate is zero, $\optMFPT{r}(x)=0$. For $\beta>1$, the optimal strategy involves non-zero bulk resetting, whose features are robust for different values of $\beta$. Profiles are split into two regions: (i) a central region where it is optimal to avoid resetting, $\optMFPT{r}(x)=0$ for $|x|<x_J$, and (ii) a region close to the boundaries, $x_J<|x|<1$, where the optimal resetting is non-zero, $\optMFPT{r}(x)>0$. This separation is in agreement with the theoretical analysis in~\eqref{eq:Ms(x_t,x)_explicit_heterogeneous}, which indicates that bulk resetting is always beneficial if the target probability is concentrated at the centre---recall that the $\beta$-family tends to a centred delta peak as $\beta\to\infty$. At $|x|=x_J$ the numerical results evidence the presence of a Dirac delta contribution to $r(x)$, which is corroborated by the use of different mesh sizes---the peak becomes narrower and higher, with a constant area below, when considering denser meshes.

One of the most remarkable features of these optimal resetting profiles is the emergence of Delta-dirac contributions, which act like separators of two regions. Although we do not have a clear physical interpretation for the emergence of these peaks, we can ensure that they are not numerical artefacts, since they are robust when varying the mesh size. Mathematically, this delta peak can be regarded, for instance, as a limit process in a narrow spatial window, inside which one has a large homogeneous resetting rate, inversely proportional to the width of the window. Interestingly, the qualitative features of the profile, \ie the vanishing window, the peak and the high heterogeneity, are robust for other families with accumulation of probability in the centre---as shown in the following. 

\subsection{Polynomial target distributions} 
\subsubsection*{Sixth-degree polynomial target distribution}

\begin{figure}
    \centering
    \includegraphics[width = 0.7\textwidth]{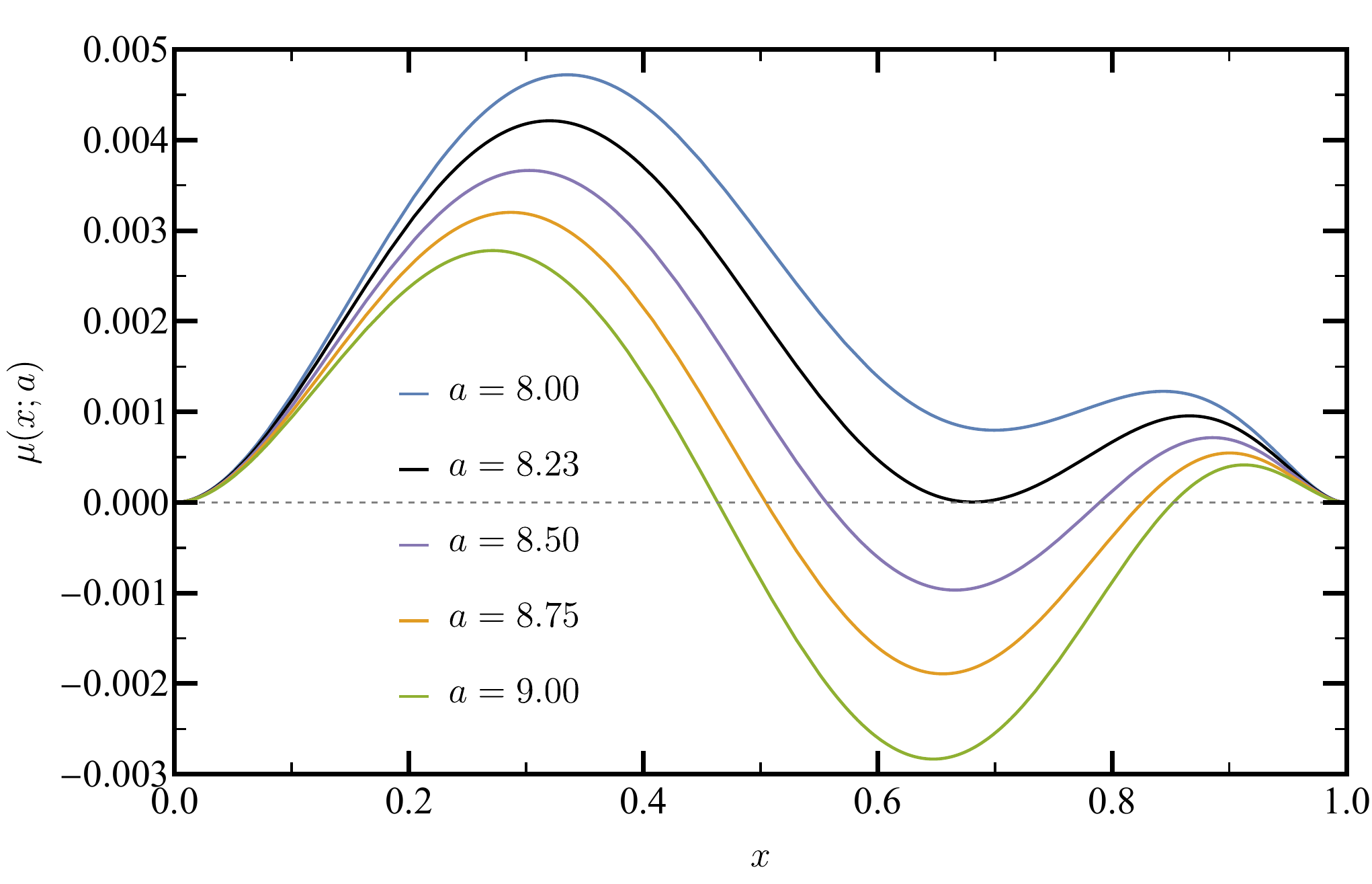}
    \caption{Functional derivative  controlling the stability of the non-resetting strategy in the bulk for the polynomial family in~\eqref{eq:polynomial_sixth_order}. The critical value $a_c$ (black solid line) demarcates when $\mu(x;a)$ becomes negative for some $x$, indicating that a non-zero resetting profile is the optimal one for $a>a_c$.}
    \label{fig:heterogeneous_sixth_polynomial}
\end{figure}

The emergence of Dirac-delta contributions in the optimal bulk resetting profiles for the $\beta$-family motivates us to explore what features of the distribution cause this phenomenon. For that purpose, we will explore some polynomial families of target distributions, in order to have a simple control of the accumulation of probability, the number of peaks, their vanishing or not at the boundaries, etc.

\begin{figure}
            \centering
            \includegraphics[width=0.49\textwidth]{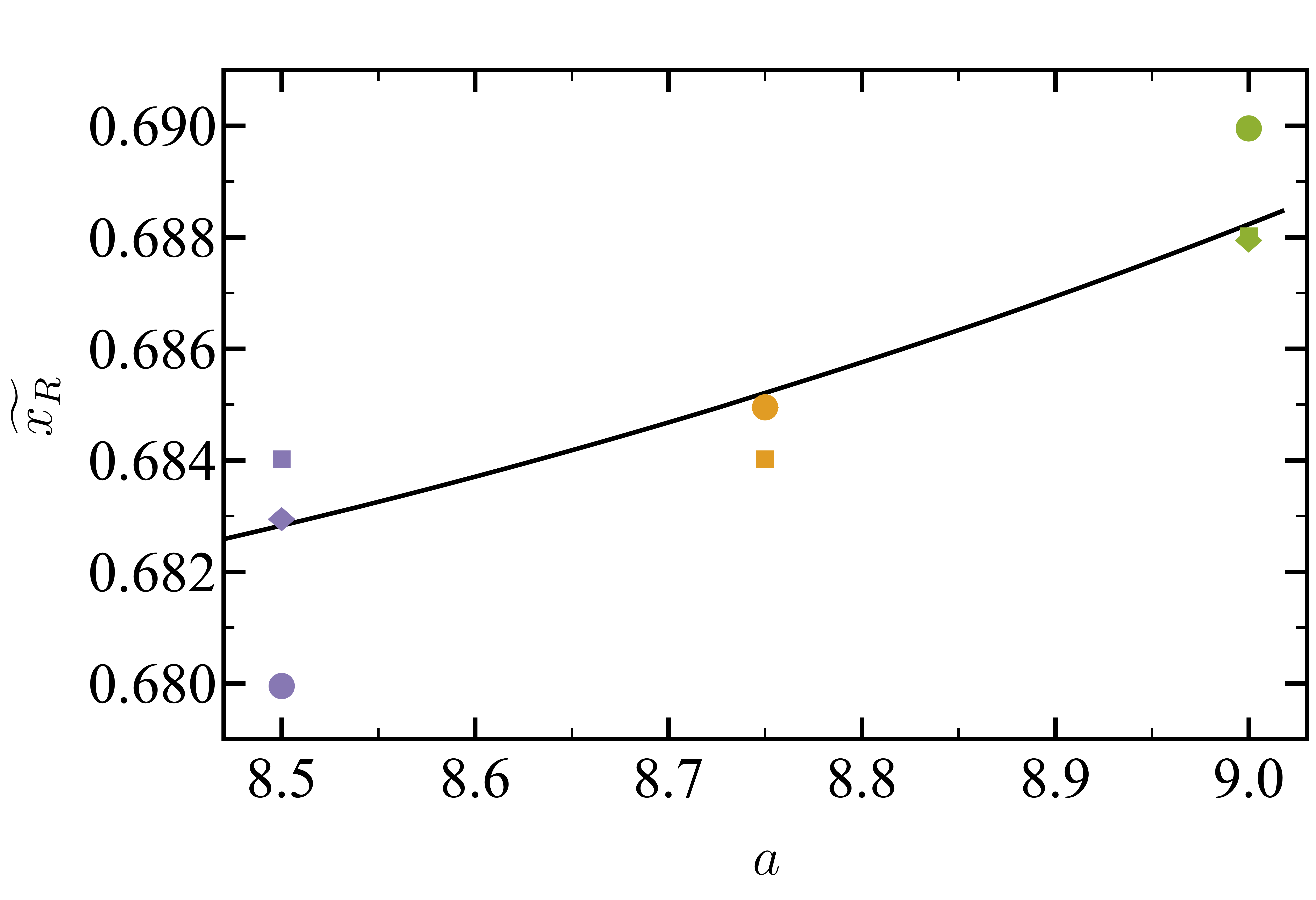}
            \includegraphics[width=0.49\textwidth]{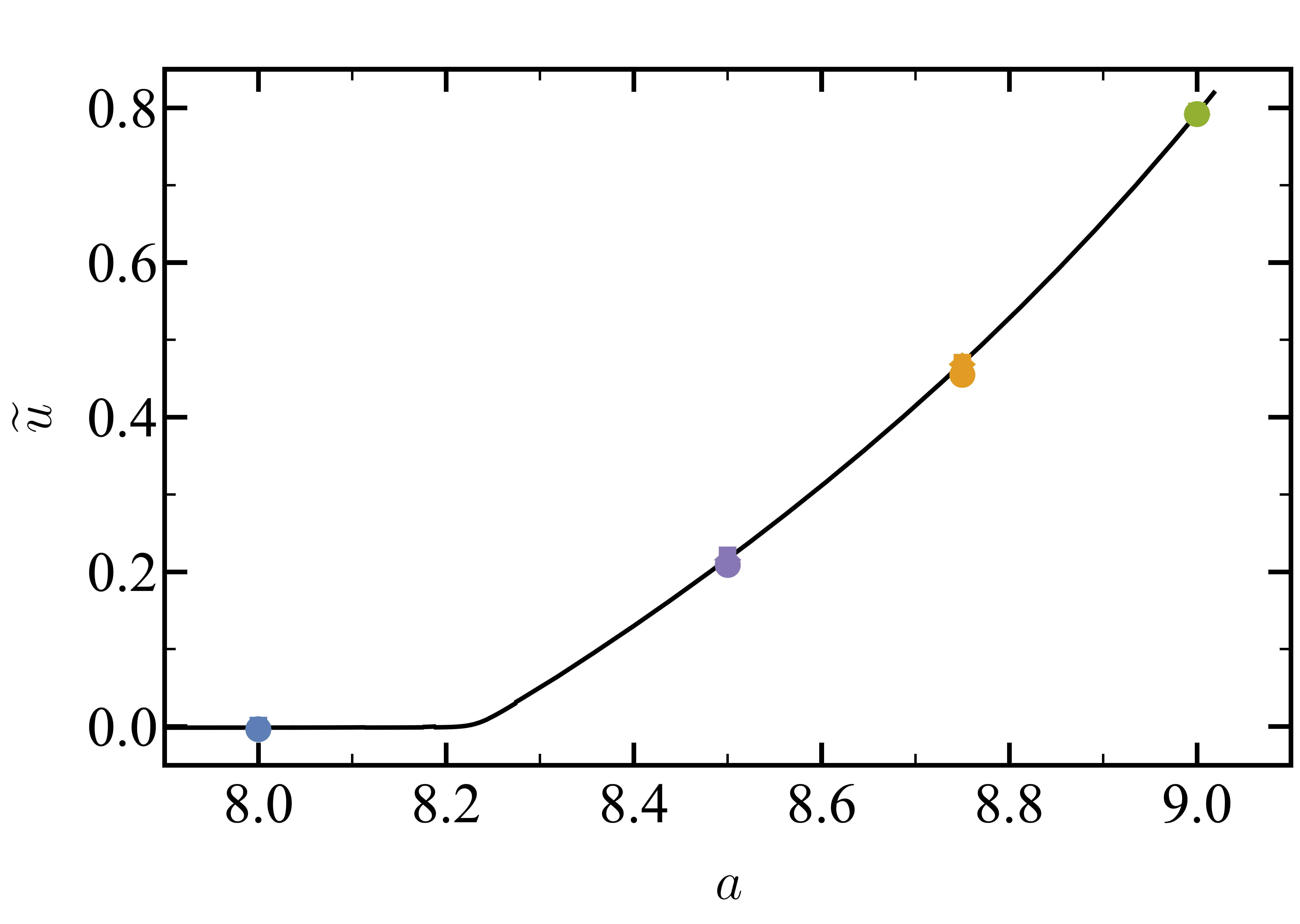}
    \caption{
        Parameters for the optimal resetting strategy for the polynomial family in~\eqref{eq:polynomial_sixth_order}. The theoretical parameters (black solid lines) are compared with the numerical parameters obtained by gradient descent optimisation, using meshes with $N=201, \, 501, \, 2001$ (circles, squares and diamonds, respectively). The color code is used to distinguish between different values of $a$, as in figure~\ref{fig:heterogeneous_sixth_polynomial}.}
    \label{fig:heterogeneous_sixth_polynomial_optimal-values}
\end{figure}

Let us consider the sixth-degree polynomial distribution
\begin{equation}
    \PDF_T(\targetx;a) = a \targetx^2 + (4-3a)\targetx^4 + \frac{7}{30}(-9+8a)\targetx^6,
    \label{eq:polynomial_sixth_order}
\end{equation}
with $0<a<(117+\sqrt{8169})/23$. The functional derivative as a function of $a$ is
\begin{align}
    \label{eq:mu(x)_sixth_polynomial}
    \mu(x;a) & = 
    \frac{x^2}{480}\brackets{171-12a+20ax^4+8(4-3a)x^6-(9-8a)x^8}
    \nonumber \\
    & \quad -\frac{|x|^3}{480}\brackets{217-4a-20ax^2-8(4-3a)x^4+(9-8a)x^6},
\end{align}
which is positive in the interval $0<x<1$ for $a < a_c$, as shown in figure~\ref{fig:heterogeneous_sixth_polynomial}. For $a<a_c\simeq 8.23$, the optimal profile is $\optMFPT{r}(x)=0$, while for $a>a_c$ we find that $\optMFPT{r}(x)$ comprises two symmetric Dirac-deltas at specific positions and with specific intensity, which can be numerically derived. 

\begin{figure}
    \centering
    \includegraphics[width = 0.7\textwidth]{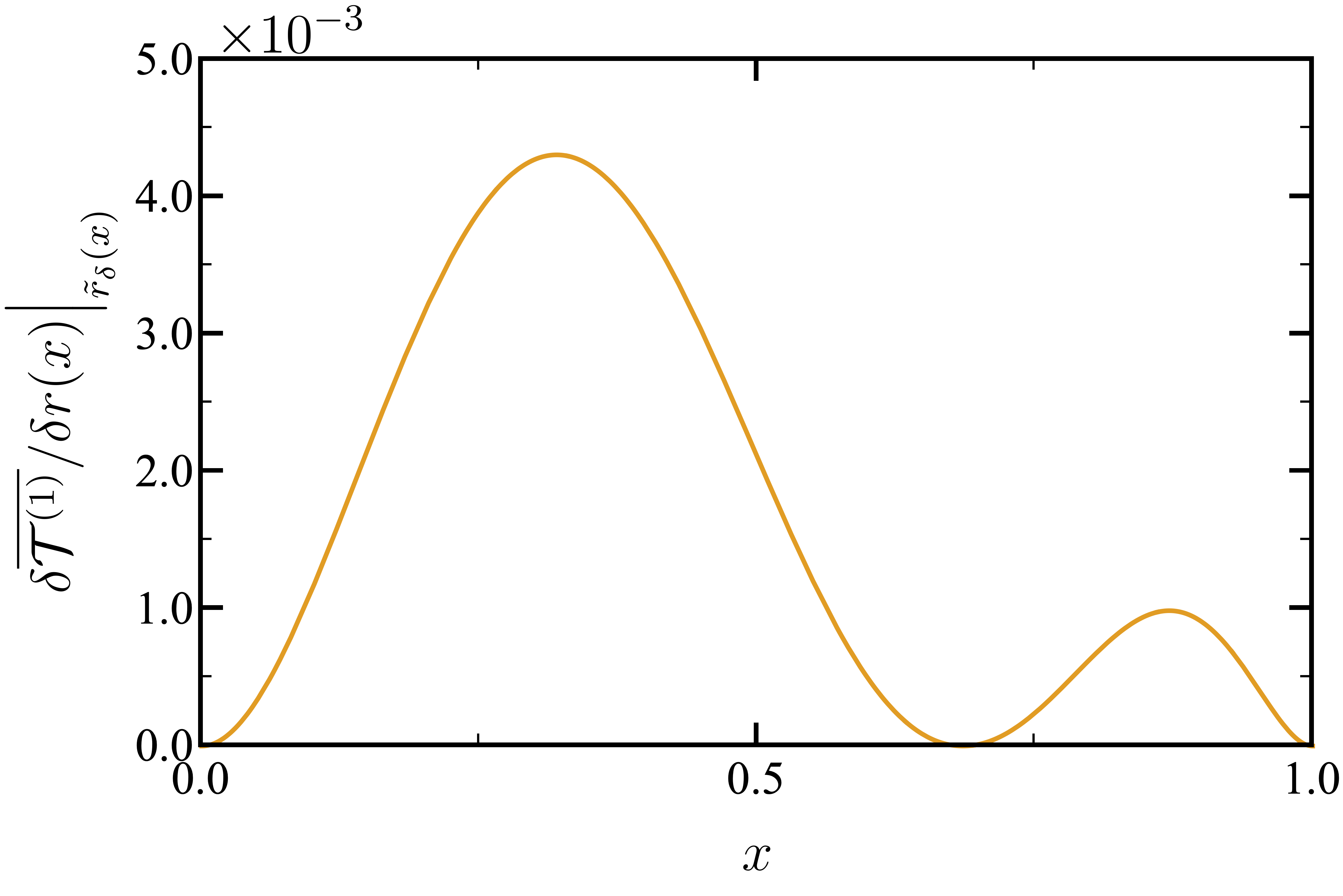}
    \caption{Functional derivative controlling the stability of the optimal resetting profile within the family~\eqref{eq:ansatz_dirac_delta} for the polynomial distribution~\eqref{eq:polynomial_sixth_order}. The curve corresponds to the parameter $a=8.75$, whose optimal values are $\optMFPT{x_R}\simeq 0.685$ and $\optMFPT{u}\simeq 0.473$.}
    \label{fig:mu(x)_ansatz_sixth_polynomial}
\end{figure}

Motivated by the just described numerical results, we propose the following ansatz for the optimal profile:
\begin{equation}
    r_{\delta}(x) = \frac{u}{x_R}\brackets{\delta(x-x_R)+\delta(x+x_R)},
    \label{eq:ansatz_dirac_delta}
\end{equation}
where $x_R$ and $u$ indicate the positions of the peaks and their intensity, respectively. The optimisation then consists of finding the parameters $\{\optMFPT{x_R},\optMFPT{u}\}$ that minimise $\overline{\FPT{(1)}}$, \ie 
\begin{equation}
    \left.\pdev{}{\overline{\FPT{(1)}}}{x_R}\right|_{\optMFPT{x_R},\optMFPT{u}}=0, \quad \left.\pdev{}{\overline{\FPT{(1)}}}{u}\right|_{\optMFPT{x_R},\optMFPT{u}}=0.
    \label{eq:optimal_values_sixth_polynomial}
\end{equation}
The expressions are intentionally omitted for the sake of brevity, though a comparison between values obtained from analytical results and simulations is presented in figure~\ref{fig:heterogeneous_sixth_polynomial_optimal-values}. Therein, the solid black line stands for the analytical curve obtained by solving~\eqref{eq:optimal_values_sixth_polynomial}, whereas the symbols correspond to the numerical results obtained by optimising the average MFPT---using the same gradient descent algorithm explained above and in appendix~\ref{app:gradient-descent}. This numerical method provides a certain resetting profile, which can be used to obtain numerical values of $\optMFPT{x_R}$ and $\optMFPT{u}$, by fitting the numerical profile with~\eqref{eq:ansatz_dirac_delta}. We observe a very good agreement between both approaches, being consistent when varying the number of nodes in the mesh---the peak becomes narrower and higher when considering denser meshes. Despite the constrained optimisation does not ensure global optimality, we have observed numerically that no other profile provides a lower average MFPT. Analytically, we are able to compute the variational derivative of the average MFPT evaluated at the optimal ansatz~\eqref{eq:ansatz_dirac_delta}, \ie $\delta \overline{\FPT{(1)}}/\delta r(x)|_{\optMFPT{r}_{\delta}(x)}$,
which tells us if any local perturbation from the optimal profiles $\optMFPT{r}_\delta (x)$ would increase the average MFPT. Clearly, if $\optMFPT{u}=0$, $\delta \overline{\FPT{(1)}}/\delta r(x)|_{\optMFPT{r}_{\delta}(x)}$ reduces to~\eqref{eq:mu(x)_sixth_polynomial}. Figure~\ref{fig:mu(x)_ansatz_sixth_polynomial} shows this quantity is positive in the whole interval for $a>a_c$, so $\optMFPT{r}_\delta (x)$ indeed minimises the average MFPT.

\subsubsection*{\texorpdfstring{$n$-th degree simple polynomial target distribution}{n-th degree simple polynomial target distribution}}

\begin{figure}
    \centering
    \includegraphics[width = 0.7\textwidth]{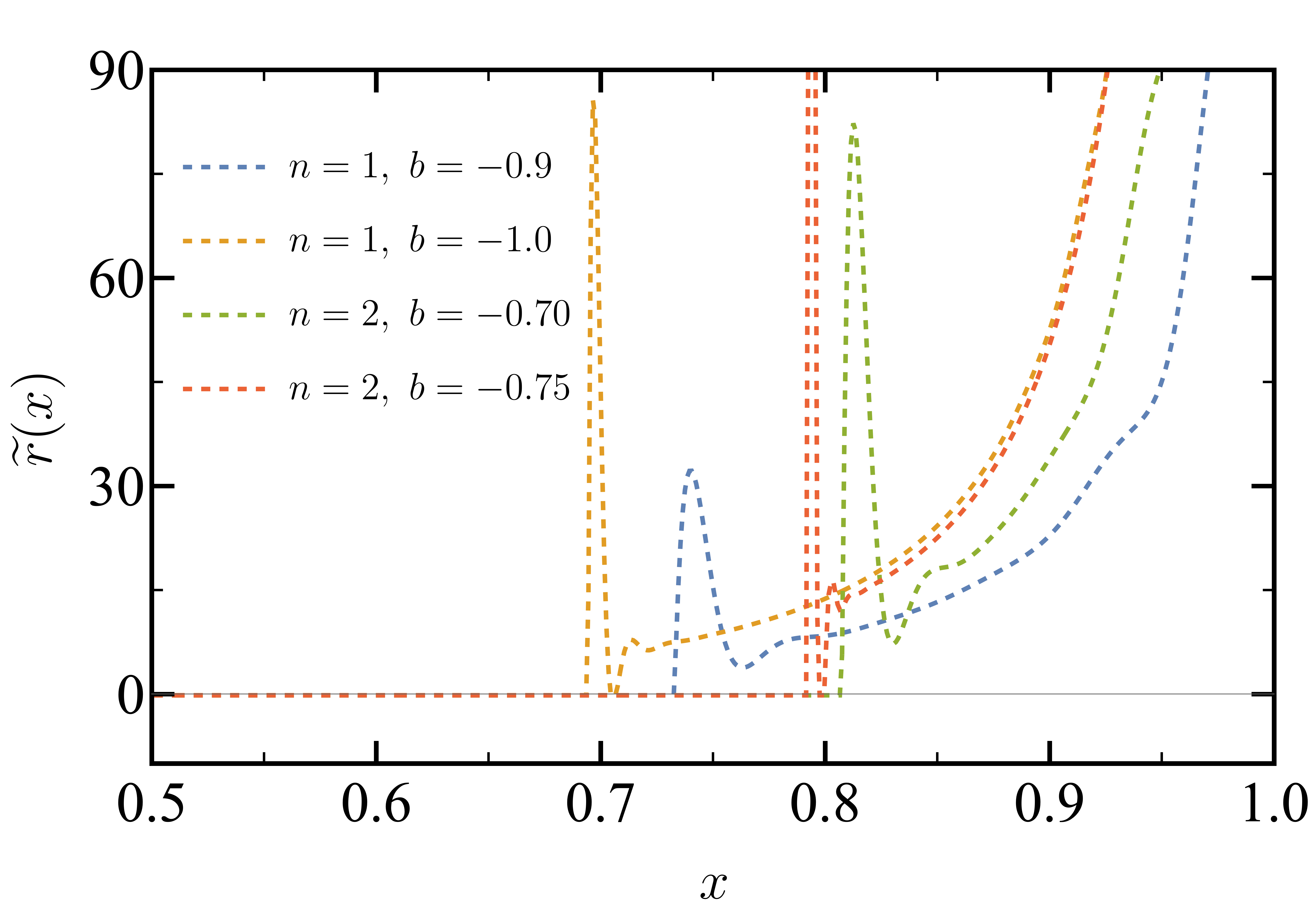}
    \caption{Optimal strategy for the polynomial family given by~\eqref{eq:polynomial_family_distribution_simple} for a mesh with $N=2001$. Resetting profiles have been obtained minimising numerically the average MFPT, given by~\eqref{eq:functional_average_MFPT}, using our gradient descent algorithm.}
    \label{fig:simple_polynomial_heterogeneous_resetting}
\end{figure}

In the same way as with the sixth-order polynomial family~\eqref{eq:polynomial_sixth_order}, we can consider the simple polynomial distribution
\begin{equation}
    \PDF_T(\targetx;b,n) = -\frac{b}{1+n} + b |x|^n, \quad n>-1,
    \label{eq:polynomial_family_distribution_simple}
\end{equation}
which is useful to explore the dependence of the optimal profiles on how the distribution behaves at the boundaries. Concretely, we are interested in the cases $n=1,2$, due to its similarities with the $\beta$-family. The coefficient $b$ indicates if the probability is accumulated at the centre ($b<0$) or at the boundaries ($b>0$).  

The optimal profiles for this distribution display a completely analogous behaviour to the ones found for the $\beta$-family, as seen in figure~\ref{fig:simple_polynomial_heterogeneous_resetting}. For $b>0$, the optimal bulk resetting rate vanishes---actually, $b=b_c=0$ is again the critical value for this family, corresponding to the flat distribution. For $b<0$, the optimal non-zero resetting profiles share the features already observed for the $\beta$-family: (i) a central region with no resetting, (ii) a peak and a strongly heterogeneous behaviour close to the wall. 

This hints at the existence of common features in the optimal resetting strategies, when the distribution is made by just one ``hill'' centred at the origin. When the probability is accumulated that way, the average functional depends weakly on the details of the resetting profiles at the boundaries. This behaviour causes that the optimal profiles share the same qualitative features, regardless of the specific target distribution, as long as the resetting strategy blows up at the boundaries.


\bookmarksetup{startatroot} 

\chapter{Conclusions}
\pagestyle{mystyle4}
\label{ch:conclusions}
This thesis attempts to provide a comprehensive understanding of some phenomena in statistical mechanics. Specifically, our approach is based on proposing minimal models that capture the essential ingredients of the phenomena under study. On the one hand, physical systems as complex as deformable solids can be effectively described by mesoscopic models, where we propose the simple interactions that are able to reproduce the observed phenomenology. On the other hand, we address the research on search problems by optimising realistic search strategies that can be performed experimentally and modelled mathematically in a simple way. 

These two research lines have been developed throughout this thesis. In this chapter, we summarise the main results, enumerating the conclusions that can be drawn from our work, and discussing some possible future research directions. For the sake of clarity, such conclusions are split into two sections, one for each of the parts of the thesis.

\subsubsection*{Part I: Buckling in low-dimensional spin-elastic models}

\begin{enumerate}
    \item We have put forward a novel rotationally invariant spin-elastic model for one- and two-dimensional solids. The discrete Hamiltonian for this model involves contributions that only depend on the discrete curvature and spin variables of the corresponding lattice. This makes the model be physically consistent from a theoretical point of view, compatible with the theory of elasticity, as it preserves the rotational invariance in the absence of external forces.
    \item A continuum description of spin-elastic models has been derived for both one- and two-dimensional lattices. To compute the continuum limit of the discrete Hamiltonian, we have taken into account how its different contributions scale with the lattice parameter. This allows us to identify the effective parameters of the continuum model in terms of the microscopic ones without considering very large size, \ie the thermodynamic limit is uncoupled from the continuum limit.
    \item In the continuum limit, we have derived an Euler-Lagrange equation describing the profiles that minimise the free energy of the system. Because of the dependence of the free energy on the Laplacian of the displacement field (\ie the continuum curvature), the Euler-Lagrange equation becomes a closed equation for the curvature.
    \item The Euler-Lagrange equation shares the same functional form for any physically reasonable topology of the spin-membrane model. This is a consequence of how the elastic potential is included in the Hamiltonian, which is written in terms of the discrete Laplacian of the displacement. The main difference between geometries lies on the partition function of the spins.
    \item The solutions of the Euler-Lagrange equation are displacement fields with a constant curvature, independent of the position, which depends on the parameters of the model: the temperature and the coupling between neighbouring spins. Then, we are able to characterise the different phases of the system by their respective curvature, which plays the role of an order parameter. 
    \item The equilibrium profiles are obtained by integrating the curvature twice, leading to parabolic shapes in one dimension and to surfaces with constant curvature in two dimensions. In the two-dimensional case, different geometries, such as circular or rectangular, have been taken into account to study the influence of the boundary conditions on the equilibrium profiles given by Poisson's equation.
    \item The solutions of the Euler-Lagrange equation allow us to give a full description of the phase diagram for the spin-string model. Three different phases have been identified: a rippled phase with zero curvature, and two buckled phases, being one of them stable and the other one unstable. A tricritical point has been found, which demarcates the change from a second-order, if the temperature is above the critical point, to a first-order phase transition, if the temperature is below it, between the flat and the stable buckled phases. 
    \item The order parameter has been computed both numerically and analytically for the one-dimensional model. On the one hand, the transcendental Euler-Lagrange equation has been solved numerically to obtain the curvature for any value of the parameters. On the other hand, we have used methods from bifurcation theory and Landau theory of phase transitions to obtain an approximate expression of the order parameter close to the transition lines. Additionally, exact results have been obtained in the low-temperature limit, providing a perfect agreement between our theoretical predictions for the transition lines and the curvature with the numerical results.
    \item The 1d spin-string model is able to qualitatively reproduce the experimental results obtained for suspended graphene sheets, due to the similarity of the obtained phase diagram with previous models. Specifically, we have found the existence of a region where rippled and buckled phases coexist. In that region, the system can be prepared in a metastable rippled state that becomes unstable when the temperature is increased.
    \item The Euler-Lagrange equation does not have a completely closed form for the 2d spin-membrane model, since we are not able to derive an explicit expression for the spin contributions to the free energy. However, their solutions can be computed in certain limits, such as the low-temperature regime or the case of uncorrelated spins. 
    \item If there is no coupling between neighbouring spins, topology plays no role and the model can be solved exactly for any temperature. In this case, the system presents a second-order phase transition between a rippled and a buckled phase. 
    \item In the low-temperature limit of the spin-membrane model, the system also presents an interval of the coupling constant where three different phases exist: a rippled phase with zero curvature, a stable buckled phase and an unstable buckled phase. The change of stability between the rippled and the stable buckled phases occurs through a first-order phase transition, analogously to the one-dimensional case.
    \item Finally, we briefly comment on some possible future research lines: 
        \begin{enumerate}
            \item It would be interesting to analyse some variant of the spin-membrane model that could be analytically solved. A possible approach could be to approximate the spin partition function, \eg using a mean-field approximation. 
            \item Including internal degrees of freedom as spin variables has been shown to be useful to model the buckling transition of low-dimensional solids. It seems worth exploring more complex internal degrees of freedom, \eg Potts-like models~\cite{journalarticle:Wu_PottsModel_Rev.Mod.Phys.82}---as a generalisation of the Ising model. Another approach could be to apply the ideas developed in this thesis to analyse phenomena in other kinds of systems, like wrinkled biofilms that have already been studied using elastic models~\cite{journalarticle:Espeso.etal_DifferentialGrowthWrinkled_Phys.Rev.E15}.
            \item So far, we have only considered equilibrium configurations of the spin-elastic models. However, we have some preliminary results regarding the dynamics of these systems, and how they relax to equilibrium. To study this problem, we will propose a Liouville-Master equation for the time evolution of the PDF of finding the system in any possible configuration at a given time. This equation will relate the dynamics of the spin variables, which follow a master equation based on Glauber dynamics, with Hamilton's equations of motion for the string.
            \item The Liouville-Master equation could be solved by integrating numerically the equations of motion for the string, while the spins are updated following a Monte Carlo algorithm. Nonetheless, we expect the dynamics of the system to strongly depend on the relative time scales of the string and spin variables. Therefore, it would be physically relevant to investigate different regimes, such as fast spins compared to the string, to derive a Fokker-Planck equation by applying a multiple scale analysis---a Chapman-Enskog expansion.
        \end{enumerate}
\end{enumerate}

\subsubsection*{Part II: Realistic implementations of stochastic resetting}

\begin{enumerate}
    \item We have studied two different aspects in realistic implementations of stochastic resetting processes. In both cases, we have considered a Brownian particle that diffuses in one dimension and is randomly reset to a given position at certain times. The particle is instantaneously relocated from any position to the resetting position in both models.
    \item In our first problem, stochastic resetting is followed by a refractory period phase, where the particle remains immobile for a certain time. The waiting times of the refractory period are also random variables. After the refractory phase is finished, the particle is free to diffuse again until the next resetting event.
    \item Resetting with refractory periods is framed within the context of intermittent search strategies. A general mathematical framework, which we term pathway formulation, has been introduced to study the evolution of every observable in this model or in any other intermittent search process with renewal properties.
    \item Particular distributions for the resetting and refractory times have been considered to obtain explicit results. Specifically, we have studied the case of Poissonian distributed waiting times for both resetting and refractory periods. In this case, we have derived exact analytical expressions for the non-equilibrium stationary state. 
    \item An asymptotic analysis of the PDF provides insight on how the system evolves towards its stationary state. The relaxation to the steady state involves a front that separates an inner region, where the PDF has already reached the stationary state, from an outer region, where the PDF is still transient and behaves as a free Brownian particle.
    \item The search problem for a fixed target is optimised by minimising the MFPT with respect to the resetting and refractory rates. A backward formulation provides an analytical expression for the MFPT, revealing a non-trivial, resonance-like, dependence with the refractory period rate.
    \item Before studying the second aspect, we discuss some perspectives for future work in this line: 
        \begin{enumerate}
            \item Motivated by the dependence on the optimal rates, the connection between resetting and resonant activation phenomena could be further explored.
            \item It would be interesting to investigate other distributions for the resetting and refractory times, such as power-law distributions, which could lead to new rich phenomenology.
            \item Refractory period has been proposed as a more realistic implementation of stochastic resetting, where the particle has to ``pay'' a time cost after each reset. Other implementations may be considered, such as a return phase, where the particle is not instantaneously relocated to the resetting position.
            \item The return phase could be modelled as a Brownian motion under the action of a harmonic potential that pulls the particle to the resetting position. It would be switched off when the reset is achieved, and the free diffusion phase would start again. In this case, the pathway formulation should be adapted to study the observables of interest.
            \item The inclusion of an external potential involves an energetic cost during the return phase. An intriguing question is related to the optimisation of the protocol for the external potential: how can we minimise the energetic cost associated with the return phase? As we reduce the cost, the return time increases, causing a trade-off in the search problem between both quantities.
        \end{enumerate}
        \item The second aspect we have focused on is related to the nature of the target. We have considered a Brownian particle under the effect of a heterogeneous resetting mechanism in quenched disordered media. Thus, the position of the target is not deterministic, instead it is drawn from a given distribution in each realisation.  
        \item Any observable of the system can be computed by averaging the results for the observables at fixed target over the target distribution. Nevertheless, the optimisation of the search becomes a mathematical challenge, since we have to minimise an involved functional.
        \item To overcome this difficulty, we have first considered a particular simple model where the target is drawn from a dichotomous distribution, and the resetting strategy belongs to a piecewise constant family of functions. 
        \item The optimisation of the search process has been performed by minimising the MFPT or the standard deviation of the first-passage time distribution. The theoretical implicit expressions for both observables have been derived, which have allowed us to find the optimal resetting rates. 
        \item In certain limits, where the target is close to be always in the same position, an interesting trade-off between minimising the MFPT and the standard deviation arises. When the target position is deterministic, both observables are minimised by the same resetting rate, which prevents the search from exploring the region opposite to the target. However, if the probability of finding the target is small, but not exactly zero, there exists very large fluctuations in the first-passage time that makes the standard deviation diverge when the MFPT is minimised. Minimising the variance seems to be a good strategy to avoid those fluctuations, while keeping the MFPT close to its optimal value.
        \item After studying the dichotomous model, the general problem has been addressed by considering the system to be bounded within a finite domain. Herein, we have put forward a new kind of boundary conditions, called resetting boundaries: the particle experiments an instantaneous reset when it hits the boundaries. 
        \item To minimise the average MFPT functional, we have analysed the optimal resetting strategy in the bulk for both homogeneous, \ie a constant resetting rate, and heterogeneous, \ie a space-dependent resetting profile. 
        \item In the case of homogeneous resetting, we have derived an exact analytical expression for the average MFPT for a specific family of target distributions. The average MFPT presents a non-trivial dependence on the parameters of the target distribution, leading to a strategy with no resetting in the bulk if the probability is accumulated close enough to the boundaries.
        \item The heterogeneous resetting strategy has been theoretically analysed by studying the perturbation with respect to the no-resetting strategy in the bulk. We have proven that the no-resetting strategy in the bulk is, at least, a local minimum if the target distribution is sufficiently peaked close to the boundaries.
        \item Optimal resetting profiles in the bulk are numerically computed by minimising the average MFPT, using a gradient descent method. The optimal non-zero profiles present similar characteristics for several target distributions: they vanish close to the resetting position, then a high peak emerges, and the resetting rate is strongly heterogeneous when approaching the boundaries, where the resetting rate takes very large values. 
        \item Certain target distributions lead to quite simple optimal profiles, as in the case of sixth-order polynomial distributions. Here, the optimal resetting function turns out to be a symmetric couple of Dirac-delta peaks close to the boundaries.
        \item In general, resetting boundaries have been proven to be more efficient than usual reflecting boundaries to minimise the MFPT in a bounded domain. In fact, numerical optimisation shows that the optimal resetting strategies in the bulk tend to very large values close to the boundaries in many scenarios, resembling the behaviour of resetting boundaries.
        \item Finally, a brief discussion of possible perspectives for this latter problem:
            \begin{enumerate}
                \item We need to have a better understanding of the physical meaning of the Dirac-delta-like peaks in the optimal resetting profiles.
                \item The concept of resetting boundaries could be further explored in other scenarios, such as higher-dimensional systems. Therein, it is not clear whether resetting boundaries will outperform reflecting boundaries or not.
                \item Another interesting research line is related to consider other kinds of disordered media. Very recently, some works have considered the predator-prey model under the effect of stochastic resetting with annealed disorder \cite{journalarticle:Evans.etal_ExactlySolvablePredator_JPhysMathTheor22}: the searcher (predator) and the target (prey) are both moving following Brownian dynamics.
            \end{enumerate}
\end{enumerate}

\clearpage


\appendix
\chapter{Variational principle for functionals with higher-order derivatives}
\label{app:variational_principle_higher_order}

Let us consider a functional $\mathcal{F}$ of some function $u(x)$ and its derivatives up to the second order: 
\begin{equation}
    \mathcal{F}[u] = \int_{x_1}^{x_2} \d{x} f\parenthesis{u,u',u'';x},
\end{equation}
where the endpoints $x_1$ and $x_2$ are fixed. In many physical situations, one is interested in finding the function $u_{\eq}(x)$ that minimises $\mathcal{F}[u]$. To accomplish that goal, we consider an arbitrary variation $\delta u$---involving the corresponding variations $\delta u'$, and $\delta u''$. Therefore, the variation of the functional is
\begin{equation}
    \delta \mathcal{F} = \int_{x_1}^{x_2} \d{x} \brackets{f_{u}\delta u + f_{u'}\delta u' + f_{u''}\delta u''},
\end{equation}
where $f_u \equiv \partial f / \partial u$, $f_{u'} \equiv \partial f / \partial u'$, and $f_{u''} \equiv \partial f / \partial u''$. Applying integration by parts to each term we get 
\begin{subequations}
    \begin{align}
        \int_{x_1}^{x_2} \d{x} f_{u'} \delta u' &= \Bigg[f_{u'} \delta u\Bigg]_{x_1}^{x_2} - \int_{x_1}^{x_2} \d{x} \cdev{}{f_{u'}}{x} \delta u, \\
        \int_{x_1}^{x_2} \d{x} f_{u''} \delta u'' &= \Bigg[f_{u''} \delta u' - \cdev{}{f_{u''}}{x} \delta u \Bigg]_{x_1}^{x_2} + \int_{x_1}^{x_2} \d{x} \cdev{2}{f_{u''}}{x} \delta u.
    \end{align}
\end{subequations}
Hence, the complete variation of the functional reads
\begin{align}
    \delta \mathcal{F} = \int_{x_1}^{x_2} \d{x} \parenthesis{
        \pdev{}{f}{u}-\cdev{}{}{x}\pdev{}{f}{u'}+\cdev{2}{}{x}\pdev{}{f}{u''}
    }\delta u + \Bigg[ 
        \parenthesis{\pdev{}{f}{u'} - \cdev{}{}{x}\pdev{}{f}{u''}}\delta u + \pdev{}{f}{u''}\delta u'
    \Bigg]_{x_1}^{x_2},
\end{align}
where we have distinguished between the integral and the boundary contributions. 

To ensure that $\delta \mathcal{F}=0$ to obtain a minimum, it is first required that the integral term vanishes for arbitrary variations $\delta u$, which leads to the Euler-Lagrange equation 
\begin{equation}
    \pdev{}{f}{u}-\cdev{}{}{x}\pdev{}{f}{u'}+\cdev{2}{}{x}\pdev{}{f}{u''}=0.
\end{equation}
Secondly, the boundary contributions must also vanish. This can be achieved by imposing appropriate boundary conditions on $u(x)$, $u'(x)$, or a combination of both at $x_1$ and $x_2$. The three most common boundary conditions in the theory of elasticity are:
\begin{enumerate}
    \item Clamped boundary conditions, where both $u(x)$ and $u'(x)$ are fixed at the boundaries, \eg
    \begin{equation}
        u(x_1)=u(x_2)=0, \quad u'(x_1)=u'(x_2)=0.
    \end{equation}
    \item Supported boundary conditions, where $u(x)$ is fixed at the boundaries, but $u'(x)$ is free. Thus, $\delta u'$ is arbitrary and the factor in front of it must vanish, \ie
    \begin{equation}
        u(x_1)=u(x_2)=0, \quad \pdev{}{f}{u''}\Bigg|_{x_1}= \pdev{}{f}{u''}\Bigg|_{x_2}=0.
    \end{equation} 
    \item Free boundary conditions, where both $u(x)$ and $u'(x)$ are free at the boundaries. Therefore, the two terms multiplying $\delta u$ and $\delta u'$ must vanish independently:
    \begin{equation}
        \pdev{}{f}{u''}\Bigg|_{x_1}= \pdev{}{f}{u''}\Bigg|_{x_2}=0, \quad \parenthesis{\pdev{}{f}{u'} - \cdev{}{}{x}\pdev{}{f}{u''}}\Bigg|_{x_1} = \parenthesis{\pdev{}{f}{u'} - \cdev{}{}{x}\pdev{}{f}{u''}}\Bigg|_{x_2}=0.
    \end{equation}
\end{enumerate}

The spin-elastic model studied in chapters~\ref{ch:materials_equilibrium} and~\ref{ch:materials_2d} is a particular case of the above formalism, where $f(u,u',u'';x)=f(u'';x)$, so we only have to take into account those terms that involve $f_{u''}$ (equivalent to $f_{u}=0$ and $f_{u'}=0$).

\chapter{Antiferromagnetic ground state for lattices without triangular loops}
\label{app:ground-state_triangular_loops}

For a general $d$-dimensional lattice, let the antiferromagnetic Ising Hamiltonian be
\begin{equation}
    \mathcal{H}(\bm{\sigma})=-h\sum_{i}\sigma_i + J\sum_{\langle i,j \rangle}\sigma_i \sigma_j, \quad h>0, \ J>0,
\end{equation}
where $\sigma_i=\pm 1$ stands for a spin variable. The subscript $i=1,\ldots,N$ runs over all the $N$ lattice sites, so it maps the $d$-dimensional index to a one-dimensional one, whereas the sum over $\langle i,j \rangle$ means that we sum over all nearest-neighbour pairs. The number of neighbours for each site is the coordination number $c$. The total energy can be also fully characterised through the number of spins aligned with the external field $n_h$, and the number of nearest-neighbour pairs with anti-aligned spins $n_J$. The total number of coupling pairs in the lattice is given by $c N/2$. Then, the number of anti-aligned or antiferromagnetic pairs is $n_J$, whereas $cN/2-n_J$ is the number of aligned pairs. Equivalently, we may use the fraction numbers $x_h\equiv n_h/N$, $x_h\in[0,1]$, and $x_J\equiv n_J/N$, $x_J\in[0,c/2]$, which are particularly adequate to analyse the thermodynamic limit $N\to\infty$. Note that $x_h$ is bounded both from below and from above by zero and one, respectively. Instead, $x_J$ is bounded by zero and $c/2$. Using this parametrisation, the Hamiltonian reads
\begin{align}
    \mathcal{H}(x_h,x_J)  
    = -N\brackets{ h (2x_h-1)+J\parenthesis{2x_J-\frac{c}{2}}}.
    \label{eq:Ising_model_Hamiltonian_xh_xJ}
\end{align}
In this way, a macrostate of the spin system is defined by a point $(x_h,x_J)$.

The partition function of the $d$-dimensional Ising model is
\begin{equation}
    Z=\sum_{\bm{\sigma}} e^{-\beta \mathcal{H} (\bm{\sigma})}=\sum_{\left(x_h,x_J\right)} \xi \left(x_h,x_J\right) e^{-\beta \mathcal{H} \left(x_h,x_J\right)},
\end{equation}
where $\xi(x_h,x_J)$ stands for the multiplicity of the macrostate $(x_h,x_J)$, \ie the number of microstates compatible with $(x_h,x_J)$. In the low-temperature limit $T\to0^+$ ($\beta\to+\infty$), the leading order of the partition function stems from the ground-state energy 
\begin{equation}
    -\frac{1}{N\beta}\zeta^{(d)} \sim e_{\gs}^{(d)}, \quad \beta\to+\infty,
\end{equation}
where 
\begin{align}
    \label{eq:ground-states_general_lowT}
    e_{\gs}^{(d)} & \equiv \frac{1}{N}\min_{\left( x_h, x_J \right)} \mathcal{H}\left( x_h, x_J \right) \nonumber \\
     &= \min_{\left( x_h, x_J \right)} \left\{-h\left(2x_h-1\right)-J\left(2 x_J-\dfrac{c}{2} \right)\right\}
\end{align}
is the minimum energy of the system per lattice site. Therefore, $e_{\gs}^{(d)}$ plays the role of the free energy of the system in the low-temperature limit. Since~\eqref{eq:Ising_model_Hamiltonian_xh_xJ} is a linear function of the two variables $(x_h,x_J)$, so 
\begin{equation}
    \pdev{}{\mathcal{H}}{x_h} = -2 N h \neq 0, \quad \pdev{}{\mathcal{H}}{x_J} = -2 N J \neq 0,
\end{equation}
the minimum energy must be found at some point belonging to the boundary of the physically available set $(x_h,x_J)$ for any non-zero $(h,J)$.

\begin{figure}
    \centering
    \includegraphics[width=0.6\textwidth]{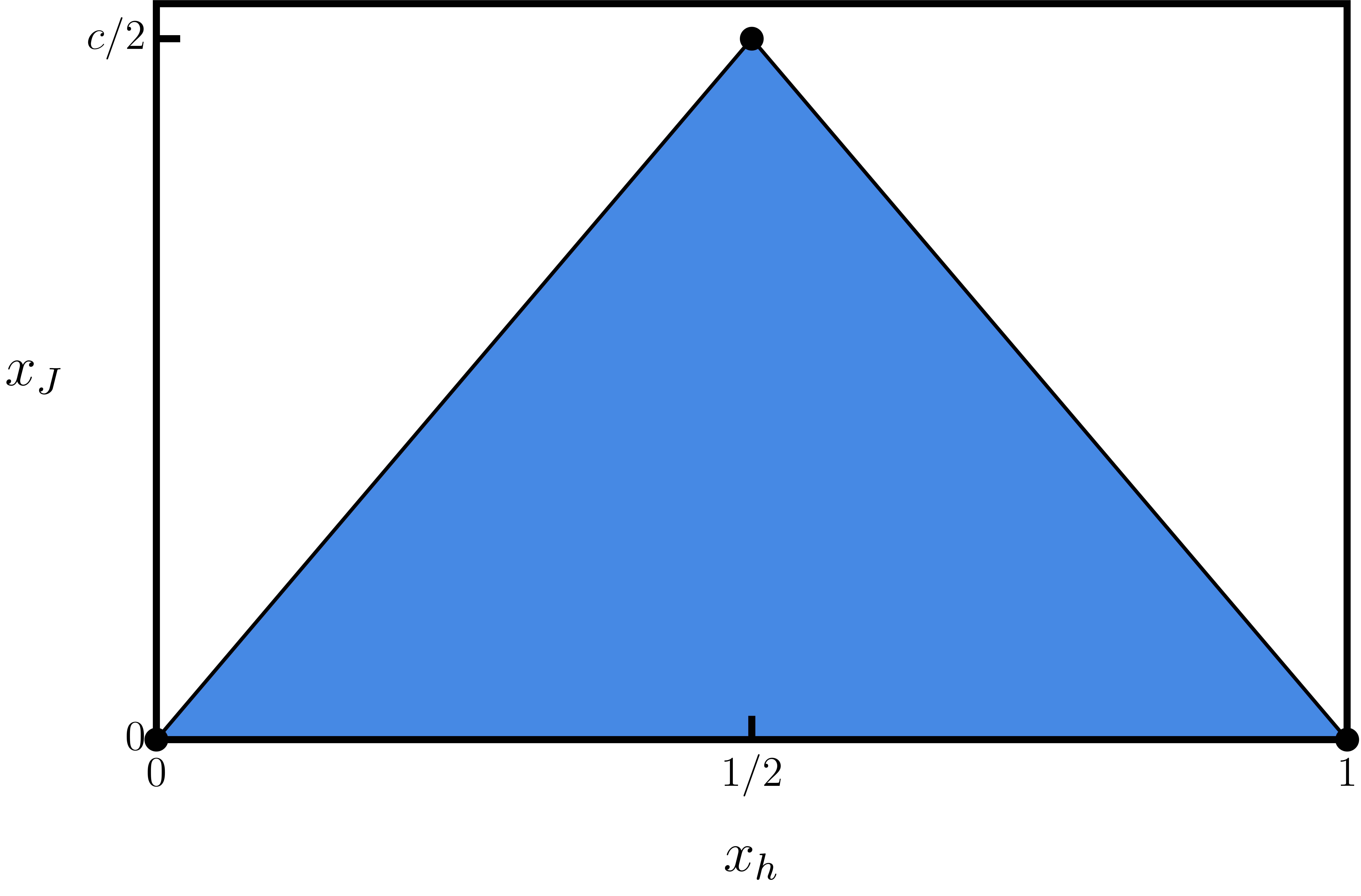}
    \caption[Antiferromagnetic Ising model parameter space $(x_h,x_J)$. The number of spins aligned with the external field is $n_h=N x_h$, whereas the number of antiferromagnetically aligned pairs is $n_J=N x_J$. The filled blue region indicates the physically available macrostates in the large system size limit, for an arbitrary $d$-dimensional lattice without triangular loops.]{Antiferromagnetic Ising model parameter space $(x_h,x_J)$. The number of spins aligned with the external field is $n_h=N x_h$, whereas the number of antiferromagnetically aligned pairs is $n_J=N x_J$. The filled blue region indicates the physically available macrostates in the large system size limit as $N\to\infty$, for an arbitrary $d$-dimensional lattice without triangular loops.}
    \label{fig:triangular_loops_solution}
\end{figure}

Not all points inside the rectangle $\left( x_h ,x_J \right) \in [0,1]\times [0,c/2]$ are physically acceptable. The specific shape of the physically available region in the $\left( x_h, x_J\right)$ space is not straightforward to derive for arbitrary lattices. Nevertheless, let us restrict ourselves to the family of lattices with no triangular loops that contains,\footnote{Loops are defined in lattice and complex systems as those pathways that start and end in the same node. The absence of triangular loops is qualitatively understood using the statement 'any neighbour of my neighbour is not my neighbour'.} for instance, all bipartite networks---such as the honeycomb lattice in figure~\ref{fig:honeycomb_lattice} or the square lattice in figure~\ref{fig:square_lattice}. The assumption of the absence of triangular loops allows us to characterise exactly the available control set in the $\left( x_h, x_J\right)$ space. If this condition holds, the number of anti-aligned couples verifies 
\begin{equation}\label{eq:AppendixB_ConditionBipartite}
    x_J\leq c\ \text{min}(x_h,1-x_h).
\end{equation}
Hence, the three boundary vertices $(0,0)$, $(1/2,c/2)$, and $(1,0)$ define a triangular region in the $\left(x_h, x_J\right)$ plane in which all physically possible macrostates are contained. Moreover, in the large system size limit as $N\to \infty$, this region is densely filled by such macrostates---both $x_h$ and $x_J$ become continuous variables in this limit. The plane $(x_h,x_J)$ is shown in figure~\ref{fig:triangular_loops_solution}, the blue region representing the physically available macrostates in the system---were we not working in the limit $N\to\infty$, only some points inside the blue region would represent acceptable macrostates.

After evaluating the energy at the boundaries, it is possible to obtain finally what the energy of the ground state is---and thus the low-temperature limit of the free energy of the system. Depending on the parameters $h$, $J$, there appear only two possibilities:
\begin{itemize}
    \item All spins are aligned with the external field for $h>cJ$. In this situation, the external field prevails over the ferromagnetic interaction. If so, $n_h=N$ and $n_J=0$, i.e. $x_h=1$ and $x_J=0$.
    \item The state is purely antiferromagnetic, \ie all spins are antiparallel to their nearest neighbours, for $cJ>h$. In this regime, it is the antiferromagnetic coupling that dominates. Such a state exists due to our assuming that triangular loops are absent---otherwise, it would not be possible to have all nearest neighbours of any site $i$ antiparallel to $\sigma_i$, some of them would be nearest neighbours among themselves.  If so, $n_h=N/2$ and $n_J=cN/2$, \ie $x_h=1/2$ and $x_J=c/2$.
\end{itemize}
Taking into account the above discussion, the energy per site in the ground state can be cast in a single equation,
\begin{equation}\label{eq:egs-expression-arbitrary-d} 
    e_{\gs} = -\left(h-c J\right) \Theta(h-cJ)-\frac{c}{2}J.
\end{equation}
This expression agrees with~\eqref{eq:ground-state_energy_1d_spin-string} and~\eqref{eq:ground-state_energy_2d_spin-membrane} for the one-dimensional and two-dimensional honeycomb lattice, where (i) the absolute value of the curvature $|\chi|$ plays the role of the external field, and (ii) the coordination number is $c=2$, for $d=1$, and $c=3$ on the honeycomb lattice for $d=2$. Still, it must be remarked that~\eqref{eq:egs-expression-arbitrary-d} holds for an arbitrary $d$-dimensional lattice---as long as it does not contain triangular loops. For instance, it is also valid for the square lattice with coordination number $c=4$.

\chapter[Asymptotic tendency to stationariness of resetting with refractory periods]{Asymptotic tendency to stationariness of resetting with refractory periods} \label{app:refractory_Laplace_method}
\pagestyle{mystyle3}
The complete derivation of the tendency to the NESS of the exploration phase of resetting with refractory periods, $\PDF^{\free}(x,t|x_0)$, requires a more careful analysis of the position of minima to properly apply Laplace's method. 
To simplify the following analysis, we introduce the dimensionless variables $t^*=r_1 t$, $x^* = |x-x_0| / \sqrt{4 D r_1^{-1}}$, then we drop the asterisks not to overload the notation.

Let us focus on the first integral of~\eqref{eq:refractory_Laplace_prop}, 
\begin{equation}
    I_1 = \int_0^1 \,\d \omega (\omega)^{-1/2} \, e^{-\omega t - x^2 / (\omega t)},
\end{equation}
The relative minimum on that interval is reached at $\widetilde{\omega}=|y|=|x|/t$, and the estimation of $I_1$ depends on whether $\widetilde{\omega}$ lies inside or outside the integration interval. Let us define $\nu = \omega/\widetilde{\omega}$, so that  
\begin{equation}
    I_1 = \widetilde{\omega}^{1/2} \int_0^{1/\widetilde{\omega}} \d{\nu} \nu^{-1/2} e^{-\widetilde{\omega}t\psi(\nu)},
\end{equation}
where $\psi(\nu) = \nu + 1/\nu$, which attains its relative minimum at $\widetilde{\nu}=1$. Now we asymptotically estimate $I_1$ for long times $ t\gg 1$, with $\widetilde{\omega}=\mathcal{O}(1)$, so $x=\mathcal{O}(t)$ (or $y=\mathcal{O}(1)$). We must discriminate between four different scenarios depending on the position of the relative minimum $\widetilde{\nu}=1$, and the boundaries of the interval $(0,1/\widetilde{\omega})$:
\begin{enumerate}[(i)]
    \item If unity lies inside $(0,1/\widetilde{\omega})$, \ie $\widetilde{\omega}<1$ or $|x|< t$, and, additionally, it is far enough the upper integration limit, in a sense that is clarified below.
    
    The Laplace's method can be applied to expand $\psi(\nu)$ around $\widetilde{\nu}=1$ up to second order, yielding
    \begin{equation}
        \widetilde{\omega}t\psi(\nu) = 2 \widetilde{\omega}t + \widetilde{\omega}t(\nu-1)^2 + \mathcal{O}((\nu-1)^3).
    \end{equation}
    Therefore, the dominant behaviour of the integral comes from a Gaussian centred at $\nu=1$ and with very small width, proportional to $(\widetilde{\omega}t)^{-1/2}\ll 1$. If the minimum is far enough from the upper limit, the integral is dominated by the Gaussian contribution from $(1-\varepsilon,1+\varepsilon)$, with $\varepsilon \ll 1$, such that
    \begin{equation}
        \delta_{\text{in}} \equiv 1/\widetilde{\omega} - 1 > \varepsilon \gg (\widetilde{\omega}t)^{-1/2}.
        \label{eq:refractory_laplace_i_condition}
    \end{equation}
    If this condition is fulfilled, the integral can be extended to $(-\infty,+\infty)$ to simplify the computation,
    \begin{equation}
        I_1\sim I_1^{(i)} =\widetilde{\omega}^{1/2} e^{-2\widetilde{\omega}t} \int_{1-\varepsilon}^{1+\varepsilon} \d{\nu} e^{-\widetilde{\omega}t(\nu-1)^2} = \frac{1}{\sqrt{t}}e^{-2\widetilde{\omega}t}\int_{-\varepsilon\sqrt{\widetilde{\omega}t}}^{+\varepsilon\sqrt{\widetilde{\omega}t}} \d{z} e^{-z^2}
        \sim \sqrt{\frac{\pi}{t}}e^{-2|x|}.
    \end{equation}

    \item If unity lies outside $(0,1/\widetilde{\omega})$, \ie $\widetilde{\omega}>1$ or $|x|> t$, and, additionally, it is far enough the upper integration limit, in a sense that is also clarified below. 
    
    In this case, the minimum of $\psi(\nu)$ within the interval is the upper limit itself, so $\widetilde{\nu} =1/\widetilde{\omega}$. Hence, the first term of the Taylor expansion does not vanish, $\psi(\nu) \simeq \widetilde{\omega} + 1/\widetilde{\omega} + (1-\widetilde{\omega}^2)(\nu-1/\widetilde{\omega})+\widetilde{\omega}^3(\nu-1/\widetilde{\omega})^2$. Applying Laplace's method, the integration interval is dominated by a small region $(1/\widetilde{\omega}-\varepsilon,1/\widetilde{\omega})$, with $\varepsilon \ll 1$. If the relative minimum outside the interval is far enough from the upper limit, in other words, 
    \begin{equation}
        \frac{\widetilde{\omega}^2-1}{\widetilde{\omega}^3} \gg \varepsilon . 
        \label{eq:refractory_laplace_ii_condition1}
    \end{equation}
    then the quadratic term can be neglected. Assuming this condition, the integral is approximated by 
    \begin{equation}
        I_1\sim I_1^{(ii)} = \widetilde{\omega} e^{-t(1+\widetilde{\omega}^2)} \int_{1/\widetilde{\omega}-\varepsilon}^{1/\widetilde{\omega}} \d{\nu} e^{\widetilde{\omega} t (\widetilde{\omega}^2-1)(\nu-1/\widetilde{\omega})}
        \sim \frac{e^{-t(1+x^2/t^2)}}{t(x^2/t^2-1)},
    \end{equation}
    provided that 
    \begin{equation}
        \widetilde{\omega}t(\widetilde{\omega}^2-1)\gg 1.
        \label{eq:refractory_laplace_ii_condition2}
    \end{equation} 
    Conditions~\eqref{eq:refractory_laplace_ii_condition1} and~\eqref{eq:refractory_laplace_ii_condition2} are satisfied when $\widetilde{\omega}-1 = \mathcal{O}(1)$, so we can choose $\varepsilon$ small but much larger than $(\widetilde{\omega}t)^{-1}\ll 1$. As $\widetilde{\omega}$ approaches unity, $1/\widetilde{\omega} = 1-\delta_{\text{out}}$, with $\delta_{\text{out}}\ll 1$, so~\eqref{eq:refractory_laplace_ii_condition1} and~\eqref{eq:refractory_laplace_ii_condition2} entail that 
    \begin{equation}
        \delta_{\text{out}} \gg \varepsilon, \quad \widetilde{\omega}t\delta_{\text{out}} \varepsilon \gg 1 \to \delta_{\text{out}} \gg (\widetilde{\omega}t)^{-1/2}.
        \label{eq:refractory_laplace_ii_condition3}
    \end{equation}
    This condition tells us the separation of the upper limit from the unity, $\widetilde{\nu} = 1$, must be much larger than the width of the Gaussian, analogously to the previous case~\eqref{eq:refractory_laplace_i_condition}.

    In original variables, this contribution has the same exponential form of the first term of~\eqref{eq:refractory_Laplace_prop}, but with a different subdominant prefactor $(x^2/t)^{-1}$, instead of $t^{-1/2}$. Since $\mathcal{O}(x/t)=1$, due to~\eqref{eq:refractory_laplace_ii_condition3}, we conclude that this contribution is subdominant and thus negligible compared to the non-resetting term.  

    \item If unity lies inside the interval $(0,1/\omega_0)$, \ie $\omega_0 < 1$ or $|x| < t$, but it is close to the upper limit,  $1/\omega_0 = 1+\delta_{\text{in}}$ with $\delta_{\text{in}} $ not fulfilling condition~\eqref{eq:refractory_laplace_ii_condition3}, \ie $\delta_{\text{in}} = \mathcal{O}((\omega_0 t)^{-1/2})$ or greater. In this case, the Gaussian integral becomes 
    \begin{equation}
        I_1\sim I_1^{(iii)} =\widetilde{\omega}^{1/2} e^{-2\widetilde{\omega}t} \int_{1-\varepsilon}^{1+\delta_{\text{in}}} \d{\nu} e^{-\widetilde{\omega}t(\nu-1)^2} \sim \frac{1}{2}\sqrt{\frac{\pi}{t}}\erfc\parenthesis{\frac{|x|-t}{\sqrt{|x|}}}e^{-2|x|},
    \end{equation}
    where, just like in case (i), we can choose $\varepsilon$ such that $\varepsilon \sqrt{\widetilde{\omega}t}\gg 1$. This expression converges to $I_1^{(i)}$ when $\delta_{\text{in}}\sqrt{\widetilde{\omega}t}\gg 1$, so we can use it to approximate $I_1$ for $|x|<t$, regardless of the value of $\delta_{\text{in}}$.

    \item If unity lies outside the interval $(0,1/\omega_0)$, \ie $\omega_0 > 1$ or $|x| > t$, but it is close to the upper limit,  $1/\omega_0 = 1-\delta_{\text{out}}$ with $\delta_{\text{out}} $ not fulfilling condition~\eqref{eq:refractory_laplace_ii_condition3}, \ie $\delta_{\text{out}} = \mathcal{O}((\omega_0 t)^{-1/2})$ or greater. As we did in (ii), we expand $\psi(\nu)$ around $1/\widetilde{\omega}=1-\delta_{\text{out}}$ in a small region $(1-\delta_{\text{out}}-\varepsilon,1-\delta_{\text{out}})$, but we have to take into account that $\varepsilon$ is at least $\mathcal{O}(\delta_{\text{out}})$, 
    \begin{equation}
        \widetilde{\omega}t\psi(\nu)\simeq \widetilde{\omega}t \left[2 +\delta_{\text{out}}^{2}+ 2\delta_{\text{out}}(\nu - 1 +\delta_{\text{out}}) + (\nu - 1 + \delta_{\text{out}})^2\right],
    \end{equation}
    where we have neglected terms of order $\mathcal{O}(\omega_{0}t\delta_{\text{out}}^{3})$, $\mathcal{O}(\omega_{0}t\delta_{\text{out}}^{2}\varepsilon)$, $\mathcal{O}(\omega_{0}t\delta_{\text{out}}\varepsilon^{2})$, and $\mathcal{O}(\omega_{0}t\varepsilon^{3})$.\footnote{Recalling that $\delta_{\text{out}}=\mathcal{O}(\omega_{0} t)^{-1/2}$, we have, on the one hand, $\mathcal{O}(\omega_{0}t\delta_{\text{out}}^{3})=\mathcal{O}(\omega_{0}t)^{-1/2}\ll 1$, $\mathcal{O}(\omega_{0}t\delta_{\text{out}}^{2}\varepsilon)=\mathcal{O}(\varepsilon)\ll 1$. On the other hand, both $\mathcal{O}(\omega_{0}t\delta_{\text{out}}\varepsilon^{2})=\mathcal{O}((\omega_{0}t)^{1/2}\varepsilon^{2})$ and $\mathcal{O}(\omega_{0}t\varepsilon^{3})$ must be much smaller than unity, so $\varepsilon\ll(\omega_{0}t)^{-1/3}$.} Substituting this expression into the integral and following the same arguments as before by choosing $\delta_{\text{out}}\widetilde{\omega}t\varepsilon\gg 1$, we finally obtain
    \begin{equation}
        I_1 \sim I_1^{(iv)} = \frac{\pi |x|}{2 t}e^{-2|x|}\erfc\parenthesis{\frac{|x|-t}{\sqrt{|x|}}}.
    \end{equation}
\end{enumerate}

The other integral that involves $\Phi_2$ in~\eqref{eq:refractory_Laplace_prop} can be also estimated using Laplace's method. However, the analysis is simpler, since $\Phi_2(\omega;y)$ is a monotonically increasing function of $\omega$. Thus, the local maximum of the exponent is reached always in the upper limit of the integral, $\widetilde{\omega}=1$. Expanding $\Phi_2(\omega;y)$ around that point, similarly to case (ii), we find that 
\begin{equation}
    I_2 = \int_0^1 \d{\omega} \omega^{-1/2} e^{\omega t- x^2/(\omega t)} \sim (x^2/t+t)^{-1} e^{t(1-x^2/t^2)},
\end{equation}
which is also negligible compared with the non-resetting contribution of~\eqref{eq:refractory_Laplace_prop}.

Summing up, if we want to refine~\eqref{eq:Approximation_NESS}, we can use the exhaustive approach for the integral terms, obtaining the most accurate approximation
\begin{eqnarray}
    \PDF^{\free} (x,t|x_0) - F(t)\PDF_0(x,t|x_0) \sim 
    \left\{
    \begin{array}{ll}       
        I_1^{(iii)}-I_2, \quad |x| < t,\\
        I_1^{(iv)}-I_2, \quad t < |x-x_0| < x_{\text{cross}}, \\
        I_1^{(ii)}-I_2, \quad x_{\text{cross}} < |x|,
    \end{array} \right.
    \label{eq:refractory_approximation_tendencyNESS}
\end{eqnarray}
which are represented in the right panel of figure~\ref{fig:refractory_RelaxationNESS}.\footnote{For the sake of compactness, we have not written the explicit expression for dimensional variables. Remember that $\PDF_0$ is the Brownian propagator~\eqref{eq:Green_function_BrownianMotion} and $F(t) = e^{-r_1 t}$.} The crossover position $x_{\text{cross}}$ indicates the change of stability where $I_1^{(iv)}=I_1^{(ii)}$, which can be obtained numerically.


\chapter{Langevin simulations} \label{app:langevin_simulations}

\pagestyle{mystyle3}
Stochastic simulations that involve Langevin equations have been produced by integrating a large number $N$ of trajectories to obtain the statistical properties at the ensemble level of description. 

The specific methods used for Langevin simulations may slightly vary from chapter to chapter, but the general procedure has the same basic ingredients. Let the overdamped Langevin equation of a one-dimensional Brownian particle be given by
\begin{equation}
    \pdev{}{x}{t} = -\frac{D}{k_B T}\frac{\partial V(x,t)}{\partial x} + \sqrt{2D}\xi(t),
\end{equation}
where $D$ is the diffusion coefficient, $k_B$ is the Boltzmann constant, $T$ is the temperature, $V(x,t)$ is the potential in which the particle is moving, and $\xi(t)$ is the unit Gaussian white noise. The former stochastic differential equation is integrated using a simple forward Euler method, which leads to 
\begin{equation}
        x(t+\Delta t) = x(t) - \frac{D}{k_B T}\frac{\partial V(x,t)}{\partial x}\Delta t + \sqrt{2D\Delta t}\mathcal{N}(0,1),
        \label{eq:langevin_integration}
\end{equation}
where $\Delta t \ll 1$ is a time step and $\mathcal{N}(0,1)$ represents a random number drawn from a Gaussian distribution with zero mean and unit variance. 

The general numerical scheme is the following:
\begin{enumerate}
    \item Set the parameters of the system: $D$, $k_B$, $T$, $\Delta t$, etc.
    \item Set the condition to end the numerical integration. It depends on the problem we are working on: it may be a maximum time $t_{\max}$, a target position $\targetx$ to be reached, etc. Set the initial trajectory $n=1$.
    \item Set the initial condition $x(0)=x_0$ at $t=0$ for the $n$-th trajectory.
    \item Let the position evolve following~\eqref{eq:langevin_integration},
    \begin{equation}
        x_{\text{new}} = x_{\text{old}} - \frac{D}{k_B T}\left.\frac{\partial V(x,t)}{\partial x}\right|_{x_{\text{old}}}\Delta t + \sqrt{2D\Delta t}\mathcal{N}(0,1).
    \end{equation}
    \item Update the dynamical variables: the time $t = t + \Delta t$ and the position $x_{\text{old}} = x_{\text{new}}$.
    \item Store the values of interest if the time coincides with the saving time step $t_{\text{save}}$, \ie if $t = j t_{\text{save}}$, $j\in \mathbb{N}$.
    \item Check the stopping condition. If it is not fulfilled, go back to step 4. If it is fulfilled and $n < N$, $n=n+1$ and go back to step 3. End the simulation when $n=N$.
\end{enumerate}

\section{Stochastic processes with resetting}
Stochastic resetting changes the dynamics of the system, making the particle return to the initial position $x_0$ at random times. Resetting events are assumed to be Poissonian, \ie the time intervals between two consecutive events are exponentially distributed with rate $r_1$, $f(t) = r_1 e^{-r_1t}$. 

In the presence of resetting, Langevin equation~\eqref{eq:langevin_integration} is integrated as
\begin{equation}
        x(t+\Delta t) = 
        \left\{
        \begin{array}{ll}
            x(t) - \dfrac{D}{k_B T}\dfrac{\partial V(x,t)}{\partial x}\Delta t + \sqrt{2D\Delta t}\mathcal{N}(0,1), & \text{with probability } 1 - r_1\Delta t, \\
            x_0, & \text{with probability } r_1\Delta t.
        \end{array}
        \right.
        \label{eq:langevin_integration_resetting}
\end{equation}
However, to simplify the numerical implementation, we are previously computing when resetting events occur. Random exponential times are generated as $-\frac{1}{r_1}\ln(1-\mathcal{U}(0,1))$, where $\mathcal{U}(0,1)$ are uniform random numbers in the interval $[0,1)$. Defining resetting times as $t_i$, $i=1,2,\ldots$, we can generate them iteratively before starting the simulation as
\begin{equation}
    t_{i+1} = t_i - \frac{1}{r_1}\ln(1-\mathcal{U}(0,1)),\quad  t_0=0. 
\end{equation}
Of course, this procedure must be done for each trajectory $n$.

In addition to the generation of resetting times a starting a counter of resetting events $i=0$ at the beginning of each trajectory, the numerical scheme is only modified in step 4, which is now 
\begin{equation}
        x_{\text{new}} = 
        \left\{
        \begin{array}{ll}
            x_{\text{old}} - \dfrac{D}{k_B T}\left.\dfrac{\partial V(x,t)}{\partial x}\right|_{x_{\text{old}}}\Delta t + \sqrt{2D\Delta t}\mathcal{N}(0,1).& t < t_{i+1}, \\
            x_0, & t \geq t_{i+1}.
        \end{array}
        \right.
\end{equation}
After this, if a reset event has occurred, the index $i$ is updated, $i=i+1$. 

\section{Additional refractory periods}
Refractory periods after resetting events are implemented in an analogous way. Assuming they also follow an exponential distribution with rate $r_2$, the whole set of time events is
\begin{subequations}
    \begin{align}
        \tau_0 = &\, 0, \\
        t_i = &\, \tau_{i-1} - \dfrac{1}{r_1}\ln(1-\mathcal{U}(0,1)), \\ 
        \tau_{i} = &\, t_i - \frac{1}{r_2}\ln(1-\mathcal{U}(0,1)).
    \end{align}
\end{subequations}

After generating the sets of times $\{t_i\}$ and $\{\tau_i\}$ for each trajectory $n$, step 4 of the numerical scheme is modified as 
\begin{equation}
        x_{\text{new}} = 
        \left\{
        \begin{array}{ll}
            x_{\text{old}} - \dfrac{D}{k_B T}\left.\dfrac{\partial V(x,t)}{\partial x}\right|_{x_{\text{old}}}\Delta t + \sqrt{2D\Delta t}\mathcal{N}(0,1).& \tau_{i} \leq t < t_{i+1}, \\
            x_0, & t_{i+1} \leq t < \tau_{i+1}.
        \end{array}
        \right.
\end{equation}
In this case, the index $i$ is updated, $i=i+1$, when $t \geq \tau_i$.


\chapter{Inverting the FPT distribution for dichotomous disordered resetting}
\label{app:dichotomous_Asymptotic_FPT}

The long-time behaviour of any function $m(t)$ can be derived from its Laplace transform $\mu(s)$ \cite{book:Schiff_LaplaceTransformTheory_99}. Assuming that the singularity of $\mu(s)$ with the highest real part $s^*$ is a simple pole, the asymptotic behaviour of $m(t)$ is 
\begin{equation}
    m(t) = \frac{1}{2\pi i}\int_{\sigma-i\infty}^{\sigma+i\infty} \d{s} \mu(s)e^{st} \sim \text{Res}\Bigg[\mu(s)e^{st};s=s^*\Bigg], \quad t\to \infty,
    \label{eq:residue_theorem_asymptotic}
\end{equation}
as a consequence of Cauchy's residue theorem. 

\begin{figure}
    \centering
    \includegraphics[width=0.7\textwidth]{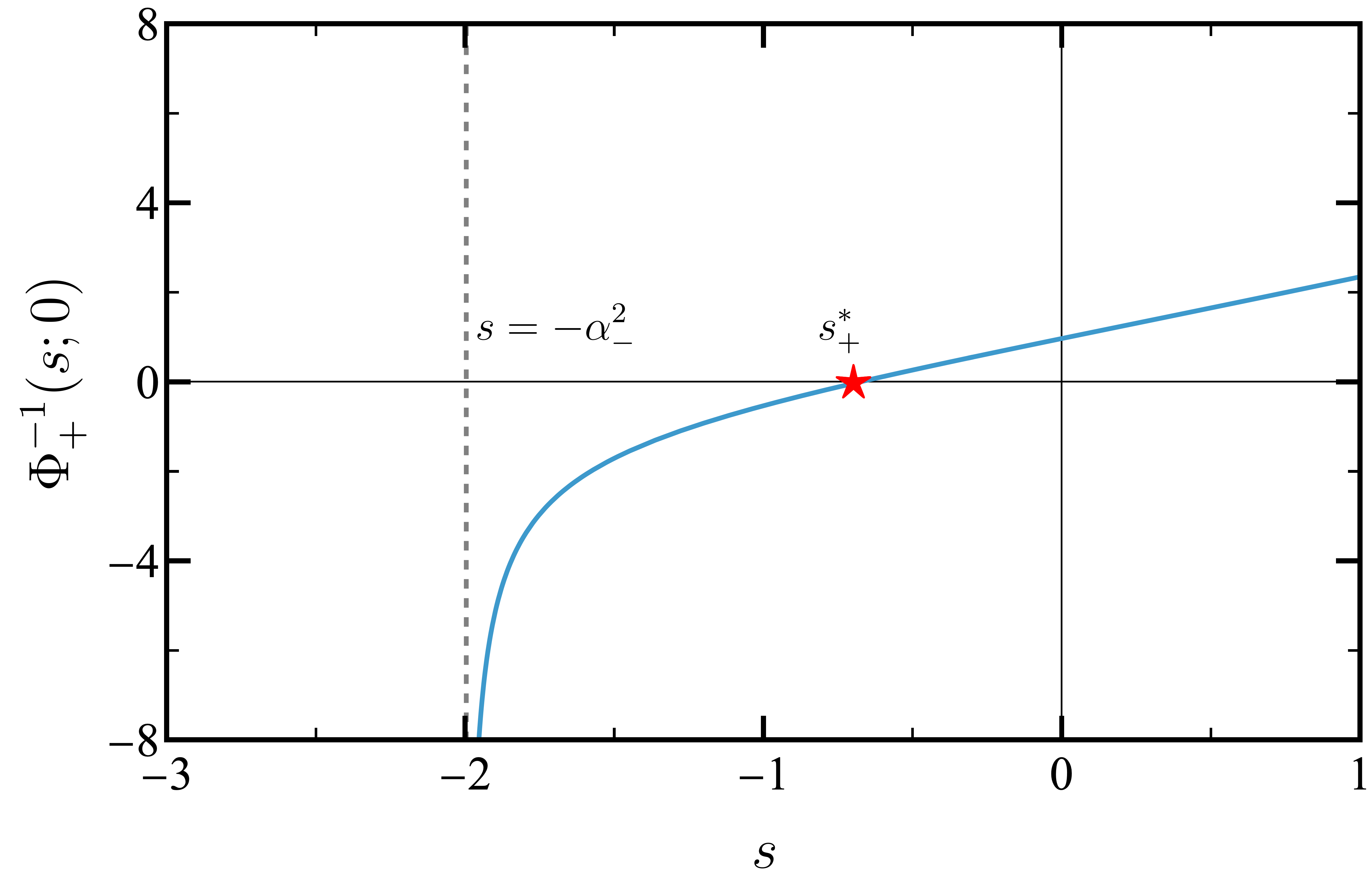}
    \caption{Localisation of the largest pole of the Laplace transform $\positiveSol{\FPTpdfLaplace}(s;0)$ of the FPT distribution. Specifically, we plot $1/\positiveSol{\FPTpdfLaplace}(s;0)$  as a function of $s$ over the real axis. The dashed line corresponds to the vertical asymptote at $s=-\alpha_-^2$, whereas the red star marks the position of the simple pole $s_+^*$. Here, we have chosen $\alpha_+=1$ and $\alpha_-=\sqrt{2}$.}
    \label{fig:dichotomous_FPT_poles}
\end{figure}

The poles of~\eqref{eq:disordered_FPT_distribution_Laplace_solution} are given by the zeros of its denominator, \ie when $\positiveSol{\FPTpdfLaplace}(s;0)^{-1}$ vanishes. First, we focus on the real axis, where we can prove that there is always a unique negative real pole. For any $(\alpha_+,\alpha_-)$,
\begin{equation}
    \lim_{s\to\parenthesis{-\alpha_-^2}^+} \positiveSol{\FPTpdfLaplace}(s;0)^{-1} = -\infty, \quad \positiveSol{\FPTpdfLaplace}^{-1}(0;0) = 1, \quad \lim_{s\to\infty} \positiveSol{\FPTpdfLaplace}(s;0)^{-1} = \infty,
\end{equation}
so it seems that $\positiveSol{\FPTpdfLaplace}(s;0)^{-1}$ is a monotonically increasing function of $s$ in the interval $(-\alpha_-^2,\infty)$, as qualitatively shown in figure~\ref{fig:dichotomous_FPT_poles}. Furthermore, the first derivative is always positive, $\partial_s\positiveSol{\FPTpdfLaplace}(s;0)^{-1}>0$, and just one simple pole $s_+^*$ exists in the interval $(-\alpha_-^2,0)$. Note that, despite $s=-\alpha_-^2$ is a branching point, the real pole is always larger than it. 

Although the real pole is always to the right of the branching point, we have to check that it is the pole with the largest real part in the whole complex plane. We do not have an analytical proof for this, but the numerical evaluation of the function gives no other pole with larger real part. Once we have shown that $s_+^*$ corresponds to a simple pole, it is fully justified to apply~\eqref{eq:residue_theorem_asymptotic}, which leads to equations~\eqref{eq:asymp+1}. Due to symmetry, an analogous result applies to $\negativeSol{\FPTpdfLaplace}(s;0)$, which has a unique real simple pole $s_-^*$.

\chapter{Numerical optimisation of the MFPT functional} \label{app:gradient-descent}

The numerical computation of the optimal resetting profile $\optMFPT{r}(x)$ involves an iterative method to decrease the average MFPT~\eqref{eq:functional_average_MFPT}. This is the so-called gradient descent method, which is based on the computation of the functional derivative $\mu(x)$, defined in~\eqref{eq:mu(x)_heterogeneous_resetting}, but with an arbitrary resetting profile $r(x)$. At fixed target position $\targetx$, the functional derivative is defined as
\begin{align}
    M(\targetx,x) \equiv  \frac{\delta \FPT{(1)}(\targetx)}{\delta r(x)}  
    = \lim_{\varepsilon \to 0 } \frac{ \FPT{(1)}(\targetx)[r(x_0)+\varepsilon \delta(x_0-x)] -\FPT{(1)}(\targetx)[r(x_0)]}{\varepsilon},
\end{align}
where our notation stresses that $\FPT{(1)}(\targetx)[r(x_0)]$ is the MFPT to the target located at $\targetx$ of our searcher with a resetting strategy given by the resetting profile $r(x_0)$. This is a generalization of~\eqref{eq:M(xt,x)_heterogeneous_resetting}, where the perturbation was made around $r(x_0)=0$, instead of around an arbitrary $r(x_0)$. 

The MFPT is obtained as $\FPT{(1)}(\targetx)=-\chi(\targetx)$, where $\chi$ satisfies~\eqref{eq:resetting_boundaries_FPT_ODE}. 
For building the solution at $x_0=\targetx$, it suffices to solve the equation in the interval with boundaries $\targetx$ and $-\sgn(\targetx)$, \ie 
\begin{align}
    \chi(-\sgn(\targetx))=0, \quad \chi(0)=0.
\end{align}
If we consider the perturbed profile, \ie $r(x_0) \to r(x_0)+ \delta(x_0-x)$, we must include the extra matching conditions
\begin{align}
    \chi(x^+)-\chi(x^-)&=0,\\
    \chi'(x^+)-\chi'(x^-)&=\varepsilon \chi(x).
\end{align}
We will denote the final perturbed solution by $\chi_{p}(\targetx;x)$, evaluated at $x_0=\targetx$ when considering the perturbation at $x_0=x$.  Note that another way to compute the functional derivative is 
\begin{equation}
    M(\targetx,x)=- \left. \pdev{}{\chi_{p}(\targetx;x)}{\varepsilon}\right|_{\varepsilon=0}.
\end{equation}
The difference $\Delta \chi(\targetx)=\chi_{p}(\targetx;x)-\chi(\targetx)$ between the perturbed and the unperturbed solutions fulfils
\begin{subequations}
    \label{eq:Delta_chi_ODE}
    \begin{align}
        \Delta \chi''(x_0)&= r(x_0)\Delta\chi(x_0), \\
        \Delta \chi(-\sgn(\targetx))&=0,\\
        \Delta \chi(0) &=0,\\
        \Delta \chi(x^+)- \Delta \chi(x^-) &=0,\\
        \Delta \chi'(x^+)- \Delta \chi'(x^-) &=\varepsilon\chi_{p}(\targetx;x). 
    \end{align}
\end{subequations}
By definition, it also fulfils
\begin{equation}
    M(\targetx,x)= - \left. \pdev{}{\Delta \chi(\targetx;x)}{\varepsilon}\right|_{\varepsilon=0}.
\end{equation}

The solution of the ODE~\eqref{eq:Delta_chi_ODE} is obtained by using an extension of the usual linear shooting method. It consists of converting a boundary-value problem---\eqref{eq:Delta_chi_ODE} with boundary conditions in $-\sgn(\targetx)$ and zero, into two initial-value problems.  Let us introduce the fundamental solutions $\chi_{1}$, $\chi_{2}$, $\chi_{3}$, and $\chi_{4}$, defined to be the solutions of the following ODEs:
\begin{subequations}
    \begin{align}
        \chi''_{1}(x_0) &= r(x_0) \chi_{1}(x_0) - 1, \quad &\chi_{1}(-\sgn(\targetx)) = 0, \quad &\chi'_{1}(-\sgn(\targetx)) = 0,
        \\
        \chi''_{2}(x_0) &= r(x_0) \chi_{2}(x_0), \quad &\chi_{2}(-\sgn(\targetx)) = 0, \quad  &\chi'_{2}(-\sgn(\targetx)) = 1,
        \\
        \chi''_{3}(x_0) &= r(x_0) \chi_{3}(x_0) - 1, \quad  &\chi_{3}(0) = 0, \quad  &\chi'_{3}(0) = 0,
        \\
        \chi''_{4}(x_0) &= r(x_0) \chi_{4}(x_0), \quad  &\chi_{4}(0) = 0, \quad  &\chi'_{4}(0) = 1.
    \end{align}
\end{subequations}

Any of these second-order differential equations can be conveniently rewritten as a two-order system of first-order differential equations. By defining $\kappa=\chi'$, our homogeneous equations can be written as
\begin{equation}
    \left(\begin{array}{c}
        \chi'(x_{0})\\
        \kappa'(x_{0})
        \end{array}
    \right)=\underbrace{\left(
        \begin{array}{cc}
            0 & 1\\
            r(x_{0}) & 0
        \end{array}\right)}_{A(x_{0})} 
        \underbrace{\left(
        \begin{array}{c}
            \chi(x_0)\\
            \kappa(x_0)
        \end{array}\right)}_{\vec{v}(x_0)} , 
\end{equation}
or, in matrix form, as
\begin{equation}
    \frac{\text{d}}{\text{d}x_0} \vec{v}(x_0) = A(x_0) \vec{v}(x_0).
\end{equation}
If one has initial conditions for $x_0=x_i$, \ie $\vec{v}(x_i)=\vec{v}_i$, the solution can be  built with the propagator $U(x_0,x_i)$, 
\begin{equation}
    \vec{v}(x_0) = U(x_0,x_i) \vec{v_i},
\end{equation}
where the propagator is an operator fulfilling  
\begin{equation}
    \frac{\partial}{\partial x_0} U(x_0,x_i) = A(x_0) U(x_0,x_i) 
\end{equation}
with the initial condition $U(x_i,x_i)=I$, being $I$ the identity operator. 

In the following, we study in detail the case $\targetx>0$. This choice makes $\sgn(\targetx)=+1$. For negative $\targetx$, the expressions are still valid under the transformation $x\to-x$.  We split the study of $M(\targetx,x)$ into different cases, depending on the relative positions between $x$ and $\targetx$.

\textbf{Case I:  $x>0$, $x>\targetx$} (general case: $x\, \targetx>0$, $|x|>|\targetx|$)

This is the easiest situation, there is no difference between $\chi_{p}(\targetx;x)$ and $\chi(\targetx)$, 
since $x>\targetx$. Consequently, $M(\targetx,x)=0$.

\textbf{Case II:  $x>0$, $x<\targetx$} (general case: $x\, \targetx>0$, $|x|<|\targetx|$)

In this case, one needs to solve the equation in the interval $x_0 \in (-1,x)$, which can be done using the conditions at $x_0=-1$ and $x_0=0$. At $x_0=x$, $\chi_{p}$ is continuous but $\chi'_{p}=\kappa$ experiments a kick stemming from the Dirac-delta perturbation, and the solution can be propagated from $x^+$ up to $\targetx$. The solution is then written in terms of $\chi_{p}(\targetx;x)$,
\begin{equation}
\left(\begin{array}{c}
    \Delta \chi(\targetx)\\
    \Delta \chi'(\targetx)
    \end{array}\right) = U(\targetx,x) \left(\begin{array}{c}
    0\\
    \varepsilon \chi_{p}(\targetx;x)
    \end{array}\right).
\end{equation}
Hence,
\begin{equation}
    M(\targetx,x)=- \left. \frac{\text{d} \Delta \chi(\targetx;x)}{\text{d}\varepsilon} \right|_{\varepsilon=0} =U_{12}(\targetx,x) \chi_{p}(\targetx;x).
\end{equation}
We have computed the functional derivative for the case $0<x<\targetx$ as a function of $\chi_{p}(\targetx;x)$, which can be conveniently written as a linear combination of the solutions $\chi_{1}$ and $\chi_2$. In the interval $x_0 \in (-1,x)$, we write the solution as
\begin{equation}
    \chi_{p}(x_0;x)= \chi_{1}(x_0)+ \alpha \chi_2(x_0).
\end{equation} 
Enforcing $\chi_{p}(0;x)=0$, we get
\begin{equation}
    \alpha=-\frac{\chi_{1}(0)}{\chi_2(0)}.
\end{equation}
So, we have
\begin{equation}
    \chi_{p}(x_0;x)= \chi_{1}(x_0) -\frac{\chi_{1}(0)}{\chi_2(0)} \chi_2(x_0),
\end{equation}
which substituted into the previous equation leads us to the final result
\begin{equation}
    M(\targetx,x)=U_{12}(\targetx,x) \left[\chi_{1}(x) -\frac{\chi_{1}(0)}{\chi_2(0)} \chi_2(x) \right].
\end{equation}

\textbf{Case III:  $x<0$} (general case: $x\, \targetx<0$)

Here, we need to solve $\chi_{p}$ for both $x_0 \in (-1,x)$ and $x_0 \in (x,\targetx)$. The solution can be written as
\begin{align}
    \chi_{p}(x_0;x)=&\left[\chi_{1}(x_0) + \beta \chi_2(x_0) \right] \Theta(x-x_0) \nonumber \\
    &+  \left[\chi_3(x_0) + \gamma \chi_4(x_0) \right] \Theta(x_0-x).
\end{align} 
Enforcing the matching conditions $\chi_{p}(x^+;x)-\chi_{p}(x^-;x)=0$ and $\chi'_{p}(x^+;x)-\chi'_{p}(x^-;x)=\varepsilon \chi_{p}(\targetx;x)$, one obtains
\begin{align}
    \beta&=\frac{[\chi_{3}-\chi_{1}]\kappa_{4}+\chi_{4}[\kappa_{1}-\kappa_{3}+\varepsilon \chi_{1}]}{\chi_{2}\kappa_{4}-\chi_{4}[\kappa_{2}+\varepsilon \chi_{2}]}, \\
    \gamma&=\frac{[\chi_{3}-\chi_{1}]\kappa_{2}+\chi_{2}[\kappa_{1}-\kappa_{3}+\varepsilon \chi_{3}]}{\chi_{2}\kappa_{4}-\chi_{4}[\kappa_{2}+\varepsilon \chi_{2}]},
\end{align}
where we will omit from now on the argument in $\chi_{n}(x)=\chi_n$ and $\kappa_{n}(x)=\kappa_n$. Now, we can evaluate $\chi_{p}(\targetx)$,
\begin{align}
    \chi_{p}(\targetx)=&\chi_{3}(\targetx) +\frac{[\chi_{3}-\chi_{1}]\kappa_{2}+\chi_{2}[\kappa_{1}-\kappa_{3}+\varepsilon \chi_{3}]}{\chi_{2}\kappa_{4}-\chi_{4}[\kappa_{2}+\varepsilon \chi_{2}]} \chi_{4}(\targetx).
\end{align}
Carrying out the derivative with respect to $\varepsilon$ and evaluating at $\varepsilon=0$, we get
\begin{align}
    M(\targetx,x)=-\frac{\chi_{2}\chi_{4}[\kappa_{1}-\kappa_{3}]+\chi_{2}\chi_{3}\kappa_{4}-\chi_{1}\chi_{4}\kappa_{2} }{\left[\chi_{4}\kappa_{2}-\chi_{2}\kappa_{4}\right]^2} \chi_{2} \,\chi_{4}(\targetx).
\end{align}

Comments on the need of the building objects for our numerical computation:
\begin{itemize}
    \item For Case II, we need $U_{12}(\targetx,x)$ for the region $0<\targetx<x<1$, which actually implies the need of $U_{22}(\targetx,x)$ too. These components are decoupled from $U_{11}$ and $U_{21}$, which are not needed. Additionally, $\chi_1(x_0)$ and $\chi_2(x_0)$ for $0<x<1$ are required, which implies the solution for $\chi_{1}$ and $\chi_{2}$ for the whole interval $x_0 \in (-1,1)$.
    
    \item For Case III, we need $\chi_{1}$, $\chi_{2}$, $\chi_{3}$, $\chi_{4}$, $\kappa_{1}$, $\kappa_{2}$, $\kappa_{3}$, and $\kappa_{4}$, for $x_0 \in (-1,0)$. Furthermore, we need $\chi_{4}$ for $x_0 \in (0,1)$.
\end{itemize}

\clearpage

\pagestyle{mystyle2}
\printbibliography[heading=bibintoc,title={Bibliography}]

@book{book:Bell_SearchingBehaviourBehavioural_12,
  title = {Searching Behaviour: The Behavioural Ecology of Finding Resources},
  author = {Bell, William J.},
  year = 2012,
  publisher = {Springer Science \& Business Media},
  address = {London},
  urldate = {2025-08-12}
}

@book{book:Bender.Orszag_AdvancedMathematicalMethods_99,
  title = {Advanced Mathematical Methods for Scientists and Engineers {{I}}: {{Asymptotic}} Methods and Perturbation Theory},
  author = {Bender, Carl M. and Orszag, Steven A.},
  year = 1999,
  publisher = {Springer},
  address = {New York},
  urldate = {2013-11-22}
}

@book{book:Feller_IntroductionProbabilityTheory_71,
  title = {An {{Introduction}} to {{Probability Theory}} and {{Its Applications}}, {{Volume}} 1},
  author = {Feller, William},
  year = 1971,
  publisher = {John Wiley \& Sons},
  address = {New York}
}

@book{book:Feller_IntroductionProbabilityTheory_91,
  title = {An {{Introduction}} to {{Probability Theory}} and {{Its Applications}}, {{Volume}} 2},
  author = {Feller, William},
  year = 1991,
  publisher = {John Wiley \& Sons},
  address = {New York}
}

@book{book:Fermi_NuclearPhysicsCourse_50,
  title = {Nuclear {{Physics}}: {{A Course Given}} by {{Enrico Fermi}} at the {{University}} of {{Chicago}}},
  author = {Fermi, Enrico},
  year = 1950,
  publisher = {University of Chicago Press},
  address = {Chicago},
  urldate = {2025-10-22}
}

@book{book:Feynman_StatisticalMechanicsSet_96,
  title = {Statistical Mechanics: A Set of Lectures},
  author = {Feynman, Richard P.},
  year = 1996,
  series = {Frontiers in Physics},
  edition = {19. printing},
  publisher = {Addison-Wesley},
  address = {Reading, Mass.},
  isbn = {978-0-8053-2509-6 978-0-8053-2508-9}
}

@book{book:Gardiner_HandbookStochasticMethods_83,
  title = {Handbook of {{Stochastic Methods}}},
  author = {Gardiner, Crispin W.},
  editor = {Haken, Hermann},
  year = 1983,
  series = {Springer {{Series}} in {{Synergetics}}},
  volume = {13},
  publisher = {Springer},
  address = {Berlin, Heidelberg},
  doi = {10.1007/978-3-662-02377-8},
  urldate = {2025-08-16},
  isbn = {978-3-662-02379-2 978-3-662-02377-8}
}

@book{book:Gelfand.Fomin_CalculusVariations_00,
  title = {Calculus of {{Variations}}},
  author = {Gelfand, I. M. and Fomin, S. V.},
  year = 2000,
  publisher = {Dover Publications},
  address = {New York}
}

@book{book:Gumbel_StatisticsExtremes_19,
  title = {Statistics of {{Extremes}}},
  author = {Gumbel, E. J.},
  year = 2019,
  publisher = {Columbia University Press},
  address = {New York},
  doi = {10.7312/gumb92958},
  urldate = {2025-08-24},
  isbn = {978-0-231-89131-8}
}

@book{book:Lanczos_VariationalPrinciplesMechanics_70,
  title = {The Variational Principles of Mechanics},
  author = {Lanczos, Cornelius},
  year = 1970,
  publisher = {Dover Publications},
  address = {New York},
  urldate = {2013-11-22}
}

@book{book:Landau.etal_TheoryElasticityVolume_86,
  title = {Theory of {{Elasticity}}: {{Volume}} 7},
  author = {Landau, L.D. and Lifshitz, E.M. and Kosevich, A.M. and Sykes, J.B. and Pitaevskii, L.P. and Reid, W.H.},
  year = 1986,
  series = {Course of Theoretical Physics},
  publisher = {Pergamon Press},
  address = {Oxford},
  isbn = {978-0-7506-2633-0},
  lccn = {86002450}
}

@book{book:Landau.Lifshitz_StatisticalPhysicsVolume_13,
  title = {Statistical {{Physics}}: {{Volume}} 5},
  author = {Landau, L.D. and Lifshitz, E.M.},
  year = 2013,
  series = {Course of Theoretical Physics},
  publisher = {Elsevier Science},
  address = {Oxford},
  isbn = {978-0-08-057046-4}
}

@book{book:Newman_NetworksIntroduction_10,
  title = {Networks: {{An Introduction}}},
  author = {Newman, M.},
  year = 2010,
  publisher = {OUP Oxford},
  address = {Oxford},
  isbn = {978-0-19-920665-0},
  lccn = {2010006011}
}

@book{book:Redner_GuideFirstpassageProcesses_08,
  title = {A Guide to First-Passage Processes},
  author = {Redner, Sidney},
  year = 2008,
  publisher = {Cambridge University Press},
  address = {Cambridge},
  isbn = {978-0-521-65248-3 978-0-521-03691-7}
}

@book{book:Schiff_LaplaceTransformTheory_99,
  title = {The {{Laplace Transform}}: {{Theory}} and {{Applications}}},
  author = {Schiff, Joel L.},
  year = 1999,
  publisher = {Springer Science \& Business Media},
  address = {New York},
  isbn = {978-0-387-98698-2}
}

@book{book:VanKampen_StochasticProcessesPhysics_92,
  title = {Stochastic Processes in {{Physics}} and {{Chemistry}}},
  author = {Van Kampen, Nicolaas Godfried},
  year = 1992,
  publisher = {North-Holland},
  address = {Amsterdam}
}

@book{book:Viswanathan.etal_PhysicsForagingIntroduction_11,
  title = {The Physics of Foraging: An Introduction to Random Searches and Biological Encounters},
  author = {Viswanathan, Gandhimohan M and Da Luz, Marcos GE and Raposo, Ernesto P and Stanley, H Eugene},
  year = 2011,
  publisher = {Cambridge University Press},
  address = {Cambridge}
}

@incollection{booksection:Chechkin.etal_IntroductionTheoryLevy_AnomalousTransport:FoundationsandApplications08,
  title = {Introduction to the {{Theory}} of {{L\'evy Flights}}},
  booktitle = {Anomalous {{Transport}}: {{Foundations}} and {{Applications}}},
  author = {Chechkin, Aleksei and Metzler, Ralf and Klafter, Joseph and Gonchar, Vsevolod},
  year = 2008,
  pages = {129--162},
  doi = {10.1002/9783527622979.ch5},
  isbn = {978-3-527-62297-9}
}

@incollection{booksection:Maida_Chapter2Cognitive_CognitiveComputing:TheoryandApplications16,
  title = {Chapter 2 - {{Cognitive Computing}} and {{Neural Networks}}: {{Reverse Engineering}} the {{Brain}}},
  booktitle = {Cognitive {{Computing}}: {{Theory}} and {{Applications}}},
  author = {Maida, A. S.},
  editor = {Gudivada, Venkat N. and Raghavan, Vijay V. and Govindaraju, Venu and Rao, C. R.},
  year = 2016,
  series = {Handbook of {{Statistics}}},
  volume = {35},
  pages = {39--78},
  publisher = {Elsevier},
  issn = {0169-7161},
  doi = {10.1016/bs.host.2016.07.011}
}

@article{journalarticle:Abramowitz.etal_HandbookMathematicalFunctions_Am.J.Phys.88,
  title = {Handbook of {{Mathematical Functions}} with {{Formulas}}, {{Graphs}}, and {{Mathematical Tables}}},
  author = {Abramowitz, Milton and Stegun, Irene A. and Romer, Robert H.},
  year = 1988,
  journal = {Am. J. Phys.},
  volume = {56},
  pages = {958--958},
  doi = {10.1119/1.15378},
  urldate = {2013-11-22}
}

@article{journalarticle:Ahmad.etal_FirstPassageParticle_Phys.Rev.E19,
  title = {First Passage of a Particle in a Potential under Stochastic Resetting: {{A}} Vanishing Transition of Optimal Resetting Rate},
  author = {Ahmad, Saeed and Nayak, Indrani and Bansal, Ajay and Nandi, Amitabha and Das, Dibyendu},
  year = 2019,
  journal = {Phys. Rev. E},
  volume = {99},
  number = {2},
  pages = {022130},
  publisher = {American Physical Society},
  doi = {10.1103/PhysRevE.99.022130}
}

@article{journalarticle:Alpern.etal_StochasticGameModel_J.R.Soc.Interface19,
  title = {A Stochastic Game Model of Searching Predators and Hiding Prey},
  author = {Alpern, Steve and Gal, Shmuel and Lee, Viciano and Casas, J{\'e}r{\^o}me},
  year = 2019,
  journal = {J. R. Soc. Interface},
  volume = {16},
  number = {153},
  pages = {20190087},
  publisher = {Royal Society},
  doi = {10.1098/rsif.2019.0087},
  urldate = {2025-09-16}
}

@article{journalarticle:Amendola.etal_SocialSearchRetrieving_OnlineSoc.Netw.Media23,
  title = {Social Search: {{Retrieving}} Information in {{Online Social}} Platforms -- {{A}} Survey},
  author = {Amendola, Maddalena and Passarella, Andrea and Perego, Raffaele},
  year = 2023,
  journal = {Online Soc. Netw. Media},
  volume = {36},
  pages = {100254},
  issn = {24686964},
  doi = {10.1016/j.osnem.2023.100254},
  urldate = {2025-08-15}
}

@article{journalarticle:Amnuanpol_BucklingInstabilityRotating_EPL21,
  title = {Buckling Instability on a Rotating Plane},
  author = {Amnuanpol, S.},
  year = 2021,
  journal = {EPL},
  volume = {135},
  number = {5},
  pages = {50003},
  issn = {0295-5075, 1286-4854},
  doi = {10.1209/0295-5075/135/50003},
  urldate = {2022-09-26}
}

@article{journalarticle:Amorim.etal_NovelEffectsStrains_Phys.Rep.16,
  title = {Novel Effects of Strains in Graphene and Other Two Dimensional Materials},
  author = {Amorim, B. and Cortijo, A. and {de Juan}, F. and Grushin, A.G. and Guinea, F. and {Guti{\'e}rrez-Rubio}, A. and Ochoa, H. and Parente, V. and Rold{\'a}n, R. and {San-Jose}, P. and Schiefele, J. and Sturla, M. and Vozmediano, M.A.H.},
  year = 2016,
  journal = {Phys. Rep.},
  volume = {617},
  pages = {1--54},
  issn = {03701573},
  doi = {10.1016/j.physrep.2015.12.006},
  urldate = {2022-05-05}
}

@article{journalarticle:Anderson_ModelCalculationsCooperative_J.Chem.Phys.70,
  title = {Model {{Calculations}} of {{Cooperative Motions}} in {{Chain Molecules}}},
  author = {Anderson, J. E.},
  year = 1970,
  journal = {J. Chem. Phys.},
  volume = {52},
  pages = {2821},
  doi = {10.1063/1.1673406},
  urldate = {2013-11-22}
}

@article{journalarticle:Bartumeus.etal_AnimalSearchStrategies_Ecology05,
  title = {Animal Search Strategies: A Quantitative Random-Walk Analysis},
  author = {Bartumeus, Frederic and {da Luz}, M. G E. and Viswanathan, G. M. and Catalan, J.},
  year = 2005,
  journal = {Ecology},
  volume = {86},
  number = {11},
  pages = {3078--3087},
  doi = {10.1890/04-1806}
}

@article{journalarticle:Basu.etal_SymmetricExclusionProcess_Phys.Rev.E19,
  title = {Symmetric Exclusion Process under Stochastic Resetting},
  author = {Basu, Urna and Kundu, Anupam and Pal, Arnab},
  year = 2019,
  journal = {Phys. Rev. E},
  volume = {100},
  number = {3},
  pages = {032136},
  publisher = {American Physical Society},
  doi = {10.1103/PhysRevE.100.032136}
}

@article{journalarticle:Benichou.etal_IntermittentSearchStrategies_Rev.Mod.Phys.11,
  title = {Intermittent Search Strategies},
  author = {B{\'e}nichou, O. and Loverdo, C. and Moreau, M. and Voituriez, R.},
  year = 2011,
  journal = {Rev. Mod. Phys.},
  volume = {83},
  number = {1},
  pages = {81--129},
  publisher = {American Physical Society},
  doi = {10.1103/RevModPhys.83.81}
}

@article{journalarticle:Benichou.etal_OptimalSearchStrategies_Phys.Rev.Lett.05,
  ids = {journalarticle:Benichou.etal_OptimalSearchStrategies_Phys.Rev.Lett.05a},
  title = {Optimal {{Search Strategies}} for {{Hidden Targets}}},
  author = {B{\'e}nichou, O. and Coppey, M. and Moreau, M. and Suet, P-H. and Voituriez, R.},
  year = 2005,
  journal = {Phys. Rev. Lett.},
  volume = {94},
  number = {19},
  pages = {198101},
  publisher = {American Physical Society},
  doi = {10.1103/PhysRevLett.94.198101}
}

@article{journalarticle:Berger-Tal.Bar-David_RecursiveMovementPatterns_Ecosphere15,
  title = {Recursive Movement Patterns: Review and Synthesis across Species},
  author = {{Berger-Tal}, Oded and {Bar-David}, Shirli},
  year = 2015,
  journal = {Ecosphere},
  volume = {6},
  number = {9},
  pages = {art149},
  issn = {2150-8925},
  doi = {10.1890/ES15-00106.1},
  urldate = {2023-05-12}
}

@article{journalarticle:Berthier.Biroli_TheoreticalPerspectiveGlass_Rev.Mod.Phys.11,
  title = {Theoretical Perspective on the Glass Transition and Amorphous Materials},
  author = {Berthier, Ludovic and Biroli, Giulio},
  year = 2011,
  journal = {Rev. Mod. Phys.},
  volume = {83},
  number = {2},
  pages = {587--645},
  issn = {0034-6861, 1539-0756},
  doi = {10.1103/RevModPhys.83.587},
  urldate = {2022-03-14}
}

@article{journalarticle:Bhat.etal_StochasticSearchPoisson_JStatMechTheoryExp16,
  title = {Stochastic Search with {{Poisson}} and Deterministic Resetting},
  author = {Bhat, Uttam and Bacco, Caterina De and Redner, S.},
  year = 2016,
  journal = {J. Stat. Mech.: Theory Exp.},
  volume = {2016},
  number = {8},
  pages = {083401},
  publisher = {{IOP Publishing and SISSA}},
  doi = {10.1088/1742-5468/2016/08/083401}
}

@article{journalarticle:Biswas.etal_RateEnhancementGated_J.Chem.Phys.23,
  title = {Rate Enhancement of Gated Drift-Diffusion Process by Optimal Resetting},
  author = {Biswas, Arup and Pal, Arnab and Mondal, Debasish and Ray, Somrita},
  year = 2023,
  journal = {J. Chem. Phys.},
  volume = {159},
  number = {5},
  pages = {054111},
  issn = {0021-9606},
  doi = {10.1063/5.0154210}
}

@article{journalarticle:Biswas.etal_TargetSearchOptimization_25,
  title = {Target Search Optimization by Threshold Resetting},
  author = {Biswas, Arup and Majumdar, Satya N. and Pal, Arnab},
  year = 2025,
  eprint = {2504.13501},
  primaryclass = {cond-mat},
  doi = {10.48550/arXiv.2504.13501},
  urldate = {2025-10-09},
  archiveprefix = {arXiv}
}

@article{journalarticle:Bodrova.etal_ScaledBrownianMotion_Phys.Rev.E19,
  title = {Scaled {{Brownian}} Motion with Renewal Resetting},
  author = {Bodrova, Anna S. and Chechkin, Aleksei V. and Sokolov, Igor M.},
  year = 2019,
  journal = {Phys. Rev. E},
  volume = {100},
  number = {1},
  pages = {012120},
  publisher = {American Physical Society},
  doi = {10.1103/PhysRevE.100.012120}
}

@article{journalarticle:Bonilla.Carpio_ModelRipplesGraphene_Phys.Rev.B12,
  title = {Model of Ripples in Graphene},
  author = {Bonilla, L. L. and Carpio, A.},
  year = 2012,
  journal = {Phys. Rev. B},
  volume = {86},
  number = {19},
  pages = {195402},
  issn = {1098-0121, 1550-235X},
  doi = {10.1103/PhysRevB.86.195402},
  urldate = {2022-09-26}
}

@article{journalarticle:Bonilla.etal_RipplesStringCoupled_Phys.Rev.E12,
  title = {Ripples in a String Coupled to {{Glauber}} Spins},
  author = {Bonilla, L. L. and Carpio, A. and Prados, A. and Rosales, R. R.},
  year = 2012,
  journal = {Phys. Rev. E},
  volume = {85},
  number = {3},
  pages = {031125},
  issn = {1539-3755, 1550-2376},
  doi = {10.1103/PhysRevE.85.031125},
  urldate = {2025-08-02}
}

@article{journalarticle:Boyer.Walsh_ModellingMobilityLiving_Philos.Trans.R.Soc.A10,
  title = {Modelling the Mobility of Living Organisms in Heterogeneous Landscapes: Does Memory Improve Foraging Success?},
  author = {Boyer, Denis and Walsh, Peter D.},
  year = 2010,
  journal = {Philos. Trans. R. Soc. A},
  volume = {368},
  number = {1933},
  pages = {5645--5659},
  issn = {1364-503X, 1471-2962},
  doi = {10.1098/rsta.2010.0275},
  urldate = {2023-05-12}
}

@article{journalarticle:Bressloff_DirectedIntermittentSearch_JPhysMathTheor20,
  title = {Directed Intermittent Search with Stochastic Resetting},
  author = {Bressloff, Paul C.},
  year = 2020,
  journal = {J. Phys. A: Math. Theor.},
  volume = {53},
  number = {10},
  pages = {105001},
  publisher = {IOP Publishing},
  doi = {10.1088/1751-8121/ab7138}
}

@article{journalarticle:Browne.etal_SurveyMonteCarlo_IEEETrans.Comp.Intell.AIGames12,
  title = {A Survey of {{Monte Carlo}} Tree Search Methods},
  author = {Browne, Cameron B. and Powley, Edward and Whitehouse, Daniel and Lucas, Simon M. and Cowling, Peter I. and Rohlfshagen, Philipp and Tavener, Stephen and Perez, Diego and Samothrakis, Spyridon and Colton, Simon},
  year = 2012,
  journal = {IEEE Trans. Comp. Intell. AI Games},
  volume = {4},
  number = {1},
  pages = {1--43},
  issn = {1943-068X},
  doi = {10.1109/TCIAIG.2012.2186810}
}

@article{journalarticle:Busiello.etal_EntropyProductionSystems_Phys.Rev.Res.20,
  title = {Entropy Production in Systems with Unidirectional Transitions},
  author = {Busiello, D. M. and Gupta, D. and Maritan, A.},
  year = 2020,
  journal = {Phys. Rev. Res.},
  volume = {2},
  number = {2},
  pages = {023011},
  publisher = {American Physical Society},
  doi = {10.1103/PhysRevResearch.2.023011}
}

@article{journalarticle:Cates_DiffusiveTransportDetailed_Rep.Prog.Phys.12,
  title = {Diffusive Transport without Detailed Balance in Motile Bacteria: Does Microbiology Need Statistical Physics?},
  author = {Cates, M E},
  year = 2012,
  journal = {Rep. Prog. Phys.},
  volume = {75},
  number = {4},
  pages = {042601},
  publisher = {IOP Publishing},
  issn = {0034-4885},
  doi = {10.1088/0034-4885/75/4/042601},
  urldate = {2025-08-15}
}

@article{journalarticle:Cavagna_SupercooledLiquidsPedestrians_Phys.Rep.09,
  title = {Supercooled Liquids for Pedestrians},
  author = {Cavagna, Andrea},
  year = 2009,
  journal = {Phys. Rep.},
  volume = {476},
  number = {4-6},
  pages = {51--124},
  issn = {03701573},
  doi = {10.1016/j.physrep.2009.03.003},
  urldate = {2022-03-14}
}

@article{journalarticle:Cea.etal_NumericalStudyRippling_Phys.Rev.B20,
  title = {Numerical Study of the Rippling Instability Driven by Electron-Phonon Coupling in Graphene},
  author = {Cea, T. and {Ruiz-Garc{\'i}a}, M. and Bonilla, L. L. and Guinea, F.},
  year = 2020,
  journal = {Phys. Rev. B},
  volume = {101},
  number = {23},
  pages = {235428},
  issn = {2469-9950, 2469-9969},
  doi = {10.1103/PhysRevB.101.235428},
  urldate = {2025-08-02}
}

@article{journalarticle:Charbonneau.etal_FractalFreeEnergy_Nat.Commun.14,
  title = {Fractal Free Energy Landscapes in Structural Glasses},
  author = {Charbonneau, Patrick and Kurchan, Jorge and Parisi, Giorgio and Urbani, Pierfrancesco and Zamponi, Francesco},
  year = 2014,
  journal = {Nat. Commun.},
  volume = {5},
  number = {1},
  pages = {3725},
  issn = {2041-1723},
  doi = {10.1038/ncomms4725}
}

@article{journalarticle:Chechkin.Sokolov_RandomSearchResetting_Phys.Rev.Lett.18,
  title = {Random {{Search}} with {{Resetting}}: {{A Unified Renewal Approach}}},
  author = {Chechkin, A. and Sokolov, I. M.},
  year = 2018,
  journal = {Phys. Rev. Lett.},
  volume = {121},
  number = {5},
  pages = {050601},
  publisher = {American Physical Society},
  doi = {10.1103/PhysRevLett.121.050601}
}

@article{journalarticle:Chen.etal_SpontaneousTiltSingleClamped_Phys.Rev.Lett.22,
  title = {Spontaneous {{Tilt}} of {{Single-Clamped Thermal Elastic Sheets}}},
  author = {Chen, Zhitao and Wan, Duanduan and Bowick, Mark J.},
  year = 2022,
  journal = {Phys. Rev. Lett.},
  volume = {128},
  number = {2},
  pages = {028006},
  issn = {0031-9007, 1079-7114},
  doi = {10.1103/PhysRevLett.128.028006},
  urldate = {2022-09-26}
}

@article{journalarticle:Chen.Huang_FirstPassageDiffusing_Phys.Rev.E22,
  title = {First Passage of a Diffusing Particle under Stochastic Resetting in Bounded Domains with Spherical Symmetry},
  author = {Chen, Hanshuang and Huang, Feng},
  year = 2022,
  journal = {Phys. Rev. E},
  volume = {105},
  number = {3},
  pages = {034109},
  publisher = {American Physical Society},
  doi = {10.1103/PhysRevE.105.034109}
}

@article{journalarticle:Christou.Schadschneider_DiffusionResettingBounded_JPhysMathTheor15,
  title = {Diffusion with Resetting in Bounded Domains},
  author = {Christou, Christos and Schadschneider, Andreas},
  year = 2015,
  journal = {J. Phys. A: Math. Theor.},
  volume = {48},
  number = {28},
  pages = {285003},
  publisher = {IOP Publishing},
  doi = {10.1088/1751-8113/48/28/285003}
}

@article{journalarticle:Chupeau.etal_CoverTimesRandom_Nat.Phys.15,
  title = {Cover Times of Random Searches},
  author = {Chupeau, Marie and B{\'e}nichou, Olivier and Voituriez, Rapha{\"e}l},
  year = 2015,
  journal = {Nat. Phys.},
  volume = {11},
  number = {10},
  pages = {844--847},
  issn = {1745-2481},
  doi = {10.1038/nphys3413}
}

@article{journalarticle:DeBruyne.etal_OptimizationFirstPassageResetting_Phys.Rev.Lett.20,
  title = {Optimization in {{First-Passage Resetting}}},
  author = {De Bruyne, B. and {Randon-Furling}, J and Redner, S.},
  year = 2020,
  journal = {Phys. Rev. Lett.},
  volume = {125},
  number = {5},
  pages = {050602},
  publisher = {American Physical Society},
  doi = {10.1103/PhysRevLett.125.050602}
}

@article{journalarticle:DeBruyne.Mori_ResettingStochasticOptimal_Phys.Rev.Research23,
  title = {Resetting in Stochastic Optimal Control},
  author = {De Bruyne, Benjamin and Mori, Francesco},
  year = 2023,
  journal = {Phys. Rev. Research},
  volume = {5},
  number = {1},
  pages = {013122},
  issn = {2643-1564},
  doi = {10.1103/PhysRevResearch.5.013122},
  urldate = {2025-08-02}
}

@article{journalarticle:Dhar.etal_DetectionQuantumParticle_Phys.Rev.A15,
  title = {Detection of a Quantum Particle on a Lattice under Repeated Projective Measurements},
  author = {Dhar, Shrabanti and Dasgupta, Subinay and Dhar, Abhishek and Sen, Diptiman},
  year = 2015,
  journal = {Phys. Rev. A},
  volume = {91},
  number = {6},
  pages = {062115},
  publisher = {American Physical Society},
  doi = {10.1103/PhysRevA.91.062115}
}

@article{journalarticle:Diz-Pita.Otero-Espinar_PredatorPreyModels_Mathematics21,
  title = {Predator--{{Prey Models}}: {{A Review}} of {{Some Recent Advances}}},
  author = {{Diz-Pita}, {\'E}rika and {Otero-Espinar}, M. Victoria},
  year = 2021,
  journal = {Mathematics},
  volume = {9},
  number = {15},
  pages = {1783},
  publisher = {Multidisciplinary Digital Publishing Institute},
  issn = {2227-7390},
  doi = {10.3390/math9151783},
  urldate = {2025-09-16}
}

@article{journalarticle:Dubey.etal_QuantumResettingContinuous_JPhysMathTheor23,
  title = {Quantum Resetting in Continuous Measurement Induced Dynamics of a Qubit},
  author = {Dubey, Varun and Chetrite, Raphael and Dhar, Abhishek},
  year = 2023,
  journal = {J. Phys. A: Math. Theor.},
  volume = {56},
  number = {15},
  pages = {154001},
  publisher = {IOP Publishing},
  doi = {10.1088/1751-8121/acc290}
}

@article{journalarticle:Durang.etal_FirstpassageStatisticsStochastic_JPhysMathTheor19,
  title = {First-Passage Statistics under Stochastic Resetting in Bounded Domains},
  author = {Durang, Xavier and Lee, Sungmin and Lizana, Ludvig and Jeon, Jae-Hyung},
  year = 2019,
  journal = {J. Phys. A: Math. Theor.},
  volume = {52},
  number = {22},
  pages = {224001},
  publisher = {IOP Publishing},
  doi = {10.1088/1751-8121/ab15f5}
}

@article{journalarticle:Eder.etal_ProbingBothSides_NanoLett.13,
  title = {Probing from {{Both Sides}}: {{Reshaping}} the {{Graphene Landscape}} via {{Face-to-Face Dual-Probe Microscopy}}},
  author = {Eder, Franz R. and Kotakoski, Jani and Holzweber, Katharina and Mangler, Clemens and Skakalova, Viera and Meyer, Jannik C.},
  year = 2013,
  journal = {Nano Lett.},
  volume = {13},
  number = {5},
  pages = {1934--1940},
  issn = {1530-6984, 1530-6992},
  doi = {10.1021/nl3042799},
  urldate = {2022-09-26}
}

@article{journalarticle:Espeso.etal_DifferentialGrowthWrinkled_Phys.Rev.E15,
  title = {Differential Growth of Wrinkled Biofilms},
  author = {Espeso, D. R. and Carpio, A. and Einarsson, B.},
  year = 2015,
  journal = {Phys. Rev. E},
  volume = {91},
  number = {2},
  pages = {022710},
  doi = {10.1103/PhysRevE.91.022710},
  urldate = {2025-10-24}
}

@article{journalarticle:Eule.Metzger_NonequilibriumSteadyStates_NewJ.Phys.16,
  title = {Non-Equilibrium Steady States of Stochastic Processes with Intermittent Resetting},
  author = {Eule, Stephan and Metzger, Jakob J.},
  year = 2016,
  journal = {New J. Phys.},
  volume = {18},
  number = {3},
  pages = {033006},
  publisher = {IOP Publishing},
  doi = {10.1088/1367-2630/18/3/033006}
}

@article{journalarticle:Evans.etal_ExactlySolvablePredator_JPhysMathTheor22,
  ids = {journalarticle:Evans.etal_ExactlySolvablePredator_J.Phys.Math.Theor.22a},
  title = {An Exactly Solvable Predator Prey Model with Resetting},
  author = {Evans, Martin R. and Majumdar, Satya N. and Schehr, Gr{\'e}gory},
  year = 2022,
  journal = {J. Phys. A: Math. Theor.},
  volume = {55},
  number = {27},
  pages = {274005},
  publisher = {IOP Publishing},
  doi = {10.1088/1751-8121/ac7269}
}

@article{journalarticle:Evans.etal_OptimalDiffusiveSearch_JPhysMathTheor13,
  title = {Optimal Diffusive Search: Nonequilibrium Resetting versus Equilibrium Dynamics},
  author = {Evans, Martin R. and Majumdar, Satya N. and Mallick, Kirone},
  year = 2013,
  journal = {J. Phys. A: Math. Theor.},
  volume = {46},
  number = {18},
  pages = {185001},
  publisher = {IOP Publishing},
  doi = {10.1088/1751-8113/46/18/185001}
}

@article{journalarticle:Evans.etal_StochasticResettingApplications_J.Phys.A:Math.Theor.20,
  title = {Stochastic Resetting and Applications},
  author = {Evans, Martin R and Majumdar, Satya N and Schehr, Gr{\'e}gory},
  year = 2020,
  journal = {J. Phys. A: Math. Theor.},
  volume = {53},
  number = {19},
  pages = {193001},
  issn = {1751-8113, 1751-8121},
  doi = {10.1088/1751-8121/ab7cfe},
  urldate = {2025-08-02}
}

@article{journalarticle:Evans.Majumdar_DiffusionOptimalResetting_J.Phys.A:Math.Theor.11,
  title = {Diffusion with Optimal Resetting},
  author = {Evans, Martin R and Majumdar, Satya N},
  year = 2011,
  journal = {J. Phys. A: Math. Theor.},
  volume = {44},
  number = {43},
  pages = {435001},
  issn = {1751-8113, 1751-8121},
  doi = {10.1088/1751-8113/44/43/435001},
  urldate = {2025-08-02}
}

@article{journalarticle:Evans.Majumdar_DiffusionStochasticResetting_Phys.Rev.Lett.11,
  title = {Diffusion with {{Stochastic Resetting}}},
  author = {Evans, Martin R. and Majumdar, Satya N.},
  year = 2011,
  journal = {Phys. Rev. Lett.},
  volume = {106},
  number = {16},
  pages = {160601},
  issn = {0031-9007, 1079-7114},
  doi = {10.1103/PhysRevLett.106.160601},
  urldate = {2025-08-02}
}

@article{journalarticle:Evans.Majumdar_EffectsRefractoryPeriod_J.Phys.A:Math.Theor.19,
  title = {Effects of Refractory Period on Stochastic Resetting},
  author = {Evans, Martin R and Majumdar, Satya N},
  year = 2019,
  journal = {J. Phys. A: Math. Theor.},
  volume = {52},
  number = {1},
  pages = {01LT01},
  issn = {1751-8113, 1751-8121},
  doi = {10.1088/1751-8121/aaf080},
  urldate = {2025-08-02}
}

@article{journalarticle:Evans.Majumdar_RunTumbleParticle_JPhysMathTheor18,
  title = {Run and Tumble Particle under Resetting: A Renewal Approach},
  author = {Evans, Martin R. and Majumdar, Satya N.},
  year = 2018,
  journal = {J. Phys. A: Math. Theor.},
  volume = {51},
  number = {47},
  pages = {475003},
  publisher = {IOP Publishing},
  doi = {10.1088/1751-8121/aae74e}
}

@article{journalarticle:Evans.Ray_StochasticResettingPrevails_Phys.Rev.Lett.25,
  title = {Stochastic {{Resetting Prevails Over Sharp Restart}} for {{Broad Target Distributions}}},
  author = {Evans, Martin R. and Ray, Somrita},
  year = 2025,
  journal = {Phys. Rev. Lett.},
  volume = {134},
  number = {24},
  pages = {247102},
  issn = {0031-9007, 1079-7114},
  doi = {10.1103/2wpq-gl2j},
  urldate = {2025-09-29}
}

@article{journalarticle:Faisant.etal_OptimalMeanFirstpassage_JPhysMathTheor21,
  title = {Optimal Mean First-Passage Time of a {{Brownian}} Searcher with Resetting in One and Two Dimensions: Experiments, Theory and Numerical Tests},
  author = {Faisant, F. and Besga, B. and Petrosyan, A. and Ciliberto, S. and Majumdar, Satya N.},
  year = 2021,
  journal = {J. Phys. A: Math. Theor.},
  volume = {2021},
  number = {11},
  pages = {113203},
  publisher = {{IOP Publishing and SISSA}},
  doi = {10.1088/1742-5468/ac2cc7}
}

@article{journalarticle:Fasolino.etal_IntrinsicRipplesGraphene_Nat.Mater.07,
  title = {Intrinsic Ripples in Graphene},
  author = {Fasolino, A. and Los, J. H. and Katsnelson, M. I.},
  year = 2007,
  journal = {Nat. Mater.},
  volume = {6},
  number = {11},
  pages = {858--861},
  issn = {1476-1122, 1476-4660},
  doi = {10.1038/nmat2011},
  urldate = {2022-09-26}
}

@article{journalarticle:Fetz.Gustafsson_RelationShapesPostsynaptic_J.Physiol.83,
  title = {Relation between Shapes of Post-Synaptic Potentials and Changes in Firing Probability of Cat Motoneurones},
  author = {Fetz, E. E. and Gustafsson, B.},
  year = 1983,
  journal = {J. Physiol.},
  volume = {341},
  number = {1},
  pages = {387--410},
  doi = {10.1113/jphysiol.1983.sp014812}
}

@article{journalarticle:Folena.etal_IntroductionDynamicsDisordered_PhysStatMechAppl22,
  title = {Introduction to the Dynamics of Disordered Systems: {{Equilibrium}} and Gradient Descent},
  author = {Folena, Giampaolo and Manacorda, Alessandro and Zamponi, Francesco},
  year = 2022,
  journal = {Phys. A: Stat. Mech. Appl.},
  pages = {128152},
  issn = {0378-4371},
  doi = {10.1016/j.physa.2022.128152}
}

@article{journalarticle:Fuchs.etal_StochasticThermodynamicsResetting_EPL16,
  title = {Stochastic Thermodynamics of Resetting},
  author = {Fuchs, Jaco and Goldt, Sebastian and Seifert, Udo},
  year = 2016,
  journal = {EPL},
  volume = {113},
  number = {6},
  pages = {60009},
  publisher = {{EDP Sciences, IOP Publishing and Societ\`a Italiana di Fisica}},
  doi = {10.1209/0295-5075/113/60009}
}

@article{journalarticle:Garcia-Valladares.etal_BucklingRotationallyInvariant_Phys.Rev.E23,
  title = {Buckling in a Rotationally Invariant Spin-Elastic Model},
  author = {{Garc{\'i}a-Valladares}, Gregorio and Plata, Carlos A. and Prados, Antonio},
  year = 2023,
  journal = {Phys. Rev. E},
  volume = {107},
  number = {1},
  pages = {014120},
  issn = {2470-0045, 2470-0053},
  doi = {10.1103/PhysRevE.107.014120},
  urldate = {2025-08-02}
}

@article{journalarticle:Garcia-Valladares.etal_OptimalResettingStrategies_NewJ.Phys.23,
  title = {Optimal Resetting Strategies for Search Processes in Heterogeneous Environments},
  author = {{Garc{\'i}a-Valladares}, Gregorio and Plata, Carlos A and Prados, Antonio and Manacorda, Alessandro},
  year = 2023,
  journal = {New J. Phys.},
  volume = {25},
  number = {11},
  pages = {113031},
  issn = {1367-2630},
  doi = {10.1088/1367-2630/ad06da},
  urldate = {2025-08-02}
}

@article{journalarticle:Garcia-Valladares.etal_StochasticResettingRefractory_Phys.Scr.24,
  title = {Stochastic Resetting with Refractory Periods: Pathway Formulation and Exact Results},
  author = {{Garc{\'i}a-Valladares}, G and Gupta, D and Prados, A and Plata, C A},
  year = 2024,
  journal = {Phys. Scr.},
  volume = {99},
  number = {4},
  pages = {045234},
  issn = {0031-8949, 1402-4896},
  doi = {10.1088/1402-4896/ad317b},
  urldate = {2025-08-02}
}

@article{journalarticle:Gikunda.etal_ArrayGrapheneVariable_Membranes22,
  title = {Array of {{Graphene Variable Capacitors}} on 100 Mm {{Silicon Wafers}} for {{Vibration-Based Applications}}},
  author = {Gikunda, Millicent N. and Harerimana, Ferdinand and Mangum, James M. and Rahman, Sumaya and Thompson, Joshua P. and Harris, Charles Thomas and Churchill, Hugh O. H. and Thibado, Paul M.},
  year = 2022,
  journal = {Membranes},
  volume = {12},
  number = {5},
  pages = {533},
  issn = {2077-0375},
  doi = {10.3390/membranes12050533},
  urldate = {2022-09-26}
}

@article{journalarticle:Golubovic.etal_DynamicsEulerBuckling_Phys.Rev.Lett.98,
  title = {Dynamics of the {{Euler Buckling Instability}}},
  author = {Golubovic, Leonardo and Moldovan, Dorel and Peredera, Anatoli},
  year = 1998,
  journal = {Phys. Rev. Lett.},
  volume = {81},
  number = {16},
  pages = {4},
  doi = {10.1103/PhysRevLett.106.045502}
}

@article{journalarticle:Gupta_StochasticResettingUnderdamped_JStatMechTheoryExp19,
  title = {Stochastic Resetting in Underdamped {{Brownian}} Motion},
  author = {Gupta, Deepak},
  year = 2019,
  journal = {J. Stat. Mech.: Theory Exp.},
  volume = {2019},
  number = {3},
  pages = {033212},
  publisher = {{IOP Publishing and SISSA}},
  doi = {10.1088/1742-5468/ab054a}
}

@article{journalarticle:Gupta.etal_ResettingStochasticReturn_JStatMechTheoryExp21,
  title = {Resetting with Stochastic Return through Linear Confining Potential},
  author = {Gupta, Deepak and Pal, Arnab and Kundu, Anupam},
  year = 2021,
  journal = {J. Stat. Mech.: Theory Exp.},
  volume = {2021},
  number = {4},
  pages = {043202},
  publisher = {{IOP Publishing and SISSA}},
  doi = {10.1088/1742-5468/abefdf}
}

@article{journalarticle:Gupta.etal_StochasticResettingStochastic_J.Phys.A:Math.Theor.21,
  title = {Stochastic Resetting with Stochastic Returns Using External Trap},
  author = {Gupta, Deepak and Plata, Carlos A and Kundu, Anupam and Pal, Arnab},
  year = 2021,
  journal = {J. Phys. A: Math. Theor.},
  volume = {54},
  number = {2},
  pages = {025003},
  issn = {1751-8113, 1751-8121},
  doi = {10.1088/1751-8121/abcf0b},
  urldate = {2025-08-02}
}

@article{journalarticle:Gupta.etal_WorkFluctuationsJarzynski_Phys.Rev.Lett.20,
  title = {Work {{Fluctuations}} and {{Jarzynski Equality}} in {{Stochastic Resetting}}},
  author = {Gupta, Deepak and Plata, Carlos A. and Pal, Arnab},
  year = 2020,
  journal = {Phys. Rev. Lett.},
  volume = {124},
  number = {11},
  pages = {110608},
  issn = {0031-9007, 1079-7114},
  doi = {10.1103/PhysRevLett.124.110608},
  urldate = {2025-08-02}
}

@article{journalarticle:Gupta.Plata_WorkFluctuationsDiffusion_NewJ.Phys.22,
  title = {Work Fluctuations for Diffusion Dynamics Submitted to Stochastic Return},
  author = {Gupta, Deepak and Plata, Carlos A.},
  year = 2022,
  journal = {New J. Phys.},
  volume = {24},
  number = {11},
  pages = {113034},
  publisher = {IOP Publishing},
  doi = {10.1088/1367-2630/aca25e}
}

@article{journalarticle:Hanakata.etal_AnomalousThermalExpansion_Phys.Rev.Lett.22,
  title = {Anomalous {{Thermal Expansion}} in {{Ising-like Puckered Sheets}}},
  author = {Hanakata, Paul Z. and Plummer, Abigail and Nelson, David R.},
  year = 2022,
  journal = {Phys. Rev. Lett.},
  volume = {128},
  number = {7},
  pages = {075902},
  issn = {0031-9007, 1079-7114},
  doi = {10.1103/PhysRevLett.128.075902},
  urldate = {2025-08-02}
}

@article{journalarticle:Hanakata.etal_ThermalBucklingSymmetry_ExtremeMech.Lett.21,
  title = {Thermal Buckling and Symmetry Breaking in Thin Ribbons under Compression},
  author = {Hanakata, Paul Z. and Bhabesh, Sourav S. and Bowick, Mark J. and Nelson, David R. and Yllanes, David},
  year = 2021,
  journal = {Extreme Mech. Lett.},
  volume = {44},
  pages = {101270},
  issn = {23524316},
  doi = {10.1016/j.eml.2021.101270},
  urldate = {2025-08-02}
}

@article{journalarticle:Harris.Touchette_PhaseTransitionsLarge_JPhysMathTheor17,
  title = {Phase Transitions in Large Deviations of Reset Processes},
  author = {Harris, Rosemary J. and Touchette, Hugo},
  year = 2017,
  journal = {J. Phys. A: Math. Theor.},
  volume = {50},
  number = {10},
  pages = {10LT01},
  publisher = {IOP Publishing},
  doi = {10.1088/1751-8121/aa5734}
}

@article{journalarticle:Helfrich_ElasticPropertiesLipid_Z.Naturforsch.CBio.Sci.73,
  title = {Elastic {{Properties}} of {{Lipid Bilayers}}: {{Theory}} and {{Possible Experiments}}},
  author = {Helfrich, Wolfgang},
  year = 1973,
  journal = {Z. Naturforsch. C Bio. Sci.},
  volume = {28},
  number = {11-12},
  pages = {693--703},
  doi = {10.1515/znc-1973-11-1209}
}

@article{journalarticle:Hollander.etal_PropertiesAdditiveFunctionals_JPhysMathTheor19,
  title = {Properties of Additive Functionals of {{Brownian}} Motion with Resetting},
  author = {den Hollander, Frank and Majumdar, Satya N. and Meylahn, Janusz M. and Touchette, Hugo},
  year = 2019,
  journal = {J. Phys. A: Math. Theor.},
  volume = {52},
  number = {17},
  pages = {175001},
  publisher = {IOP Publishing},
  doi = {10.1088/1751-8121/ab0efd}
}

@article{journalarticle:Hua.Shen_LowdimensionalNanostructuresMonolithic_Chem.Soc.Rev.24,
  title = {Low-Dimensional Nanostructures for Monolithic {{3D-integrated}} Flexible and Stretchable Electronics},
  author = {Hua, Qilin and Shen, Guozhen},
  year = 2024,
  journal = {Chem. Soc. Rev.},
  volume = {53},
  number = {3},
  pages = {1316--1353},
  publisher = {Royal Society of Chemistry},
  doi = {10.1039/D3CS00918A},
  urldate = {2025-09-25}
}

@article{journalarticle:Huang.etal_GrapheneBasedSensorsHuman_Front.Chem.19,
  title = {Graphene-{{Based Sensors}} for {{Human Health Monitoring}}},
  author = {Huang, Haizhou and Su, Shi and Wu, Nan and Wan, Hao and Wan, Shu and Bi, Hengchang and Sun, Litao},
  year = 2019,
  journal = {Front. Chem.},
  volume = {7},
  pages = {399},
  issn = {2296-2646},
  doi = {10.3389/fchem.2019.00399},
  urldate = {2022-09-26}
}

@article{journalarticle:Jain.etal_CompressioncontrolledDynamicBuckling_Phys.Rev.E21,
  title = {Compression-Controlled Dynamic Buckling in Thin Soft Sheets},
  author = {Jain, Harsh and Ghosh, Shankar and Sahu, Kirti Chandra},
  year = 2021,
  journal = {Phys. Rev. E},
  volume = {104},
  number = {3},
  pages = {L033001},
  issn = {2470-0045, 2470-0053},
  doi = {10.1103/PhysRevE.104.L033001},
  urldate = {2022-09-26}
}

@article{journalarticle:Jolakoski.etal_FirstPassageResetting_ChaosSolitonFract.23,
  title = {A First Passage under Resetting Approach to Income Dynamics},
  author = {Jolakoski, Petar and Pal, Arnab and Sandev, Trifce and Kocarev, Ljupco and Metzler, Ralf and Stojkoski, Viktor},
  year = 2023,
  journal = {Chaos Soliton Fract.},
  volume = {175},
  pages = {113921},
  issn = {0960-0779},
  doi = {10.1016/j.chaos.2023.113921}
}

@article{journalarticle:Kunihiro.etal_NewComputationalApproach_Nanoscale25,
  title = {A New Computational Approach for Evaluating Bending Rigidity of Graphene Sheets Incorporating Disclinations},
  author = {Kunihiro, Yushi and Lei, Xiao-Wen and Uneyama, Takashi and Fujii, Toshiyuki},
  year = 2025,
  journal = {Nanoscale},
  volume = {17},
  number = {31},
  pages = {18112--18126},
  issn = {2040-3364, 2040-3372},
  doi = {10.1039/D5NR01102G},
  urldate = {2025-10-14}
}

@article{journalarticle:LeDoussal.Radzihovsky_ThermalBucklingTransition_Phys.Rev.Lett.21,
  title = {Thermal {{Buckling Transition}} of {{Crystalline Membranes}} in a {{Field}}},
  author = {Le Doussal, Pierre and Radzihovsky, Leo},
  year = 2021,
  journal = {Phys. Rev. Lett.},
  volume = {127},
  number = {1},
  pages = {015702},
  issn = {0031-9007, 1079-7114},
  doi = {10.1103/PhysRevLett.127.015702},
  urldate = {2025-08-02}
}

@article{journalarticle:Lenzi.etal_TransientAnomalousDiffusion_PhysStatMechAppl22,
  title = {Transient Anomalous Diffusion in Heterogeneous Media with Stochastic Resetting},
  author = {Lenzi, M. K. and Lenzi, E. K. and Guilherme, L. M. S. and Evangelista, L. R. and Ribeiro, H. V.},
  year = 2022,
  journal = {Phys. A: Stat. Mech. Appl.},
  volume = {588},
  pages = {126560},
  issn = {0378-4371},
  doi = {10.1016/j.physa.2021.126560}
}

@article{journalarticle:Lindahl.etal_DeterminationBendingRigidity_NanoLett.12,
  title = {Determination of the {{Bending Rigidity}} of {{Graphene}} via {{Electrostatic Actuation}} of {{Buckled Membranes}}},
  author = {Lindahl, Niklas and Midtvedt, Daniel and Svensson, Johannes and Nerushev, Oleg A. and Lindvall, Niclas and Isacsson, Andreas and Campbell, Eleanor E. B.},
  year = 2012,
  journal = {Nano Lett.},
  volume = {12},
  number = {7},
  pages = {3526--3531},
  issn = {1530-6984, 1530-6992},
  doi = {10.1021/nl301080v},
  urldate = {2022-09-26}
}

@article{journalarticle:Lipowsky_ConformationMembranes_Nature91,
  title = {The Conformation of Membranes},
  author = {Lipowsky, Reinhard},
  year = 1991,
  journal = {Nature},
  volume = {349},
  number = {6309},
  pages = {475--481},
  publisher = {Nature Publishing Group},
  doi = {10.1038/349475a0}
}

@article{journalarticle:Majumdar.etal_DynamicalTransitionTemporal_Phys.Rev.E15,
  title = {Dynamical Transition in the Temporal Relaxation of Stochastic Processes under Resetting},
  author = {Majumdar, Satya N. and Sabhapandit, Sanjib and Schehr, Gr{\'e}gory},
  year = 2015,
  journal = {Phys. Rev. E},
  volume = {91},
  number = {5},
  pages = {052131},
  publisher = {American Physical Society},
  doi = {10.1103/PhysRevE.91.052131}
}

@article{journalarticle:Mamun.etal_RecentReviewElectrospun_Membranes23,
  title = {A {{Recent Review}} of {{Electrospun Porous Carbon Nanofiber Mats}} for {{Energy Storage}} and {{Generation Applications}}},
  author = {Mamun, Al and Kiari, Mohamed and Sabantina, Lilia},
  year = 2023,
  journal = {Membranes},
  volume = {13},
  number = {10},
  pages = {830},
  publisher = {Multidisciplinary Digital Publishing Institute},
  issn = {2077-0375},
  doi = {10.3390/membranes13100830},
  urldate = {2025-10-15}
}

@article{journalarticle:Mangum.etal_MechanismsSpontaneousCurvature_Membranes21,
  title = {Mechanisms of {{Spontaneous Curvature Inversion}} in {{Compressed Graphene Ripples}} for {{Energy Harvesting Applications}} via {{Molecular Dynamics Simulations}}},
  author = {Mangum, James M. and Harerimana, Ferdinand and Gikunda, Millicent N. and Thibado, Paul M.},
  year = 2021,
  journal = {Membranes},
  volume = {11},
  number = {7},
  pages = {516},
  issn = {2077-0375},
  doi = {10.3390/membranes11070516},
  urldate = {2025-08-02}
}

@article{journalarticle:Marion.etal_UnderstandingForagingBehaviour_J.Theor.Biol.05,
  title = {Understanding Foraging Behaviour in Spatially Heterogeneous Environments},
  author = {Marion, Glenn and Swain, David L. and Hutchings, Mike R.},
  year = 2005,
  journal = {J. Theor. Biol.},
  volume = {232},
  number = {1},
  pages = {127--142},
  issn = {0022-5193},
  doi = {10.1016/j.jtbi.2004.08.005}
}

@article{journalarticle:Maso-Puigdellosas.etal_StochasticMovementSubject_J.Stat.Mech.19,
  title = {Stochastic Movement Subject to a Reset-and-Residence Mechanism: Transport Properties and First Arrival Statistics},
  author = {{Mas{\'o}-Puigdellosas}, Axel and Campos, Daniel and M{\'e}ndez, Vicen{\c c}},
  year = 2019,
  journal = {J. Stat. Mech.},
  volume = {2019},
  number = {3},
  pages = {033201},
  issn = {1742-5468},
  doi = {10.1088/1742-5468/ab02f3},
  urldate = {2025-08-02}
}

@article{journalarticle:McAdams.Arkin_StochasticMechanismsGene_Proc.Natl.Acad.Sci.U.S.A.97,
  title = {Stochastic Mechanisms in Gene\,Expression},
  author = {McAdams, Harley H. and Arkin, Adam},
  year = 1997,
  journal = {Proc. Natl. Acad. Sci. U.S.A.},
  volume = {94},
  number = {3},
  pages = {814--819},
  publisher = {Proceedings of the National Academy of Sciences},
  doi = {10.1073/pnas.94.3.814},
  urldate = {2025-09-16}
}

@article{journalarticle:Mendez.Campos_CharacterizationStationaryStates_Phys.Rev.E16,
  title = {Characterization of Stationary States in Random Walks with Stochastic Resetting},
  author = {M{\'e}ndez, Vicen{\c c} and Campos, Daniel},
  year = 2016,
  journal = {Phys. Rev. E},
  volume = {93},
  number = {2},
  pages = {022106},
  publisher = {American Physical Society},
  doi = {10.1103/PhysRevE.93.022106}
}

@article{journalarticle:Mercado-Vasquez.Boyer_FirstHittingTimes_Phys.Rev.Lett.19,
  title = {First {{Hitting Times}} to {{Intermittent Targets}}},
  author = {{Mercado-V{\'a}squez}, Gabriel and Boyer, Denis},
  year = 2019,
  journal = {Phys. Rev. Lett.},
  volume = {123},
  number = {25},
  pages = {250603},
  issn = {0031-9007, 1079-7114},
  doi = {10.1103/PhysRevLett.123.250603},
  urldate = {2023-05-12}
}

@article{journalarticle:Mercado-Vasquez.Boyer_LotkaVolterraSystems_JPhysMathTheor18,
  title = {Lotka--{{Volterra}} Systems with Stochastic Resetting},
  author = {{Mercado-V{\'a}squez}, Gabriel and Boyer, Denis},
  year = 2018,
  journal = {J. Phys. A: Math. Theor.},
  volume = {51},
  number = {40},
  pages = {405601},
  publisher = {IOP Publishing},
  doi = {10.1088/1751-8121/aadbc0}
}

@article{journalarticle:Mercado-Vasquez.etal_IntermittentResettingPotentials_JStatMechTheoryExp20,
  title = {Intermittent Resetting Potentials},
  author = {{Mercado-V{\'a}squez}, Gabriel and Boyer, Denis and Majumdar, Satya N. and Schehr, Gr{\'e}gory},
  year = 2020,
  journal = {J. Stat. Mech.: Theory Exp.},
  volume = {2020},
  number = {11},
  pages = {113203},
  publisher = {{IOP Publishing and SISSA}},
  doi = {10.1088/1742-5468/abc1d9}
}

@article{journalarticle:Mermin.Wagner_AbsenceFerromagnetismAntiferromagnetism_Phys.Rev.Lett.66,
  title = {Absence of {{Ferromagnetism}} or {{Antiferromagnetism}} in {{One-}} or {{Two-Dimensional Isotropic Heisenberg Models}}},
  author = {Mermin, N. D. and Wagner, H.},
  year = 1966,
  journal = {Phys. Rev. Lett.},
  volume = {17},
  number = {22},
  pages = {1133--1136},
  issn = {0031-9007},
  doi = {10.1103/PhysRevLett.17.1133},
  urldate = {2022-09-26}
}

@article{journalarticle:Metzler.Klafter_RandomWalksGuide_Phys.Rep.00,
  title = {The Random Walk's Guide to Anomalous Diffusion: A Fractional Dynamics Approach},
  author = {Metzler, Ralf and Klafter, Joseph},
  year = 2000,
  journal = {Phys. Rep.},
  volume = {339},
  number = {1},
  pages = {1--77},
  issn = {03701573},
  doi = {10.1016/S0370-1573(00)00070-3},
  urldate = {2025-08-22}
}

@article{journalarticle:Metzler.Klafter_RestaurantEndRandom_J.Phys.A:Math.Gen.04,
  title = {The Restaurant at the End of the Random Walk: Recent Developments in the Description of Anomalous Transport by Fractional Dynamics},
  author = {Metzler, Ralf and Klafter, Joseph},
  year = 2004,
  journal = {J. Phys. A: Math. Gen.},
  volume = {37},
  number = {31},
  pages = {R161},
  issn = {0305-4470},
  doi = {10.1088/0305-4470/37/31/R01},
  urldate = {2025-08-22}
}

@article{journalarticle:Meyer.etal_StructureSuspendedGraphene_Nature07,
  title = {The Structure of Suspended Graphene Sheets},
  author = {Meyer, Jannik C. and Geim, A. K. and Katsnelson, M. I. and Novoselov, K. S. and Booth, T. J. and Roth, S.},
  year = 2007,
  journal = {Nature},
  volume = {446},
  number = {7131},
  pages = {60--63},
  doi = {10.1038/nature05545},
  urldate = {2013-11-22}
}

@article{journalarticle:Meylahn.etal_LargeDeviationsMarkov_Phys.Rev.E15,
  title = {Large Deviations for {{Markov}} Processes with Resetting},
  author = {Meylahn, Janusz M. and Sabhapandit, Sanjib and Touchette, Hugo},
  year = 2015,
  journal = {Phys. Rev. E},
  volume = {92},
  number = {6},
  pages = {062148},
  publisher = {American Physical Society},
  doi = {10.1103/PhysRevE.92.062148}
}

@article{journalarticle:Mohammed.etal_TwodimensionalPureBromine_Comp.Cond.Mat.23,
  title = {Two-Dimensional Pure and Bromine Doped {{MoTe2}} and {{WSe2}} as Electron Transport Materials for Photovoltaic Application: {{A DFT}} Approach},
  author = {Mohammed, A. and Shu'aibu, A. and Abdu, Sadiq G. and Aliyu, Muhammed M.},
  year = 2023,
  journal = {Comp. Cond. Mat.},
  volume = {37},
  pages = {e00855},
  issn = {2352-2143},
  doi = {10.1016/j.cocom.2023.e00855},
  urldate = {2025-09-25}
}

@article{journalarticle:Montero.etal_ValuingDistantFuture_JPhysMathTheor22,
  title = {Valuing the Distant Future under Stochastic Resettings: The Effect on Discounting},
  author = {Montero, Miquel and Perell{\'o}, Josep and Masoliver, Jaume},
  year = 2022,
  journal = {J. Phys. A: Math. Theor.},
  volume = {55},
  number = {46},
  pages = {464001},
  publisher = {IOP Publishing},
  doi = {10.1088/1751-8121/ac9f8a}
}

@article{journalarticle:Monthus_LargeDeviationsMarkov_JStatMechTheoryExp21,
  title = {Large Deviations for {{Markov}} Processes with Stochastic Resetting: Analysis via the Empirical Density and Flows or via Excursions between Resets},
  author = {Monthus, C{\'e}cile},
  year = 2021,
  journal = {J. Stat. Mech.: Theory Exp.},
  volume = {2021},
  number = {3},
  pages = {033201},
  publisher = {{IOP Publishing and SISSA}},
  doi = {10.1088/1742-5468/abdeaf}
}

@article{journalarticle:Moreau.etal_IntermittentSearchProcesses_EPL07,
  title = {Intermittent Search Processes in Disordered Medium},
  author = {Moreau, M. and B{\'e}nichou, O. and Loverdo, C. and Voituriez, R.},
  year = 2007,
  journal = {EPL},
  volume = {77},
  number = {2},
  pages = {20006},
  doi = {10.1209/0295-5075/77/20006}
}

@article{journalarticle:Mukherjee.etal_QuantumDynamicsStochastic_Phys.Rev.B18,
  title = {Quantum Dynamics with Stochastic Reset},
  author = {Mukherjee, B. and Sengupta, K. and Majumdar, Satya N.},
  year = 2018,
  journal = {Phys. Rev. B},
  volume = {98},
  number = {10},
  pages = {104309},
  publisher = {American Physical Society},
  doi = {10.1103/PhysRevB.98.104309}
}

@article{journalarticle:Neek-Amal.etal_ThermalMirrorBuckling_Nat.Commun.14,
  title = {Thermal Mirror Buckling in Freestanding Graphene Locally Controlled by Scanning Tunnelling Microscopy},
  author = {{Neek-Amal}, M. and Xu, P. and Schoelz, J.K. and Ackerman, M.L. and Barber, S.D. and Thibado, P.M. and Sadeghi, A. and Peeters, F.M.},
  year = 2014,
  journal = {Nat. Commun.},
  volume = {5},
  number = {1},
  pages = {4962},
  issn = {2041-1723},
  doi = {10.1038/ncomms5962},
  urldate = {2022-05-05}
}

@article{journalarticle:Neto.etal_ElectronicPropertiesGraphene_Rev.Mod.Phys.09,
  title = {The Electronic Properties of Graphene},
  author = {Neto, AH Castro and Guinea, F. and Peres, N. M. R. and Novoselov, Kostya S. and Geim, Andre K.},
  year = 2009,
  journal = {Rev. Mod. Phys.},
  volume = {81},
  number = {1},
  pages = {109},
  doi = {10.1103/RevModPhys.81.109},
  urldate = {2013-11-22}
}

@article{journalarticle:Novoselov.etal_ElectricFieldEffect_Science04,
  title = {Electric {{Field Effect}} in {{Atomically Thin Carbon Films}}},
  author = {Novoselov, K. S. and Geim, A. K. and Morozov, S. V. and Jiang, D. and Zhang, Y. and Dubonos, S. V. and Grigorieva, I. V. and Firsov, A. A.},
  year = 2004,
  journal = {Science},
  volume = {306},
  number = {5696},
  pages = {666--669},
  issn = {0036-8075, 1095-9203},
  doi = {10.1126/science.1102896},
  urldate = {2022-09-26}
}

@article{journalarticle:OBrien.etal_SearchStrategiesForaging_Am.Sci.90,
  title = {Search Strategies of Foraging Animals},
  author = {O'Brien, W. John and Browman, Howard I. and Evans, Barbara I.},
  year = 1990,
  journal = {Am. Sci.},
  volume = {78},
  number = {2},
  pages = {152--160},
  urldate = {2025-08-12}
}

@article{journalarticle:Olsen.etal_ThermodynamicCostFinitetime_Phys.Rev.Res.24,
  title = {Thermodynamic Cost of Finite-Time Stochastic Resetting},
  author = {Olsen, Kristian St{\o}levik and Gupta, Deepak and Mori, Francesco and Krishnamurthy, Supriya},
  year = 2024,
  journal = {Phys. Rev. Res.},
  volume = {6},
  number = {3},
  pages = {033343},
  publisher = {American Physical Society},
  doi = {10.1103/PhysRevResearch.6.033343}
}

@article{journalarticle:Onsager_StatisticalHydrodynamics_NuovoCimento49,
  title = {Statistical {{Hydrodynamics}}},
  author = {Onsager, L.},
  year = 1949,
  journal = {Nuovo Cimento},
  volume = {6},
  number = {Suppl. 2},
  pages = {279},
  doi = {10.1007/BF02780991}
}

@article{journalarticle:Oshanin.etal_IntermittentRandomWalks_J.Phys.Condens.Matter07,
  title = {Intermittent Random Walks for an Optimal Search Strategy: One-Dimensional Case},
  author = {Oshanin, G. and Wio, H. S. and Lindenberg, K. and Burlatsky, S. F.},
  year = 2007,
  journal = {J. Phys. Condens. Matter},
  volume = {19},
  number = {6},
  pages = {065142},
  doi = {10.1088/0953-8984/19/6/065142}
}

@article{journalarticle:Pal.etal_DiffusionTimedependentResetting_JPhysMathTheor16,
  title = {Diffusion under Time-Dependent Resetting},
  author = {Pal, Arnab and Kundu, Anupam and Evans, Martin R.},
  year = 2016,
  journal = {J. Phys. A: Math. Theor.},
  volume = {49},
  number = {22},
  pages = {225001},
  publisher = {IOP Publishing},
  doi = {10.1088/1751-8113/49/22/225001}
}

@article{journalarticle:Pal.etal_InvariantsMotionStochastic_NewJ.Phys.19,
  title = {Invariants of Motion with Stochastic Resetting and Space-Time Coupled Returns},
  author = {Pal, Arnab and Ku{\'s}mierz, {\L}ukasz and Reuveni, Shlomi},
  year = 2019,
  journal = {New J. Phys.},
  volume = {21},
  number = {11},
  pages = {113024},
  publisher = {IOP Publishing},
  doi = {10.1088/1367-2630/ab5201}
}

@article{journalarticle:Pal.etal_SearchHomeReturns_Phys.Rev.Res.20,
  title = {Search with Home Returns Provides Advantage under High Uncertainty},
  author = {Pal, Arnab and Ku{\'s}mierz, {\L}ukasz and Reuveni, Shlomi},
  year = 2020,
  journal = {Phys. Rev. Res.},
  volume = {2},
  number = {4},
  pages = {043174},
  publisher = {American Physical Society},
  doi = {10.1103/PhysRevResearch.2.043174}
}

@article{journalarticle:Pal.etal_ThermodynamicUncertaintyRelation_Phys.Rev.Res.21,
  title = {Thermodynamic Uncertainty Relation for Systems with Unidirectional Transitions},
  author = {Pal, Arnab and Reuveni, Shlomi and Rahav, Saar},
  year = 2021,
  journal = {Phys. Rev. Res.},
  volume = {3},
  number = {1},
  pages = {013273},
  publisher = {American Physical Society},
  doi = {10.1103/PhysRevResearch.3.013273}
}

@article{journalarticle:Pal.Prasad_FirstPassageStochastic_Phys.Rev.E19,
  title = {First Passage under Stochastic Resetting in an Interval},
  author = {Pal, Arnab and Prasad, V. V.},
  year = 2019,
  journal = {Phys. Rev. E},
  volume = {99},
  number = {3},
  pages = {032123},
  publisher = {American Physical Society},
  doi = {10.1103/PhysRevE.99.032123}
}

@article{journalarticle:Pal.Prasad_LandaulikeExpansionPhase_Phys.Rev.Res.19,
  title = {Landau-like Expansion for Phase Transitions in Stochastic Resetting},
  author = {Pal, Arnab and Prasad, V. V.},
  year = 2019,
  journal = {Phys. Rev. Res.},
  volume = {1},
  number = {3},
  pages = {032001},
  issn = {2643-1564},
  doi = {10.1103/PhysRevResearch.1.032001},
  urldate = {2025-08-27}
}

@article{journalarticle:Pal.Rahav_IntegralFluctuationTheorems_Phys.Rev.E17,
  title = {Integral Fluctuation Theorems for Stochastic Resetting Systems},
  author = {Pal, Arnab and Rahav, Saar},
  year = 2017,
  journal = {Phys. Rev. E},
  volume = {96},
  number = {6},
  pages = {062135},
  publisher = {American Physical Society},
  doi = {10.1103/PhysRevE.96.062135}
}

@article{journalarticle:Pinsky_OptimizingDriftDiffusive_Electron.J.Probab.19,
  title = {Optimizing the Drift in a Diffusive Search for a Random Stationary Target},
  author = {Pinsky, Ross G.},
  year = 2019,
  journal = {Electron. J. Probab.},
  volume = {24},
  number = {none},
  pages = {1--22},
  publisher = {{Institute of Mathematical Statistics and Bernoulli Society}},
  doi = {10.1214/19-EJP335}
}

@article{journalarticle:Plata.etal_AsymmetricStochasticResetting_Phys.Rev.E20,
  title = {Asymmetric Stochastic Resetting: {{Modeling}} Catastrophic Events},
  author = {Plata, Carlos A. and Gupta, Deepak and Azaele, Sandro},
  year = 2020,
  journal = {Phys. Rev. E},
  volume = {102},
  number = {5},
  pages = {052116},
  issn = {2470-0045, 2470-0053},
  doi = {10.1103/PhysRevE.102.052116},
  urldate = {2025-08-02}
}

@article{journalarticle:Plummer.Nelson_BucklingMetastabilityMembranes_Phys.Rev.E20,
  title = {Buckling and Metastability in Membranes with Dilation Arrays},
  author = {Plummer, Abigail and Nelson, David R.},
  year = 2020,
  journal = {Phys. Rev. E},
  volume = {102},
  number = {3},
  pages = {033002},
  issn = {2470-0045, 2470-0053},
  doi = {10.1103/PhysRevE.102.033002},
  urldate = {2025-08-02}
}

@article{journalarticle:Poincloux.etal_BendingResponseBook_Phys.Rev.Lett.21,
  title = {Bending {{Response}} of a {{Book}} with {{Internal Friction}}},
  author = {Poincloux, Samuel and Chen, Tian and Audoly, Basile and Reis, Pedro M.},
  year = 2021,
  journal = {Phys. Rev. Lett.},
  volume = {126},
  number = {21},
  pages = {218004},
  issn = {0031-9007, 1079-7114},
  doi = {10.1103/PhysRevLett.126.218004},
  urldate = {2025-08-02}
}

@article{journalarticle:Qamar.etal_CarbonNanotubesPerovskite_Synth.Met.24,
  title = {Carbon Nanotubes in Perovskite Solar Cells: {{A}} Comprehensive Review of Recent Developments and Future Directions},
  author = {Qamar, Muhammad Azam and Aroosh, Komal and Nawaz, Aqsa and Almashnowi, Majed Y. A. and Alnasir, M. Hisham},
  year = 2024,
  journal = {Synth. Met.},
  volume = {307},
  pages = {117651},
  issn = {0379-6779},
  doi = {10.1016/j.synthmet.2024.117651},
  urldate = {2025-10-15}
}

@article{journalarticle:Radice_DiffusionProcessesGammadistributed_JPhysMathTheor22,
  title = {Diffusion Processes with {{Gamma-distributed}} Resetting and Non-Instantaneous Returns},
  author = {Radice, Mattia},
  year = 2022,
  journal = {J. Phys. A: Math. Theor.},
  volume = {55},
  number = {22},
  pages = {224002},
  publisher = {IOP Publishing},
  doi = {10.1088/1751-8121/ac654f}
}

@article{journalarticle:Ray.etal_PecletNumberGoverns_J.Phys.A:Math.Theor.19,
  title = {P\'eclet Number Governs Transition to Acceleratory Restart in Drift-Diffusion},
  author = {Ray, Somrita and Mondal, Debasish and Reuveni, Shlomi},
  year = 2019,
  journal = {J. Phys. A: Math. Theor.},
  volume = {52},
  number = {25},
  pages = {255002},
  publisher = {IOP Publishing},
  issn = {1751-8121},
  doi = {10.1088/1751-8121/ab1fcc},
  urldate = {2025-10-10}
}

@article{journalarticle:Reuveni_OptimalStochasticRestart_Phys.Rev.Lett.16,
  title = {Optimal {{Stochastic Restart Renders Fluctuations}} in {{First Passage Times Universal}}},
  author = {Reuveni, Shlomi},
  year = 2016,
  journal = {Phys. Rev. Lett.},
  volume = {116},
  number = {17},
  pages = {170601},
  publisher = {American Physical Society},
  doi = {10.1103/PhysRevLett.116.170601}
}

@article{journalarticle:Reuveni.etal_RoleSubstrateUnbinding_Proc.Natl.Acad.Sci.U.S.A.14,
  title = {Role of Substrate Unbinding in {{Michaelis}}--{{Menten}} Enzymatic Reactions},
  author = {Reuveni, Shlomi and Urbakh, Michael and Klafter, Joseph},
  year = 2014,
  journal = {Proc. Natl. Acad. Sci. U.S.A.},
  volume = {111},
  number = {12},
  pages = {4391--4396},
  doi = {10.1073/pnas.1318122111}
}

@article{journalarticle:Rojo.etal_IntermittentSearchStrategies_JPhysMathTheor10,
  title = {Intermittent Search Strategies Revisited: Effect of the Jump Length and Biased Motion},
  author = {Rojo, F. and Revelli, J. and Budde, C. E. and Wio, H. S. and Oshanin, G. and Lindenberg, Katja},
  year = 2010,
  journal = {J. Phys. A: Math. Theor.},
  volume = {43},
  number = {34},
  pages = {345001},
  doi = {10.1088/1751-8113/43/34/345001}
}

@article{journalarticle:Roldan.etal_StochasticResettingBacktrack_Phys.Rev.E16,
  title = {Stochastic Resetting in Backtrack Recovery by {{RNA}} Polymerases},
  author = {Rold{\'a}n, {\'E}dgar and Lisica, Ana and {S{\'a}nchez-Taltavull}, Daniel and Grill, Stephan W.},
  year = 2016,
  journal = {Phys. Rev. E},
  volume = {93},
  number = {6},
  pages = {062411},
  publisher = {American Physical Society},
  doi = {10.1103/PhysRevE.93.062411}
}

@article{journalarticle:Roldan.Gupta_PathintegralFormalismStochastic_Phys.Rev.E17,
  title = {Path-Integral Formalism for Stochastic Resetting: {{Exactly}} Solved Examples and Shortcuts to Confinement},
  author = {Rold{\'a}n, {\'E}dgar and Gupta, Shamik},
  year = 2017,
  journal = {Phys. Rev. E},
  volume = {96},
  number = {2},
  pages = {022130},
  issn = {2470-0045, 2470-0053},
  doi = {10.1103/PhysRevE.96.022130},
  urldate = {2025-08-02}
}

@article{journalarticle:Ros.Fyodorov_HighdLandscapesParadigm_22,
  title = {The High-d Landscapes Paradigm: Spin-Glasses, and Beyond},
  author = {Ros, Valentina and Fyodorov, Yan V.},
  year = 2022,
  doi = {arXiv.2209.07975},
  urldate = {2025-08-13}
}

@article{journalarticle:Rose.etal_SpectralPropertiesSimple_Phys.Rev.E18,
  title = {Spectral Properties of Simple Classical and Quantum Reset Processes},
  author = {Rose, Dominic C. and Touchette, Hugo and Lesanovsky, Igor and Garrahan, Juan P.},
  year = 2018,
  journal = {Phys. Rev. E},
  volume = {98},
  number = {2},
  pages = {022129},
  publisher = {American Physical Society},
  doi = {10.1103/PhysRevE.98.022129}
}

@article{journalarticle:Rotbart.etal_MichaelisMentenReactionScheme_Phys.Rev.E15,
  title = {Michaelis-{{Menten}} Reaction Scheme as a Unified Approach towards the Optimal Restart Problem},
  author = {Rotbart, Tal and Reuveni, Shlomi and Urbakh, Michael},
  year = 2015,
  journal = {Phys. Rev. E},
  volume = {92},
  number = {6},
  pages = {060101},
  publisher = {American Physical Society},
  doi = {10.1103/PhysRevE.92.060101}
}

@article{journalarticle:Ruiz-Garcia.etal_BifurcationAnalysisPhase_Phys.Rev.E17,
  title = {Bifurcation Analysis and Phase Diagram of a Spin-String Model with Buckled States},
  author = {{Ruiz-Garcia}, M. and Bonilla, L. L. and Prados, A.},
  year = 2017,
  journal = {Phys. Rev. E},
  volume = {96},
  number = {6},
  pages = {062147},
  issn = {2470-0045, 2470-0053},
  doi = {10.1103/PhysRevE.96.062147},
  urldate = {2025-08-02}
}

@article{journalarticle:Ruiz-Garcia.etal_RipplesHexagonalLattices_J.Stat.Mech.15,
  title = {Ripples in Hexagonal Lattices of Atoms Coupled to {{Glauber}} Spins},
  author = {{Ruiz-Garc{\'i}a}, M and Bonilla, L L and Prados, A},
  year = 2015,
  journal = {J. Stat. Mech.},
  volume = {2015},
  number = {5},
  pages = {P05015},
  issn = {1742-5468},
  doi = {10.1088/1742-5468/2015/05/P05015},
  urldate = {2025-08-02}
}

@article{journalarticle:Ruiz-Garcia.etal_STMdrivenTransitionRippled_Phys.Rev.B16,
  title = {{{STM-driven}} Transition from Rippled to Buckled Graphene in a Spin-Membrane Model},
  author = {{Ruiz-Garc{\'i}a}, M. and Bonilla, L. L. and Prados, A.},
  year = 2016,
  journal = {Phys. Rev. B},
  volume = {94},
  number = {20},
  pages = {205404},
  issn = {2469-9950, 2469-9969},
  doi = {10.1103/PhysRevB.94.205404},
  urldate = {2025-08-02}
}

@article{journalarticle:Samy.etal_ReviewMoS2Properties_Crystals21,
  title = {A {{Review}} on {{MoS2 Properties}}, {{Synthesis}}, {{Sensing Applications}} and {{Challenges}}},
  author = {Samy, Omnia and Zeng, Shuwen and Birowosuto, Muhammad Danang and El Moutaouakil, Amine},
  year = 2021,
  journal = {Crystals},
  volume = {11},
  number = {4},
  pages = {355},
  issn = {2073-4352},
  doi = {10.3390/cryst11040355},
  urldate = {2022-09-26}
}

@article{journalarticle:Samykano_ProgressOnedimensionalNanostructures_Mater.Charact.21,
  title = {Progress in One-Dimensional Nanostructures},
  author = {Samykano, M.},
  year = 2021,
  journal = {Mater. Charact.},
  volume = {179},
  pages = {111373},
  issn = {10445803},
  doi = {10.1016/j.matchar.2021.111373},
  urldate = {2025-10-15}
}

@article{journalarticle:San-Jose.etal_ElectroninducedRipplingGraphene_Phys.Rev.Lett.11,
  title = {Electron-Induced Rippling in Graphene},
  author = {{San-Jose}, Pablo and Gonz{\'a}lez, J. and Guinea, F.},
  year = 2011,
  journal = {Phys. Rev. Lett.},
  volume = {106},
  number = {4},
  pages = {045502},
  doi = {10.1103/PhysRevLett.106.045502},
  urldate = {2013-11-22}
}

@article{journalarticle:Sandev.etal_HeterogeneousDiffusionStochastic_JPhysMathTheor22,
  title = {Heterogeneous Diffusion with Stochastic Resetting},
  author = {Sandev, Trifce and Domazetoski, Viktor and Kocarev, Ljupco and Metzler, Ralf and Chechkin, Aleksei},
  year = 2022,
  journal = {J. Phys. A: Math. Theor.},
  volume = {55},
  number = {7},
  pages = {074003},
  publisher = {IOP Publishing},
  doi = {10.1088/1751-8121/ac491c}
}

@article{journalarticle:Sangchap.etal_ExploringPromiseOnedimensional_IntJHydrog.Energy24,
  title = {Exploring the Promise of One-Dimensional Nanostructures: {{A}} Review of Hydrogen Gas Sensors},
  author = {Sangchap, Mohammad and Hashtroudi, Hanie and Thathsara, Thilini and Harrison, Christopher J. and Kingshott, Peter and Kandjani, Ahmad E. and Trinchi, Adrian and Shafiei, Mahnaz},
  year = 2024,
  journal = {Int. J. Hydrogen Energy},
  volume = {50},
  pages = {1443--1457},
  issn = {03603199},
  doi = {10.1016/j.ijhydene.2023.11.115},
  urldate = {2025-10-15}
}

@article{journalarticle:Santra_EffectTaxDynamics_EPL22,
  title = {Effect of Tax Dynamics on Linearly Growing Processes under Stochastic Resetting: {{A}} Possible Economic Model},
  author = {Santra, Ion},
  year = 2022,
  journal = {EPL},
  volume = {137},
  number = {5},
  pages = {52001},
  publisher = {{EDP Sciences, IOP Publishing and Societ\`a Italiana di Fisica}},
  doi = {10.1209/0295-5075/ac5e53}
}

@article{journalarticle:Schmidhuber_DeepLearningNeural_NeuralNetw.15,
  title = {Deep Learning in Neural Networks: {{An}} Overview},
  author = {Schmidhuber, J{\"u}rgen},
  year = 2015,
  journal = {Neural Netw.},
  volume = {61},
  pages = {85--117},
  issn = {0893-6080},
  doi = {10.1016/j.neunet.2014.09.003},
  urldate = {2025-08-15}
}

@article{journalarticle:Schoelz.etal_GrapheneRipplesRealization_Phys.Rev.B15,
  title = {Graphene Ripples as a Realization of a Two-Dimensional {{Ising}} Model: {{A}} Scanning Tunneling Microscope Study},
  author = {Schoelz, J. K. and Xu, P. and Meunier, V. and Kumar, P. and {Neek-Amal}, M. and Thibado, P. M. and Peeters, F. M.},
  year = 2015,
  journal = {Phys. Rev. B},
  volume = {91},
  number = {4},
  pages = {045413},
  doi = {10.1103/PhysRevB.91.045413},
  urldate = {2017-05-12}
}

@article{journalarticle:Seo.etal_SinglechiralitySinglewallCarbon_Phys.Chem.Chem.Phys.25,
  title = {Single-Chirality Single-Wall Carbon Nanotubes for Electrochemical Biosensing},
  author = {Seo, Ju-Yeon and Mostafiz, Bahar and Tu, Xiaomin and Khripin, Constantine Y. and Zheng, Ming and Li, Han and Peltola, Emilia},
  year = 2025,
  journal = {Phys. Chem. Chem. Phys.},
  volume = {27},
  number = {9},
  pages = {4959--4967},
  publisher = {The Royal Society of Chemistry},
  issn = {1463-9084},
  doi = {10.1039/D4CP04206A},
  urldate = {2025-10-15}
}

@article{journalarticle:Sevilla.Valdes-Hernandez_DynamicsClosedQuantum_JPhysMathTheor23,
  title = {Dynamics of Closed Quantum Systems under Stochastic Resetting},
  author = {Sevilla, Francisco J. and {Vald{\'e}s-Hern{\'a}ndez}, Andrea},
  year = 2023,
  journal = {J. Phys. A: Math. Theor.},
  volume = {56},
  number = {3},
  pages = {034001},
  publisher = {IOP Publishing},
  doi = {10.1088/1751-8121/acb29d}
}

@article{journalarticle:Shankar.Nelson_ThermalizedBucklingIsotropically_Phys.Rev.E21,
  title = {Thermalized Buckling of Isotropically Compressed Thin Sheets},
  author = {Shankar, Suraj and Nelson, David R.},
  year = 2021,
  journal = {Phys. Rev. E},
  volume = {104},
  number = {5},
  pages = {054141},
  issn = {2470-0045, 2470-0053},
  doi = {10.1103/PhysRevE.104.054141},
  urldate = {2025-08-02}
}

@article{journalarticle:Shanmughan.etal_ExploringFuture2D_IntJHydrog.Energy23,
  title = {Exploring the Future of {{2D}} Catalysts for Clean and Sustainable Hydrogen Production},
  author = {Shanmughan, Bhavana and Nighojkar, Amrita and Kandasubramanian, Balasubramanian},
  year = 2023,
  journal = {Int. J. Hydrogen Energy},
  volume = {48},
  number = {74},
  pages = {28679--28693},
  issn = {0360-3199},
  doi = {10.1016/j.ijhydene.2023.04.053},
  urldate = {2025-09-25}
}

@article{journalarticle:Singh.etal_RipplingBucklingMelting_Phys.Rev.B15,
  title = {Rippling, Buckling, and Melting of Single- and Multilayer {{MoS}} 2},
  author = {Singh, Sandeep Kumar and {Neek-Amal}, M. and Costamagna, S. and Peeters, F. M.},
  year = 2015,
  journal = {Phys. Rev. B},
  volume = {91},
  number = {1},
  pages = {014101},
  issn = {1098-0121, 1550-235X},
  doi = {10.1103/PhysRevB.91.014101},
  urldate = {2022-09-26}
}

@article{journalarticle:Smith.Majumdar_CondensationTransitionLarge_J.Stat.Mech.TheoryExp.22,
  title = {Condensation Transition in Large Deviations of Self-Similar {{Gaussian}} Processes with Stochastic Resetting},
  author = {Smith, Naftali R. and Majumdar, Satya N.},
  year = 2022,
  journal = {Journal of Statistical Mechanics: Theory and Experiment},
  volume = {2022},
  number = {5},
  pages = {053212},
  publisher = {{IOP Publishing and SISSA}},
  doi = {10.1088/1742-5468/ac6f04}
}

@article{journalarticle:Stojkoski.etal_AutocorrelationFunctionsErgodicity_JPhysMathTheor22,
  title = {Autocorrelation Functions and Ergodicity in Diffusion with Stochastic Resetting},
  author = {Stojkoski, Viktor and Sandev, Trifce and Kocarev, Ljupco and Pal, Arnab},
  year = 2022,
  journal = {J. Phys. A: Math. Theor.},
  volume = {55},
  number = {10},
  pages = {104003},
  publisher = {IOP Publishing},
  doi = {10.1088/1751-8121/ac4ce9}
}

@article{journalarticle:Stojkoski.etal_GeneralisedGeometricBrownian_Entropy20,
  title = {Generalised {{Geometric Brownian Motion}}: {{Theory}} and {{Applications}} to {{Option Pricing}}},
  author = {Stojkoski, Viktor and Sandev, Trifce and Basnarkov, Lasko and Kocarev, Ljupco and Metzler, Ralf},
  year = 2020,
  journal = {Entropy},
  volume = {22},
  number = {12},
  pages = {1432},
  publisher = {Multidisciplinary Digital Publishing Institute},
  issn = {1099-4300},
  doi = {10.3390/e22121432},
  urldate = {2025-08-15}
}

@article{journalarticle:Stojkoski.etal_GeometricBrownianMotion_Phys.Rev.E21,
  title = {Geometric {{Brownian}} Motion under Stochastic Resetting: {{A}} Stationary yet Nonergodic Process},
  author = {Stojkoski, Viktor and Sandev, Trifce and Kocarev, Ljupco and Pal, Arnab},
  year = 2021,
  journal = {Phys. Rev. E},
  volume = {104},
  number = {1},
  pages = {014121},
  publisher = {American Physical Society},
  doi = {10.1103/PhysRevE.104.014121}
}

@article{journalarticle:Stojkoski.etal_IncomeInequalityMobility_Philos.Trans.R.Soc.A22,
  title = {Income Inequality and Mobility in Geometric {{Brownian}} Motion with Stochastic Resetting: Theoretical Results and Empirical Evidence of Non-Ergodicity},
  author = {Stojkoski, Viktor and Jolakoski, Petar and Pal, Arnab and Sandev, Trifce and Kocarev, Ljupco and Metzler, Ralf},
  year = 2022,
  journal = {Philos. Trans. R. Soc. A},
  volume = {380},
  number = {2224},
  pages = {20210157},
  doi = {10.1098/rsta.2021.0157}
}

@article{journalarticle:Thibado.etal_FluctuationinducedCurrentFreestanding_Phys.Rev.E20,
  title = {Fluctuation-Induced Current from Freestanding Graphene},
  author = {Thibado, P. M. and Kumar, P. and Singh, Surendra and {Ruiz-Garcia}, M. and Lasanta, A. and Bonilla, L. L.},
  year = 2020,
  journal = {Phys. Rev. E},
  volume = {102},
  number = {4},
  pages = {042101},
  issn = {2470-0045, 2470-0053},
  doi = {10.1103/PhysRevE.102.042101},
  urldate = {2022-09-26}
}

@article{journalarticle:Tian.Burrage_StochasticModelsRegulatory_Proc.Natl.Acad.Sci.U.S.A.06,
  title = {Stochastic Models for Regulatory Networks of the Genetic Toggle Switch},
  author = {Tian, Tianhai and Burrage, Kevin},
  year = 2006,
  journal = {Proc. Natl. Acad. Sci. U.S.A.},
  volume = {103},
  number = {22},
  pages = {8372--8377},
  publisher = {Proceedings of the National Academy of Sciences},
  doi = {10.1073/pnas.0507818103},
  urldate = {2025-09-16}
}

@article{journalarticle:Toledo-Marin.etal_PredatorpreyDynamicsChasing_19,
  title = {Predator-Prey Dynamics: {{Chasing}} by Stochastic Resetting},
  author = {{Toledo-Marin}, J. Quetzalcoatl and Boyer, Denis and Sevilla, Francisco J.},
  year = 2019,
  eprint = {1912.02141},
  primaryclass = {cond-mat},
  doi = {arXiv.1912.02141},
  urldate = {2025-10-06},
  archiveprefix = {arXiv}
}

@article{journalarticle:Touchette_LargeDeviationApproach_Phys.Rep.09,
  title = {The Large Deviation Approach to Statistical Mechanics},
  author = {Touchette, Hugo},
  year = 2009,
  journal = {Phys. Rep.},
  volume = {478},
  number = {1},
  pages = {1--69},
  doi = {10.1016/j.physrep.2009.05.002},
  urldate = {2013-11-22}
}

@article{journalarticle:Vinod.etal_TimeaveragingNonergodicityReset_Phys.Rev.E22,
  title = {Time-Averaging and Nonergodicity of Reset Geometric {{Brownian}} Motion with Drift},
  author = {Vinod, Deepak and Cherstvy, Andrey G. and Metzler, Ralf and Sokolov, Igor M.},
  year = 2022,
  journal = {Phys. Rev. E},
  volume = {106},
  number = {3},
  pages = {034137},
  publisher = {American Physical Society},
  doi = {10.1103/PhysRevE.106.034137}
}

@article{journalarticle:Viswanathan.etal_OptimizingSuccessRandom_Nature99,
  title = {Optimizing the Success of Random Searches},
  author = {Viswanathan, G. M. and Buldyrev, Sergey V. and Havlin, Shlomo and {da Luz}, M. G. E. and Raposo, E. P. and Stanley, H. Eugene},
  year = 1999,
  journal = {Nature},
  volume = {401},
  number = {6756},
  pages = {911--914},
  issn = {1476-4687},
  doi = {10.1038/44831}
}

@article{journalarticle:Wald.Bottcher_ClassicalQuantumWalks_Phys.Rev.E21,
  title = {From Classical to Quantum Walks with Stochastic Resetting on Networks},
  author = {Wald, Sascha and B{\"o}ttcher, Lucas},
  year = 2021,
  journal = {Phys. Rev. E},
  volume = {103},
  number = {1},
  pages = {012122},
  publisher = {American Physical Society},
  doi = {10.1103/PhysRevE.103.012122}
}

@article{journalarticle:Wang.etal_RandomWalksComplex_Chaos21,
  title = {Random Walks on Complex Networks with Multiple Resetting Nodes: {{A}} Renewal Approach},
  author = {Wang, Shuang and Chen, Hanshuang and Huang, Feng},
  year = 2021,
  journal = {Chaos},
  volume = {31},
  number = {9},
  pages = {093135},
  doi = {10.1063/5.0064791}
}

@article{journalarticle:Wang.etal_TimeAveragingEmerging_Phys.Rev.E21,
  title = {Time Averaging and Emerging Nonergodicity upon Resetting of Fractional {{Brownian}} Motion and Heterogeneous Diffusion Processes},
  author = {Wang, Wei and Cherstvy, Andrey G. and Kantz, Holger and Metzler, Ralf and Sokolov, Igor M.},
  year = 2021,
  journal = {Phys. Rev. E},
  volume = {104},
  number = {2},
  pages = {024105},
  publisher = {American Physical Society},
  doi = {10.1103/PhysRevE.104.024105}
}

@article{journalarticle:Wei.etal_BendingRigidityGaussian_NanoLett.13,
  title = {Bending Rigidity and {{Gaussian}} Bending Stiffness of Single-Layered Graphene},
  author = {Wei, Yujie and Wang, Baoling and Wu, Jiangtao and Yang, Ronggui and Dunn, Martin L},
  year = 2013,
  journal = {Nano Lett.},
  volume = {13},
  number = {1},
  pages = {26--30},
  publisher = {ACS Publications},
  doi = {10.1021/nl303168w}
}

@article{journalarticle:Wu_PottsModel_Rev.Mod.Phys.82,
  title = {The {{Potts}} Model},
  author = {Wu, F. Y.},
  year = 1982,
  journal = {Rev. Mod. Phys.},
  volume = {54},
  number = {1},
  pages = {235--268},
  publisher = {American Physical Society},
  doi = {10.1103/RevModPhys.54.235},
  urldate = {2025-10-24}
}

@article{journalarticle:Wu.etal_EvolutionSearchThree_ACMTrans.Manag.Inf.Syst.22,
  title = {The {{Evolution}} of {{Search}}: {{Three Computing Paradigms}}},
  author = {Wu, Xindong and Zhu, Xingquan and Wu, Minghui},
  year = 2022,
  journal = {ACM Trans. Manag. Inf. Syst.},
  volume = {13},
  number = {2},
  pages = {20:1--20:20},
  issn = {2158-656X},
  doi = {10.1145/3495214},
  urldate = {2025-08-15}
}

@article{journalarticle:Ye.Chen_RandomWalksComplex_JStatMechTheoryExp22,
  title = {Random Walks on Complex Networks under Node-Dependent Stochastic Resetting},
  author = {Ye, Yanfei and Chen, Hanshuang},
  year = 2022,
  journal = {J. Stat. Mech.: Theory Exp.},
  volume = {2022},
  number = {5},
  pages = {053201},
  publisher = {{IOP Publishing and SISSA}},
  doi = {10.1088/1742-5468/ac625b}
}

@article{journalarticle:Zamparo_StatisticalFluctuationsResetting_JPhysMathTheor22,
  title = {Statistical Fluctuations under Resetting: Rigorous Results},
  author = {Zamparo, Marco},
  year = 2022,
  journal = {J. Phys. A: Math. Theor.},
  volume = {55},
  number = {48},
  pages = {484001},
  publisher = {IOP Publishing},
  doi = {10.1088/1751-8121/aca452}
}

@article{journalarticle:Zhang.etal_ScalablyNanomanufacturedAtomically_SmallStruct.22,
  title = {Scalably {{Nanomanufactured Atomically Thin Materials}}-{{Based Wearable Health Sensors}}},
  author = {Zhang, Ruifang and Jiang, Jing and Wu, Wenzhuo},
  year = 2022,
  journal = {Small Struct.},
  volume = {3},
  number = {1},
  pages = {2100120},
  issn = {2688-4062, 2688-4062},
  doi = {10.1002/sstr.202100120},
  urldate = {2022-09-26}
}

@article{journalarticle:Zwanzig_RateProcessesDynamical_Acc.Chem.Res.90,
  title = {Rate Processes with Dynamical Disorder},
  author = {Zwanzig, Robert},
  year = 1990,
  journal = {Acc. Chem. Res.},
  volume = {23},
  number = {5},
  pages = {148--152},
  issn = {0001-4842, 1520-4898},
  doi = {10.1021/ar00173a005},
  urldate = {2025-08-02}
}

@phdthesis{thesis:Pandelov_InvestigationStructureReactivity_07,
  title = {Investigation of the Structure and Reactivity of Nanostructured Surfaces},
  author = {Pandelov, Stanislav},
  year = 2007,
  school = {Technical University of Munich}
}


{\clearpage \thispagestyle{empty}}
\chapter*{List of acronyms}
\addcontentsline{toc}{chapter}{List of acronyms}
\begin{labeling}{MFPT}
\item[BM]   Brownian motion
\item[FPE]  Fokker-Planck equation
\item[FPT]  First-passage time
\item[FVT]  Final value theorem
\item[KME]  Kramers-Moyal expansion
\item[lhs]  Left-hand side
\item[MFPT] Mean first-passage time
\item[NESS] Nonequilibrium steady state
\item[ODE]  Ordinary differential equation
\item[PDE]  Partial differential equation 
\item[PDF]  Probability density function 
\item[rhs]  Right-hand side
\item[SDE]  Stochastic differential equation
\item[SSR]  Standard stochastic resetting
\item[STM]  Scanning tunneling microscopy
\end{labeling}


{\clearpage \thispagestyle{empty}}

\renewcommand{\listfigurename}{List of figures}
\listoffigures

\end{document}